\documentclass{elsart} 

\usepackage{amsmath,amsfonts,amssymb}
\usepackage{array}
\usepackage{bm}
\usepackage{cancel}
\usepackage{dcolumn}
\usepackage{graphicx}
\usepackage{picins}
\usepackage[scanall]{psfrag}
\usepackage{times}

\journal{Physics Reports}

\newcommand{\psfragstyle}[1]{\small{#1}}

\newcommand{\figref}[1]{Fig.~\ref{#1}}

\newcommand{\tableref}[1]{Table~\ref{#1}}

\newlength{\tempfboxsep}
\newcommand{\sidebar}[1]{%
	\setlength{\fboxsep}{0pt}%
	\setlength{\temptabcolsep}{\tabcolsep}%
	\setlength{\tabcolsep}{0pt}%
	\begin{tabular}{rcl}%
		\vrule width 0.03\columnwidth &%
		\hspace*{0.05\columnwidth} &%
		\begin{minipage}{0.92\columnwidth}%
			#1%
		\end{minipage}%
	\end{tabular}%
	\setlength{\tabcolsep}{\temptabcolsep}%
	\setlength{\fboxsep}{\tempfboxsep}%
}
\newlength{\temptabcolsep}

\newcommand{\zero}{\ensuremath{\text{0}}}
\newcommand{\one}{\ensuremath{\text{1}}}
\newcommand{\two}{\ensuremath{\text{2}}}
\newcommand{\three}{\ensuremath{\text{3}}}
\newcommand{\four}{\ensuremath{\text{4}}}
\newcommand{\five}{\ensuremath{\text{5}}}
\newcommand{\ten}{\ensuremath{\text{10}}}

\newcommand{\ith}{\ensuremath{i^{\text{th}}}~}

\newcommand{\trademark}{\ensuremath{^{\text{\tiny{TM}}}}~}

\begin{document}

\newlength{\figurewidth}
\setlength{\figurewidth}{\columnwidth}
\newlength{\halffigurewidth}
\setlength{\halffigurewidth}{0.47\columnwidth}
\newlength{\figureseparation}
\setlength{\figureseparation}{0.2cm}

\setlength{\parindent}{0pt}
\setlength{\parskip}{1ex plus 0.5ex minus 0.2ex}

\begin{frontmatter}

\title{
  Cellular Automata Models of Road Traffic\\
  \vspace*{0.3cm}
  \normalsize{
  	\emph{
  		Accepted for publication in Physics Reports\\
  	}
	  \vspace*{0.25cm}
 		A high-quality version of this paper can be found at\\
 		\texttt{http://phdsven.dyns.cx}\\
  }
}

\author{Sven Maerivoet, Bart De Moor}
\ead{$\lbrace$sven.maerivoet,bart.demoor$\rbrace$@esat.kuleuven.be}
\ead[url]{http://www.esat.kuleuven.be/scd}

\address{
	Department of Electrical Engineering ESAT-SCD (SISTA)\\
	Katholieke Universiteit Leuven\\
	Kasteelpark Arenberg 10, 3001 Leuven, Belgium\\
	Phone: +32 (0) 16 32 17 09 Fax: +32 (0) 16 32 19 70
}

\setlength{\parskip}{0pt}

\tableofcontents

\begin{abstract}
	In this paper, we give an elaborate and understandable review of traffic 
	cellular automata (TCA) models, which are a class of computationally efficient 
	microscopic traffic flow models. TCA models arise from the physics discipline 
	of statistical mechanics, having the goal of reproducing the correct 
	macroscopic behaviour based on a minimal description of microscopic 
	interactions. After giving an overview of cellular automata (CA) models, their 
	background and physical setup, we introduce the mathematical notations, show 
	how to perform measurements on a TCA model's lattice of cells, as well as how 
	to convert these quantities into real-world units and vice versa. The majority 
	of this paper then relays an extensive account of the behavioural aspects of 
	several TCA models encountered in literature. Already, several reviews of TCA 
	models exist, but none of them consider all the models exclusively from the 
	behavioural point of view. In this respect, our overview fills this void, as 
	it focusses on the behaviour of the TCA models, by means of time-space and 
	phase-space diagrams, and histograms showing the distributions of vehicles' 
	speeds, space, and time gaps. In the report, we subsequently give a concise 
	overview of TCA models that are employed in a multi-lane setting, and some of 
	the TCA models used to describe city traffic as a two-dimensional grid of 
	cells, or as a road network with explicitly modelled intersections. The final 
	part of the paper illustrates some of the more common analytical 
	approximations to single-cell TCA models.
\end{abstract}

\clearpage

\begin{keyword}
	cellular automata, traffic flow modelling, tempo-spatial behaviour, phase-space diagrams

	\PACS 
	02.50.-r,45.70.Vn,89.40.-a
\end{keyword}

\end{frontmatter}

\setlength{\parskip}{1ex plus 0.5ex minus 0.2ex}

%

In the field of traffic flow modelling, microscopic traffic simulation has 
always been regarded as a time consuming, complex process involving detailed 
models that describe the behaviour of individual vehicles. Approximately a 
decade ago, however, new microscopic models were being developed, based on the 
\emph{cellular automata} programming paradigm from \emph{statistical physics}. 
The main advantage was an \emph{efficient and fast performance} when used in 
computer simulations, due to their rather low accuracy on a microscopic scale. 
These so-called \emph{traffic cellular automata} (TCA) are dynamical systems 
that are discrete in nature, in the sense that time advances with discrete steps 
and space is coarse-grained (e.g., the road is discretised into cells of 7.5 
metres wide, each cell being empty or containing a vehicle). This 
coarse-graininess is fundamentally different from the usual microscopic models, 
which adopt a semi-continuous space, formed by the usage of IEEE floating-point 
numbers \cite{MAERIVOET:05c}. True to the spirit of statistical mechanics, all 
the TCA models discussed in this report do not have a realistic microscopic 
description of traffic flows as their primary intent, but are rather aimed at 
obtaining a correct macroscopic behaviour through their crude microscopic 
description. Such an approach would involve more human-oriented aspects such as 
those found in socio-economic, behavioural, and psychological sciences. Due to 
large lack of knowledge about the manner in which human beings operate in a 
traffic system, traffic engineers currently stick with this higher-level 
scientific approach. As such, they are able to positively capture the first- and 
second-order macroscopic effects of traffic streams. TCA models are very 
flexible and powerful, in that they are also able to capture all previously 
mentioned basic phenomena that occur in traffic flows 
\cite{BARLOVIC:99,CHOWDHURY:00}. In a larger setting, these models describe 
\emph{self-driven, many-particle systems, operating far from equilibrium}. And 
in contrast to strictly gaseous analogies, the particles in these systems are 
intelligent and able to learn from past experience, thereby opening the door to 
the incorporation of behavioural and psychological aspects 
\cite{CHOWDHURY:99b,WOLF:99,HELBING:01}.

The cellular automata approach proved to be quite useful, not only in the field 
of vehicular traffic flow modelling, but also in other fields such as pedestrian 
behaviour, escape and panic dynamics, the spreading of forest fires, population 
growth and migrations, cloud formation, material properties (corrosion, cracks, 
creases, peeling et cetera), ant colonies and pheromone trails, \ldots 
\cite{HELBING:99,KARAFYLLIDIS:97,NAGEL:92b,GOBRON:01,NISHINARI:03}. It is now 
feasible to simulate large systems containing many `intelligent particles', such 
that is it possible to observe their interactions, collective behaviour, 
self-organisation, \ldots 
\cite{IMMERS:98b,VANZUYLEN:99,HELBING:99,HELBING:01,HELBING:04,NAGEL:02b,NAGEL:02c,CHOWDHURY:04}

In this report, we provide a detailed description of the methodology of cellular 
automata applied to traffic flows. We first discuss their background and 
physical setup, followed by an account of the mathematical notations we adopt. 
The remaining majority of this report extensively discusses the behavioural 
aspects of several state-of-the-art TCA models encountered in literature (our 
overview distinguishes between single-cell and multi-cell models). The report 
concludes with a concise overview of TCA models in a multi-lane setting, and TCA 
models used to describe two-dimensional traffic (e.g., a grid for city traffic). 
We end with a description of several common analytical approximations to 
single-cell TCA models.\\

\sidebar{
	Note that aside from our phenomenological discussion of different TCA models, 
	we refer the reader to the work of Chowdhury et al. \cite{CHOWDHURY:00}, Santen 
	\cite{SANTEN:99}, and Knospe et al. \cite{KNOSPE:04} for more theoretically- 
	and quantitatively-oriented overviews.
}\\

	\section{Background and physical setup for road traffic}
	\label{sec:TCA:BackgroundPhysicalSetupForRoadTraffic}

In this section, we give a brief overview of the historic origins of cellular 
automata, as they were conceived around 1950. We subsequently describe which 
main ingredients constitute a cellular automaton: the physical environment, the 
cells' states, their neighbourhoods, and finally a local transition rule. We 
then move on to a general description on how cellular automata are applied to 
vehicular road traffic, discussing their physical environment and the 
accompanying rule set that describes the vehicles' physical propagation.

		\subsection{Historic origins of cellular automata}

The mathematical concepts of cellular automata (CA) models can be traced back as 
far as 1948, when Johann Louis von Neumann introduced them to study (living) 
biological systems \cite{VONNEUMANN:48}. Central to von Neumann's work, was the 
notion of \emph{self-reproduction} and theoretical machines (called 
\emph{kinematons}) that could accomplish this. As his work progressed, von 
Neumann started to cooperate with Stanislaw Marcin Ulam, who introduced him to 
the concept of \emph{cellular spaces}. These described the physical structure of 
a cellular automaton, i.e., a grid of cells which can be either `on' or `off' 
\cite{WOLFRAM:83,DELORME:98}. Interestingly, Alan Mathison Turing proposed in 
1952 a model that illustrated reaction-diffusion in the context of 
\emph{morphogenesis} (e.g., to explain the patterns of spots on giraffes, of 
stripes on zebras, \ldots). His model can be seen as a type of continuous CA, in 
which the cells have a direct analogy with a simplified biological organism 
\cite{TURING:52}.

\begin{figure}[!htbp]
	\centering
	\includegraphics[width=\figurewidth]{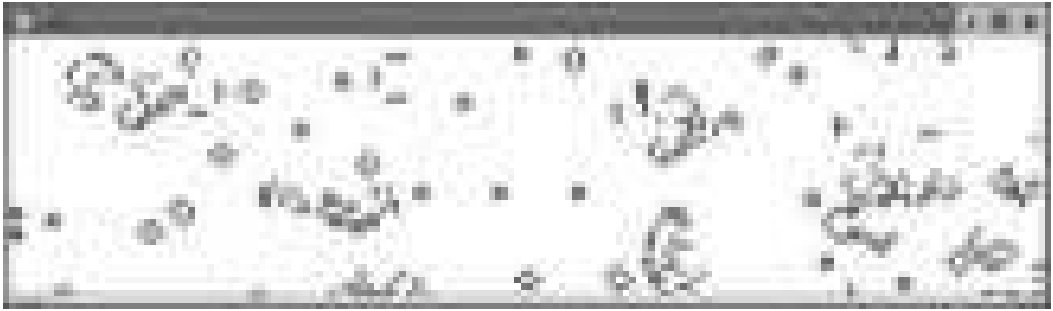}
	\caption{
		An example of the Game of Life, with a rectangular grid of cells. Live cells 
		are coloured black, whereas dead cells remain white. The image shows a 
		snapshot during the game's course, illustrating many different shapes to 
		either die out, or live indefinitely by remaining stationary or moving 
		around (image adapted from \cite{GEORGET:02}).
	}
	\label{fig:TCA:GameOfLife}
\end{figure}

In the seventies, CA models found their way to one of the most popular 
applications called `simulation games', of which John Horton Conway's 
\emph{``Game of Life''} \cite{GARDNER:70} is probably the most famous. The game 
found its widespread fame due to Martin Gardner who, at that time, devoted a 
Scientific American column, called \emph{``Mathematical Games''}, to it. Life, 
as it is called for short, is traditionally `played' on an infinitely large grid 
of cells. Each cell can either be `alive' or `dead'. The game evolves by 
considering a cell's all surrounding neighbours, deciding whether or not the 
cell should live or die, leading to phenomenon called `birth', `survival', and 
`overcrowding' (or `loneliness'). An example of a Life game board can be seen in 
\figref{fig:TCA:GameOfLife}. Typical of Life, is the spawning of a whole 
plethora of patterns or shapes, having illustrious names such as gliders, guns, 
space ships, puffers, beehives, oscillators, \ldots The Game of Life is now all 
about how these shapes evolve, and whether or not they die out or live 
indefinitely (either by remaining stationary or moving around).

The widespread popularisation of CA models was achieved in the eighties through 
the work of Stephen Wolfram. Based on empirical experiments using computers, he 
gave an extensive classification of CA models as mathematical models for 
self-organising statistical systems \cite{WOLFRAM:83,WOLFRAM:02}. Wolfram's work 
culminated in his mammoth monograph, called \emph{A New Kind of Science} 
\cite{WOLFRAM:02}. In this book, Wolfram related cellular automata to all 
disciplines of science (e.g., sociology, biology, physics, mathematics, \ldots). 
Despite the broad range of science areas touched upon, Wolfram's book has 
received its share of criticism. As an example of this, we mention the comments 
of Gray, who points out that Wolfram's results suffer from a rigourous 
mathematical test. As a consequence, the physical examples in his book are 
deemed either uncheckable or unconvincing. Gray's final critique is that 
\emph{``\ldots he [Wolfram] has helped to popularise a relatively little-known 
mathematical area (CA theory), and he has unwittingly provided several highly 
instructive examples of the pitfalls of trying to dispense with mathematical 
rigour''} \cite{GRAY:03}. However, with respect to their computational power, CA 
models can emulate universal Turing machines within the theories of computation 
and complexity. Recently, Chua took Wolfram's empirical observations one step 
further, proving that some of the CA models are capable of Turing universal 
computations. He furthermore introduced the paradigm of \emph{cellular neural 
networks} (CNN), which provide a very efficient method for performing massive 
parallel computations, and are a generalisation of cellular automata 
\cite{CHUA:05}.

Finally, an important step in this direction, is Bill Gosper's proof that the 
Game of Life is computationally universal, i.e., it can mimic arbitrary 
algorithms \cite{GOSPER:74}. Notably, one of the most profound testimonies 
related to this concept, is the work of Konrad Zuse and Edward Fredkin at the 
end of the sixties. Their Zuse-Fredkin thesis states that \emph{``The Universe 
is a cellular automaton''}, and is based on the assumption that the Universe's 
physical laws are discrete in nature \cite{ZUSE:67,ZUSE:69,FREDKIN:90}. This 
latter statement was also conveyed by Wolfram in his famous CA compendium 
\cite{WOLFRAM:02}.

		\subsection{Ingredients of a cellular automaton}
		\label{sec:TCA:IngredientsOfACellularAutomaton}

\begin{figure}[!htbp]
	\centering
	\includegraphics[width=\figurewidth]{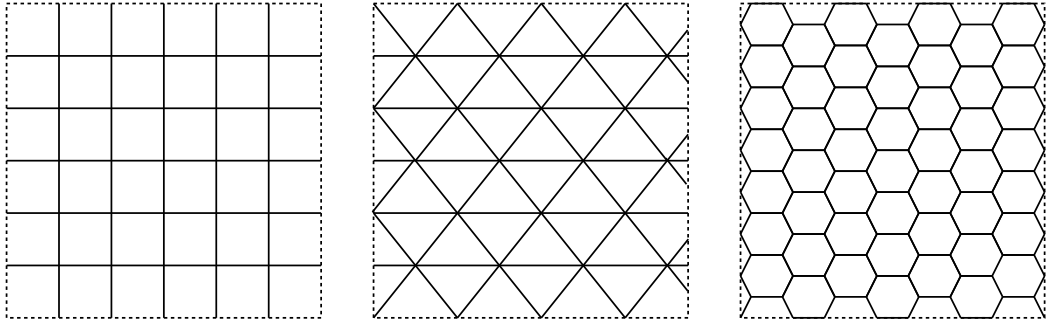}
	\caption{
		Some examples of different Euclidean lattice topologies for a cellular 
		automaton in two dimensions. \emph{Left:} rectangular. \emph{Middle:} 
		triangular/isometric. \emph{Right:} hexagonal.
	}
	\label{fig:TCA:LatticeTopologies}
\end{figure}

From a theoretical point of view, four main ingredients play an important role 
in cellular automata models \cite{GUTOWITZ:96,DELORME:98,SARKAR:00}:

\textbf{(1) The physical environment}\\
This defines the \emph{universe} on which the CA is computed. This underlying 
structure consists of a \emph{discrete lattice of cells} with a rectangular, 
hexagonal, or other topology (see \figref{fig:TCA:LatticeTopologies} for some 
examples). Typically, these cells are all equal in size; the lattice itself can 
be finite or infinite in size, and its dimensionality can be 1 (a linear string 
of cells called an \emph{elementary cellular automaton} or ECA), 2 (a grid), or 
even higher dimensional. In most cases, a common --- but often neglected --- 
assumption, is that the CA's lattice is embedded in a \emph{Euclidean space}.\\

\textbf{(2) The cells' states}\\
Each cell can be in a certain state, where typically an integer represents the 
number of distinct states a cell can be in, e.g., a binary state. Note that a 
cell's state is not restricted to such an integer domain (e.g., 
$\mathbb{Z}_{\two}$), as a continuous range of values is also possible (e.g., 
$\mathbb{R}^{+}$), in which case we are dealing with \emph{coupled map lattices} 
(CML) \cite{CRUTCHFIELD:87,KANEKO:90}. We call the states of all cells 
collectively a CA's \emph{global configuration}. This convention asserts that 
states are local and refer to cells, while a configuration is global and refers 
to the whole lattice.\\

\begin{figure}[!htbp]
	\centering
	\psfrag{C}[c][c]{\scriptsize{C}}
	\psfrag{N}[c][c]{\scriptsize{N}}
	\psfrag{E}[c][c]{\scriptsize{E}}
	\psfrag{S}[c][c]{\scriptsize{S}}
	\psfrag{W}[c][c]{\scriptsize{W}}
	\psfrag{NE}[c][c]{\scriptsize{NE}}
	\psfrag{SE}[c][c]{\scriptsize{SE}}
	\psfrag{SW}[c][c]{\scriptsize{SW}}
	\psfrag{NW}[c][c]{\scriptsize{NW}}
	\includegraphics[width=\figurewidth]{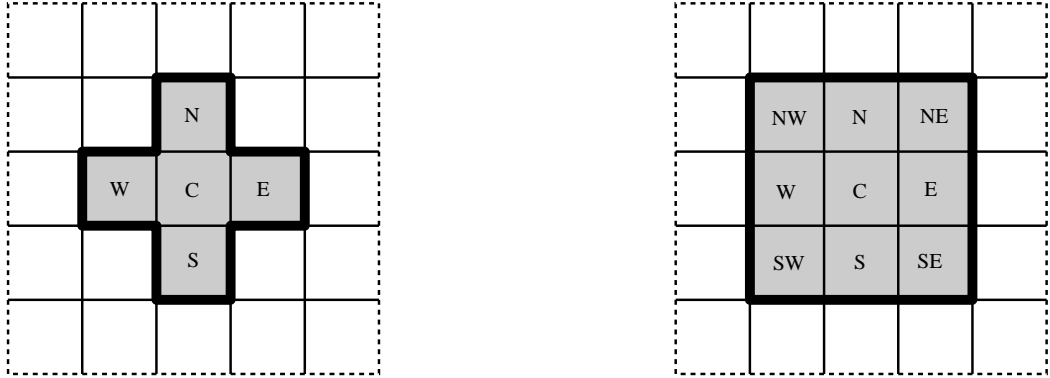}
	\caption{
		Two commonly used two dimensional CA neighbourhoods with a radius of 1: the 
		von Neumann neighbourhood (left) consisting of the central cell itself plus 
		4 adjacent cells, and the Moore neighbourhood (right) where there are 8 
		adjacent cells. Note that for one dimensional CAs, both types of 
		neighbourhoods are the same.
	}
	\label{fig:TCA:CellNeighbourhoods}
\end{figure}

\textbf{(3) The cells' neighbourhoods}\\
For each cell, we define a neighbourhood that locally determines the 
evolution of the cell. The size of neighbourhood is the same for each cell 
in the lattice. In the simplest case, i.e., a 1D lattice, the neighbourhood 
consists of the cell itself plus its adjacent cells. In a 2D rectangular 
lattice, there are several possibilities, e.g., with a radius of 1 there 
are, besides the cell itself, the four north, east, south, and west adjacent 
cells (\emph{von Neumann neighbourhood)}, or the previous five cells as well 
as the four north-east, south-east, south-west, and north-west diagonal 
cells (\emph{Moore neighbourhood)}; see \figref{fig:TCA:CellNeighbourhoods} 
for an example of both types of neighbourhoods. Note that as the 
dimensionality of the lattice increases, the number of direct neighbours of 
a cell increases exponentially.\\

\textbf{(4) A local transition rule}\\
This rule (also called function) acts upon a cell and its direct neighbourhood, 
such that the cell's state changes from one \emph{discrete time step} to another 
(i.e., the system's iterations). The CA evolves in time and space as the rule is 
subsequently applied to all the cells \emph{in parallel}. Typically, the same 
rule is used for all the cells (if the converse is true, then the term 
\emph{hybrid} CA is used). When there are no stochastic components present in 
this rule, we call the model a \emph{deterministic} CA, as opposed to a 
\emph{stochastic} (also called \emph{probabilistic}) CA.

As the local transition rule is applied to all the cells in the CA's lattice, 
the global configuration of the CA changes. This is also called the CA's 
\emph{global map}, which transforms one global configuration into another. This 
corresponds to the notion of \emph{computing a function} in automata theory, see 
also section \ref{sec:TCA:ClassicNotation}. Sometimes, the CA's evolution can be 
reversed by computing past states out of future states. By evolving the CA 
backwards in time in this manner, the CA's \emph{inverse global map} is 
computed. If this is possible, the CA is called \emph{reversible}, but if there 
are states for which no precursive state exists, these states are called 
\emph{Garden of Eden} (GoE) states and the CA is said to be \emph{irreversible}.

Finally, when the local transition rule is applied to all cells, its global map 
is computed. In the context of the theory of dynamical systems, this phenomenon 
of \emph{local simple interactions} that lead to a \emph{global complex 
behaviour} (i.e., the spontaneous development of order in a system due to 
\emph{internal} interactions), is termed \emph{self-organisation} or 
\emph{emergence}.

Whereas the previous paragraphs discussed the classic approach to CA models, the 
following sections will exclusively focus on vehicular traffic flows, leading to 
traffic cellular automata (TCA) models: section 
\ref{sec:TCA:RoadLayoutAndThePhysicalEnvironment} discusses the physical 
environment on which these TCA models are based, and section 
\ref{sec:TCA:VehicleMovementsAndTheRuleSet} deals with their accompanying rule 
set that determines the vehicular motion.

		\subsection{Road layout and the physical environment}
		\label{sec:TCA:RoadLayoutAndThePhysicalEnvironment}

When applying the cellular automaton analogy to vehicular road traffic flows, 
the physical environment of the system represents the road on which the vehicles 
are driving. In a classic single-lane setup for traffic cellular automata, this 
layout consists of a one-dimensional lattice that is composed of individual 
cells (our description here thus focuses on unidirectional, single-lane 
traffic). Each cell can either be empty, or is occupied by \emph{exactly} one 
vehicle; we use the term \emph{single-cell models} to describe these systems. 
Another possibility is to allow a vehicle to span several consecutive cells, 
resulting in what we call \emph{multi-cell models}. Because vehicles move from 
one cell to another, TCA models are also called \emph{particle-hopping models} 
\cite{NAGEL:96}.

\begin{figure}[!htbp]
	\centering
	\psfrag{t}[][]{\psfragstyle{$t$}}
	\psfrag{t+1}[][]{\psfragstyle{$t + \one$}}
	\psfrag{i}[][]{\psfragstyle{$i$}}
	\psfrag{j}[][]{\psfragstyle{$j$}}
	\psfrag{deltat}[][]{\psfragstyle{$\Delta T$}}
	\psfrag{deltax}[][]{\psfragstyle{$\Delta X$}}
	\includegraphics[width=0.95\figurewidth]{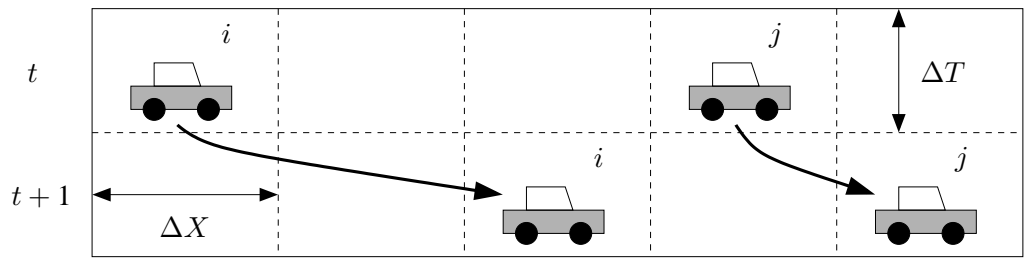}
	\caption{
		Schematic diagram of the operation of a single-lane traffic cellular 
		automaton (TCA); here, the time axis is oriented downwards, the space axis 
		extends to the right. The TCA's configuration is shown for two consecutive 
		time steps $t$ and $t + \one$, during which two vehicles $i$ and $j$ 
		propagate through the lattice.
	}
	\label{fig:TCA:ClassicLattice}
\end{figure}

An example of the tempo-spatial dynamics of such a system is depicted in 
\figref{fig:TCA:ClassicLattice}, where two consecutive vehicles $i$ and $j$ are 
driving on a 1D lattice. A typical discretisation scheme assumes $\Delta T 
=$~1~s and $\Delta X =$~7.5~m, corresponding to speed increments of $\Delta V = 
\Delta X / \Delta T =$~27~km/h. The spatial discretisation corresponds to the 
average length a conventional vehicle occupies in a closely packed jam (and as 
such, its width is neglected), whereas the temporal discretisation is based on a 
typical driver's reaction time and we implicitly assume that a driver does not 
react to events between two consecutive time steps \cite{NAGEL:92}.

With respect to the layout of the system, we can distinguish two main cases: 
closed versus open systems. They correspond to periodic (or cyclic) versus open 
boundary conditions. The former is usually implemented as a closed ring of 
cells, sometimes called the \emph{Indianapolis scenario}, while the latter 
considers an open road. This last type of system, is also called the 
\emph{bottleneck scenario}. The name is derived from the fact that this 
situation can be seen as the outflow from a jam, where vehicles are placed at 
the left boundary whenever there is a vacant spot. Note that, in closed systems, 
the number of vehicles is always conserved, leading to the description of 
\emph{number conserving cellular automata} (NCCA) \cite{MOREIRA:03}.

		\subsection{Vehicle movements and the rule set}
		\label{sec:TCA:VehicleMovementsAndTheRuleSet}

The propagation of the individual vehicles in a traffic stream, is described by 
means of a rule set that reflects the car-following and lane-changing behaviour 
of a traffic cellular automaton evolving in time and space. The TCA's local 
transition rule actually comprises this set of rules. They are consecutively 
applied to all vehicles in parallel (called a \emph{parallel update}). So in a 
classic setup, the system's state is changed through \emph{synchronous position 
updates} of all the vehicles: for each vehicle, the new speed is computed, after 
which its position is updated according to this speed and a possible lane-change 
manoeuvre. Note that there are other ways to perform this update procedure, 
e.g., a random sequential update (see section \ref{sec:TCA:TASEP}). Because time 
is discretised in units of $\Delta T$ seconds, an \emph{implicit reaction time} 
is assumed in TCA models. It is furthermore assumed that a driver does not react 
to events between consecutive time steps.

For single-lane traffic, we assume that vehicles act as \emph{anisotropic 
particles}, i.e., they only respond to frontal stimuli. So typically, the 
car-following part of a rule set only considers the direct frontal neighbourhood 
of the vehicle to which the rules are applied. The radius of this neighbourhood 
should be taken large enough such that vehicles are able to drive 
collision-free. Typically, this radius is equal to the maximum speed a vehicle 
can achieve, expressed in cells per time step.

From a microscopic point of view, the process of a vehicle following its 
predecessor is typically expressed using a \emph{stimulus-response relation} 
\cite{MAERIVOET:05c}. Typically, this response is the speed or the acceleration 
of a vehicle; in TCA models, a vehicle's stimulus is mainly composed of its 
speed and the distance to its leader, with the response directly being a new 
(adjusted) speed of the vehicle. In a strict sense, this only leads to the 
avoidance of accidents. Some models however, incorporate more detailed stimuli, 
such as anticipation terms. These forms of `anticipation' only take leaders' 
reactions into account, \emph{without predicting} them. When these effects are 
taken into account together with a safety distance, strong accelerations and 
abrupt braking can be avoided. Hence, as the speed variance is decreased, this 
results in a more stable traffic stream 
\cite{KNOSPE:02b,EISSFELDT:03,LARRAGA:04}.

To conclude this section, we note that a TCA model can also be derived from a 
so-called Gipps car-following model. All speeds in this Gipps model are directly 
computed from one discrete time step to another \cite{MAERIVOET:05c}. If now the 
spatial dimension is also discretised (a procedure called \emph{coarse 
graining}), then this will result in a TCA model.

	\section{Mathematical notation}

In this section, we give an overview of the mathematical notation adopted 
throughout this report. The focus will be on the variables in TCA models, the 
measurements that can be done on a TCA model's lattice, and their conversion to 
real-world units. We first take a look at the notation that is commonly used in 
automata theory, from which cellular automata sprung.

		\subsection{Classic notation based on automata theory}
		\label{sec:TCA:ClassicNotation}

Let us first briefly present the notation for cellular automata models, adopted 
in spirit of \emph{automata theory}. As mentioned in section 
\ref{sec:TCA:BackgroundPhysicalSetupForRoadTraffic}, a CA model represents a 
discrete dynamic system, consisting of four ingredients:

\begin{equation}
	\text{CA} = (\mathcal{L},\Sigma,\mathcal{N},\delta),
\end{equation}

where the physical environment is represented by the discrete lattice 
$\mathcal{L}$ and the set of possible states denoted by $\Sigma$. Each \ith cell 
of the lattice, has at time step $t$ a state $\sigma_{i}(t) \in \Sigma$. 
Furthermore, the associated neighbourhood with this cell is represented by 
$\mathcal{N}_{i}(t)$, i.e., a (partially) ordered set of cells. Finally, the 
local transition rule is represented as:

\begin{equation}
\label{eq:TCA:TransitionRule}
	\delta: \Sigma^{| \mathcal{N} |} \longrightarrow \Sigma: \bigcup_{j \in \mathcal{N}_{i}(t)} \sigma_{j}(t) \longmapsto \sigma_{i}(t + \one).
\end{equation}

Equation \eqref{eq:TCA:TransitionRule} shows that the state of the $i$th cell at 
the next time step $t + \one$ is computed by $\delta$ based on the states of all 
the cells in its neighbourhood at the current time step $t$. In the previous 
equation, $| \mathcal{N} |$ represents the number of cells in this 
neighbourhood, which is taken to be invariant with respect to time and space. 
Note that the local transition rule is commonly given by a \emph{rule table}, 
where the output state is listed for each possible input configuration of the 
neighbourhood. Given the sizes of $\Sigma$ and $\mathcal{N}$, the total number 
of possible rules equals:

\begin{equation}
\label{eq:TCA:NumberOfPossibleRules}
	| \Sigma^{\Sigma^{\mathcal{N}}} |,
\end{equation}

where each of the $| \Sigma^{\mathcal{N}} |$ possible configurations of a cell's 
neighbourhood is mapped to the number of possible states a cell can be in.

Considering the ordered set of all the states of all cells collectively at time 
step $t$, a CA's global configuration is obtained as:

\begin{equation}
	\mathcal{C}(t) = \bigcup_{j \in \mathcal{L}} \sigma_{j}(t),
\end{equation}

with $\mathcal{C}(t) \in \Sigma^{\mathcal{L}}$ where the latter refers to the 
set of all possible global configurations a CA can be in (also called its 
\emph{phase space}). Sometimes, such a global configuration $\mathcal{C}(t)$ is 
also represented by its characteristic polynomial (i.e., generating function) 
\cite{WOLFRAM:84b}:

\begin{equation}
	\mathcal{C}(t) = \sum_{j = \zero}^{| \mathcal{L} |} \sigma_{j}(t)~x^{j}.
\end{equation}

If we now apply the local transition rule to all the cells in the CA's lattice, 
the next configuration of the CA can be computed by its induced global map:

\begin{equation}
	G : \Sigma^{\mathcal{L}} \longrightarrow \Sigma^{\mathcal{L}} : \mathcal{C}(t) \longmapsto \mathcal{C}(t + \one).
\end{equation}

Note that if the CA is reversible, the inverse global map $G^{\text{-\one}}$ can 
be computed. As the CA evolves in time and space, the global map is iterated 
from a certain initial configuration $\mathcal{C}(\zero)$ at $t = \zero$, 
leading to the following sequence of configurations:

\begin{equation}
	\mathcal{C}(\zero) \rightarrow G(\mathcal{C}(\zero)) \rightarrow G^{\two}(\mathcal{C}(\zero)) \rightarrow G^{\text{\three}}(\mathcal{C}(\zero)) \rightarrow \cdots
\end{equation}

The above sequence is called the \emph{trajectory} of the initial configuration 
$\mathcal{C}(\zero)$ under the global map $G$, and we denote it by:

\begin{equation}
	\mathcal{T}_{\mathcal{C}(\zero) | G} = \lbrace G^{n}(\mathcal{C}(\zero))~|~n \in \mathbb{N} \rbrace.
\end{equation}

When this trajectory is periodic or chaotic, we use the terminology 
\emph{forward orbit} and denote it by $\mathcal{O}_{\mathcal{C}(\zero) | 
G}^{+}$. Similarly, the \emph{backward orbit} (i.e., the reverse trajectory) is 
denoted by $\mathcal{O}_{\mathcal{C}(t) | G^{\text{-1}}}^{-}$, where we specify 
a certain global configuration $\mathcal{C}(t)$ at time step $t$ under the 
inverse global map $G^{\text{-\one}}$.

			\subsubsection{Classification of CA rules}

Computing the global map $G$ is rather difficult, as it may require many or even 
an infinite amount of iterations in order to obtain the trajectories. In 
practice, the system's lattice size should be taken infinitely large, but even 
only considering 1000 cells of a binary elementary cellular automaton (ECA) 
would increase the size of the search space of global configurations to 
$\two^{\text{1000}} \approx \text{10}^{\text{300}}$.

A more intuitive methodology, is to observe a CA's tempo-spatial behaviour, 
i.e., its evolution on the lattice in the course of time. To this end, Stephen 
Wolfram empirically studied many configurations of binary ECA rules, with a 
neighbourhood of three cells. According to equation 
\eqref{eq:TCA:NumberOfPossibleRules}, this amounts to $\two^{\two^{\three}} 
=$~256 different rules. In 1984, based on this research, Wolfram conjectured 
four distinct \emph{universality classes} \cite{WOLFRAM:84}:

\begin{quote}
\begin{description}
	\item[Class I]~\\
		These CA evolve after a finite number of iterations to a unique homogeneous 
		state, i.e., a \emph{limit point}.
	\item[Class II]~\\
		These CA generate regular, periodic patterns, i.e., entering a \emph{limit 
		cycle}.
	\item[Class III]~\\
		CAs in this class evolve to aperiodic patterns, independent of the initial 
		configuration; their trajectories in the configuration space lie on a 
	\emph{chaotic attractor}.
	\item[Class IV]~\\
		This class encompasses all the CAs that seem to behave in a \emph{complex} 
		way, with features such as propagating structures, long transients; they are 
		thought to have the capability of universal computation.
\end{description}
\end{quote}

Although Wolfram's classification scheme is widely adopted, it still remains a 
tentative result as he himself states \cite{WOLFRAM:02}. Note that the type of 
classification he provides is \emph{phenotypic}, in the sense that it is based 
on observed behaviour, whereas a \emph{genotypic} classification would be based 
on the intrinsic structure of the rules in each class.

Despite these observations, classification still remains a difficult task as is 
evidenced by the ongoing research in dynamical systems. Other attempts at 
classification of ECA rules include the following. Firstly, \v Culik and Yu gave 
a formalisation of Wolfram's classes \cite{CULIK:88}. Secondly, Li and Packard 
studied the structure of the ECA rule space according to a certain distance 
metric, resulting in five classes \cite{LI:90}. Then, Braga et al. identified 
three classes based on the growth of patterns observed in CA models 
\cite{BRAGA:95}. Next, Wuensche used a whole arsenal of local measures to 
automatically create complex rules, thereby classifying the rule space for the 
CAs' dynamics \cite{WUENSCHE:99}. Furthermore, Dubacq et al. classified CA 
models based on their algorithmic complexity by measuring the information 
content of the local transition rule \cite{DUBACQ:01}. And finally, Fat\`es who 
used a macroscopic parameter, i.e., the density of 1's, to separate chaotic ECA 
rules from non-chaotic ones \cite{FATES:03}.

			\subsubsection{An example of a CA}
			\label{sec:TCA:AnExampleOfACA}

To end this section, let us give some definitions of a one dimensional, 
infinitely large, binary state CA with a neighbourhood of radius 1:

\begin{eqnarray}
	\mathcal{L}       & = & \mathbb{Z}^{d} \quad \text{(with $d = \one$)},\\
	\Sigma            & = & \mathbb{Z}_{\two} = \lbrace \zero,\one \rbrace,\\
	\mathcal{N}_{i}   & = & \lbrace i - \one, i, i + \one \rbrace,\label{eq:TCA:ExampleNeighbourhood}\\
	\delta(i,t)       & : & \mathbb{Z}_{\two}^{\three} \longrightarrow \mathbb{Z}_{\two}\nonumber\\
	                  & : & \lbrace \sigma_{i - \one}(t), \sigma_{i}(t), \sigma_{i + \one}(t) \rbrace \longmapsto \sigma_{i}(t + \one),\label{eq:TCA:ExampleLocalRule}\\
	G(\mathcal{C}(t)) & : & \mathbb{Z}_{\two}^{\mathbb{Z}} \longrightarrow \mathbb{Z}_{\two}^{\mathbb{Z}}\nonumber\\
	                  & : & \mathcal{C}(t) \longmapsto \mathcal{C}(t + \one).\label{eq:TCA:ExampleGlobalMap}
\end{eqnarray}

Note that in equation \eqref{eq:TCA:ExampleNeighbourhood}, we assume that the 
\ith cell's neighbourhood is represented by integer indices (i.e., the cells 
form a totally ordered set). This alleviates the need for an explicit 
representation of the cells themselves, as it is now sufficient to work with the 
cells' indices and states. The transition rule $\delta$ in equation 
\eqref{eq:TCA:ExampleLocalRule} takes as its arguments a cell's index $i$ and 
current time step $t$, but operates on the states of this cell's neighbourhood. 
The global map in equation \eqref{eq:TCA:ExampleGlobalMap} operates on the 
global configuration of the CA at time step $t$.

		\subsection{Basic variables and conventions}
		\label{sec:TCA:BasicVariablesAndConventions}

\begin{figure}[!htbp]
	\centering
	\psfrag{i}[][]{\psfragstyle{$i$}}
	\psfrag{Li}[][]{\psfragstyle{$l_{i}$}}
	\psfrag{gsi}[][]{\psfragstyle{$g_{s_{i}}$}}
	\psfrag{hsi}[][]{\psfragstyle{$h_{s_{i}}$}}
	\psfrag{gsilf}[][]{\psfragstyle{$g_{s_{i}}^{l,f}$}}
	\psfrag{gsilb}[][]{\psfragstyle{$g_{s_{i}}^{l,b}$}}
	\psfrag{gsirf}[][]{\psfragstyle{$g_{s_{i}}^{r,f}$}}
	\psfrag{gsirb}[][]{\psfragstyle{$g_{s_{i}}^{r,b}$}}
	\includegraphics[width=\figurewidth]{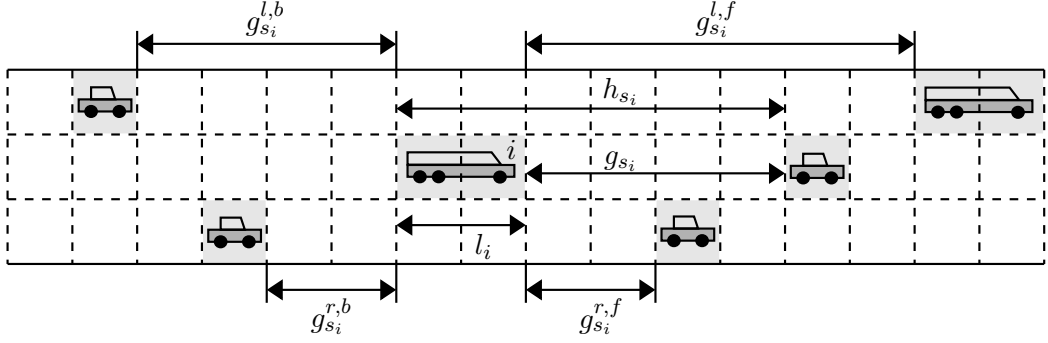}
	\caption{
		A portion of the lattice $\mathcal{L}$ at a certain time step; it has 
		$L_{\mathcal{L}} =$~3 lanes, containing six vehicles. The central vehicle 
		$i$ has a space headway $h_{s_{i}} =$~6~cells, consisting of a space gap 
		$g_{s_{i}} =$~4~cells and its length $l_{i} =$~2~cells. There are four other 
		space gaps to be considered when the neighbouring lanes are taken into 
		account: $g_{s_{i}}^{l,f}$ (left-front), $g_{s_{i}}^{l,b}$ (left-back), 
		$g_{s_{i}}^{r,f}$ (right-front), and $g_{s_{i}}^{r,b}$ (right-back), 
		equalling 6, 4, 2 and 2 cells, respectively.
	}
	\label{fig:TCA:LatticeLayout}
\end{figure}

Conforming to the setup and notation discussed in the previous sections, we 
denote a TCA's discrete lattice by $\mathcal{L}$ (for the remainder of this 
section, we assume a \emph{rectangular lattice}). This lattice physically 
represents the road on which vehicles will drive in a TCA model. It consists of 
$L_{\mathcal{L}}$ lanes, each of which has $K_{\mathcal{L}}$ cells, so in total 
there are $L_{\mathcal{L}} \times K_{\mathcal{L}}$ cells in the lattice 
($L_{\mathcal{L}}, K_{\mathcal{L}} \in \mathbb{N}_{\zero}$). Each cell can 
either be empty, or occupied with a single vehicle that spans one or more 
consecutive cells. An example of a lattice containing several vehicles, can be 
seen in \figref{fig:TCA:LatticeLayout}.

Based on the microscopic vehicle characteristics of a vehicle's space headway, 
space gap, length, time headway, time gap, and occupancy time, we propose to use 
the following set of definitions for \emph{multi-lane} vehicular road traffic 
flows that are \emph{heterogeneous} (in the sense of having different vehicle 
lengths) \cite{MAERIVOET:05d}:

\begin{eqnarray}
	g_{s_{i}}^{l,f} & = & x_{i}^{l,f} - x_{i} - l_{i},\label{eq:TCA:SpaceGapLeftFront}\\
	g_{s_{i}}^{r,f} & = & x_{i}^{r,f} - x_{i} - l_{i},\\
	g_{s_{i}}^{l,b} & = & x_{i} - x_{i}^{l,b} - l_{i}^{l,b},\\
	g_{s_{i}}^{r,b} & = & x_{i} - x_{i}^{r,b} - l_{i}^{r,b}\label{eq:TCA:SpaceGapRightBack},
\end{eqnarray}

for which we assume that a vehicle's position is denoted by the cell that 
contains its rear bumper. For the example in \figref{fig:TCA:LatticeLayout}, the 
left and right frontal and backward space gaps of the central vehicle $i$ are 6, 
4, 2 and 2 cells, respectively (all these space gaps thus represent effective 
distances, corresponding to the number of empty cells between vehicles). Similar 
definitions hold for the space headways $h_{s_{i}}^{l,f}$, $h_{s_{i}}^{r,f}$, 
$h_{s_{i}}^{l,b}$, and $h_{s_{i}}^{r,b}$, i.e., the vehicle lengths in the 
right-hand sides of equations \eqref{eq:TCA:SpaceGapLeftFront} -- 
\eqref{eq:TCA:SpaceGapRightBack} are dropped. Derivations for the time gaps 
$g_{t_{i}}^{l,f}$, $g_{t_{i}}^{r,f}$, $g_{t_{i}}^{l,b}$, and $g_{t_{i}}^{r,b}$, 
and time headways $h_{t_{i}}^{l,f}$, $h_{t_{i}}^{r,f}$, $h_{t_{i}}^{l,b}$, and 
$h_{t_{i}}^{r,b}$ are analogous.

Discriminating between frontal an backward neighbours in the adjacent lanes to 
the \ith vehicle, is done based on their positions, i.e.:

\begin{equation}
\label{eq:TCA:NeighboursPositionDiscrimination}
	\lbrace x_{i}^{l,b}, x_{i}^{r,b} \rbrace < x_{i} \leq \lbrace x_{i}^{l,f}, x_{i}^{r,f} \rbrace.
\end{equation}

According to equation \eqref{eq:TCA:NeighboursPositionDiscrimination}, a vehicle 
that is driving alongside in an adjacent lane to the \ith vehicle, will be 
considered as a backward neighbour as long as its rear bumper is located 
strictly behind the rear bumper of the \ith vehicle (even if this neighbour has 
a large length that `sticks out' in front of the \ith vehicle).

Under the above set of assumptions, we can now write the conditions for a 
successful lane change (i.e., a possible gap acceptance) as the following 
constraints:

\begin{eqnarray}
	g_{s_{i}}^{l,f} \geq \zero \quad \wedge \quad g_{s_{i}}^{l,b} \geq \zero & & \text{(left lane change)},\\
	g_{s_{i}}^{r,f} \geq \zero \quad \wedge \quad g_{s_{i}}^{r,b} \geq \zero & & \text{(right lane change)}.
\end{eqnarray}

With respect to the domains of all variables, we note that all vehicle lengths, 
space gaps, and space headways are expressed as integers, or more specifically:

\begin{eqnarray}
	l_{i}, h_{s_{i}}, h_{s_{i}}^{l,b}, h_{s_{i}}^{r,b}                 & \in & \mathbb{N}_{\zero},\nonumber\\
	g_{s_{i}}, h_{s_{i}}^{l,f}, h_{s_{i}}^{r,f}                        & \in & \mathbb{N},\nonumber\\
	g_{s_{i}}^{l,b}, g_{s_{i}}^{r,b}, g_{s_{i}}^{l,f}, g_{s_{i}}^{r,f} & \in & \mathbb{Z}.\nonumber
\end{eqnarray}

In contrast to this, the occupancy times, time headways, and time gaps are not 
restricted to the domain of integers, i.e.:

\begin{eqnarray}
	\rho_{i}, h_{t_{i}}, h_{t_{i}}^{l,b}, h_{t_{i}}^{r,b}                    & \in & \mathbb{R}_{\zero}^{+},\nonumber\\
	g_{t_{i}}, h_{t_{i}}^{l,f}, h_{t_{i}}^{r,f}                              & \in & \mathbb{R}^{+},\nonumber\\
	g_{t_{i}}^{l,b}, g_{t_{i}}^{r,b}, g_{t_{i}}^{l,f}, g_{t_{i}}^{r,f} & \in & \mathbb{R}.\nonumber
\end{eqnarray}

For example, the occupancy time $\rho_{i}$ as defined by $\rho_{i} = l_{i} / 
v_{i}$ \cite{MAERIVOET:05d}, corresponds to the time a vehicle `spends' in its 
own cells.

To conclude, each vehicle $i$ in the lattice has an associated speed $v_{i} \in 
\mathbb{N}$ (expressed in cells per time step $\Delta T$), which is bounded by a 
maximum speed $v_{\text{max}} \in \mathbb{N}_{\zero}$. For example, if we set 
$\Delta T =$~1.2 s, $\Delta X =$~7.5~m, and $v_{\text{max}} = \five$ cells/time 
step, then $v_{i} \in \lbrace \zero, \ldots, \five \rbrace$ which corresponds to 
a maximum of $\five \times \Delta X / \Delta T = \five \times \text{7.5~m/s} 
\div \text{1.2~s} = \text{31.25~m/s} = \text{112.5~km/h}$. As can be seen in 
this derivation, we only consider positive speeds in our models, i.e., vehicles 
always move forward.

		\subsection{Performing macroscopic measurements}

The previously discussed quantities are all microscopic traffic stream 
characteristics. In this section, we reconsider the macroscopic quantities 
densities, flows, and mean speeds \cite{MAERIVOET:05d}. As we now have to 
measure these quantities on a TCA's lattice $\mathcal{L}$, we present three 
possibilities for obtaining the data points:

\begin{itemize}
	\item by performing local measurements with an artificial loop detector of 
	finite length (open and closed systems),
	\item by performing global measurements on the entire lattice (closed system), 
	\item and by performing local measurements with an artificial loop detector of 
	unit length (open and closed systems).
\end{itemize}

In the following three sections, we give detailed derivations of each of these 
measurement techniques. Locally measured quantities are indicated by a `$l$' 
subscript, whereas globally measured ones are indicated by an `$g$' subscript. A 
temporal and spatial discretisation of respectively $\Delta T$ (in seconds) and 
$\Delta X$ (in metres) is implicitly assumed.\\

\sidebar{
	For all following techniques, we assume an integer measurement period of 
	$T_{\text{mp}}$ time steps. Thus, aggregating data into intervals of 60 
	seconds with $\Delta T =$~1.2~s, requires a measurement period of:

	\begin{equation} 
		T_{\text{mp}} = \left [ \frac{\text{60}}{\text{1.2}} \right ] = \text{50 time steps}.
	\end{equation} 

	Furthermore, densities are expressed in vehicles per cell, flows in vehicles 
	per time step, and space-mean speeds in cells per time step.
}\\

			\subsubsection{Local measurements with a detector of finite length} 
			\label{sec:TCA:LocalMeasurementsFiniteLength}

In this section, we deal with an artificial loop detector of finite length 
$K_{\text{ld}} \in \mathbb{N_{\zero}}$, located in a single lane. Note that 
typically, $K_{\text{ld}} \geq v_{\text{max}}$, so as to ensure that no vehicles 
can `skip' the detector between consecutive time steps. The first step in our 
approach for performing these measurements, is based on obtaining local 
measurements of the density and flow for such a spatial measurement region at a 
certain time step $t$ \cite{MAERIVOET:05d}. Once these are known, the space-mean 
speed can be derived using the fundamental relation of traffic flow theory $q = 
k~\overline v_{s}$ \cite{MAERIVOET:05d}:

\begin{eqnarray}
	k_{l}(t) & = &
		\frac{N(t)}{K_{\text{ld}}},\\
	q_{l}(t) & = &
		\frac{\one}{K_{\text{ld}}} \sum_{i = \one}^{N(t)} v_{i}(t),\\
	         & \Downarrow &\nonumber\\
	\overline v_{s_{l}}(t) & = &
		\frac{q_{l}(t)}{k_{l}(t)} =
		\frac{\one}{N(t)} \sum_{i = \one}^{N(t)} v_{i}(t),
\end{eqnarray}

where we assumed $N(t)$ vehicles are present at time $t$ in the loop detector's 
segment. The density and flow measurements of consecutive time steps are now 
temporally averaged over subsequent spatial measurement regions. In similar 
fashion as before, the space-mean speed is derived using the previously 
mentioned fundamental relation:

\begin{eqnarray}
	k_{l} & = &
		\frac{\one}{T_{\text{mp}}} \sum_{t = \one}^{T_{\text{mp}}} k_{l}(t) =
		\frac{\one}{T_{\text{mp}}~K_{\text{ld}}} \sum_{t = \one}^{T_{\text{mp}}} N(t),\\
	q_{l} & = &
		\frac{\one}{T_{\text{mp}}} \sum_{t = \one}^{T_{\text{mp}}} q_{l}(t) =
		\frac{\one}{T_{\text{mp}}~K_{\text{ld}}} \sum_{t = \one}^{T_{\text{mp}}} \sum_{i = \one}^{N(t)} v_{i}(t),\\
	         & \Downarrow &\nonumber\\
	\overline v_{s_{l}} & = &
		\frac{q_{l}}{k_{l}} =
		\sum_{t = \one}^{T_{\text{mp}}} \sum_{i = \one}^{N(t)} v_{i}(t) ~ \left / ~ \sum_{t = \one}^{T_{\text{mp}}} N(t) \right.,\\
		& = &
		\sum_{t = \one}^{T_{\text{mp}}} N(t) \frac{\one}{N(t)} \sum_{i = \one}^{N(t)} v_{i}(t) ~ \left / ~ \sum_{t = \one}^{T_{\text{mp}}} N(t) \right.,\nonumber\\
		& = &
		\sum_{t = \one}^{T_{\text{mp}}} N(t) ~ \overline v_{s_{l}}(t) ~ \left / ~ \sum_{t = \one}^{T_{\text{mp}}} N(t) \right..\label{eq:TCA:MeasurementsLocalSMSFiniteDetector}
\end{eqnarray}

Our derivations for $k_{l}$ and $q_{l}$ as outlined above, also correspond to 
the generalised definitions of density and flow, defined as the total time 
spent, respectively the total distance travelled, divided by the area of the 
measurement region (which corresponds to $T_{\text{mp}} \times K_{\text{ld}}$). 
Furthermore, note that the last equation 
\eqref{eq:TCA:MeasurementsLocalSMSFiniteDetector} essentially is a weighted mean 
of the local space-mean speeds $\overline v_{s_{l}}(t)$ at each time step $t$, 
with the number of vehicles $N(t)$ as weights.

			\subsubsection{Global measurements on the entire lattice}
			\label{sec:TCA:GlobalMeasurements}

For the global measurements, we consider $N$ vehicles that are driving in a 
closed single-lane system, i.e., with a length of $K_{\mathcal{L}}$ cells (the 
extension to multi-lane traffic is straightforward). As a consequence, the 
global density $k_{g}$ remains constant during the entire measurement period. 
The derivations of the equations for $k_{g}$, $q_{g}$, and $\overline 
v_{s_{g}}$, are completely equivalent to those of the previous section 
\ref{sec:TCA:LocalMeasurementsFiniteLength}, but now with $K_{\text{ld}} = 
K_{\mathcal{L}}$:

\begin{eqnarray}
	k_{g} & = &
		\frac{N}{K_{\mathcal{L}}},\label{eq:TCA:MeasurementsGlobalDensity}\\
	q_{g} & = &
		\frac{\one}{T_{\text{mp}}~K_{\mathcal{L}}} \sum_{t = \one}^{T_{\text{mp}}} \sum_{i = \one}^{N} v_{i}(t),\\
	& \Downarrow & \nonumber\\
	\overline v_{s_{g}} & = &
		\frac{q_{g}}{k_{g}} =
		\frac{\one}{T_{\text{mp}}~N} \sum_{t = \one}^{T_{\text{mp}}} \sum_{i = \one}^{N} v_{i}(t),\\
		& = &
		\frac{\one}{T_{\text{mp}}~N} \sum_{t = \one}^{T_{\text{mp}}} N \frac{\one}{N} \sum_{i = \one}^{N} v_{i}(t),\nonumber\\
		& = &
		\frac{\one}{T_{\text{mp}}} \sum_{t = \one}^{T_{\text{mp}}} \overline v_{s_{g}}(t).\label{eq:TCA:MeasurementsGlobalSMS}
\end{eqnarray}

Note that, for single-cell TCA models, the global density computed with equation 
\eqref{eq:TCA:MeasurementsGlobalDensity} actually corresponds to the macroscopic 
characteristic called occupancy $\rho$ \cite{MAERIVOET:05d}. For multi-cell 
models, the number of vehicles is in general less than the number of occupied 
cells.

			\subsubsection{Local measurements with a detector of unit length}
			\label{sec:TCA:LocalMeasurementsUnitLength}

The third technique for measuring macroscopic traffic flow characteristics on a 
TCA's lattice, bears perhaps the closest resemblance to reality: it is based on 
an artificial loop detector with unit length, i.e., $K_{\text{ld}} = \one$~cell. 
The loop detector now explicitly counts all the vehicles that pass it at each 
time step $\Delta T$ during the measurement period $T_{\text{mp}}$.

This type of measurement corresponds to a point measurement in a temporal 
measurement region. Because of this, the appropriate method for computation is 
different from the one used in the previous two sections: we now first compute 
the local flow, and the local space-mean speed, both for single-lane traffic. 
The local density is then derived according to the previously mentioned 
fundamental relation, resulting in the following set of equations:

\begin{eqnarray}
	q_{l} & = &
		\displaystyle \frac{N}{T_{\text{mp}}},\label{eq:TCA:MeasurementsLocalPointFlow}\\
	\overline v_{s_{l}} & = &
		\displaystyle \left ( \frac{\one}{N} \sum_{i = \one}^{N} \frac{\one}{v_{i}} \right )^{-\one},\\
	& \Downarrow & \nonumber\\
	k_{l} & = &
		\displaystyle \frac{q_{l}}{\overline v_{s_{l}}},\label{eq:TCA:MeasurementsLocalPointDensity}
\end{eqnarray}

in which $N$ now denotes the number of vehicles that have passed the detector 
during the measurement period $T_{\text{mp}}$. Because the detector physically 
occupies one cell and because a vehicle has to `drive by' in order to get 
counted, this means that stopped vehicles are ignored: \emph{only moving 
vehicles are counted}. Note that, as opposed to the previous two techniques, the 
above measurements no longer denote temporal averages. And because we are 
working with a temporal measurement region, we have to take the harmonic average 
of the vehicles' speeds $v_{i}$ in order to obtain the local space-mean speed 
$\overline v_{s_{l}}$ \cite{MAERIVOET:05d}.

		\subsection{Conversion to real-world units}
		\label{sec:TCA:ConversionToRealWorldUnits}

Converting between TCA and real-world units seems straightforward, as we only 
need to suitably multiply with or divide by the temporal and spatial 
discretisations $\Delta T$ and $\Delta X$, respectively. However, problems arise 
due to the discrete nature of a TCA model, involving some intricacies with 
respect to coordinate systems and their associated units. For example, as 
defined in section \ref{sec:TCA:BasicVariablesAndConventions}, a vehicle $i$'s 
space headway $h_{s_{i}}$ is always an integer, expressing the number of cells. 
The same holds true for its space gap $g_{s_{i}}$ and length $l_{i}$. The 
difficulty now lies in the fact that fractions of cells are not representable in 
our definition of a TCA model. Keeping in mind that $h_{s_{i}} = g_{s_{i}} + 
l_{i}$ \cite{MAERIVOET:05d}, and noting that $h_{s_{i}} \in \mathbb{N}_{\zero}$, 
it follows that $g_{s_{i}} + l_{i} > \zero$, which means that either $g_{s_{i}} 
\neq \zero$ and/or $l_{i} \neq \zero$.

As a solution, we therefore adopt throughout this report the convention that, 
without loss of generality, a vehicle's length $l_{i} \geq \one$ cell (which 
agrees perfectly with our earlier definitions in section 
\ref{sec:TCA:BasicVariablesAndConventions}). Consequently, when a vehicle $i$ is 
residing in a compact jam (i.e., `bumper-to-bumper' traffic), its space headway 
$h_{s_{i}} = l$ cells and its space gap $g_{s_{i}} = \zero$ cells. Our 
convention thus gives a rigourous justification to formulate the TCA's update 
rules more intuitively using space gaps, because as already stated in section 
\ref{sec:TCA:VehicleMovementsAndTheRuleSet}, the rules in a TCA rule set are 
typically not expressed in terms of space headways, but rather in terms of 
speeds and space gaps (i.e., the distance to the leading vehicle).

In a similar fashion, time headways, time gaps, and occupancy times represent 
multiples of the temporal discretisation $\Delta T$. But note that, as explained 
before in section \ref{sec:TCA:BasicVariablesAndConventions}, these are however 
no longer constrained to integer values.

In the following two sections, we explain how to convert between coordinate 
systems of TCA models and the real world. All common variables (e.g., 
$h_{s_{i}}$) are expressed in \emph{TCA units}, except for their `primed' 
counterparts (e.g., $h_{s_{i}}'$), which are expressed in \emph{real-world 
units}. The conversions will be done with respect to the following conventions:

\begin{itemize}
	\item \textbf{TCA model}
	\begin{itemize}
		\item $h_{s_{i}}$, $g_{s_{i}}$, and $l_{i}$ are dimensionless integers, 
		denoting a number of cells,
		\item $h_{t_{i}}$, $g_{t_{i}}$, and $\rho_{i}$ are dimensionless real 
		numbers, denoting a fractional multiple of a time step,
		\item $k_{l}$ and $k_{g}$ are real numbers, expressed in vehicles/cell,
		\item $q_{l}$ and $q_{g}$ are real numbers, expressed in vehicles/time step,
		\item and $v_{i}$, $\overline v_{s_{l}}$, and $\overline v_{s_{g}}$ are real 
		numbers, expressed in cells/time step.
	\end{itemize}

	\item \textbf{Real world}
	\begin{itemize}
		\item $\Delta X$, $h_{s_{i}}'$, $g_{s_{i}}'$, and $l_{i}'$ are real numbers, 
		expressed in metres,
		\item $\Delta T$, $h_{t_{i}}'$, $g_{t_{i}}'$, and $\rho_{i}'$ are real 
		numbers, expressed in seconds,
		\item $k_{l}'$ and $k_{g}'$ are real numbers, expressed in 
		vehicles/kilometre,
		\item $q_{l}'$ and $q_{g}'$ are real numbers, expressed in vehicles/hour,
		\item and $v_{i}'$, $\overline v_{s_{l}}'$, and $\overline v_{s_{g}}'$ are 
		real numbers, expressed in kilometres/hour.
	\end{itemize}
\end{itemize}

			\subsubsection{From a TCA model to the real world}
			\label{sec:TCA:TCAModelRealWorldConversion}

Under the previously mentioned convention that $l_{i} \in \mathbb{N}_{\zero}$, 
we can write the conversions of the microscopic characteristics related to the 
space and time headways and gaps, and the vehicle lengths and occupancy times, 
in a straightforward manner:

\begin{equation}
\label{eq:TCA:ConvertTCARealWorldMicro}
	\left \lbrace
		\begin{array}{lcl}
				h_{s_{i}}' = h_{s_{i}} \cdot \Delta X, \quad
					g_{s_{i}}' = g_{s_{i}} \cdot \Delta X, \quad
					l_{i}' = l_{i} \cdot \Delta X,\\
				\\
				h_{t_{i}}' = h_{t_{i}} \cdot \Delta T, \quad
					g_{t_{i}}' = g_{t_{i}} \cdot \Delta T, \quad
					\rho_{i}' = \rho_{i} \cdot \Delta T.
		\end{array}
	\right.
\end{equation}

Related to equations \eqref{eq:TCA:ConvertTCARealWorldMicro}, there is a small 
but important detail that is easily overlooked: we can not just convert between 
$g_{s_{i}}$, $g_{s_{i}}'$, $l_{i}$, and $l_{i}'$ without making some 
assumptions. Because we adopted the convention that $l_{i} \geq \one$ cell, it 
follows that $l_{i}' \geq \Delta X$. So it is not possible to take the real 
length of a vehicle smaller than the spatial discretisation, because we assumed 
that the spatial units of a TCA model are all integer values.

The conversions for the macroscopic traffic stream characteristics densities, 
flows, and space-mean speeds, as well as the microscopic vehicle speed, are as 
follows:

\begin{equation}
\label{eq:TCA:ConvertTCARealWorldMacro}
	\left \lbrace
		\begin{array}{lcl}
				k' & = &
					\displaystyle k \cdot \frac{\text{1000}}{\Delta X},\\
				\\
				q' & = &
					\displaystyle q \cdot \frac{\text{3600}}{\Delta T},\\
				\\
				\overline v_{s}' & = &
					\displaystyle \overline v_{s} \cdot \text{3.6} \cdot \frac{\Delta X}{\Delta T}.\\
		\end{array}
	\right.
\end{equation}

To keep the previous equations clear, we have dropped the subscripts denoting 
global and local measurements.\\

\sidebar{
	It is interesting to see what happens at the jam density, i.e., the maximum 
	density when all cells in the lattice are occupied. As all vehicles are 
	standing still bumper-to-bumper, the associated space gap at this density, 
	equals zero. Computing the space headway, results in $h_{s_{i}} = \zero + 
	l_{i}$. By virtue of the fact that density is inversely proportional to the 
	average space headway \cite{MAERIVOET:05d}, we can cast this space headway 
	into a density, e.g., for a single-cell TCA model: $k_{j} = {\overline 
	h_{s_{j}}}^{-\one} = {\overline l}^{-\one} = l_{i}^{-\one} = \one$. Applying 
	the conversion by means of equations \eqref{eq:TCA:ConvertTCARealWorldMacro} 
	and assuming a spatial discretisation $\Delta X = $~7.5~m, results in a 
	real-world jam density $k_{j}' = \text{1000} \div \text{7.5~m} \approx 
	\text{133}$~vehicles/kilometre. Conversely, if we know $k_{j}'$, then we can 
	derive $k_{j}$ (see section \ref{sec:TCA:RealWorldTCAModelConversion}) and 
	hence we have a method to pick a $\Delta X$.\\

	If we were to consider multi-cell traffic, e.g., vehicles with different 
	lengths, then the jam density would be inversely proportional to the average 
	vehicle length. A solution here is to assume a common unit for all vehicle 
	lengths, e.g., passenger car units (PCU) \cite{MAERIVOET:05d}. Even though the 
	jam density can be defined for each vehicle class separately, it would be more 
	correct to speak of an \emph{average jam density} at this point due to the 
	temporal and spatial variations in traffic flows.
}\\

			\subsubsection{From the real world to a TCA model}
			\label{sec:TCA:RealWorldTCAModelConversion}

Based on equations \eqref{eq:TCA:ConvertTCARealWorldMicro}, we can write the 
reverse conversion of the microscopic characteristics in the following manner:

\begin{equation}
\label{eq:TCA:ConvertRealWorldTCAMicro}
	\left \lbrace
		\begin{array}{lcl}
				h_{s_{i}} = \displaystyle \frac{h_{s_{i}}'}{\Delta X}, \quad
					g_{s_{i}} = \displaystyle \frac{g_{s_{i}}'}{\Delta X},\quad
					l_{i} = \displaystyle \frac{l_{i}'}{\Delta X},\\
				\\
				h_{t_{i}} = \displaystyle \frac{h_{t_{i}}'}{\Delta T}, \quad
					g_{t_{i}} = \displaystyle \frac{g_{t_{i}}'}{\Delta T}, \quad
					\rho_{i} = \displaystyle \frac{\rho_{i}'}{\Delta T}.
		\end{array}
	\right.
\end{equation}

In order to agree with our previously stated convention, i.e., all spatial 
microscopic characteristics in a TCA model are integers, equations 
\eqref{eq:TCA:ConvertRealWorldTCAMicro} implicitly assume that the real-world 
spatial variables are multiples of the spatial discretisation (e.g., $h_{s_{i}}' 
= m \cdot \Delta X$ with $m \in \mathbb{N}_{\zero}$).

Another possible approach to the spatial conversion to TCA model units, is to 
\emph{approximate} the real-world values as best as possible, whilst keeping our 
adopted convention. As $l_{i} \geq \one$ cell, this leads to the following 
scheme where we use upward rounding (i.e., ceiling):

\begin{equation}
\label{eq:TCA:ConvertRealWorldTCAMicroApproximation}
	\left \lbrace
		\begin{array}{lcl}
			h_{s_{i}} = \displaystyle \left \lceil \frac{h_{s_{i}}'}{\Delta X} \right \rceil, \quad
				l_{i} = \displaystyle \left \lceil \frac{l_{i}'}{\Delta X} \right \rceil,\\
			\\
			\Longrightarrow g_{s_{i}} = h_{s_{i}} - l_{i}.
		\end{array}
	\right.
\end{equation}

For example, if $\Delta X =$~7.5~m, $l_{i}' =$~4.5~m, and $g_{s_{i}}' = 
\five$~m, then $h_{s_{i}}' = \text{4.5} + \five = \text{9.5~m}$, and from 
equation \eqref{eq:TCA:ConvertRealWorldTCAMicroApproximation} it follows that 
$h_{s_{i}} = \two$ cells, $l_{i} = \one$ cell, and $g_{s_{i}} = \two - \one = 
\one$ cell. Because equation 
\eqref{eq:TCA:ConvertRealWorldTCAMicroApproximation} is only an approximation, 
it more than often occurs that the computed space headway `exceeds' the 
real-world space headway.

In similar spirit, the conversion for the macroscopic characteristics can be 
easily derived from equations \eqref{eq:TCA:ConvertTCARealWorldMacro}. However, 
as opposed to equations \eqref{eq:TCA:ConvertRealWorldTCAMicro} and 
\eqref{eq:TCA:ConvertRealWorldTCAMicroApproximation}, there is no need for an 
approximation by means of rounding, because these quantities are real numbers, 
as mentioned in the introduction of section 
\ref{sec:TCA:ConversionToRealWorldUnits}.

	\section{Single-cell models}
	\label{sec:TCA:SingleCellModels}

Having discussed the mathematical and physical aspects of cellular automata and 
TCA models in particular, we now focus on single-cell models. As explained 
before in section \ref{sec:TCA:RoadLayoutAndThePhysicalEnvironment}, each cell 
can either be empty, or is occupied by exactly one vehicle; all vehicles have 
the same length $l_{i} = \one$ cell. Traffic is also considered to be 
homogeneous, so all vehicles' characteristics are assumed to be the same. In the 
subsequent sections, we take a look at the following TCA models (accompanied by 
their suggested abbreviations):

\begin{itemize}
	\item \emph{Deterministic models}~
		\begin{itemize}
			\item Wolfram's rule 184 (CA-184)
			\item Deterministic Fukui-Ishibashi TCA (DFI-TCA)
		\end{itemize}

	\item \emph{Stochastic models}~
		\begin{itemize}
			\item Nagel-Schreckenberg TCA (STCA)
			\item STCA with cruise control (STCA-CC)
			\item Stochastic Fukui-Ishibashi TCA (SFI-TCA)
			\item Totally asymmetric simple exclusion process (TASEP)
			\item Emmerich-Rank TCA (ER-TCA)
		\end{itemize}

	\item \emph{Slow-to-start models}~
		\begin{itemize}
			\item Takayasu-Takayasu TCA (T$^{\two}$-TCA)
			\item Benjamin, Johnson, and Hui TCA (BJH-TCA)
			\item Velocity-dependent randomisation TCA (VDR-TCA)
			\item Time-oriented TCA (TOCA)
			\item TCA models incorporating anticipation
			\item Ultra discretisation, slow-to-accelerate, and driver's perspective
		\end{itemize}
\end{itemize}

For other excellent overviews of TCA models, we refer the reader to the works of 
Chowdhury et al. \cite{CHOWDHURY:00}, Knospe et al. \cite{KNOSPE:04}, Nagel 
\cite{NAGEL:96}, Nagel et al. \cite{NAGEL:03}, Schadschneider 
\cite{SCHADSCHNEIDER:00,SCHADSCHNEIDER:02}, and Schreckenberg et al. 
\cite{SCHRECKENBERG:01}.

All following TCA models will be empirically studied using simulations that are 
performed on a \emph{unidirectional, single-lane lattice} with periodic boundary 
conditions, i.e., a closed loop with ${L}_\mathcal{L} = \one$. The length of 
this lattice equals $K_\mathcal{L} = \ten^{\three}$ cells, which is taken large 
enough in order to reduce most unwanted \emph{finite-size effects}. Our own 
experiments indicate that larger lattice sizes do not render any significant 
advantage, aside from the burden of a larger computation time.\\

\sidebar{
	\textbf{The importance of studying closed-loop, single-lane traffic}\\

	There is often a criticism expressed as to why it is important to study the 
	behaviour of traffic flows in such a simplified system. After all, can such a 
	basic system capture all the dynamics of real-life traffic flows, or be even 
	representative of them ? The answer to this question is that, in our opinion, 
	the dynamics of these constrained systems play an important, non-negligible 
	role. For example, when considering traffic flows on most unidirectional 
	two-lane European motorways, drivers are by law obliged to drive on the right 
	shoulder lane, unless when performing overtaking manoeuvres. A frequently 
	observed phenomenon is then that under light traffic conditions (e.g., 10 
	vehicles/km/lane), a slower moving vehicle (e.g., a truck) is located on the 
	right lane, and is acting as a \emph{moving bottleneck}. As a result, all 
	faster vehicles will line up on the left lane (overtaking on the right lane is 
	prohibited by law), thereby causing a \emph{density} or \emph{lane inversion} 
	\cite{NAGEL:98,NEWELL:98,WOLF:99,KERNER:04}. It is under these circumstances 
	that the stability of the car-following behaviour plays an important role. 
	Similarly, in densely congested traffic, e.g., the synchronised-flow regime, 
	the same stability may govern the fact whether or not a traffic breakdown is 
	likely to be induced (see our work in \cite{MAERIVOET:05c} for a discussion on 
	the nature of this breakdown). Even for multi-lane traffic, we believe its 
	dynamics are essentially those of parallel single lanes when considering 
	densely congested traffic flows. Another argument for the necessity of 
	studying these simplified systems, is the one given by Nagel and Nelson. They 
	state that this is the easiest way to determine whether or not internal 
	effects of a traffic flow model play a role in e.g., the spontaneous breakdown 
	of traffic, as all external effects (i.e., the boundary conditions) are 
	eliminated \cite{NAGEL:05}. Nevertheless, when applying these models to 
	real-life traffic networks, closed-loop traffic is not very representative, as 
	the behaviour near bottlenecks plays a far more important role 
	\cite{HELBING:01}.
}\\

All measurements on the TCA models' lattices are based on two possible initial 
conditions: depending on the nature of the study, we will either use 
\emph{homogeneous initial conditions} (the default), or a \emph{compact 
superjam} to start with. In the former case, all vehicles are uniformly 
distributed over the lattice, implying equal space headways. In the latter case, 
all vehicles are `bunched up' behind each other, with zero space gaps. When 
going from one global density to another, an equivalent method would be to 
\emph{adiabatically} add (or remove) vehicles to an already homogeneous or 
jammed state. In our experiments however, we always reset the initial 
conditions, corresponding to the first method. The simulations ran each time for 
$\ten^{\four}$ time steps, after an initial period of $\ten^{\three}$ time steps 
was discarded in order to let transients from the initial conditions in the 
system die out. Global densities, flows, and space-mean speeds are computed by 
means of equations \eqref{eq:TCA:MeasurementsGlobalDensity} -- 
\eqref{eq:TCA:MeasurementsGlobalSMS}, whereas we use a point detector, i.e., 
equations \eqref{eq:TCA:MeasurementsLocalPointFlow} -- 
\eqref{eq:TCA:MeasurementsLocalPointDensity}, to compute their local variants. 
In this latter case, the data points were collected with a measurement period 
$T_{\text{mp}} =$~60 time steps. Based on these results, we can construct 
($k_{g}$,$\overline v_{s_{g}}$), ($k_{g}$,$q_{g}$), ($k_{l}$,$\overline 
v_{s_{l}}$), and ($k_{l}$,$q_{l}$) diagrams. To keep a clear formulation, we 
will however from now on drop the subscripts denoting global and local 
measurements. All simulations were performed by means of our \emph{Traffic 
Cellular Automata +} software (developed for the Java\trademark Virtual 
Machine); more information can be found in appendix 
\ref{sec:TCA+Software:Appendix}.

For a deeper insight into the behaviour of the space-mean speed $\overline 
v_{s}$, the average space gap $\overline g_{s}$, and the median time gap 
$\overline g_{t}$, detailed histograms showing their \emph{distributions} are 
provided. Note that with respect to the time gaps and time headways, we will 
work in the remainder of this report with the \emph{median} instead of the 
arithmetic mean. The median gives more robust results when $h_{t_{i}}, g_{t_{i}} 
\rightarrow +\infty$, which occurs when a vehicle $i$ stops. These histograms 
are interesting because in the existing literature (e.g., 
\cite{CHOWDHURY:98,SCHADSCHNEIDER:00,HELBING:01}) these distributions are only 
considered at several distinct global densities, whereas we show them for 
\emph{all} densities. Each of our histograms is constructed by varying the 
global density $k$ between 0 and 1, computing the space-mean speed, the average 
space gap and the median time gap for each simulation run. A simulation run 
consists of $\five \times \ten^{\four}$ time steps (with a transient period of 
500 time steps) on systems of 300 cells, varying the density in 150 steps. Note 
that a larger size of the system's lattice, has no significant effects on the 
results, except for an increase of the variance \cite{MAERIVOET:04e}.\\

\sidebar{
	Before giving an elaborate discussion of some of the classic TCA models, it is 
	worthwhile to mention the first historical and practical implementations of 
	traffic cellular automata. Cremer and Ludwig conceived an implementation of 
	traffic flows based on \emph{lattice gas automata} (LGA), which are a special 
	case of cellular automata typically employed when simulating viscous fluids 
	\cite{CREMER:86}. Their seminal work, using individual bits to represent 
	vehicles, was extended by Sch\"utt, who provided a simulation package for 
	heterogeneous traffic, multi-lane motorways, and network and city traffic 
	\cite{SCHUTT:91}. Unfortunately, the developed models were quite inefficient 
	when they were used in setting that called for large scale Monte Carlo 
	simulations \cite{NAGEL:95c}.
}\\

		\subsection{Deterministic models}

In this section, we discuss Wolfram's original rule 184, and its generalisation 
to higher speeds as proposed by Fukui and Ishibashi's deterministic model. We 
abbreviate these two TCA models as CA-184 and DFI-TCA, respectively.

			\subsubsection{Wolfram's rule 184 (CA-184)}
			\label{sec:TCA:CA184}

The first deterministic model we consider, is a one-dimensional TCA model with 
binary states. As $L_{\mathcal{L}} = \one$, this model is called an elementary 
cellular automaton (ECA), according to the terminology introduced in section 
\ref{sec:TCA:IngredientsOfACellularAutomaton}. If we furthermore assume a local 
neighbourhood of three cells wide (i.e., a radius of 1), then there are 
$\two^{\two^{\three}} =$~256 different rules possible, according to equation 
\eqref{eq:TCA:NumberOfPossibleRules}. Around 1983, Stephen Wolfram classified 
all these 256 binary ECAs \cite{WOLFRAM:83}. One of these is called \emph{rule 
184}, who's name is derived from Wolfram's naming scheme.

\begin{figure}[!htbp]
	\centering
	\psfrag{i-1}[][]{\psfragstyle{$i - 1$}}
	\psfrag{i}[][]{\psfragstyle{$i$}}
	\psfrag{i+1}[][]{\psfragstyle{$i + 1$}}
	\psfrag{t}[][]{\psfragstyle{$t$}}
	\psfrag{t+1}[][]{\psfragstyle{$t + 1$}}
	\includegraphics[scale=0.75]{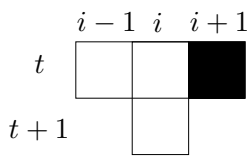}
	\caption{
		An illustrative method for representing the evolution of a cell's state in 
		time, based on its local neighbourhood. We can see the state $\sigma_{i}(t)$ 
		of a central cell $i$ at time step $t$, together with the states $\sigma_{i 
		- \one}(t)$ and $\sigma_{i + \one}(t)$ of its two direct neighbours $i - 
		\one$ and $i + \one$, respectively. This local neighbourhood is mapped onto 
		a new state $\sigma_{i}(t + \one)$. For binary states, we use a black square 
		to represent a state of 1 (e.g., state $\sigma_{i + \one}(t)$), and an empty 
		(white) square for a state of 0. The depicted transition maps the triplet 
		$(\zero\zero\one)_{\two}$ onto the state $\sigma_{i}(t + \one) = \zero$.
	}
	\label{fig:TCA:WolframsRuleNamingScheme}
\end{figure}

Wolfram's scheme is based on the representation of how a cell's state evolves in 
time, depending on its local neighbourhood. In 
\figref{fig:TCA:WolframsRuleNamingScheme}, we have provided a convenient 
visualisation for the evolution of the states in a binary ECA. Here, we can see 
the state $\sigma_{i}(t)$ of a central cell $i$ at time step $t$, together with 
the states $\sigma_{i - \one}(t)$ and $\sigma_{i + \one}(t)$ of its two direct 
neighbours $i - \one$ and $i + \one$, respectively. All three of them constitute 
the local neighbourhood $\mathcal{N}_{i}(t)$ of radius 1 (see also our example 
of a CA in section \ref{sec:TCA:AnExampleOfACA}). Because states are binary, we 
can indicate them with a colour, i.e., a black square represents a state of 1 
(e.g., state $\sigma_{i + \one}(t)$ in 
\figref{fig:TCA:WolframsRuleNamingScheme}), whereas an empty (white) square 
represents a state of 0. According to the local transition rule $\delta(i,t)$, 
the local neighbourhood $\mathcal{N}_{i}(t)$ is then mapped from $t$ to $t + 
\one$ onto a new state $\sigma_{i}(t + \one)$. The graphical representation in 
\figref{fig:TCA:WolframsRuleNamingScheme} thus provides us with an illustrative 
method to indicate the evolution of $\lbrace \sigma_{i - \one}(t), 
\sigma_{i}(t), \sigma_{i + \one}(t) \rbrace \longmapsto \sigma_{i}(t + \one)$.

\begin{figure}[!htbp]
	\centering
	\psfrag{111}[][]{\psfragstyle{1~~~1~~~1}}
	\psfrag{110}[][]{\psfragstyle{1~~~1~~~0}}
	\psfrag{101}[][]{\psfragstyle{1~~~0~~~1}}
	\psfrag{100}[][]{\psfragstyle{1~~~0~~~0}}
	\psfrag{011}[][]{\psfragstyle{0~~~1~~~1}}
	\psfrag{010}[][]{\psfragstyle{0~~~1~~~0}}
	\psfrag{001}[][]{\psfragstyle{0~~~0~~~1}}
	\psfrag{000}[][]{\psfragstyle{0~~~0~~~0}}
	\psfrag{1}[][]{\psfragstyle{1}}
	\psfrag{0}[][]{\psfragstyle{0}}
	\includegraphics[width=\figurewidth]{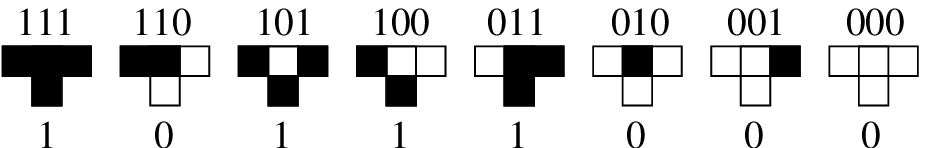}
	\caption{
		A graphical representation of Wolfram's rule 184, which is written as 
		$(\text{10111000)}_{\two}$ in base 2. All 8 possible configurations for the 
		local neighbourhood are sorted in descending order, expressing the local 
		transition rule $\delta(i,t)$ as explained by  
		\figref{fig:TCA:WolframsRuleNamingScheme}. For example, the local 
		neighbourhood $(\one\zero\zero)_{\two}$ gets mapped onto a state of 1. This 
		has the physical meaning that a particle (black square) moves to the right 
		if its neighbouring cell is empty.
	}
	\label{fig:TCA:CA184Rule}
\end{figure}

Considering the transition depicted in 
\figref{fig:TCA:WolframsRuleNamingScheme}, we can see that a complete 
neighbourhood contains three cells, each of which can be in a 0 (white) or 1 
(black) state. So in total, there are $\two^{\three} =$~8 possible 
configurations for such a local neighbourhood. Wolfram's naming scheme for the 
binary ECAs is now based on an integer coding of this neighbourhood. Indeed, the 
local transition rule $\delta(i,t)$ is given by a table lookup containing eight 
entries, one for each of the possible local neighbourhoods. If we binary sort 
these eight configurations in the descending order (111), (110), (101), (100), 
(011), \ldots, then we obtain a graphic scheme such as the one in 
\figref{fig:TCA:CA184Rule}. As can be seen, for each of the local 
configurations, a resulting 0 or 1 state is returned for cell $i$ at time step 
$t + \one$. Collecting all resulting states, and writing them in base 2, results 
in the number $(\text{10111000)}_{\two}$. Converting this code to base 10, we 
obtain the number 184. Wolfram now coded all 256 possible binary ECAs by a 
unique number in the range from 0 to 255, resulting in 256 rules for these CAs.

Rule 184 (which we abbreviate as CA-184) is an \emph{asymmetrical} rule because 
$\delta((\one\one\zero)_{\two},t) = \zero \neq \delta((\zero\one\one)_{\two},t) 
= \one$. It is also called a \emph{quiescent} rule because 
$\delta((\zero\zero\zero)_{\two},t) = \zero$ (so all zero-initial conditions 
remain zero). As an example of the rule's evolution, \figref{fig:TCA:CA184Rule} 
shows that the local neighbourhood $(\one\zero\zero)_{\two}$ gets mapped onto a 
state of 1. If we consider these 1 states as \emph{particles} (i.e., vehicles), 
and the 0 states as \emph{holes}, then rule 184 dictates that all particles move 
one cell to the right, on the condition that this right neighbour cell is empty. 
Equivalently, all holes have the tendency to move to the left for each particle 
that moves to the right, a phenomenon which is termed the \emph{particle-hole 
symmetry}.

For a TCA model, we can rewrite the previous actions as a set of rules that are 
consecutively applied to all vehicles in the lattice, as explained in section 
\ref{sec:TCA:VehicleMovementsAndTheRuleSet}. For the CA-184, we have the 
following two rules:

\begin{quote}
	\textbf{R1}: \emph{acceleration and braking}\\
		\begin{equation}
		\label{eq:TCA:CA184R1}
			v_{i}(t) \leftarrow \text{min} \lbrace g_{s_{i}}(t - \one),\one \rbrace,
		\end{equation}

	\textbf{R2}: \emph{vehicle movement}\\
		\begin{equation}
		\label{eq:TCA:CA184R2}
			x_{i}(t) \leftarrow x_{i}(t - \one) + v_{i}(t).
		\end{equation}
\end{quote}

Rule R1, equation \eqref{eq:TCA:CA184R1}, sets the speed of the \ith vehicle, 
for the current updated configuration of the system; it states that a vehicle 
always strives to drive at a speed of 1 cell/time step, unless its impeded by 
its direct leader, in which case $g_{s_{i}}(t - \one) = \zero$ and the vehicle 
consequently stops in order to avoid a collision. The second rule R2, equation 
\eqref{eq:TCA:CA184R2}, is not actually a `real' rule; it just allows the 
vehicles to advance in the system.

\begin{figure}[!htbp]
	\centering
	\includegraphics[width=\halffigurewidth]{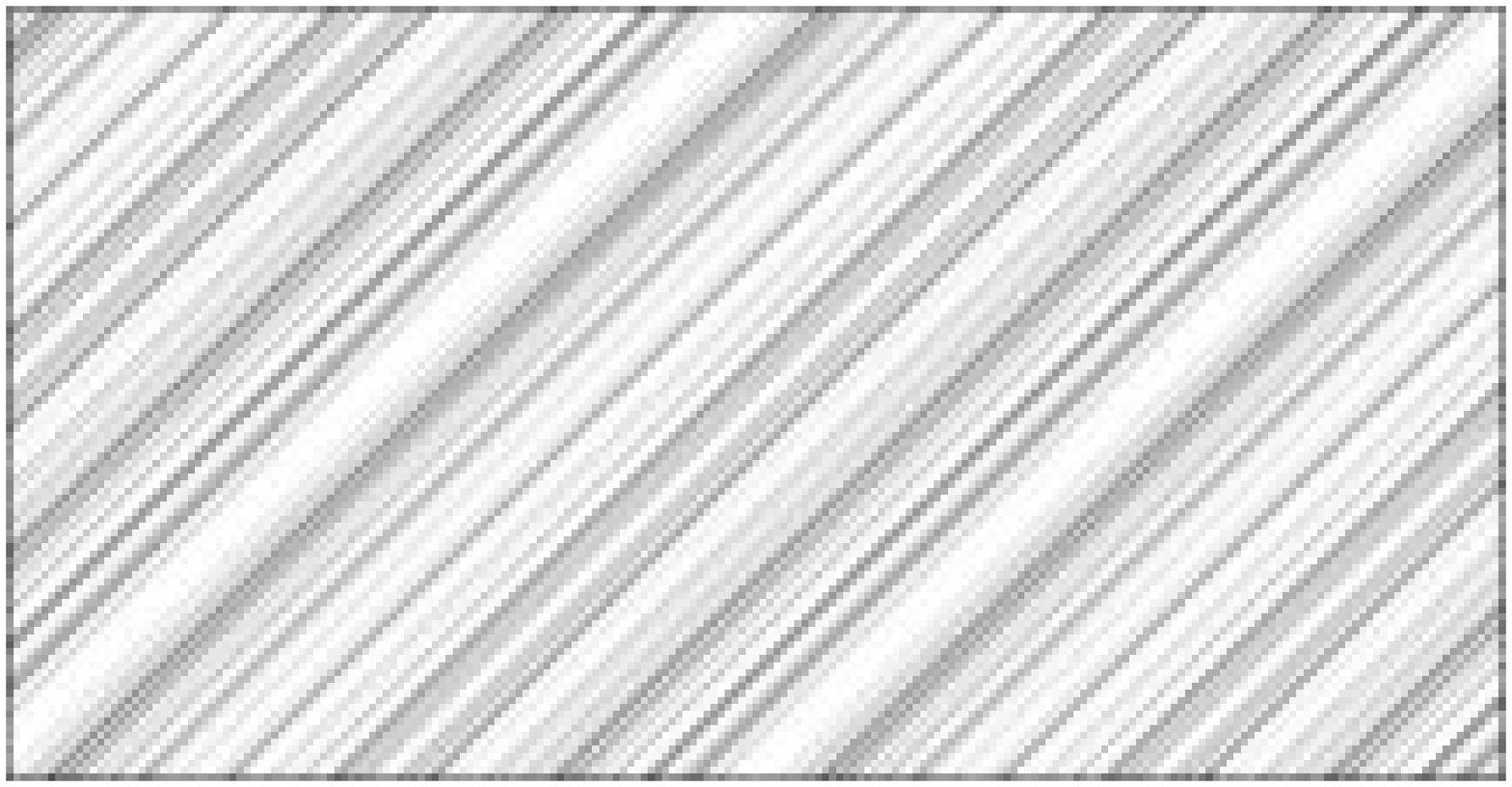}
	\hspace{\figureseparation}
	\includegraphics[width=\halffigurewidth]{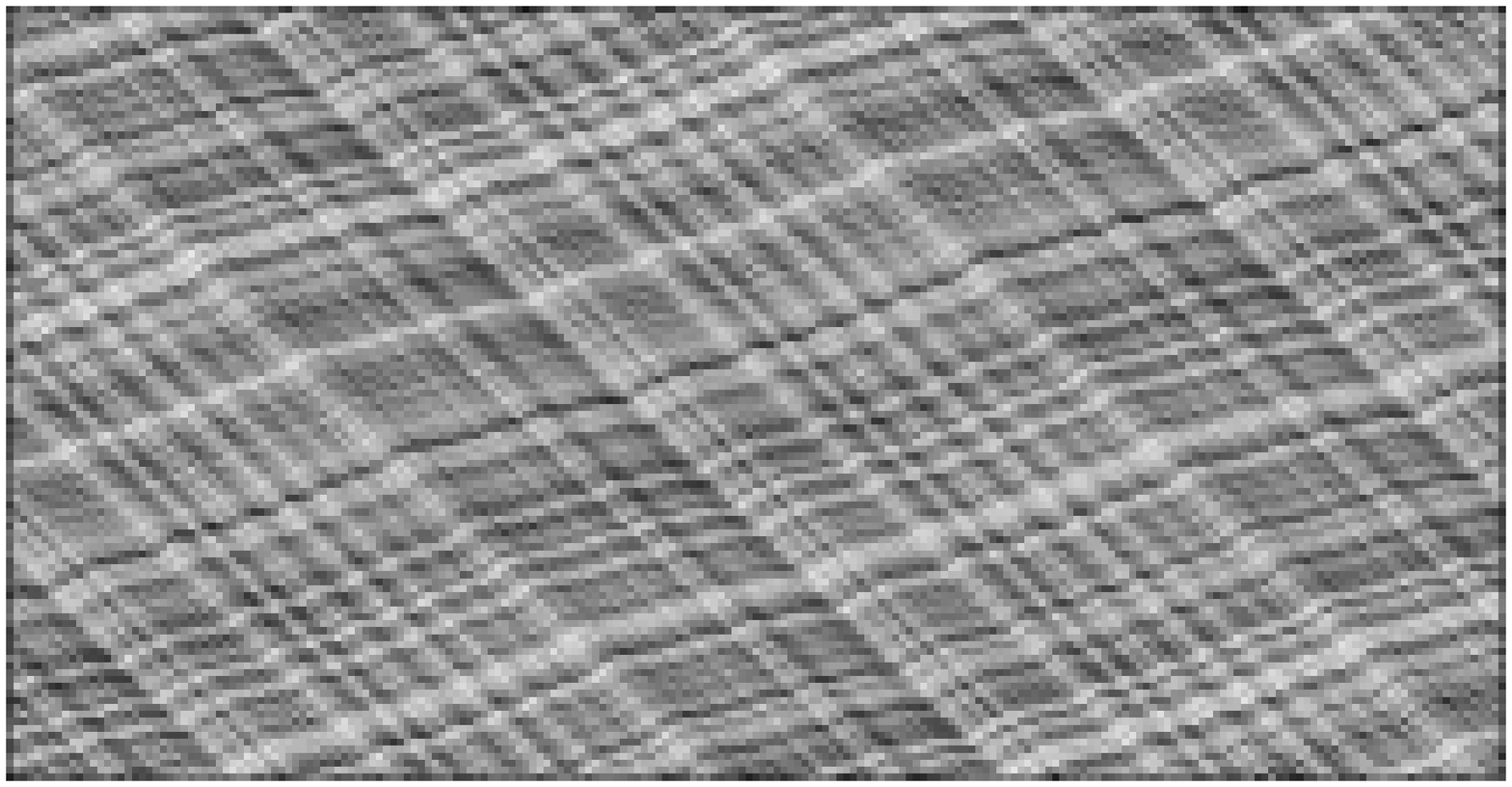}
	\caption{
		Typical time-space diagrams of the CA-184 TCA model. The shown closed-loop 
		lattices each contain 300 cells, with a visible period of 580 time steps 
		(each vehicle is represented as a single coloured dot). \emph{Left:} 
		vehicles driving a free-flow regime with a global density $k = $~0.2 
		vehicles/cell. \emph{Right:} vehicles driving in a congested regime with $k 
		= $~0.75 vehicles/cell. The congestion waves can be seen as propagating in 
		the opposite direction of traffic; they have an eternal life time in the 
		system. Both time-space diagrams show a fully deterministic system that 
		continuously repeats itself.
	}
	\label{fig:TCA:CA184TimeSpaceDiagrams}
\end{figure}

In \figref{fig:TCA:CA184TimeSpaceDiagrams}, we have applied these rules to a 
lattice consisting of 300 cells (closed loop), showing the evolution over a 
period of 580 time steps. The time and space axes are oriented from left to 
right, and bottom to top, respectively. In the left part, we show a free-flow 
regime with a global density $k = $~0.2 vehicles/cell, in the right part we have 
a congested regime with $k = $~0.75 vehicles/cell. Each vehicle is represented 
as a single coloured dot; as time advances, vehicles move to the upper right 
corner, whereas congestion waves move to the lower right corner, i.e., backwards 
in space. From both parts of \figref{fig:TCA:CA184TimeSpaceDiagrams}, we can see 
that the CA-184 TCA model constitutes a fully deterministic system that 
continuously repeats itself. A characteristic of the encountered congestion 
waves is that they have an eternal life time in the system.

\begin{figure}[!htbp]
	\centering
	\includegraphics[width=\halffigurewidth]{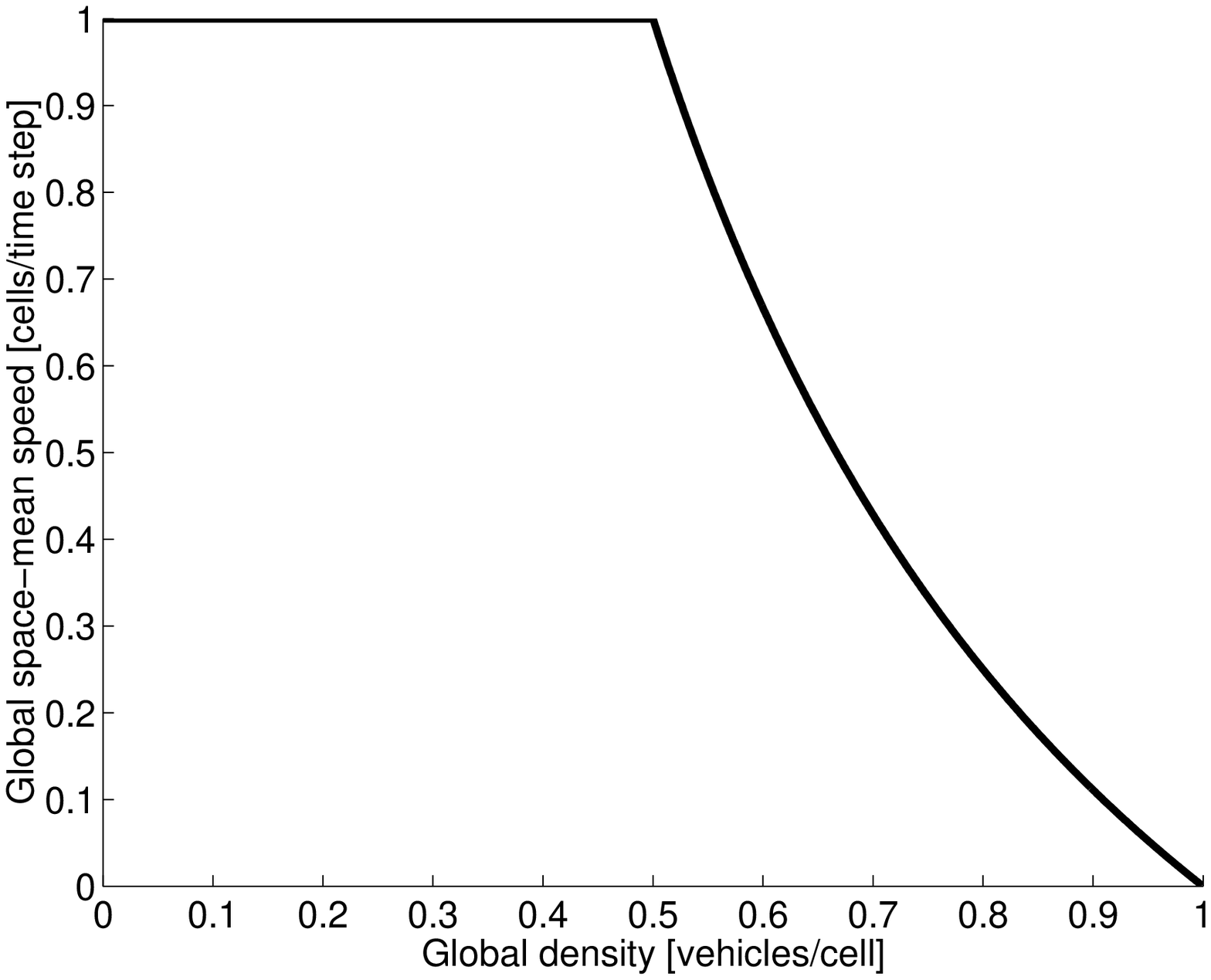}
	\hspace{\figureseparation}
	\includegraphics[width=\halffigurewidth]{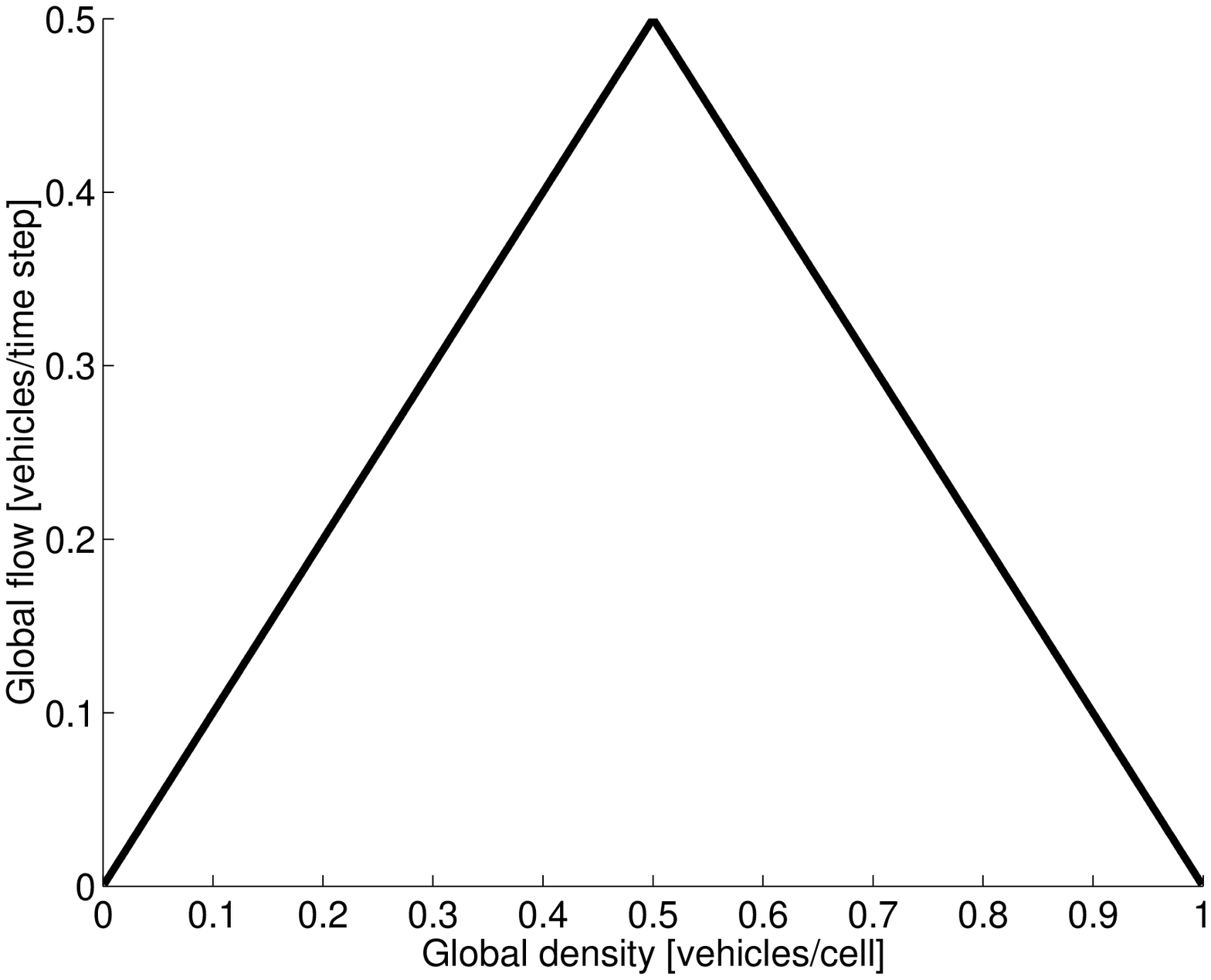}
	\caption{
		\emph{Left:} the ($k$,$\overline v_{s}$) diagram for the CA-184, based on 
		global measurements on the lattice. The global space-mean speed remains 
		constant at $\overline v_{s} = \one$ cell/time step, until the critical 
		density $k_{c} = $~0.5 is reached, at which point $\overline v_{s}$ will 
		start to diminish towards zero. \emph{Right:} the CA-184's ($k$,$q$) 
		diagram, with its characteristic isosceles triangular shape. The 
		transition between the free-flowing and the congested regime is of a 
		first-order nature.
	}
	\label{fig:TCA:CA184DensityFlowSMSFundamentalDiagrams}
\end{figure}

In \figref{fig:TCA:CA184DensityFlowSMSFundamentalDiagrams}, we have plotted both 
the ($k$,$\overline v_{s}$) and ($k$,$q$) diagrams. As can be seen from the left 
part, the global space-mean speed remains constant at $\overline v_{s} = \one$ 
cell/time step, until the critical density $k_{c} = $~0.5 is reached, at which 
point $\overline v_{s}$ will start to diminish towards zero where the critical 
density $k_{j} = \one$ is reached. Similarly, the global flow first increases 
and then decreases linearly with the density, below and respectively above, the 
critical density. Here, the capacity flow $q_{\text{cap}} =$~0.5 vehicles/time 
step is reached. The transition from the free-flowing to the congested regime is 
characterised by a first-order phase transition. As is evidenced by the 
\emph{isosceles triangular shape} of the CA-184's resulting ($k$,$q$) 
fundamental diagram, there are only two possible kinematic wave speeds, i.e., +1 
and -1 cell/time step. Both speeds are also clearly visible in the left, 
respectively right, time-space diagrams of 
\figref{fig:TCA:CA184TimeSpaceDiagrams}. More analytical details on these values 
will be provided in the following section \ref{sec:TCA:DeterministicFITCA}.

			\subsubsection{Deterministic Fukui-Ishibashi TCA (DFI-TCA)}
			\label{sec:TCA:DeterministicFITCA}

In 1996, Fukui and Ishibashi constructed a generalisation of the prototypical 
CA-184 TCA model \cite{FUKUI:96}. Although their model is essentially a 
stochastic one (see section \ref{sec:TCA:SFITCA}), we will first discuss its 
deterministic version. Fukui and Ishibashi's idea was two-fold: on the one hand, 
the maximum speed was increased from 1 to $v_{\text{max}}$ cells/time step, and 
on the other hand, vehicles would accelerate \emph{instantaneously} to the 
highest possible speed. Corresponding to the definitions of the rule set of a 
TCA model, the CA-184's rule R1, equation \eqref{eq:TCA:CA184R1}, changes as 
follows:

\begin{quote}
	\textbf{R1}: \emph{acceleration and braking}\\
		\begin{equation}
		\label{eq:TCA:DFITCAR1}
			v_{i}(t) \leftarrow \text{min} \lbrace g_{s_{i}}(t - \one),v_{\text{max}} \rbrace.
		\end{equation}
\end{quote}

Just as before, a vehicle will now avoid a collision by taking into account the 
size of its space gap. To this end, it will apply an instantaneous deceleration: 
for example, a fast-moving vehicle might have to come to a complete stop when 
nearing the end of a jam, thereby \emph{abruptly} dropping its speed from 
$v_{\text{max}}$ to 0 in one time step.

\begin{figure}[!htbp]
	\centering
	\includegraphics[width=\halffigurewidth]{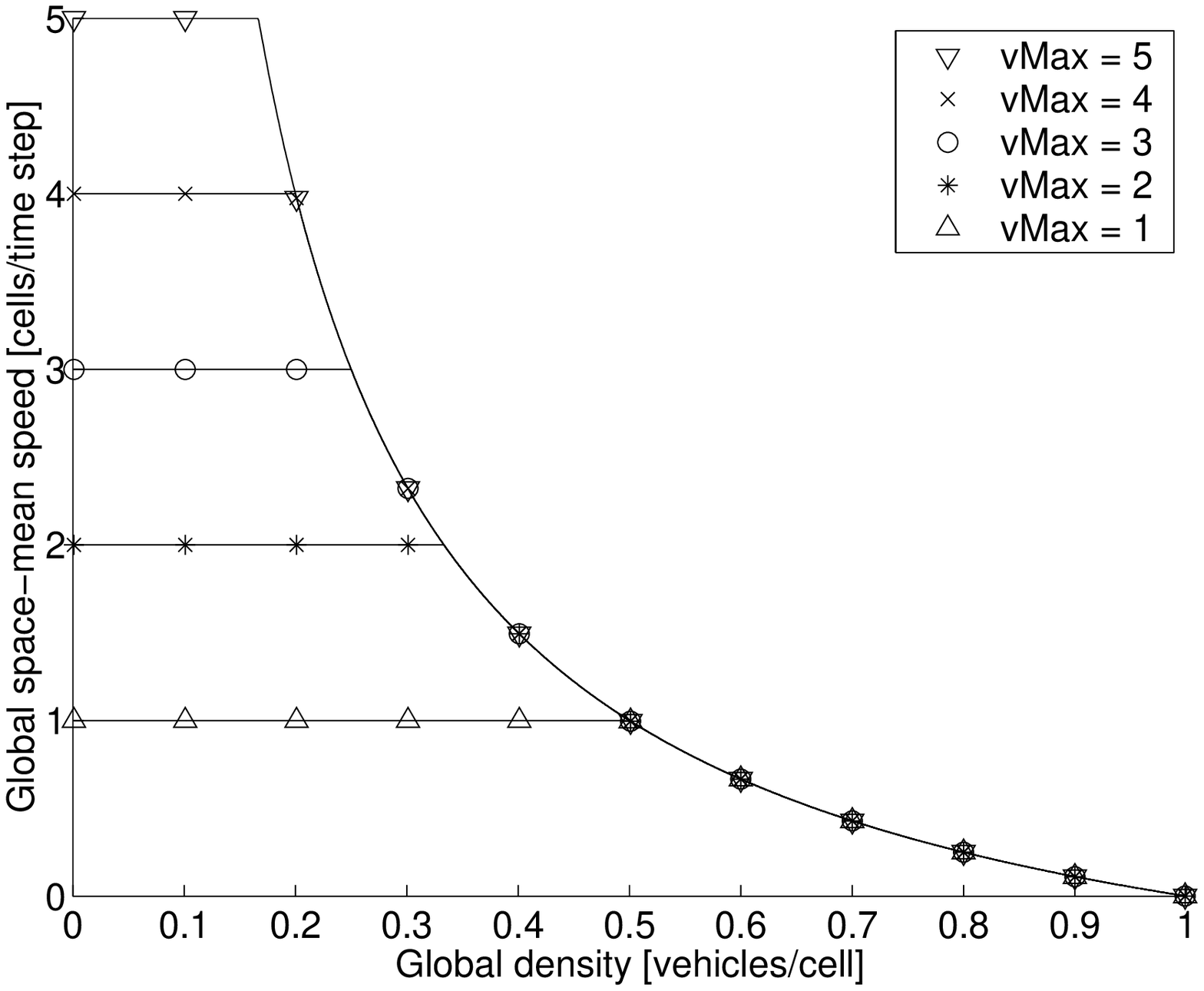}
	\hspace{\figureseparation}
	\includegraphics[width=\halffigurewidth]{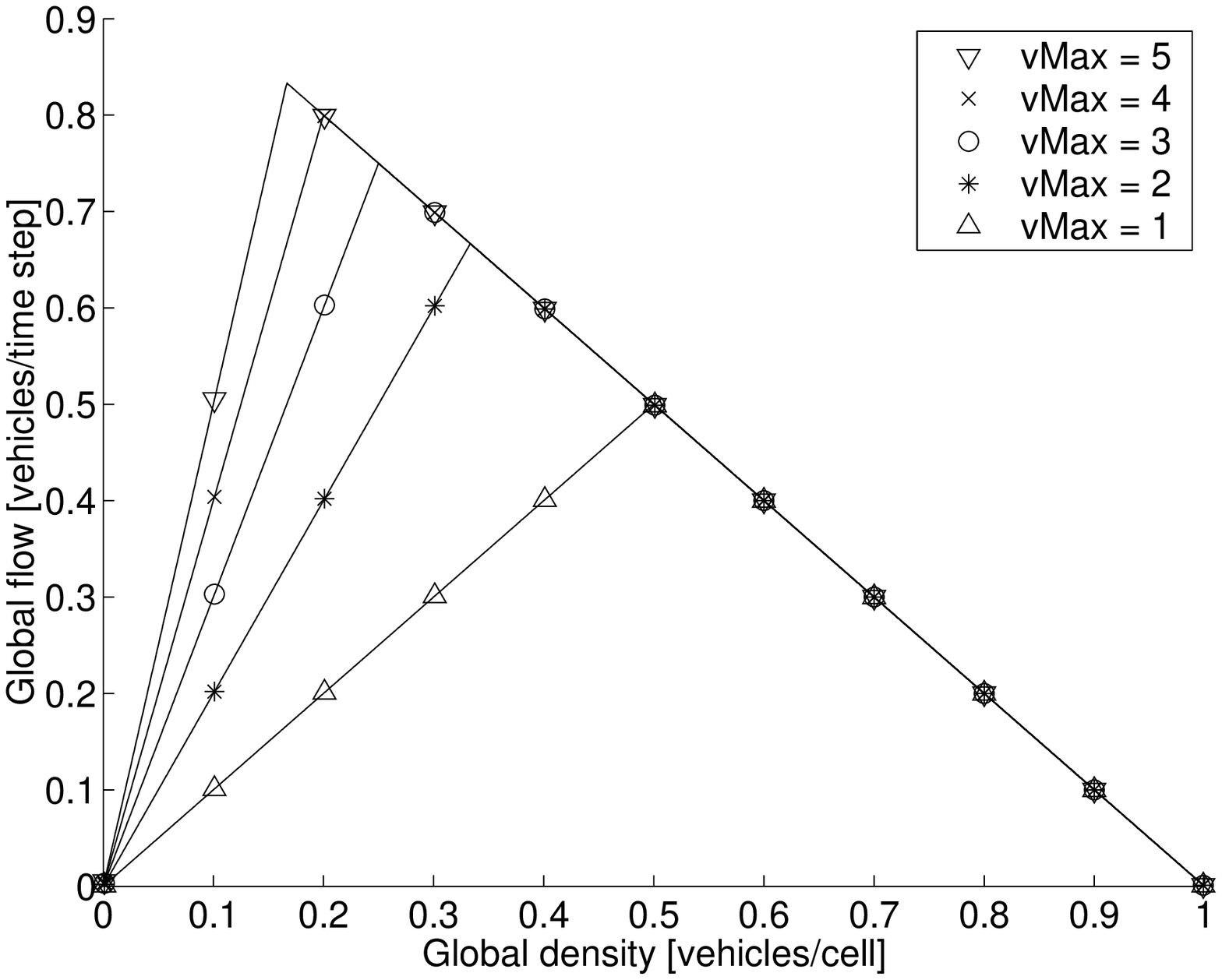}
	\caption{
		\emph{Left:} several ($k$,$\overline v_{s}$) diagrams for the deterministic 
		DFI-TCA, each for a different $v_{\text{max}} \in \lbrace \one, \ldots, 
		\five \rbrace$. Similarly to the CA-184, the global space-mean speed remains 
		constant, until the critical density is reached, at which point $\overline 
		v_{s}$ will start to diminish towards zero. \emph{Right:} several of the 
		DFI-TCA's ($k$,$q$) diagrams, each having a triangular shape (with the slope 
		of the congestion branch invariant for the different $v_{\text{max}}$).
	}
	\label{fig:TCA:DFITCADensityFlowSMSFundamentalDiagrams}
\end{figure}

Due to the strictly deterministic behaviour of the system, the time-space 
diagrams of the DFI-TCA do not differ much from those of the CA-184. The only 
difference is the speed of the vehicles in the free-flow regime, leading to 
steeper trajectories. It is however interesting to study the ($k$,$\overline 
v_{s}$) and ($k$,$q$) diagrams in 
\figref{fig:TCA:DFITCADensityFlowSMSFundamentalDiagrams}. Here we can see that 
increasing the maximum speed $v_{\text{max}}$ creates --- as expected --- a 
steeper free-flow branch in the ($k$,$q$) diagram. Interestingly, the slope of 
the congested branch does not change, logically implying that the kinematic wave 
speed for jams remains constant, i.e., -1 cell/time step. This can be confirmed 
with an analytical kinematic wave analysis, as explained by Nagel 
\cite{NAGEL:03}.

Based on the behaviour of the vehicles near the critical density, we can 
analytically compute the capacity flow as follows: in the free-flow regime, all 
vehicles move with a constant speed of $v_{\text{max}}$ cells/time step. When 
the critical density is reached, all vehicles drive collision-free at this 
maximum speed, which implies that $g_{s_{i}} = v_{\text{max}}$ cells. The space 
headway $h_{s_{i}} = v_{\text{max}} + \one$ (because $l_{i} = \one$ for 
single-cell models). Consequently, the value for the critical density as 
\cite{MAERIVOET:05d}:

\begin{equation}
\label{eq:TCA:DFITCACriticalDensity}
	k_{c} = \frac{\one}{\overline h_{s_{c}}} = \frac{\one}{v_{\text{max}} + \one}.
\end{equation}

The capacity flow is now computed by means of the fundamental relation, i.e., 
$q_{\text{cap}} = k_{c}~v_{\text{max}}$:

\begin{equation}
\label{eq:TCA:DFITCACapacityFlow}
	q_{\text{cap}} = \frac{v_{\text{max}}}{v_{\text{max}} + \one}.
\end{equation}

Applying equations \eqref{eq:TCA:DFITCACriticalDensity} and 
\eqref{eq:TCA:DFITCACapacityFlow}, for e.g., $v_{\text{max}} = \five$ cells/time 
step, results in $k_{c} \approx$~0.167 vehicles/cell and $q_{\text{cap}} 
\approx$~0.83 vehicles/time step. If we furthermore assume $\Delta X = $~7.5~m 
and $\Delta T = $~1~s, then both values correspond to 22 vehicles/kilometre and 
3000 vehicles/hour, respectively.

As opposed to the instantaneous acceleration in rule R1, equation 
\eqref{eq:TCA:DFITCAR1}, we can also assume a \emph{gradual acceleration} of one 
cell per time step (the braking remains instantaneous):

\begin{quote}
	\textbf{R1}: \emph{acceleration and braking}\\
		\begin{equation}
			v_{i}(t) \leftarrow \text{min} \lbrace v_{i}(t - \one) + \one,g_{s_{i}}(t - \one),v_{\text{max}} \rbrace.
		\end{equation}
\end{quote}

However, our experimental observations have indicated that there is no 
difference in global system dynamics, with respect to either adopting gradual or 
instantaneous vehicle accelerations.\\

\sidebar{
	There exists a strong relation between the previously discussed deterministic 
	TCA models, and the macroscopic first-order LWR model with a triangular 
	$q_{e}(k)$ fundamental diagram \cite{MAERIVOET:05c}. Some of the finer results 
	in this case, are the work of Nagel who extensively discusses some analytical 
	results of both deterministic and stochastic TCA models \cite{NAGEL:96}, and 
	the work of Daganzo who explicitly proves an equivalency between two TCA 
	models and the kinematic wave model with a triangular $q_{e}(k)$ fundamental 
	diagram \cite{DAGANZO:04b}. More details with respect to such analytical 
	relations, are given in sections \ref{sec:TCA:TASEP} and 
	\ref{sec:TCA:AnalyticalTCA}.
}\\

\begin{figure}[!htbp]
	\centering
	\includegraphics[width=\halffigurewidth]{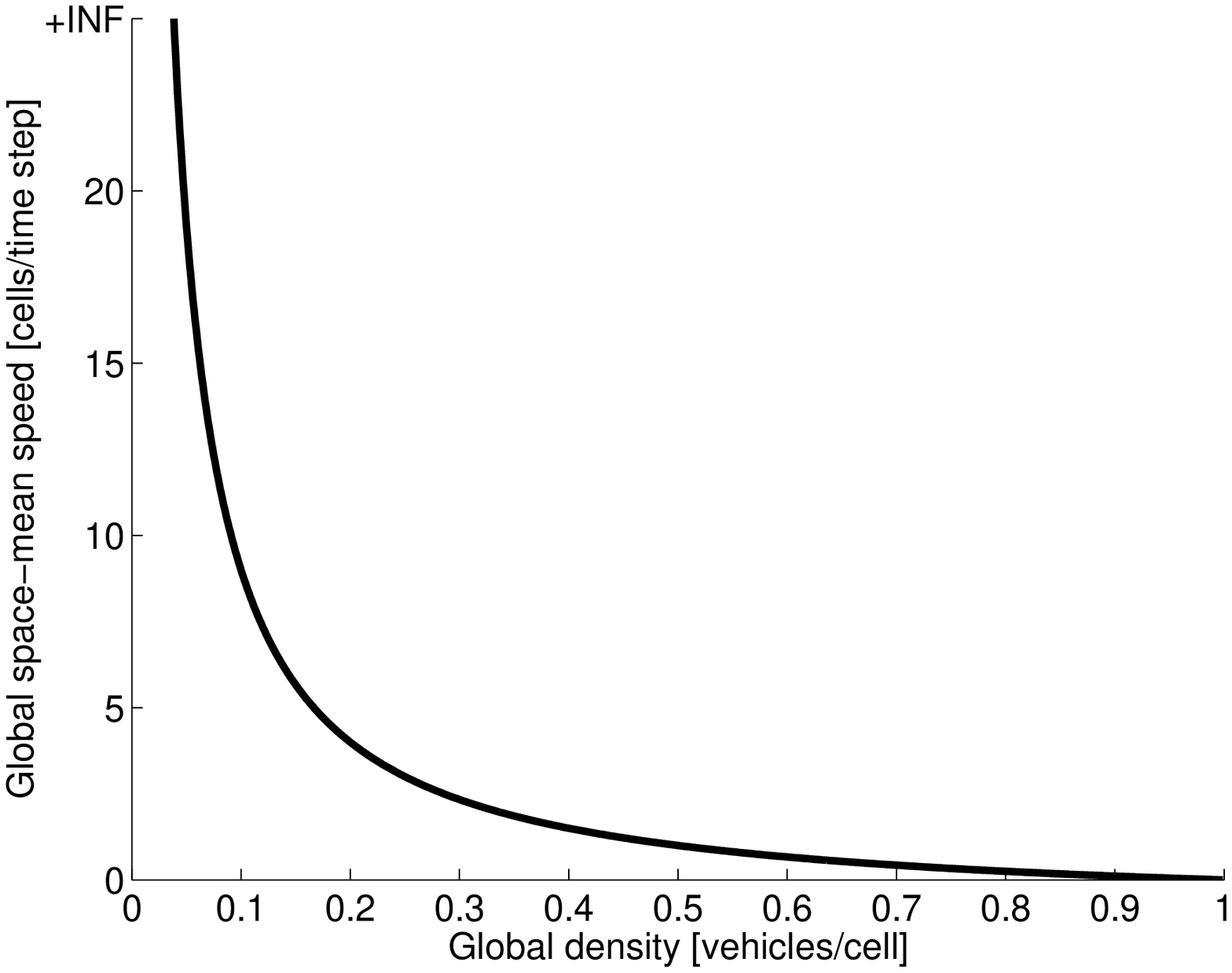}
	\hspace{\figureseparation}
	\includegraphics[width=\halffigurewidth]{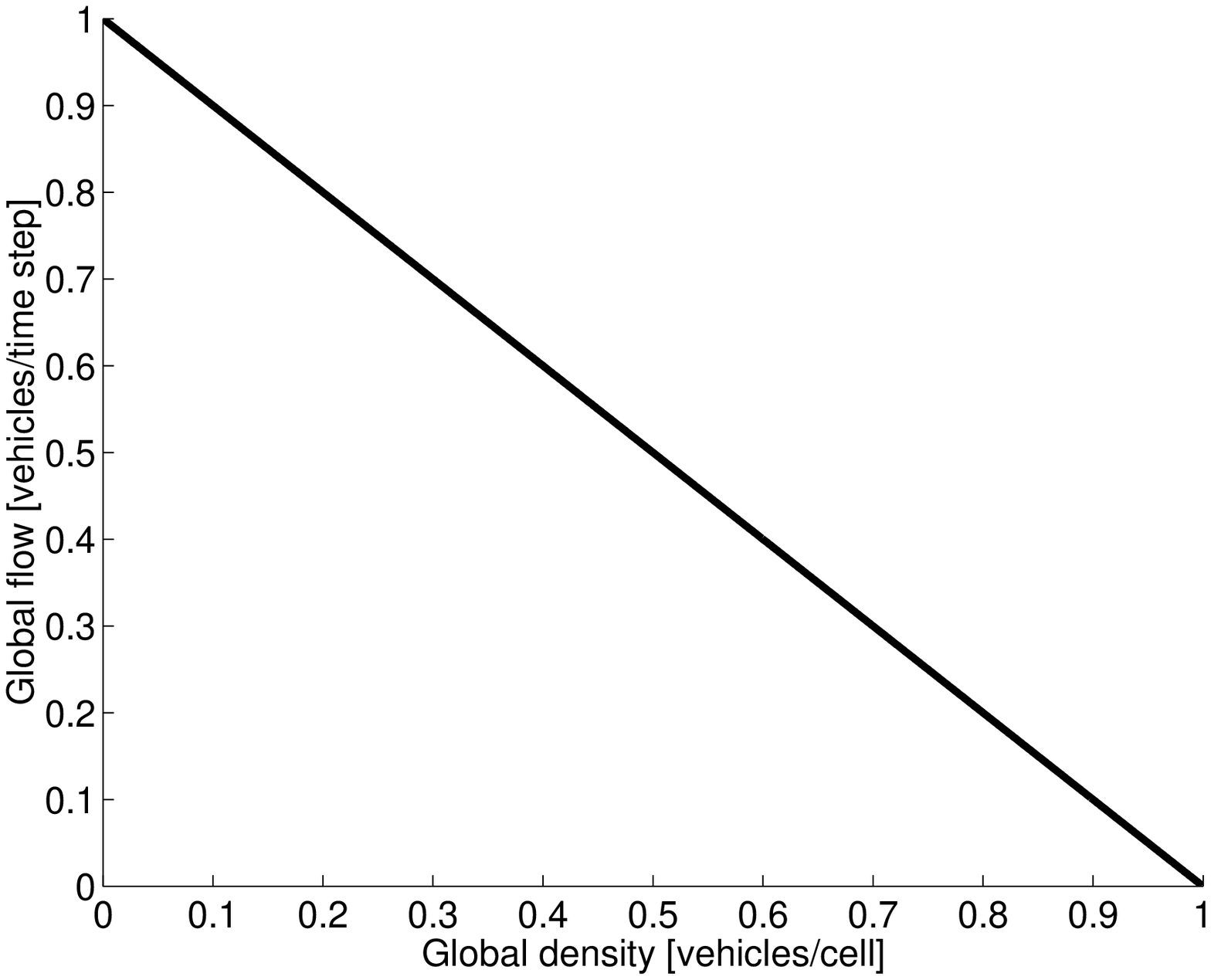}
	\caption{
		\emph{Left:} the ($k$,$\overline v_{s}$) diagram for the deterministic 
		CA-184, with now $v_{\text{max}} \rightarrow +\infty$. \emph{Right:} the 
		($k$,$q$) diagram for the same TCA model, resulting in a critical density 
		$k_{c} = \zero$, with a capacity flow $q_{\text{cap}} = \one$. This type of 
		diagram corresponds to a simplified triangular $q_{e}(k)$ fundamental 
		diagram that is expressed in a moving coordinate system.
	}
	\label{fig:TCA:CA184VMAXINFDensityFlowSMSFundamentalDiagrams}
\end{figure}

To conclude our discussion of deterministic models, we take a look at what 
happens in the limiting case where $v_{\text{max}} \rightarrow +\infty$. As can 
be seen in \figref{fig:TCA:CA184VMAXINFDensityFlowSMSFundamentalDiagrams}, the 
congested branches in both ($k$,$\overline v_{s}$) and ($k$,$q$) diagrams grow, 
at the cost of the free-flow branches which disappear. Interestingly, these 
diagrams correspond one-to-one with a triangular $q_{e}(k)$ fundamental diagram 
that is now expressed in a \emph{moving coordinate system}, as explained by 
Newell \cite{NEWELL:99}. In such a simplified system, the critical density 
$k_{c} = \zero$, with a capacity flow $q_{\text{cap}} = \one$.

		\subsection{Stochastic models}

The encountered models in the previous section were all deterministic in nature, 
implying that there can be no spontaneous formation of jam structures. All 
congested conditions produced in those models, essentially stemmed from the 
assumed initial conditions. In contrast to this, we now discuss stochastic TCA 
models (i.e., these are probabilistic CAs) that allow for the spontaneous 
emergence of phantom jams. As will be shown, all these models explicitly 
incorporate a stochastic term in their equations, in order to accomplish this 
kind of real-life behaviour \cite{NAGEL:93}.

			\subsubsection{Nagel-Schreckenberg TCA (STCA)}
			\label{sec:TCA:STCA}

In 1992, Nagel and Schreckenberg proposed a TCA model that was able to reproduce 
several characteristics of real-life traffic flows, e.g., the spontaneous 
emergence of traffic jams \cite{NAGEL:92,NAGEL:95c}. Their model is called the 
\emph{NaSch TCA}, but is more commonly known as the \emph{stochastic traffic 
cellular automaton} (STCA). It explicitly includes a stochastic noise term in 
one of its rules, which we present in the same fashion as those of the 
previously discussed deterministic TCA models. The STCA then comprises the 
following three rules (note that in Nagel and Schreckenberg's original 
formulation, they decoupled acceleration and braking, resulting in four rules):

\begin{quote}
	\textbf{R1}: \emph{acceleration and braking}\\
		\begin{equation}
		\label{eq:TCA:STCAR1}
			v_{i}(t) \leftarrow \text{min} \lbrace v_{i}(t - \one) + \one,g_{s_{i}}(t - \one),v_{\text{max}} \rbrace,
		\end{equation}

	\textbf{R2}: \emph{randomisation}\\
		\begin{equation}
		\label{eq:TCA:STCAR2}
			\xi(t) < p \Longrightarrow v_{i}(t) \leftarrow \max \lbrace \zero,v_{i}(t) - \one \rbrace,
		\end{equation}

	\textbf{R3}: \emph{vehicle movement}\\
		\begin{equation}
		\label{eq:TCA:STCAR3}
			x_{i}(t) \leftarrow x_{i}(t - \one) + v_{i}(t).
		\end{equation}
\end{quote}

Like in both CA-184 and DFI-TCA deterministic TCA models (see sections 
\ref{sec:TCA:CA184} and \ref{sec:TCA:DeterministicFITCA}), the STCA contains a 
rule for increasing the speed of a vehicle and braking to avoid collisions, 
i.e., rule R1, equation \eqref{eq:TCA:STCAR1}, as well as rule R3, equation 
\eqref{eq:TCA:STCAR3}, for the actual vehicle movement. However, the STCA also 
contains an additional rule R2, equation \eqref{eq:TCA:STCAR2}, which introduces 
stochasticity in the system. At each time step $t$, a random number $\xi(t) \in 
[\zero,\one[$ is drawn from a uniform distribution. This number is then compared 
with a stochastic noise parameter $p \in [\zero,\one]$ (called the 
\emph{slowdown probability}); as a result, there is a probability of $p$ that a 
vehicle will slow down to $v_{i}(t) - \one$ cells/time step. The STCA model is 
called a \emph{minimal model}, in the sense that all these rules are a necessity 
for mimicking the basic features of real-life traffic flows.\\

\sidebar{
	According to Nagel and Schreckenberg, the randomisation of rule R2 captures 
	natural speed fluctuations due to human behaviour or varying external 
	conditions. The rule introduces overreactions of drivers when braking, 
	providing the key to the formation of spontaneously emerging jams.\\

	Although the above rationale is widely agreed upon, much criticism was however 
	expressed due to this second rule. For example, Brilon and Wu believe that 
	this rule has no theoretical background and is in fact introduced quite 
	heuristically \cite{BRILON:99b}.
}\\

\begin{figure}[!htbp]
	\centering
	\includegraphics[width=\halffigurewidth]{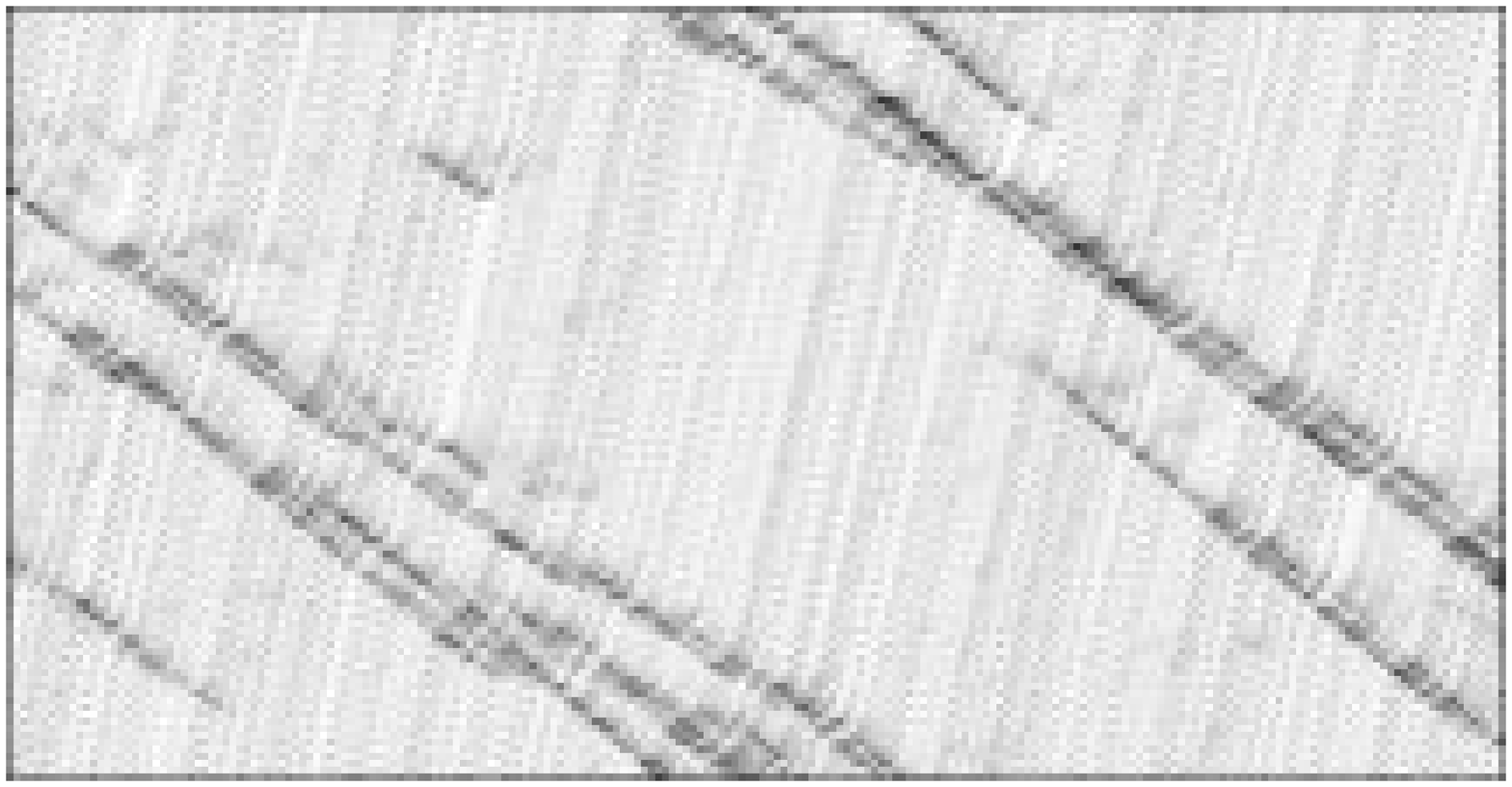}
	\hspace{\figureseparation}
	\includegraphics[width=\halffigurewidth]{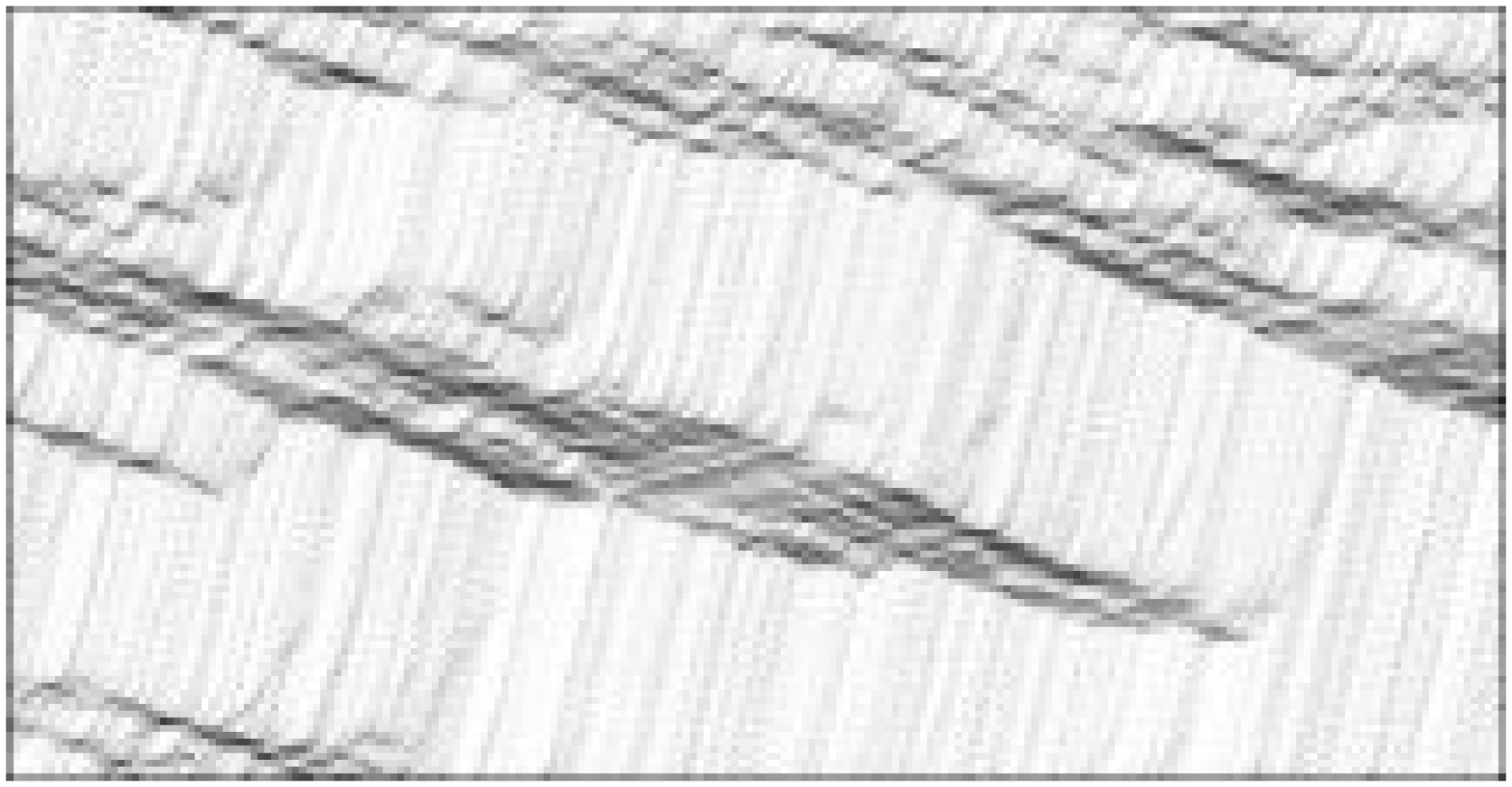}
	\caption{
		Typical time-space diagrams of the STCA model (similar setup as for the 
		CA-184 TCA model in \figref{fig:TCA:CA184TimeSpaceDiagrams}). Both diagrams 
		have a global density of $k =$~0.2 vehicles/cell. \emph{Left:} the evolution 
		of the system for $p =$~0.1. \emph{Right:} the evolution of the system, but 
		now for $p = $~0.5. The effects of the randomisation rule R2 are clearly 
		visible in both diagrams, as there occur many unstable artificial phantom 
		mini-jams. Furthermore, the speed $w$ of the backward propagating kinematic 
		waves decreases with an increasing $p$.
	}
	\label{fig:TCA:STCATimeSpaceDiagrams}
\end{figure}

To get an intuitive feeling for the STCA's system dynamics, we have provided two 
time-space diagrams in \figref{fig:TCA:STCATimeSpaceDiagrams}. Both diagrams 
show the evolution for a global density of $k =$~0.2 vehicles/cell, but with $p$ 
set to 0.1 for the left diagram, and $p = $~0.5 for the right diagram. As can be 
seen in both diagrams, the randomisation in the model gives rise to many 
unstable artificial phantom mini-jams. The downstream fronts of these jams smear 
out, forming \emph{unstable interfaces} \cite{NAGEL:03}. This is a direct result 
of the fact that the intrinsic noise (as embodied by $p$) in the STCA model is 
too strong: a jam can always form at \emph{any} density, meaning that breakdown 
can (and will) occur, even in the free-flow traffic regime. For low enough 
densities however, these jams can vanish as they are absorbed by vehicles with 
sufficient space headways, or by new jams in the system \cite{KRAUSS:99}. It has 
been experimentally shown that below the critical density, these jams have 
finite life times with a cut-off that is about $\five \times \ten^{\five}$ time 
steps and independent of the lattice size. When the critical density is crossed, 
these long-lived jams evolve into jams with an infinite life time, i.e., they 
will survive for an infinitely long time 
\cite{NAGEL:94b,NAGEL:95c,SCHADSCHNEIDER:99b}.

\begin{figure}[!htbp]
	\centering
	\includegraphics[width=\halffigurewidth]{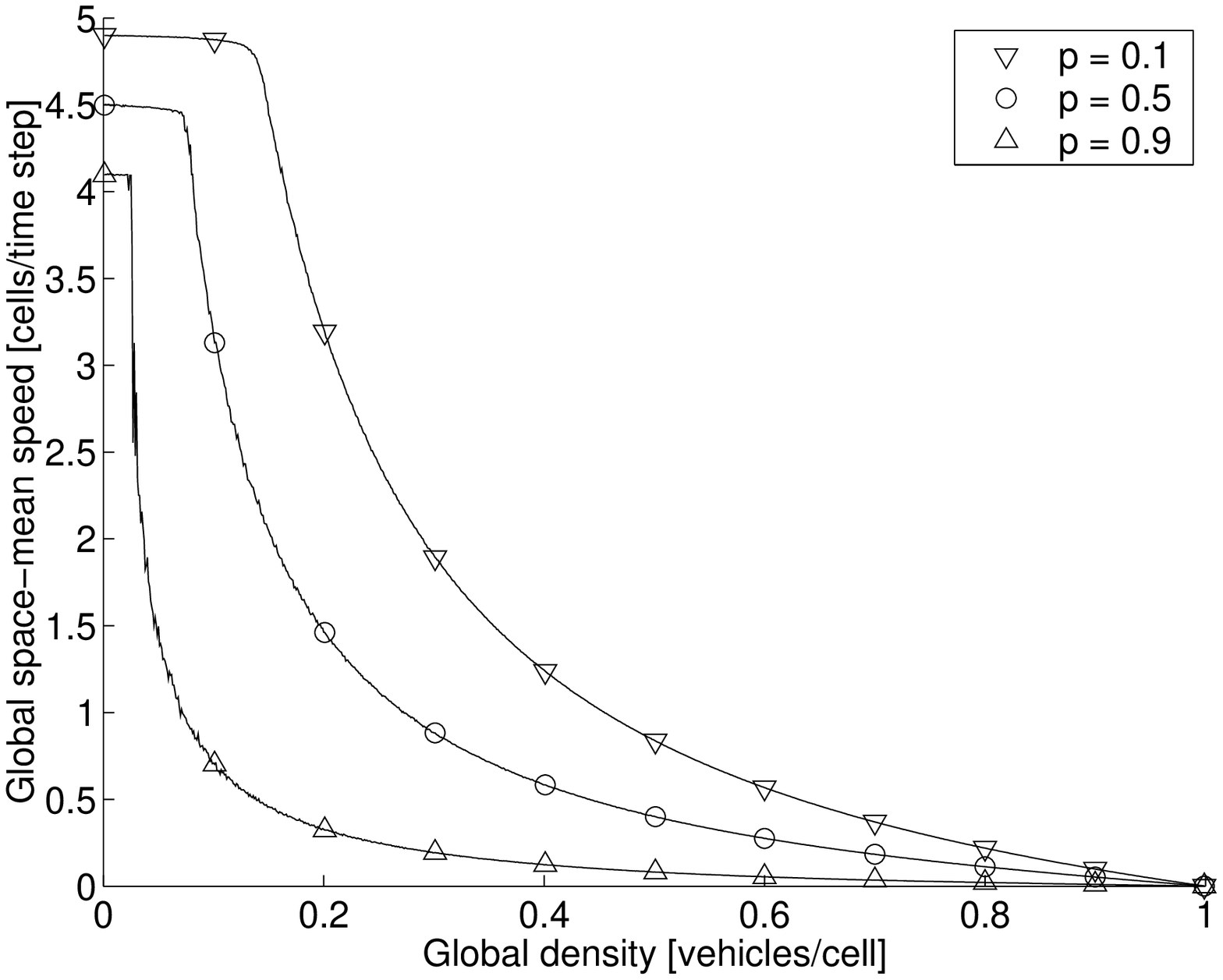}
	\hspace{\figureseparation}
	\includegraphics[width=\halffigurewidth]{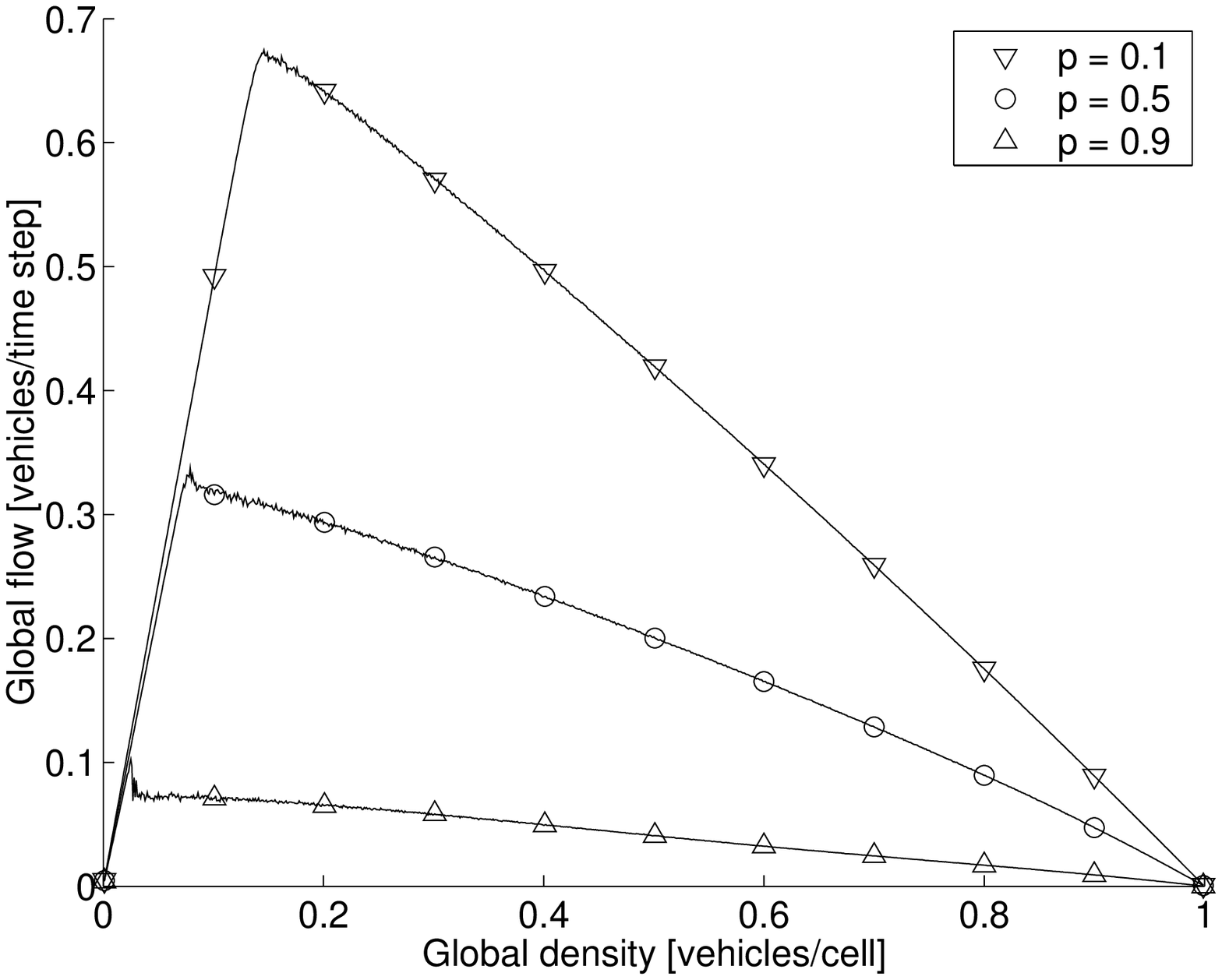}
	\caption{
		\emph{Left:} several ($k$,$\overline v_{s}$) diagrams for the STCA, each for 
		a different $p \in \lbrace \text{0.1}, \text{0.5}, \text{0.9} \rbrace$. It 
		is clear from the diagram, that a vehicle will drive with an average 
		free-flow speed of $\overline v_{\text{ff}} = v_{\text{max}} - p$. 
		\emph{Right:} several ($k$,$q$) diagrams for the same STCA models as before. 
		The slope of the congested branch tends toward zero for an increasing 
		slowdown probability $p$. Note that the seemingly small capacity drops at 
		the critical density in the right part, are actually finite-size effects 
		\cite{NAGEL:95b,KRAUSS:97b}.
	}
	\label{fig:TCA:STCADensityFlowSMSFundamentalDiagrams}
\end{figure}

In free-flow traffic, a vehicle's speed will fluctuate between $v_{\text{max}}$ 
and $v_{\text{max}} - \one$, due to the randomisation rule R2. We can compute 
the space-mean speed in the free-flow regime by means of a weighted average. 
This average corresponds to the probability $\one - p$ for driving with the 
speed $v_{\text{max}}$ and the probability $p$ for slowing down to the speed 
$v_{\text{max}} - \one$. As such, we get $\overline v_{s_{\text{ff}}} = \sum 
w_{i} v_{i} / \sum w_{i} = [(\one - p) v_{\text{max}} + p (v_{\text{max}} - 
\one)] / [(\one - p) + p] = v_{\text{max}} - p$. In agreement with the 
space-mean speed observed in the left ($k$,$\overline v_{s}$) diagram of 
\figref{fig:TCA:STCADensityFlowSMSFundamentalDiagrams}, we can state that a 
vehicle will drive with an average free-flow speed of $\overline v_{\text{ff}} = 
v_{\text{max}} - p$.

As mentioned in section \ref{sec:TCA:DeterministicFITCA}, the slope of the 
free-flow branch in a ($k$,$q$) diagram can be changed by adjusting 
$v_{\text{max}}$. Similarly, the slope of the congested branch can be changed by 
tuning the slowdown probability $p$ (note that this also affects the average 
free-flow speed). Looking at the ($k$,$q$) diagram in the right part of 
\figref{fig:TCA:STCADensityFlowSMSFundamentalDiagrams}, we note that an increase 
of $p$ will on the one hand result in a smaller $\overline v_{\text{ff}}$, and 
on the other hand the congested branch will lie lower, with a smaller critical 
density $k_{c}$. In this latter case, the speed $w$ of the backward propagating 
kinematic waves will decrease, an effect that is also visible in the time-space 
diagrams of \figref{fig:TCA:STCATimeSpaceDiagrams}. Note that the presence of 
noise in the STCA model causes both free-flow and congested branches of the 
($k$,$q$) diagram to be slightly curved, as opposed to the perfectly linear 
branches of the deterministic models. 

If we set $p = \zero$, then the STCA model becomes deterministic; additionally, 
setting $v_{\text{max}}$ will recover the CA-184 TCA model. In the other 
deterministic case, when $p = \one$, the system behaves differently: in the 
congested state, all vehicles will come to a full stop, thereby reducing the 
global flow in the system to zero. As a result, the congested branch in the 
($k$,$q$) regime will coincide with the horizontal axis. This implies that the 
behaviour of a system with $v_{\text{max}}$ and $p = \one$ is totally different 
than that of a system with $v_{\text{max}} - \one$ and $p = \zero$.

\begin{figure}[!htbp]
	\centering
	\includegraphics[width=0.32\figurewidth]{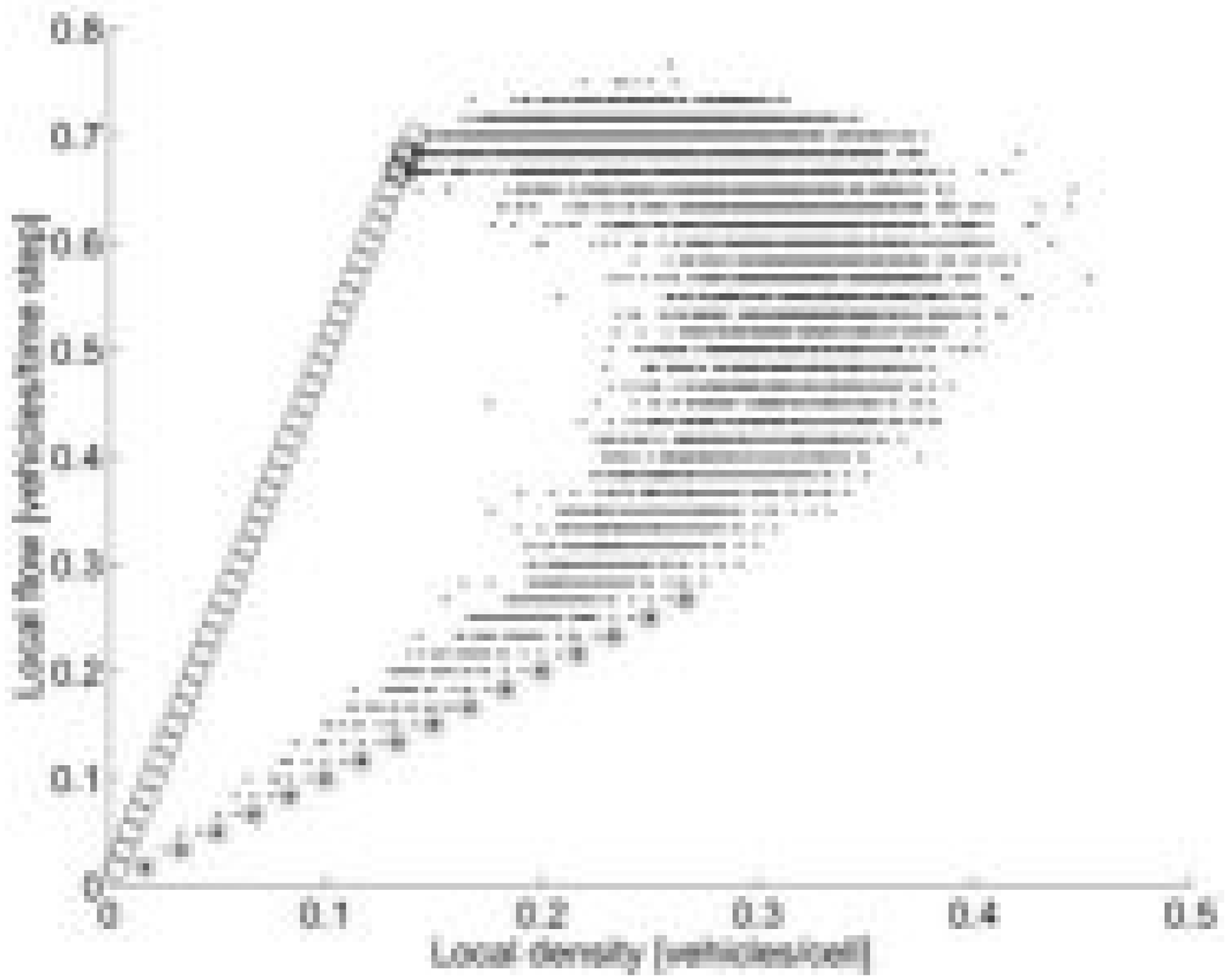}
	\includegraphics[width=0.32\figurewidth]{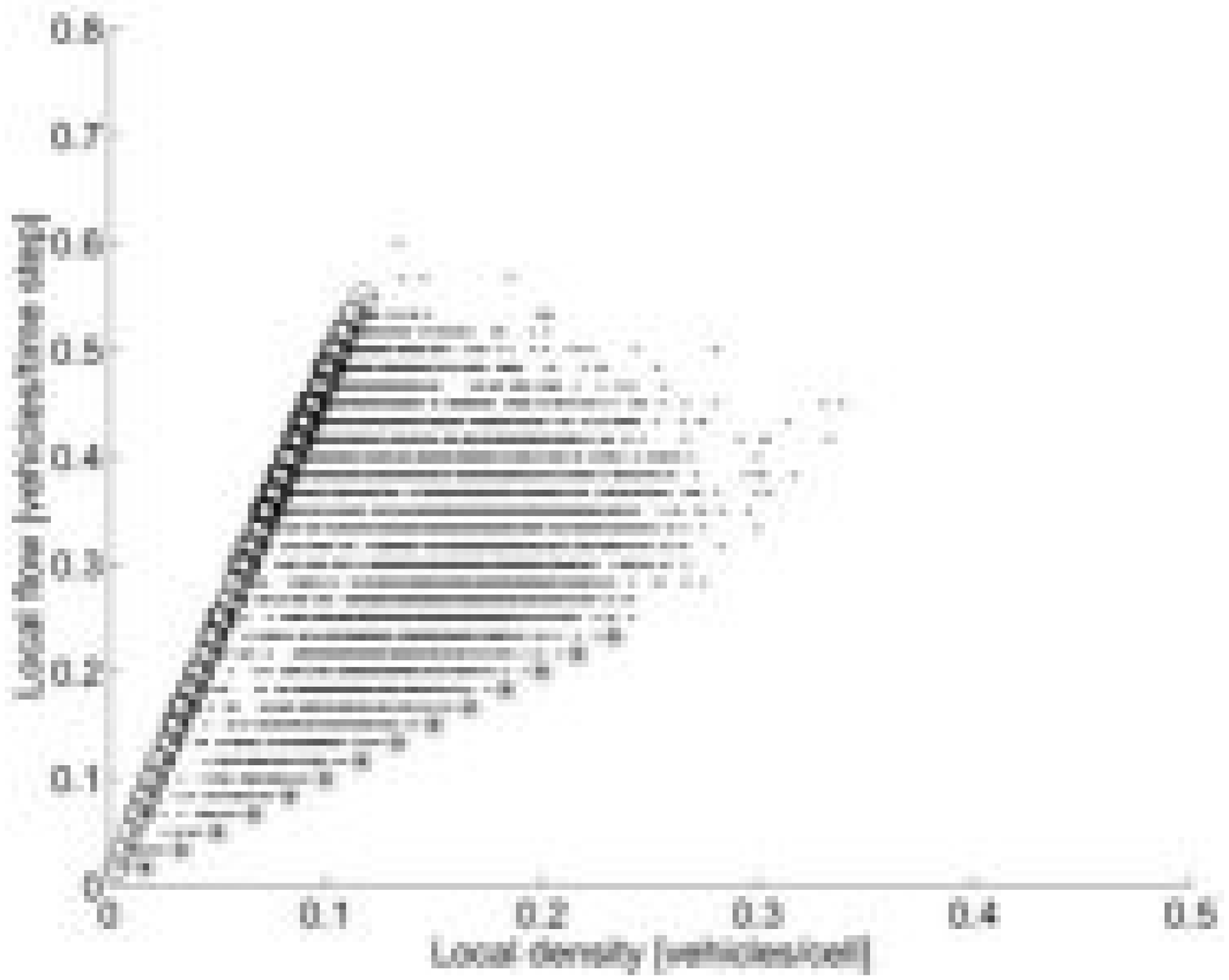}
	\includegraphics[width=0.32\figurewidth]{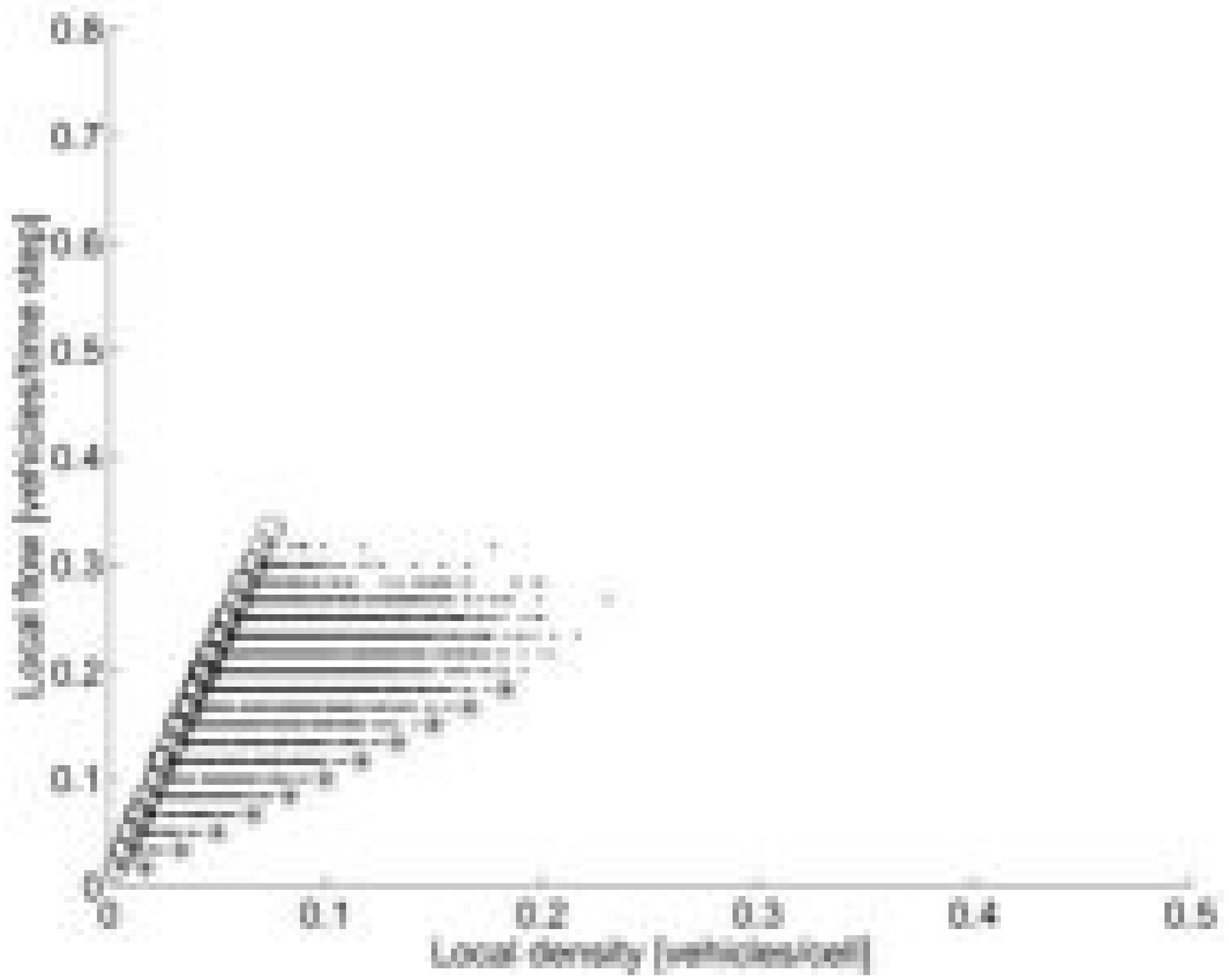}
	\caption{
		Three ($k$,$q$) diagrams based on local measurements in the STCA model with 
		$v_{\text{max}} = \five$ cells/time step. \emph{Left:} $p = \zero$. 
		\emph{Middle:} $p = \frac{\one}{\three}$. \emph{Right:} $p = 
		\frac{\two}{\three}$. Points obtained in the free-flow regime (i.e., for 
		$\overline v_{s} \approx v_{\text{max}}$ cells/time step) are marked with a 
		$\circ$, points obtained in the congested regime with a $\cdot$, and points 
		that imply heavy congestion (i.e., for $\overline v_{s} < \one$ cell/time 
		step) with a $\star$. Note that for these local diagrams, the slopes of the 
		congested branches (indicated by the points marked as $\star$) are the 
		negative of its corresponding slope in a global diagram.
	}
	\label{fig:TCA:STCADensityFlowSMSLocalFundamentalDiagrams}
\end{figure}

Considering local measurements of the density, flow, and space-mean speed, the 
($k$,$q$) diagrams in 
\figref{fig:TCA:STCADensityFlowSMSLocalFundamentalDiagrams} reveal that an 
increasing slowdown probability $p$, results in (i) a lower value for the 
critical density, (ii) a lower capacity flow, and (iii) a more localised scatter 
of the data points.

\begin{figure}[!htbp]
	\centering
	\begin{tabular}{l}
		\includegraphics[width=\halffigurewidth]{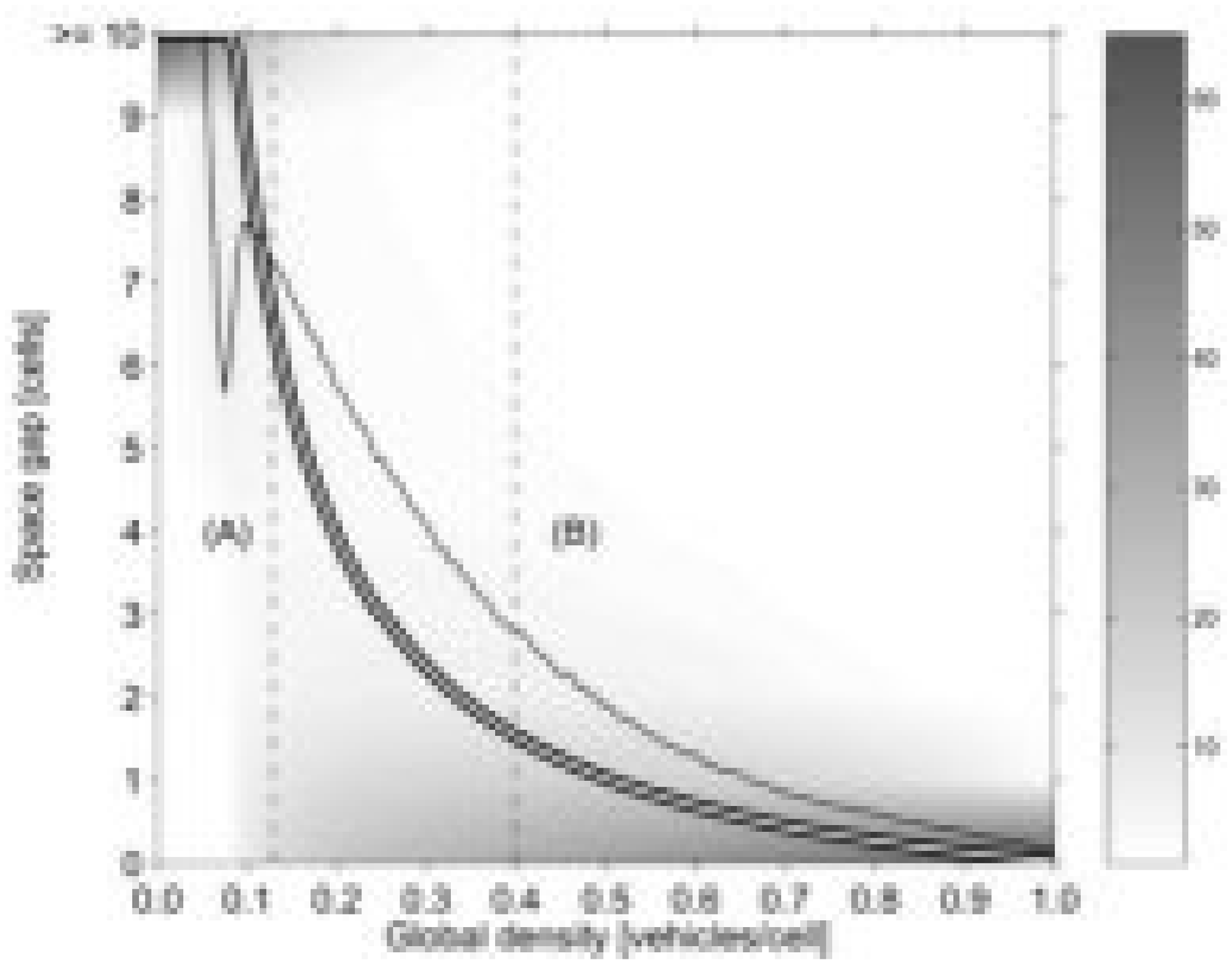}
	\end{tabular}
	\begin{tabular}{l}
		\includegraphics[width=0.9\halffigurewidth,height=2.5cm]{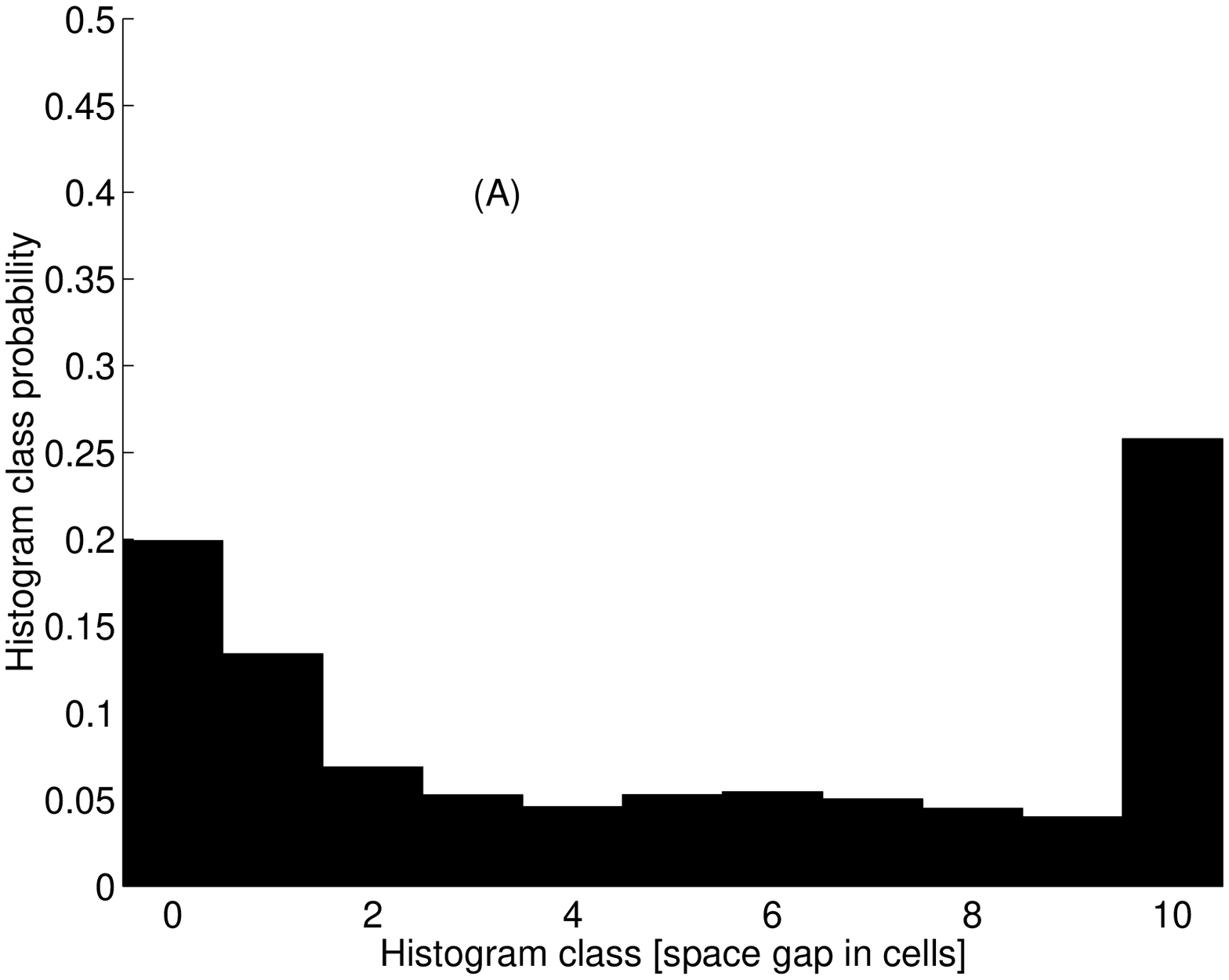}\\
		\includegraphics[width=0.9\halffigurewidth,height=2.5cm]{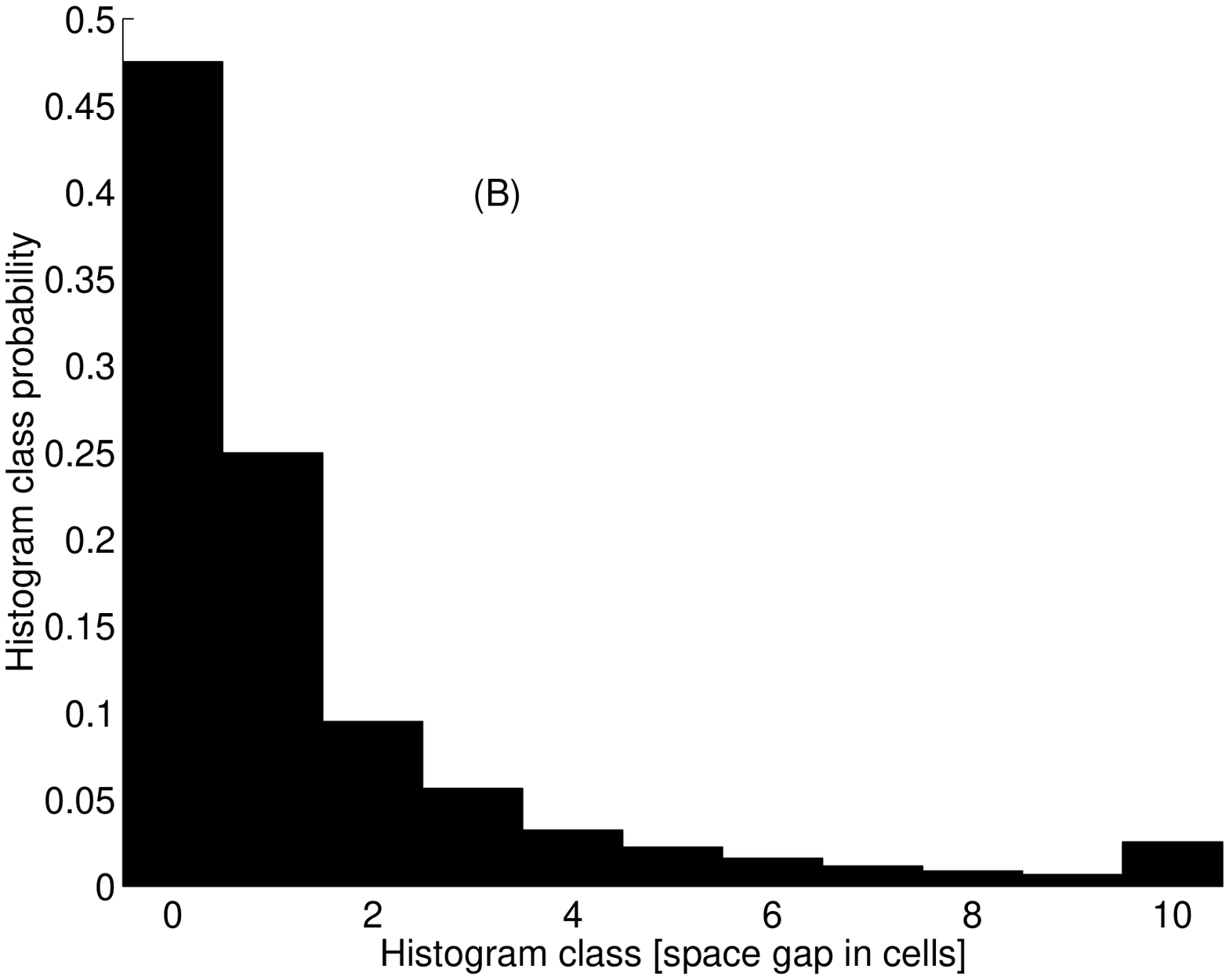}
	\end{tabular}
	\caption{
		A histogram of the distributions of the vehicles' space gaps $g_{s}$, as a 
		function of the global density $k$ in the STCA (with $v_{\text{max}} = 
		\five$ cells/time step and $p = $~0.5). In the contour plot to the left, the 
		thick solid line denotes the average space gap, whereas the thin solid line 
		shows its standard deviation. The grey regions denote the probability 
		densities. The histograms ($A$) and ($B$) to the right, show two cross 
		sections made in the left contour plot at $k =$~0.1325 and $k =$~0.4000 
		respectively: for example, in ($B$), the distribution exhibits a distinct 
		unique maximum at the histogram class $g_{s} = \zero$ cells, corresponding 
		to the dark region in the lower right corner of the contour plot where high 
		global densities occur.
	}
	\label{fig:TCA:STCASpaceGapHistogram}
\end{figure}

In \figref{fig:TCA:STCASpaceGapHistogram}, we have plotted a histogram of the 
distributions of the STCA's vehicles' space gaps, for all global densities $k 
\in [\zero,\one]$. For very low densities, the distributions have a distinct 
maximum, indicating that all vehicles travel with very large space gaps. At 
higher densities, the maxima of the distributions shift toward smaller space 
gaps, as more and more vehicles encounter jams, even leading to a reduction of 
their space gap to zero. Around the critical density however, the distributions 
are smeared out across consecutive densities, but for each of those densities 
they exhibit a bimodal structure. Because the STCA contains many jams, the 
system now contains both vehicles in free-flow traffic, as well as vehicles that 
are in a congested state (i.e., driving closer to each other) 
\cite{CHOWDHURY:98,CHOWDHURY:99b,HELBING:01,SCHADSCHNEIDER:99b}.

\begin{figure}[!htbp]
	\centering
	\includegraphics[width=\halffigurewidth]{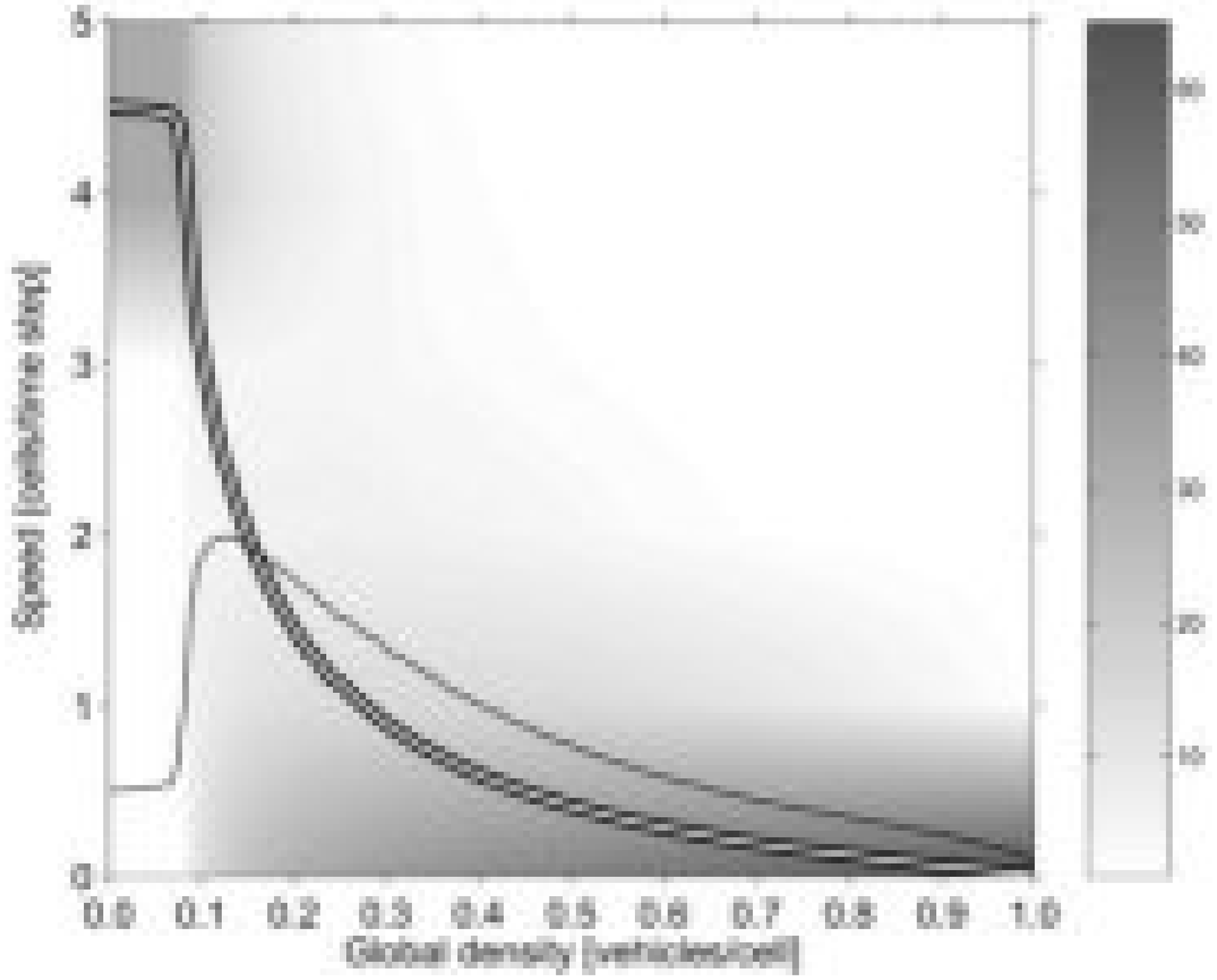}
	\hspace{\figureseparation}
	\includegraphics[width=\halffigurewidth]{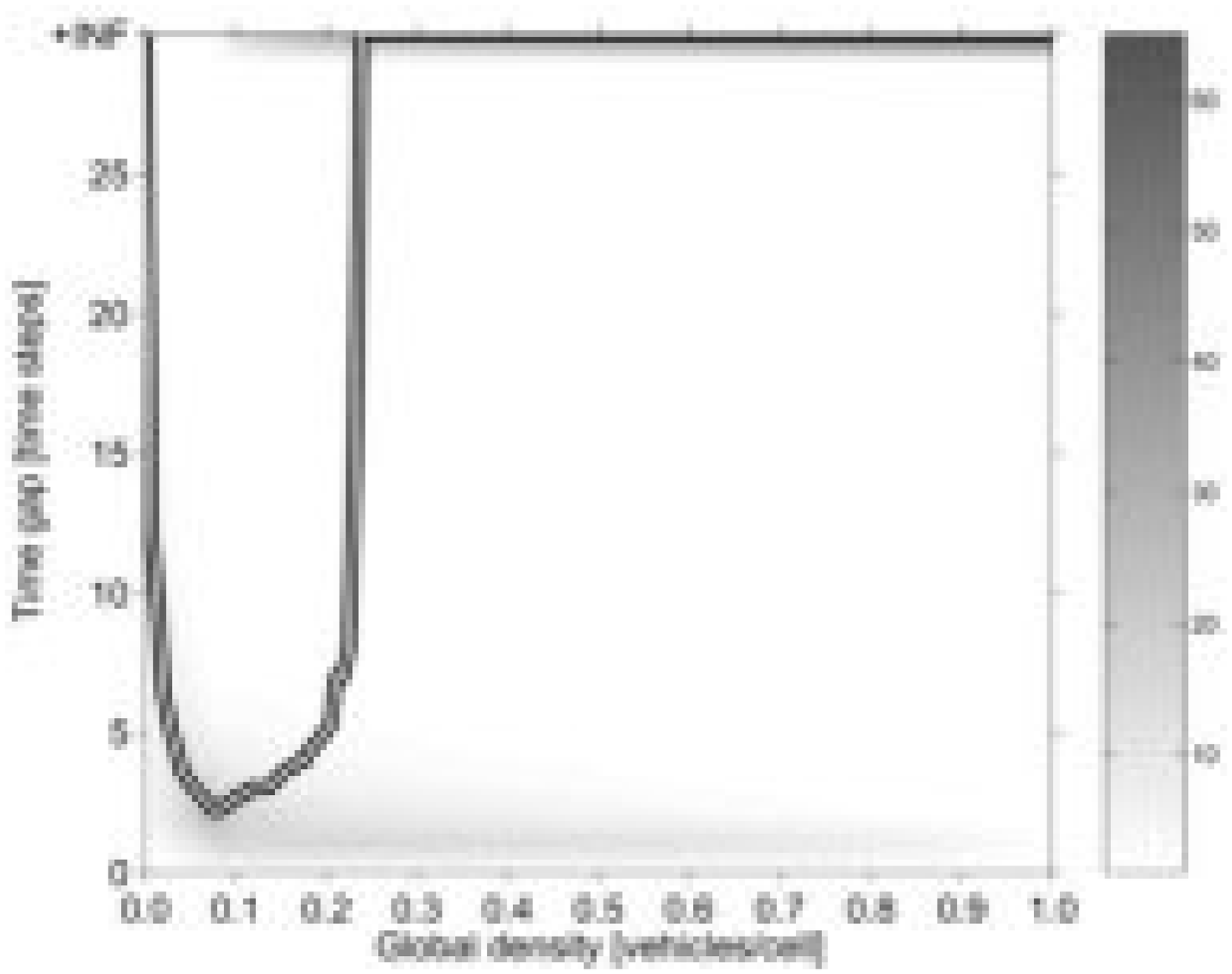}
	\caption{
		Histograms of the distributions of the vehicles' speeds $v$ (\emph{left}) 
		and time gaps $g_{t}$ (\emph{right}), as a function of the global density 
		$k$ in the STCA (with $v_{\text{max}} = \five$ cells/time step and $p = 
		$~0.5). The thick solid lines denote the space-mean speed and median time 
		gap, whereas the thin solid line shows the former's standard deviation. The 
		grey regions denote the probability densities.
	}
	\label{fig:TCA:STCASpaceMeanSpeedTimeGapHistogram}
\end{figure}

In similarly spirit, \figref{fig:TCA:STCASpaceMeanSpeedTimeGapHistogram} shows 
the distribution of the vehicles' speeds and time gaps. Corresponding with our 
observations of the ($k$,$\overline v_{s}$) diagrams in 
\figref{fig:TCA:STCADensityFlowSMSFundamentalDiagrams}, the left part of 
\figref{fig:TCA:STCASpaceMeanSpeedTimeGapHistogram} shows a distinct cluster of 
probability mass at the histogram class $v_{\text{max}} - p$ for very low global 
densities. In this region, the standard deviation of the space-mean speed is 
more or less constant and equal to $p$. At higher global densities, the 
distributions become temporarily bimodal, after which they again tend to a 
unique maximum of 0 cells/time step, corresponding to severely congested 
traffic; the standard deviation drastically encounters a maximum at the critical 
density, after which it declines steadily. With respect to the distributions of 
the time gaps, the right part of 
\figref{fig:TCA:STCASpaceMeanSpeedTimeGapHistogram} shows an rapidly decreasing 
median time gap as the critical density is approached. At this density, the time 
gaps settle around a local cluster at the minimum of 1 time step. Going to 
higher global densities, the number of stopped vehicles increases rapidly, 
frequently resulting in infinite time gaps. From the critical density on, all 
distributions exhibit a bimodal structure, corresponding to vehicles that are 
caught inside a jam, and other vehicles that are able to move freely (possibly 
at a lower speed) \cite{CHOWDHURY:98,GHOSH:98,SCHADSCHNEIDER:99b}.

			\subsubsection{STCA with cruise control (STCA-CC)}
			\label{sec:TCA:STCACC}

\begin{figure}[!htbp]
	\centering
	\includegraphics[width=\figurewidth]{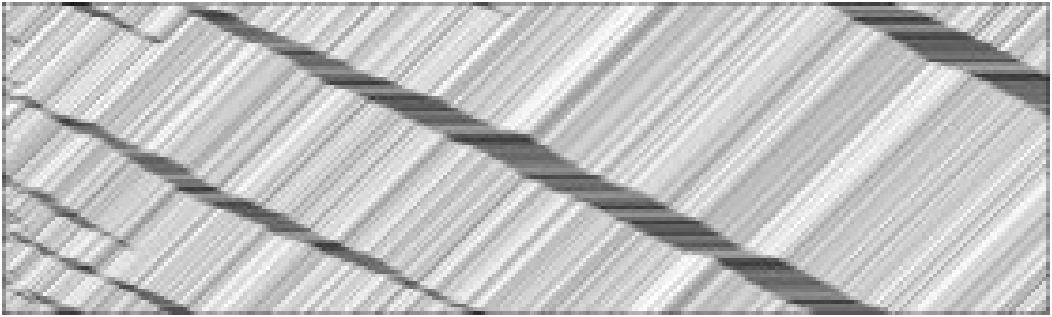}
	\caption{
		A time-space diagram of the STCA-CC model for $v_{\text{max}} = \one$ 
		cell/time step and a global density of $k =$~0.4 vehicles/cell. The shown 
		lattice contains 300 cells, with a visible period of 1000 time steps. We can 
		see over ten initial jams evolving, coalescing over time into one superjam. 
		The system exhibits two distinct phases, i.e., a free-flow and a congested 
		regime with $\overline v_{s} = \one$ and $\overline v_{s} = \zero$ 
		cells/time step respectively.
	}
	\label{fig:TCA:STCACCTimeSpaceDiagram}
\end{figure}

As mentioned in the previous section \ref{sec:TCA:STCA}, a typical artifact of 
the STCA model is that it gives rise to many unstable artificial jams. Due to 
the noise inherent in the model, a jam can always form at any density, even 
inducing a local breakdown of traffic in the free-flow traffic regime. One way 
to remedy this, is by stabilising the free-flow branch of the ($k$,$q$) diagram. 
This can be done by inhibiting the randomisation for high-speed vehicles. To 
this end, Nagel and Paczuski considered again the rules R1 -- R3 of the STCA, 
i.e., equations \eqref{eq:TCA:STCAR1} -- \eqref{eq:TCA:STCAR3}, but now 
complemented with a rule R0 \cite{NAGEL:95b}:

\begin{quote}
	\textbf{R0}: \emph{determine stochastic noise}\\
		\begin{equation}
			\left \lbrace
				\begin{array}{lcl}
					v_{i}(t - \one) = v_{\text{max}} & \Longrightarrow & p'(t) \leftarrow \zero, \\
					v_{i}(t - \one) < v_{\text{max}} & \Longrightarrow & p'(t) \leftarrow p,
				\end{array}
			\right.
		\end{equation}
\end{quote}

with now $p$ replaced by $p'(t)$ in the STCA's randomisation rule R2, i.e., 
equation \eqref{eq:TCA:STCAR2}. This new rule effectively turns off the 
randomisation for high-speed vehicles, as only `jammed' vehicles will now have 
stochastic behaviour. The resulting TCA model, is called the STCA in the 
\emph{cruise-control limit}, or STCA-CC for short. If we set the maximum speed 
$v_{\text{max}} = \one$ cell/time step, then all jams initially present in the 
system will coalesce with each other, giving rise to one superjam as depicted in 
\figref{fig:TCA:STCACCTimeSpaceDiagram}. This superjam has been found to follow 
a \emph{random walk} in the time-space diagram \cite{NAGEL:95b,NAGEL:96}. Note 
that $v_{\text{max}} > \one$ cell/time step does not alter the critical 
behaviour of the model, even though jam clusters are now branching, having 
regions of free-flow traffic in between them \cite{NAGEL:96}.

\begin{figure}[!htbp]
	\centering
	\includegraphics[width=\halffigurewidth]{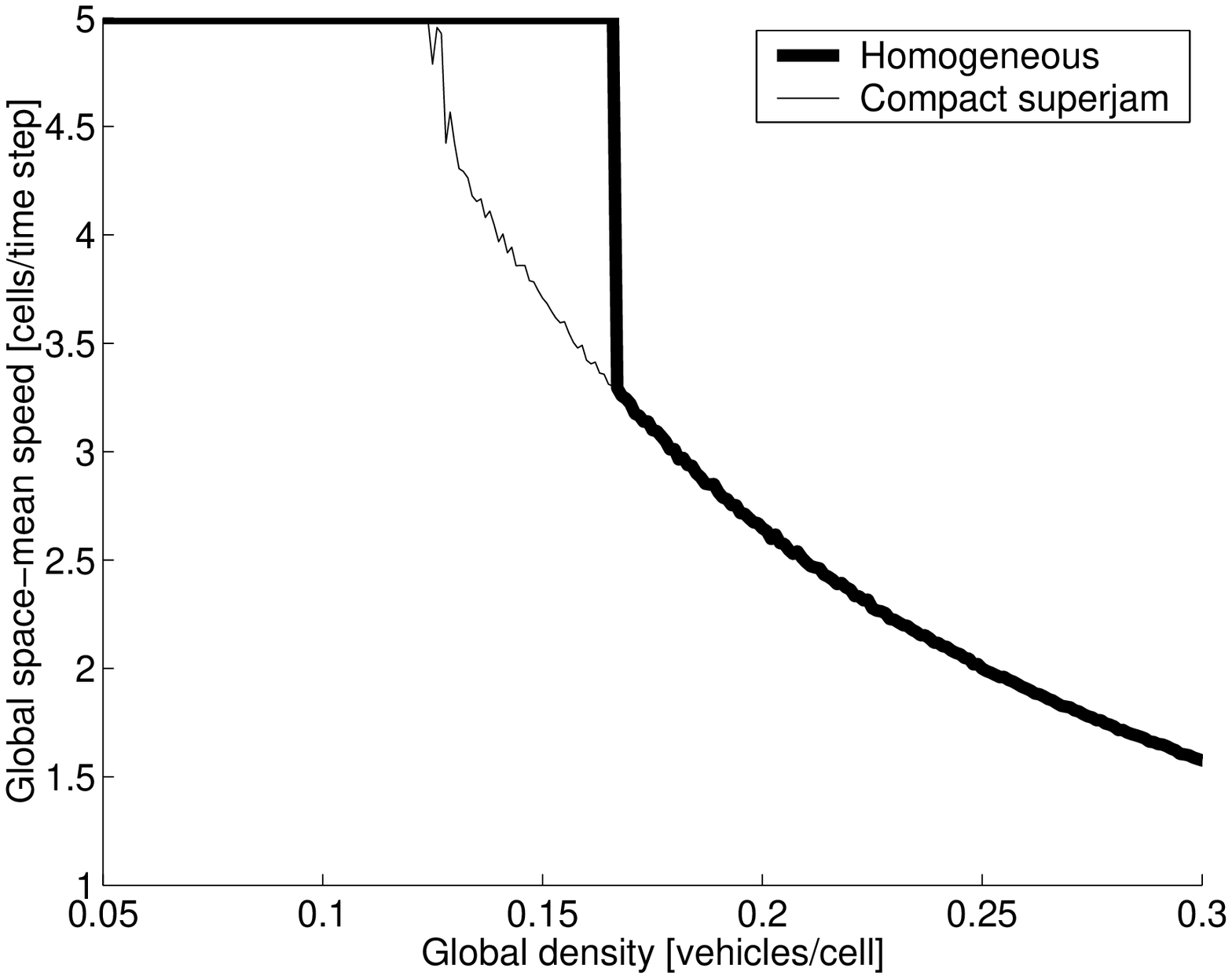}
	\hspace{\figureseparation}
	\includegraphics[width=\halffigurewidth]{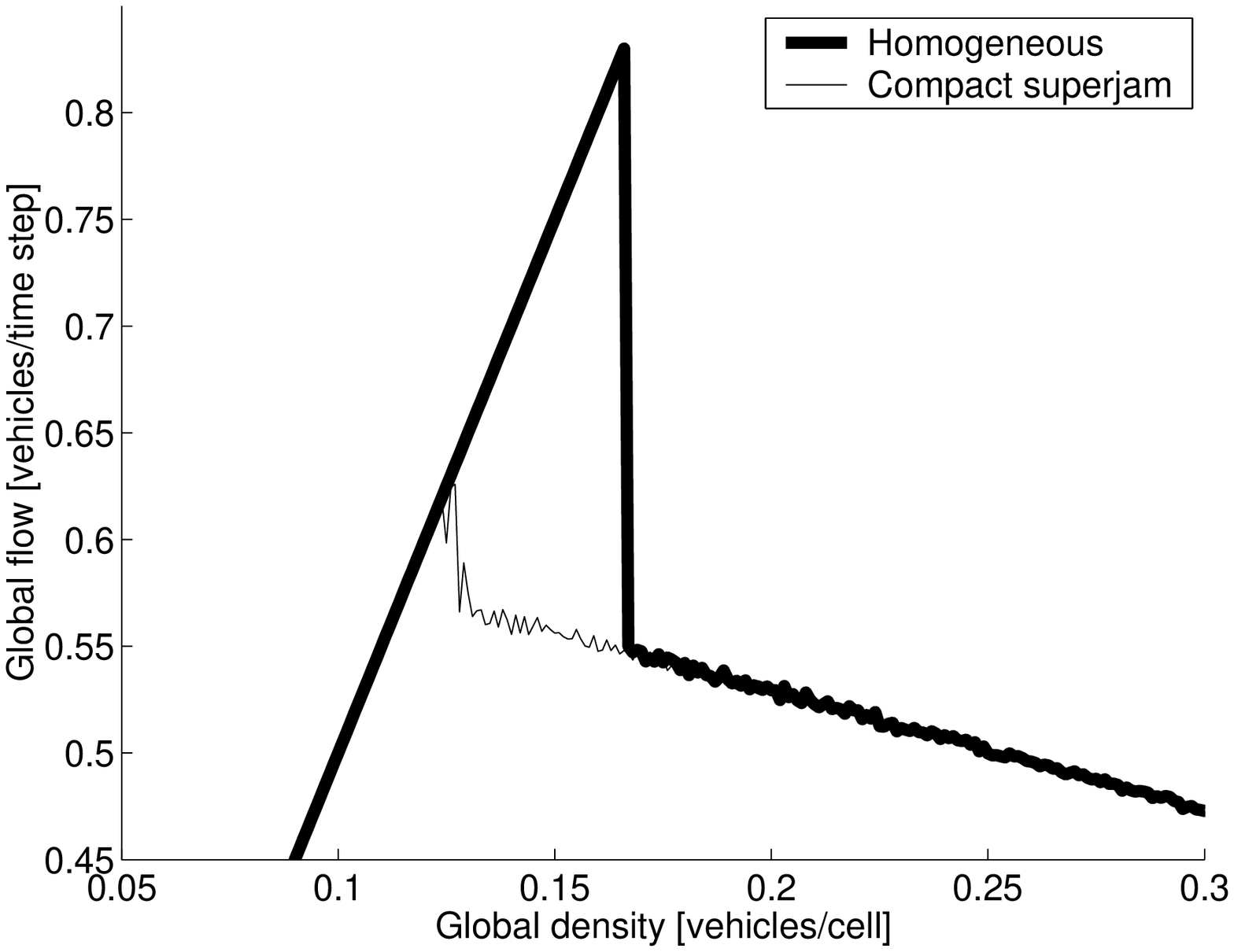}
	\caption{
		Two ($k$,$\overline v_{s}$) (\emph{left}) and ($k$,$q$) (\emph{right}) 
		diagrams for the STCA-CC model, with $v_{\text{max}} = \five$ cells/time 
		step and $p =$~0.2. The thick solid line denotes global measurements that 
		were obtained when starting from homogeneous initial conditions; the thin 
		solid line is based on a compact superjam as the initial condition (see 
		section \ref{sec:TCA:SingleCellModels} for an explanation of these 
		conditions). The right part clearly shows a typical reversed $\lambda$ 
		shape, which indicates a capacity drop. Note that the observed smaller drop 
		in flow for the compact superjam, is actually a finite-size effect 
		\cite{NAGEL:95b,KRAUSS:97b}.
	}
	\label{fig:TCA:STCACCDensityFlowSMSFundamentalDiagrams}
\end{figure}

In \figref{fig:TCA:STCACCDensityFlowSMSFundamentalDiagrams}, we show the 
($k$,$\overline v_{s}$) and ($k$,$q$) diagram of the STCA-CC with 
$v_{\text{max}} = \five$ cells/time step and $p =$~0.2. As can be seen in the 
right part, the ($k$,$q$) diagram has a typical inverted $\lambda$ shape (see 
also our discussion in \cite{MAERIVOET:05c} about the hysteresis and capacity 
drop phenomena). The STCA-CC is said to be \emph{bistable}, in that both the 
free-flow as well as the congested branches of the ($k$,$q$) diagram are stable 
(the former because it is noise-free). Vehicles going from the free-flow to the 
congested regime encounter at the critical density a phenomenon much like a 
capacity drop. The reverse transition to the free-flow branch proceeds via a 
lower density and, correspondingly, a lower flow (which is the outflow 
$q_{\text{out}}$ of a jam). Comparing the right parts of 
\figref{fig:TCA:STCADensityFlowSMSFundamentalDiagrams} and 
\figref{fig:TCA:STCACCDensityFlowSMSFundamentalDiagrams}, it is evident that a 
destabilisation of the free-flow branch forms the main reason for a lower 
capacity flow, reached at a lower critical density.

To conclude our discussion of the STCA-CC, we note that the use of cruise 
control as an ADAS can have unintended consequences. The traffic system can be 
perceived as having an underlying critical point, at which the life times of 
jams switch from finite to infinite (see our discussion at the beginning of 
section \ref{sec:TCA:STCA}). The existence of this point is closely tied to the 
\emph{self-organised criticality} (SOC) of the STCA model: the outflow from an 
infinite jam automatically self-organises to a state of maximum attainable flow 
\cite{BAK:88,NAGEL:93,NAGEL:95c,TURCOTTE:99}. Stabilising the free-flow branch 
with cruise-control measures, results on the one hand in traffic higher 
achievable flows which is beneficial, but on the other hand the system is driven 
closer to its critical point which is more dangerous. At this stage, travel 
times will experience a high degree of variability, thereby reducing its 
predictability \cite{NAGEL:94,NAGEL:95b,NAGEL:95c}.

			\subsubsection{Stochastic Fukui-Ishibashi TCA (SFI-TCA)}
			\label{sec:TCA:SFITCA}

In section \ref{sec:TCA:DeterministicFITCA}, we discussed the deterministic 
FI-TCA which is a generalisation of the CA-184 TCA model. In their original 
formulation, Fukui and Ishibashi also introduced stochasticity, but now only for 
vehicles driving at the highest possible speed of $v_{\text{max}}$ cells/time 
step \cite{FUKUI:96}. We can express the rules of this model, by considering the 
rules R2 and R3 of the STCA, i.e., equations \eqref{eq:TCA:STCAR2} and 
\eqref{eq:TCA:STCAR3}, but now complemented with the DFI-TCA's rule R1 for 
instantaneous accelerations, i.e., equation \eqref{eq:TCA:DFITCAR1} of section 
\ref{sec:TCA:DeterministicFITCA}, and, as in the STCA-CC model, an extra rule 
R0:

\begin{quote}
	\textbf{R0}: \emph{determine stochastic noise}\\
		\begin{equation}
			\left \lbrace
				\begin{array}{lcl}
					v_{i}(t - \one) = v_{\text{max}} & \Longrightarrow & p'(t) \leftarrow p, \\
					v_{i}(t - \one) < v_{\text{max}} & \Longrightarrow & p'(t) \leftarrow \zero,
				\end{array}
			\right.
		\end{equation}
\end{quote}

with now $p$ replaced by $p'(t)$ in the randomisation rule R2. It can be seen 
that for $v_{\text{max}} = \one$, the SFI-TCA and STCA models are the same. 
Furthermore, for $p = \zero$ the SFI-TCA becomes fully deterministic, and in 
contrast to the STCA's zero-flow behaviour (see section \ref{sec:TCA:STCA}), the 
SFI-TCA's $p = \one$ case corresponds to the STCA with $p = \zero$ and 
$v_{\text{max}} - \one$.

The rationale behind the specific randomisation in the SFI-TCA model, is that 
drivers who are moving at a high speed, are not able to focus their attention 
indefinitely. As a consequence, there will be fluctuations at these high speeds. 
As such, this corresponds to the opposite of a cruise-control limit, e.g., the 
STCA-CC model. There will be no capacity drop, but the effect on the 
($k$,$\overline v_{s}$) diagram is that its free-flow branch will become 
slightly downward curving, starting at $\overline v_{s} = v_{\text{max}} - p$ 
for $k = \zero$.

To conclude, we mention the related work of Wang et al., who studied the SFI-TCA 
both analytically and numerically, providing an exact result for $p = \zero$, 
and a close approximation for the model with $p \neq \zero$ \cite{WANG:98}. 
Based on the SFI-TCA, Wang et al. developed a model that is subtly different. 
They assumed that drivers do not suffer from concentration lapses at high 
speeds, but are instead only subjected to the random deceleration when they are 
driving close enough to their direct frontal leaders \cite{WANG:01}. And 
finally, we mention the work of Lee et al., who incorporate anticipation with 
respect to a vehicle's changing space gap $g_{s}$ as its leader is driving away. 
This results in a higher capacity flow, as well as the appearance of a 
synchronised-traffic regime, in which vehicles have a lower speed, but are 
\emph{all} moving \cite{LEE:02}.

			\subsubsection{Totally asymmetric simple exclusion process (TASEP)}
			\label{sec:TCA:TASEP}

The simple exclusion process is a simplified well-known particle transport model 
from non-equilibrium statistical mechanics, defined on a one-dimensional 
lattice. In the case of open boundary conditions (i.e., the bottleneck 
scenario), particles enter the system from the left side at an \emph{entry rate} 
$\alpha$, move through the lattice, and leave it at an \emph{exit rate} $\beta$. 
The term `simple exclusion' refers to the fact that a cell in the lattice can 
only be empty, or occupied by one particle. When moving through the lattice, 
particles move one cell to the left with probability $\gamma$, and one cell to 
the right with probability $\delta$. When $\gamma = \delta$, the process is 
called the \emph{symmetric simple exclusion process} (SSEP); if $\gamma \neq 
\delta$, then it is called the \emph{asymmetric simple exclusion process} (ASEP) 
\cite{DERRIDA:92}. Finally, if we set $\gamma = \zero$ and $\delta = \one$, the 
system is called the \emph{totally asymmetric simple exclusion process} (TASEP). 
If we consider the TASEP as a TCA model, then all vehicles move with 
$v_{\text{max}} = \one$ cell/time step to their direct right-neighbouring cell, 
on the condition that this cell is empty.

\begin{figure*}[!htbp]
	\centering
		\framebox{\includegraphics[width=0.17\textwidth]{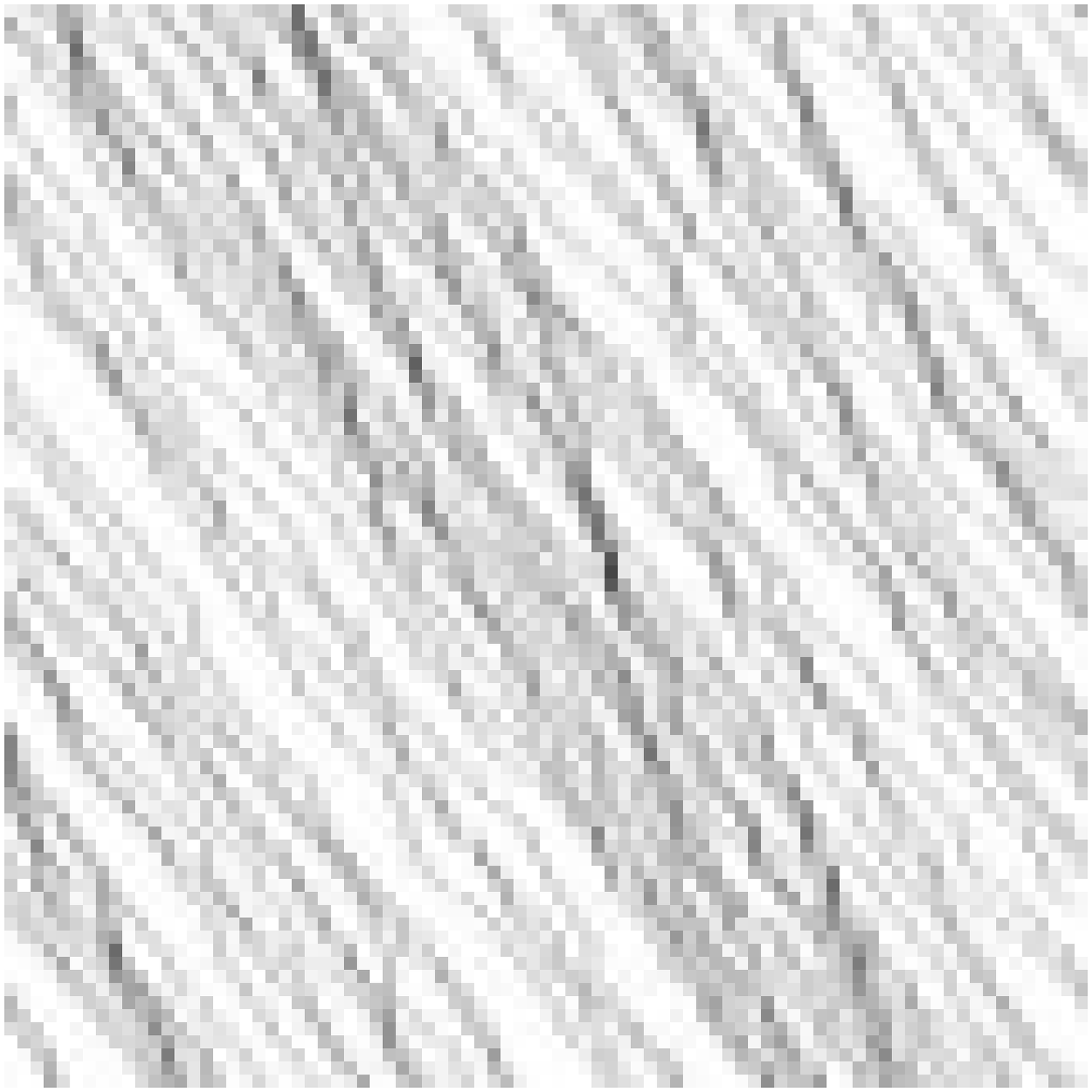}}
		\hspace{0.05cm}
		\framebox{\includegraphics[width=0.17\textwidth]{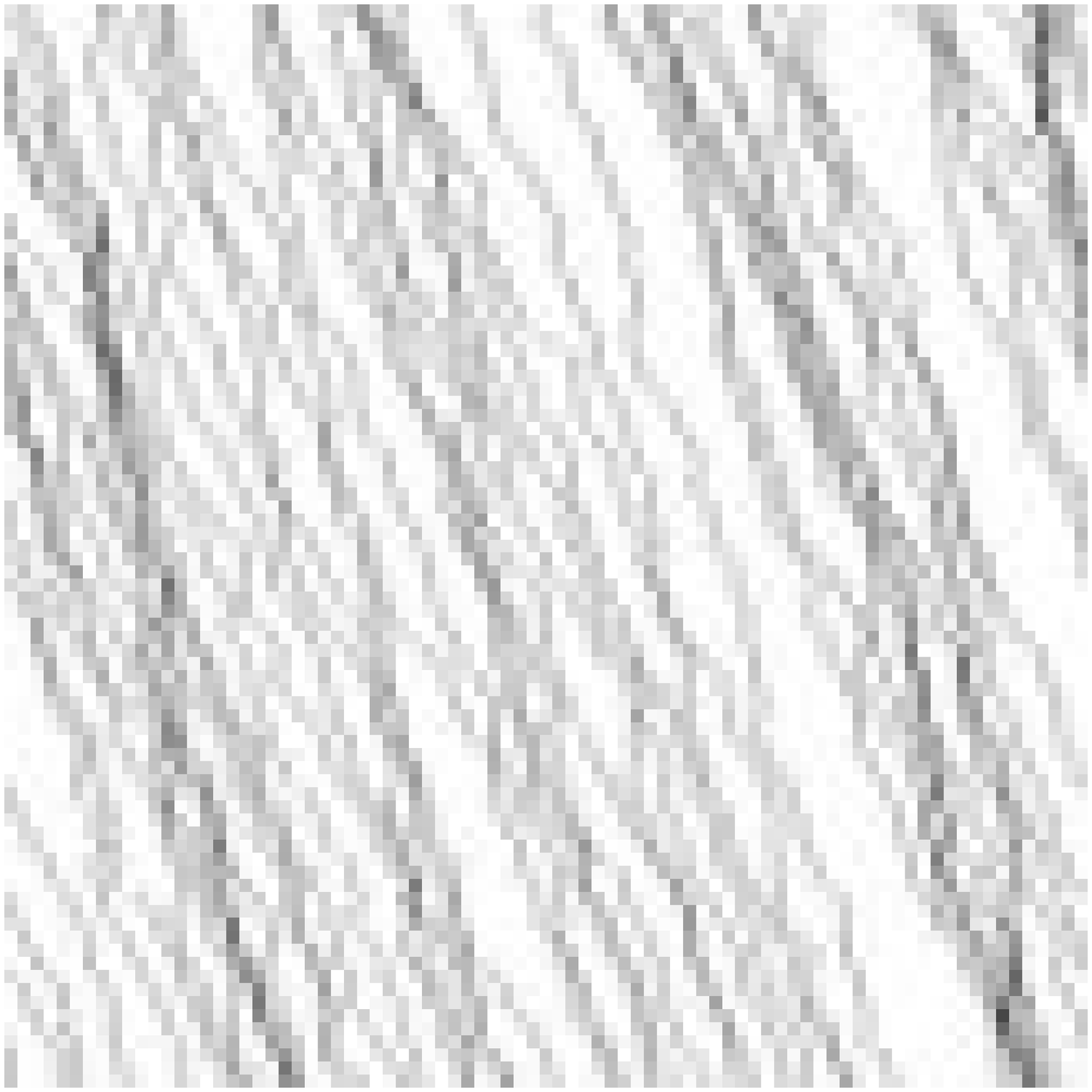}}
		\hspace{0.05cm}
		\framebox{\includegraphics[width=0.17\textwidth]{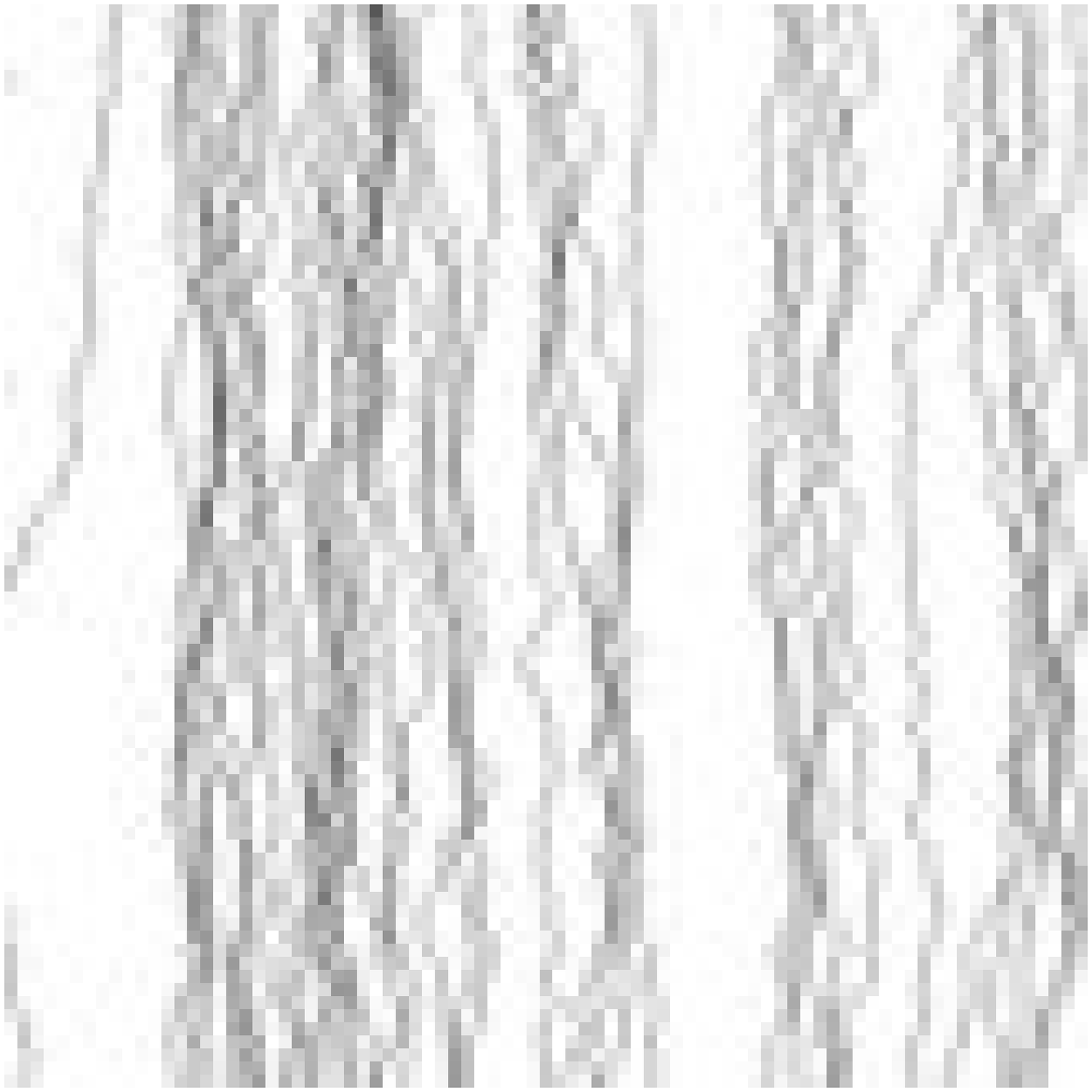}}
		\hspace{0.05cm}
		\framebox{\includegraphics[width=0.17\textwidth]{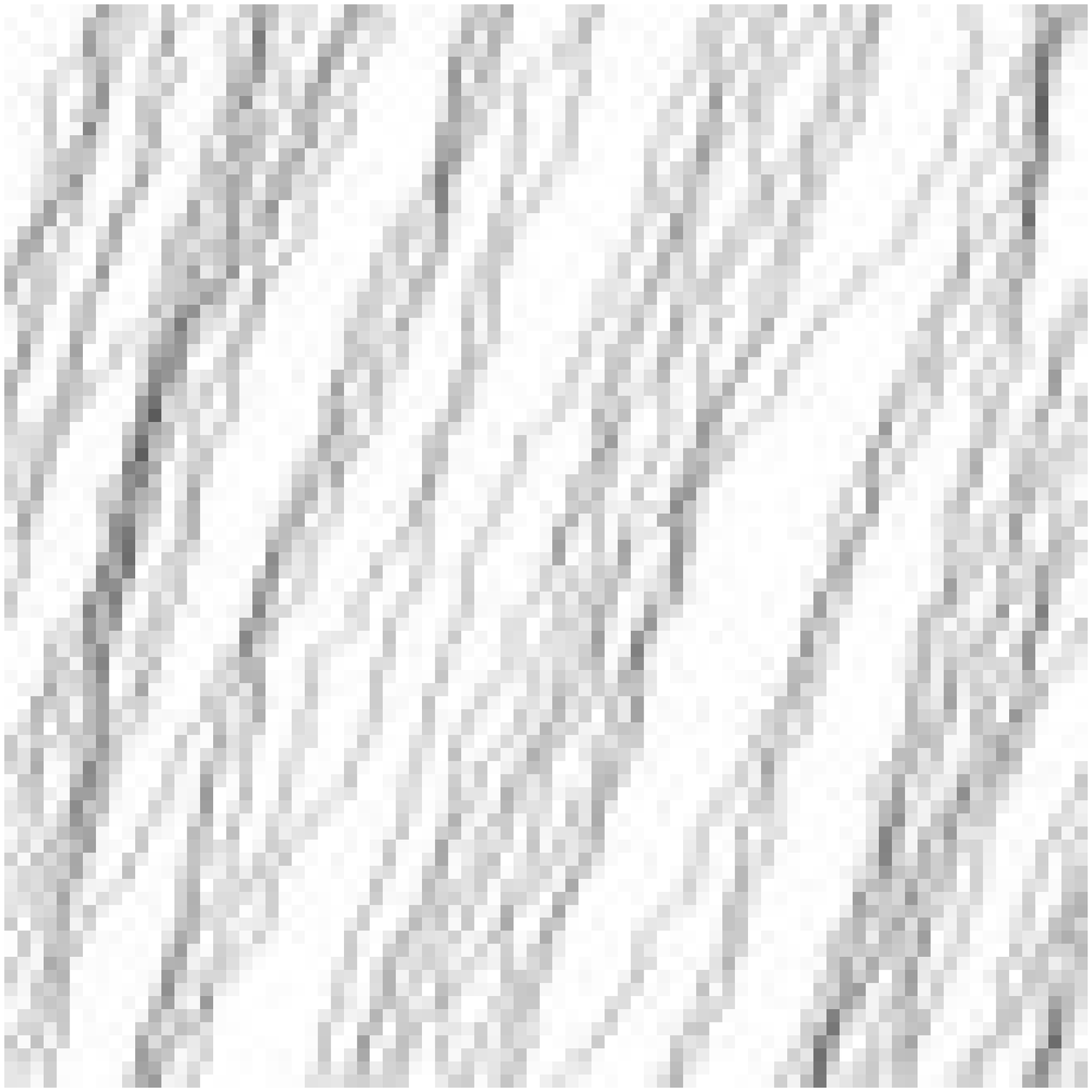}}
		\hspace{0.05cm}
		\framebox{\includegraphics[width=0.17\textwidth]{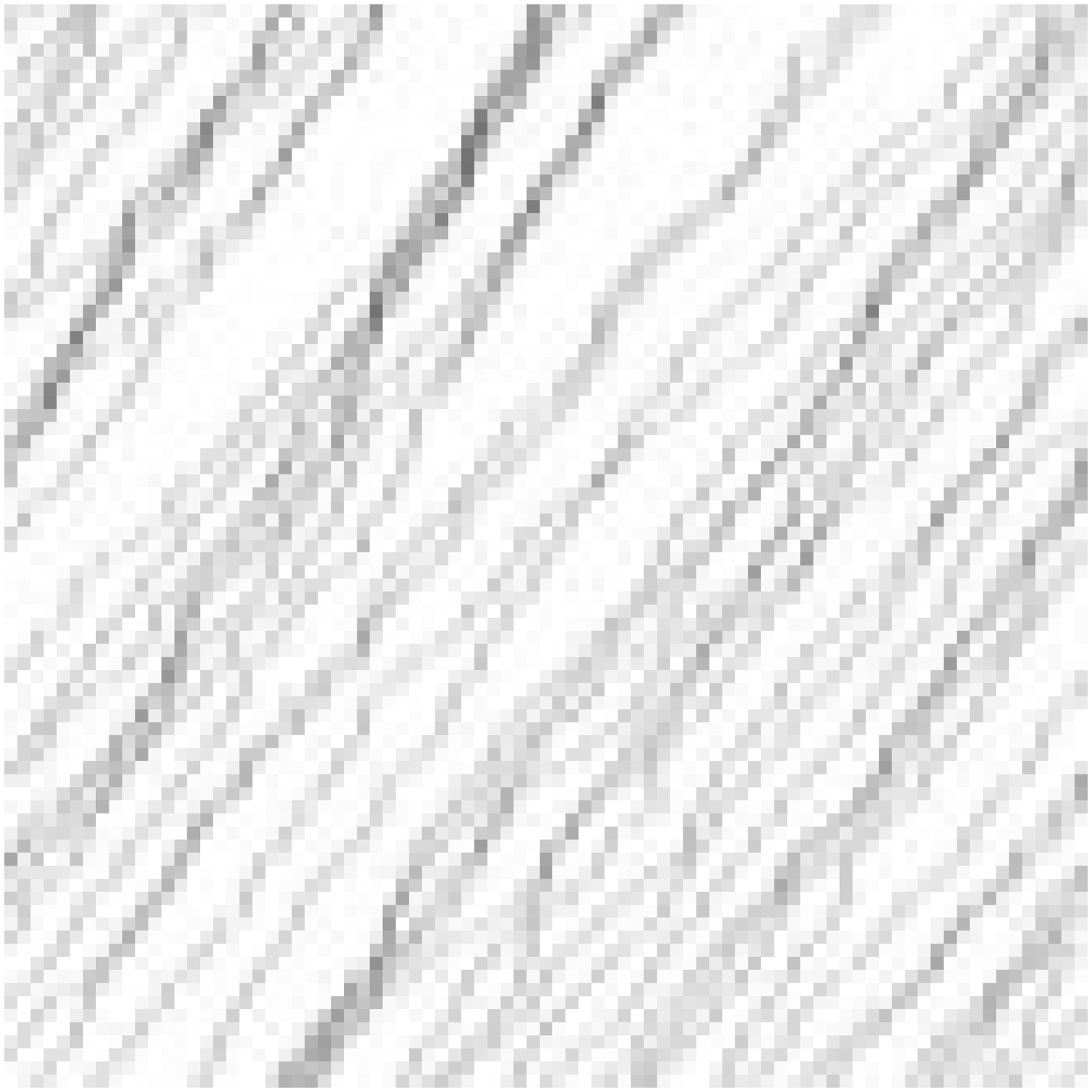}}\\
	\vspace{0.15cm}
		\framebox{\includegraphics[width=0.17\textwidth]{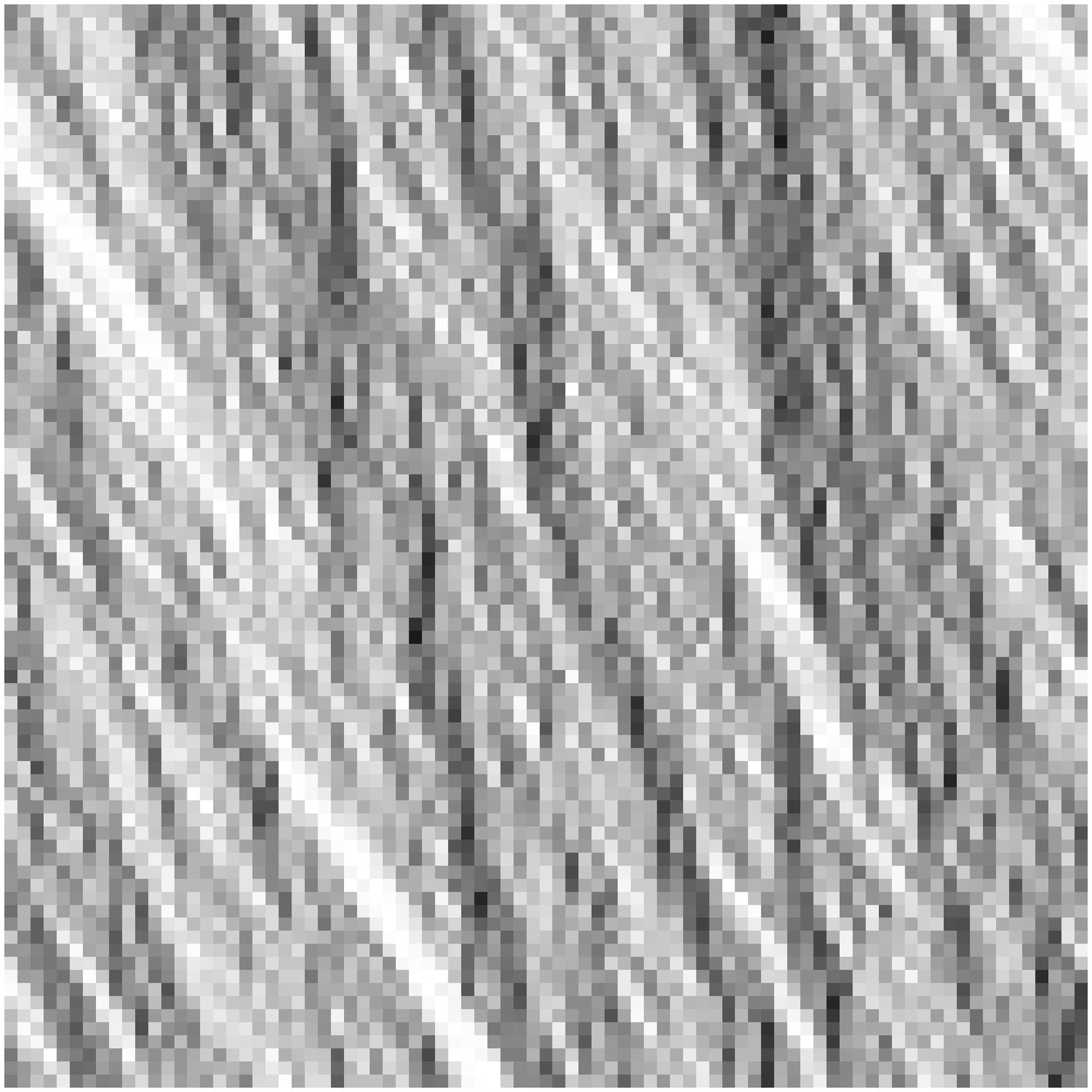}}
		\hspace{0.05cm}
		\framebox{\includegraphics[width=0.17\textwidth]{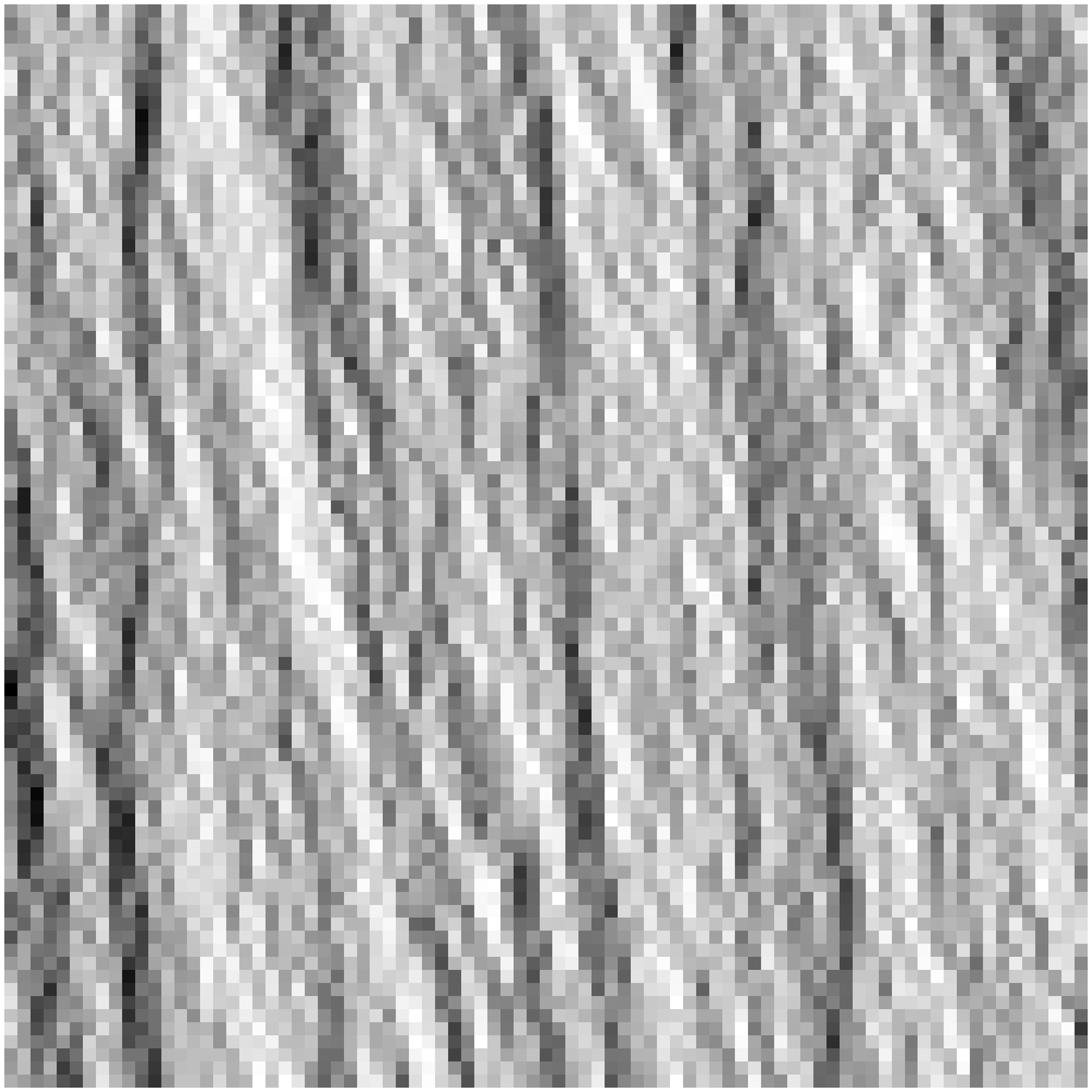}}
		\hspace{0.05cm}
		\framebox{\includegraphics[width=0.17\textwidth]{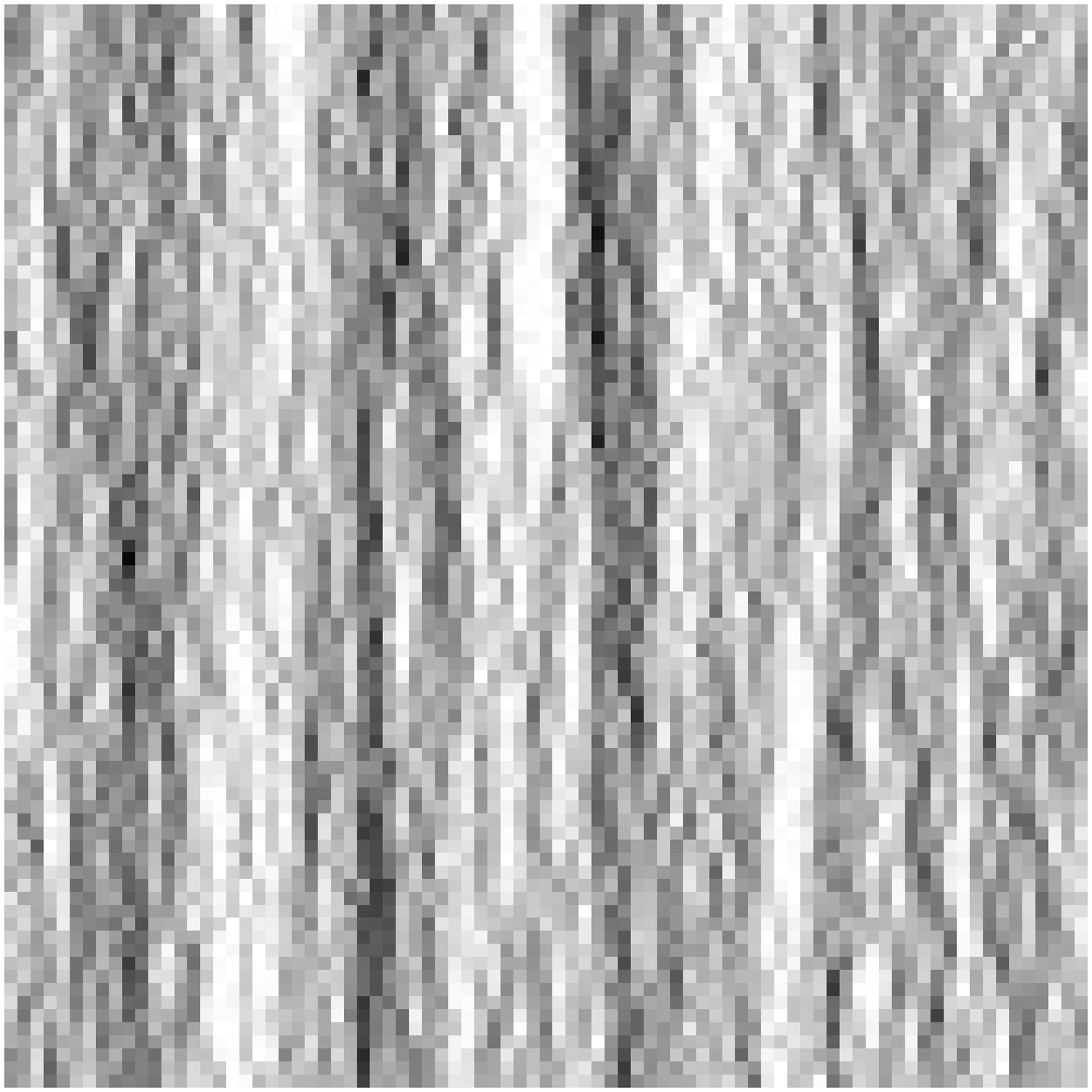}}
		\hspace{0.05cm}
		\framebox{\includegraphics[width=0.17\textwidth]{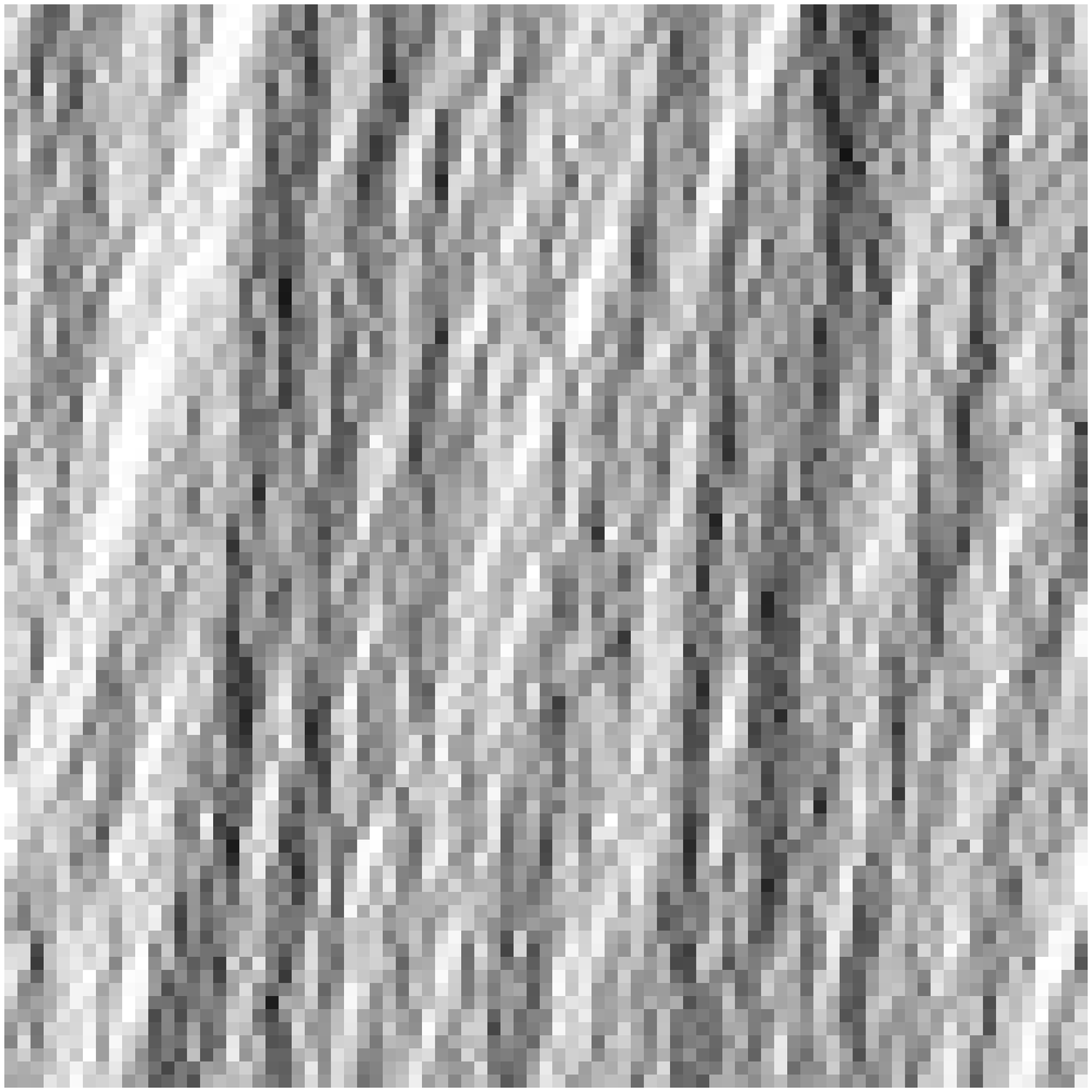}}
		\hspace{0.05cm}
		\framebox{\includegraphics[width=0.17\textwidth]{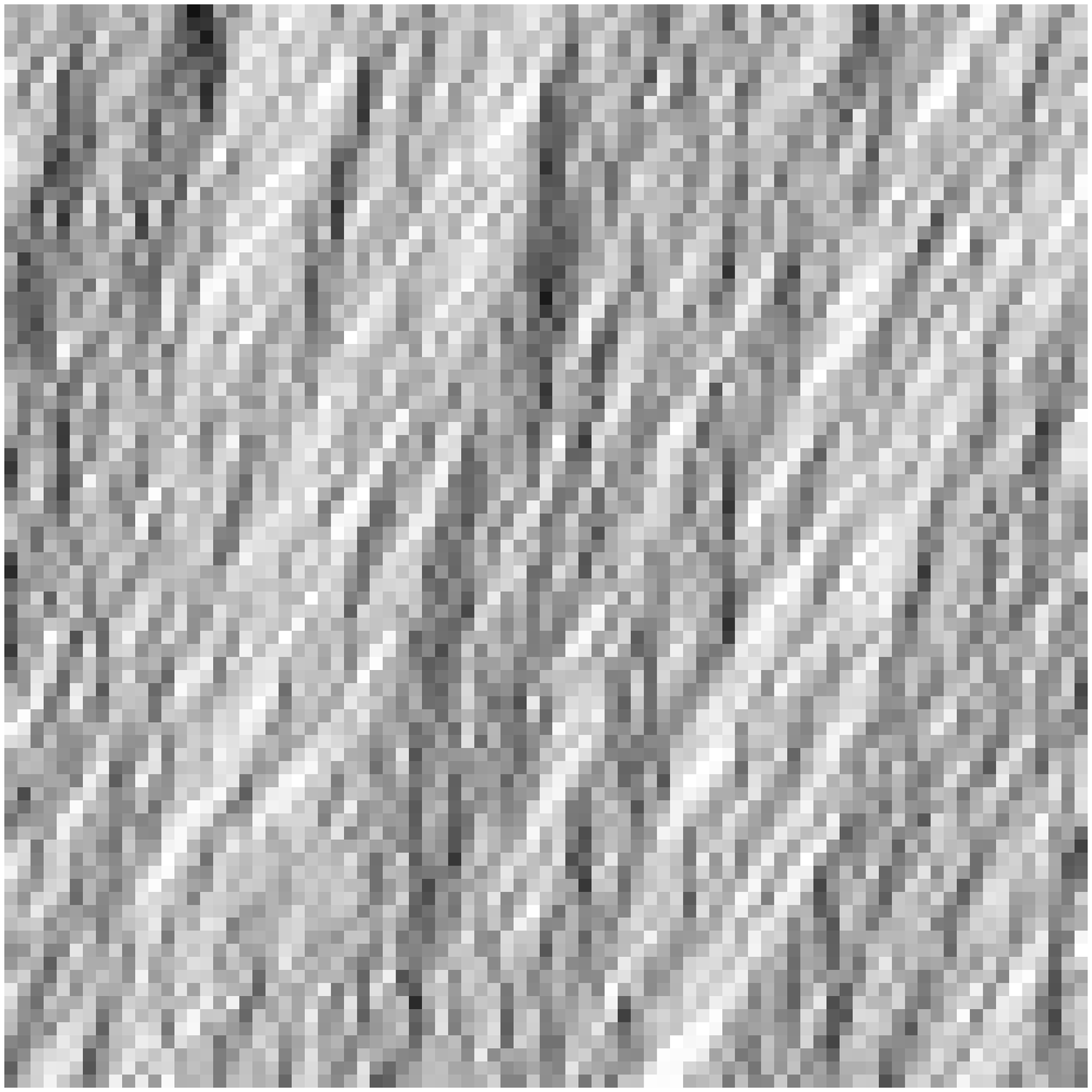}}\\
	\vspace{0.15cm}
		\framebox{\includegraphics[width=0.17\textwidth]{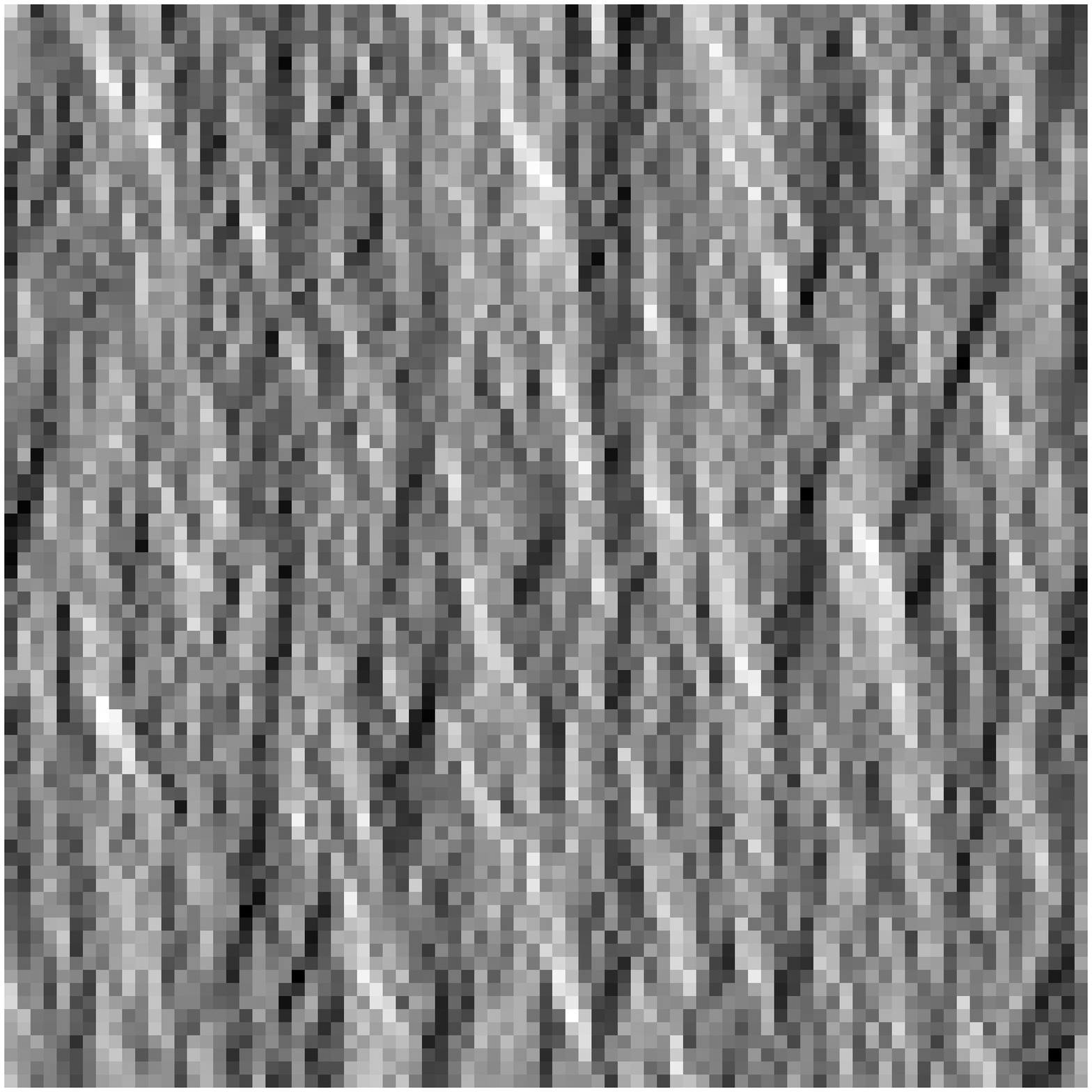}}
		\hspace{0.05cm}
		\framebox{\includegraphics[width=0.17\textwidth]{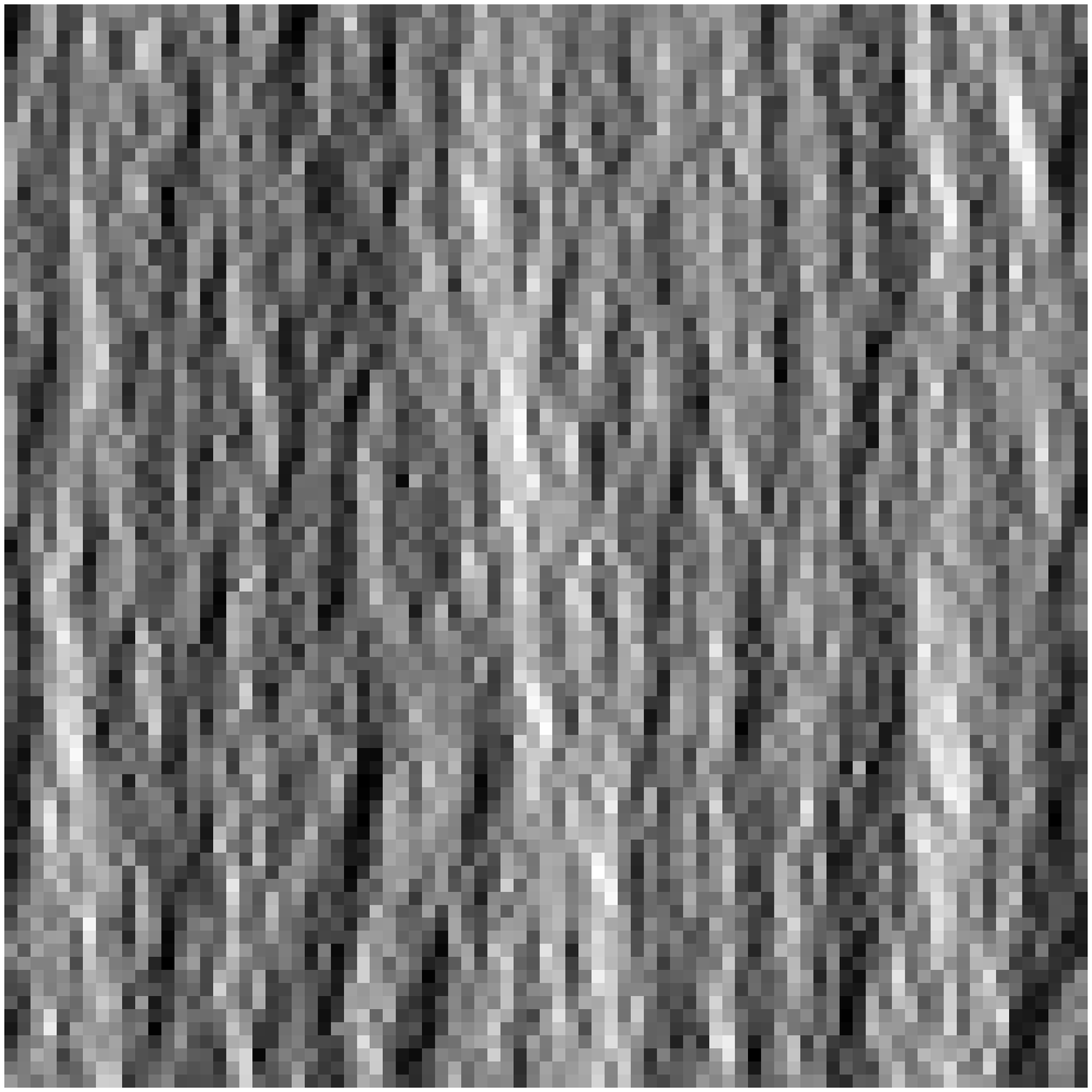}}
		\hspace{0.05cm}
		\framebox{\includegraphics[width=0.17\textwidth]{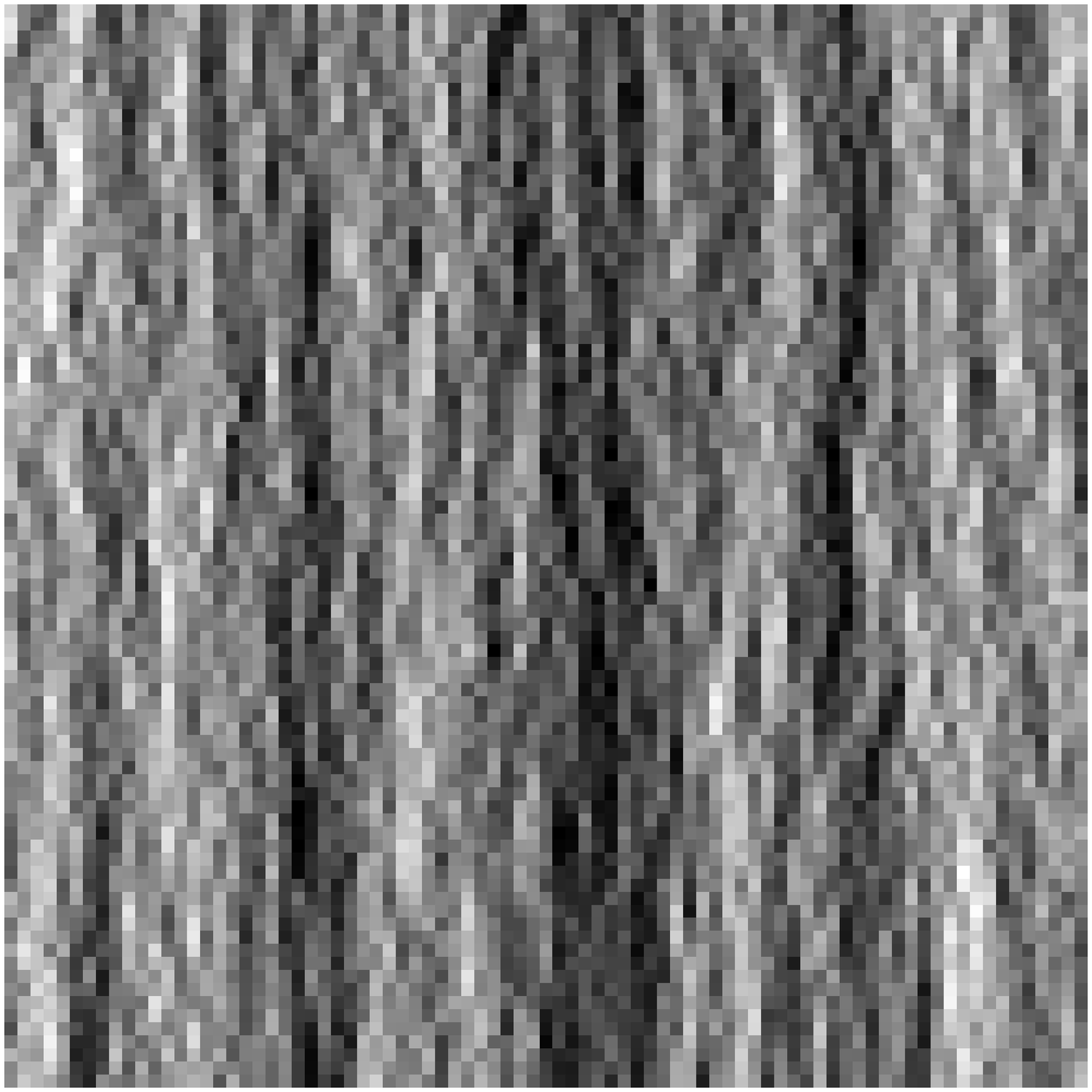}}
		\hspace{0.05cm}
		\framebox{\includegraphics[width=0.17\textwidth]{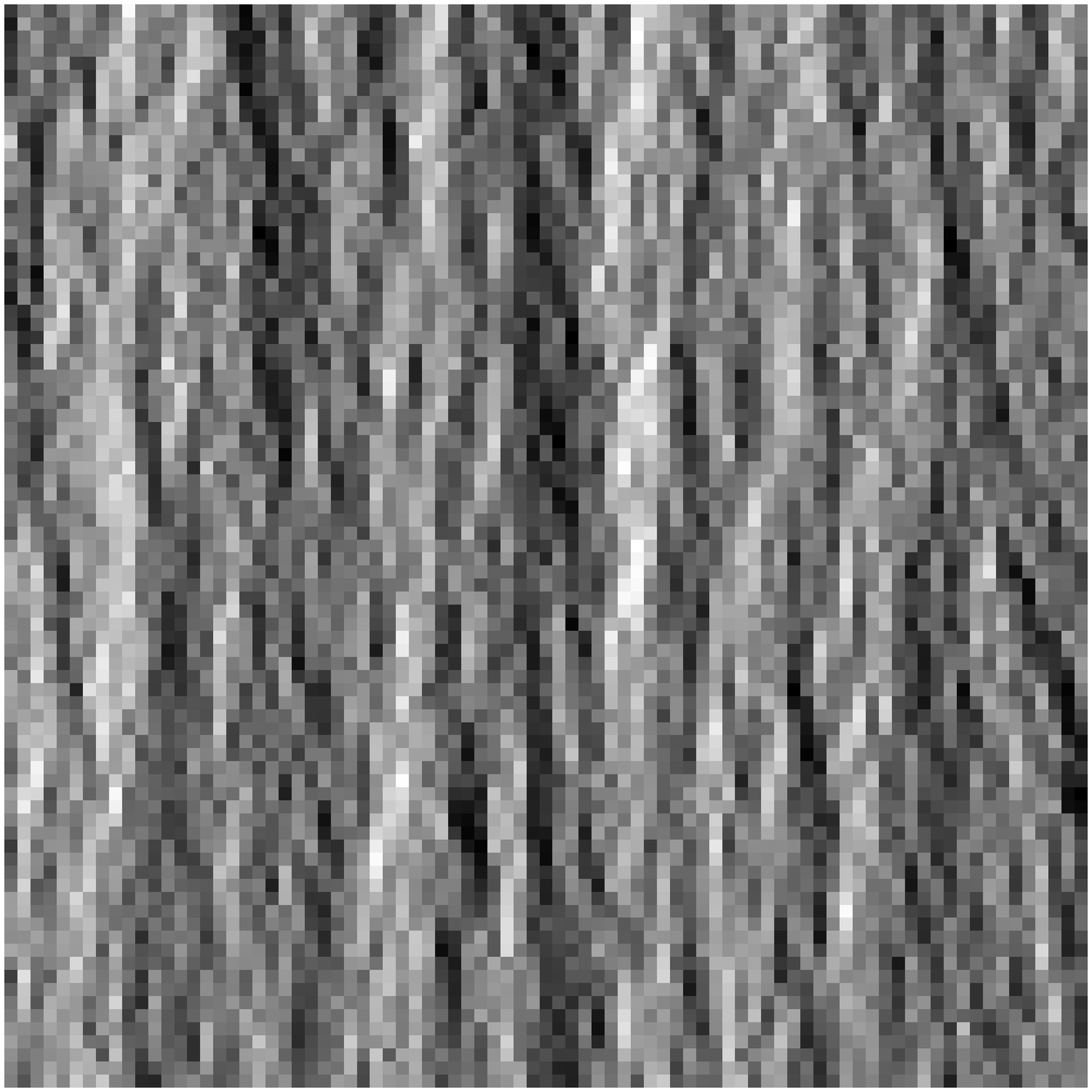}}
		\hspace{0.05cm}
		\framebox{\includegraphics[width=0.17\textwidth]{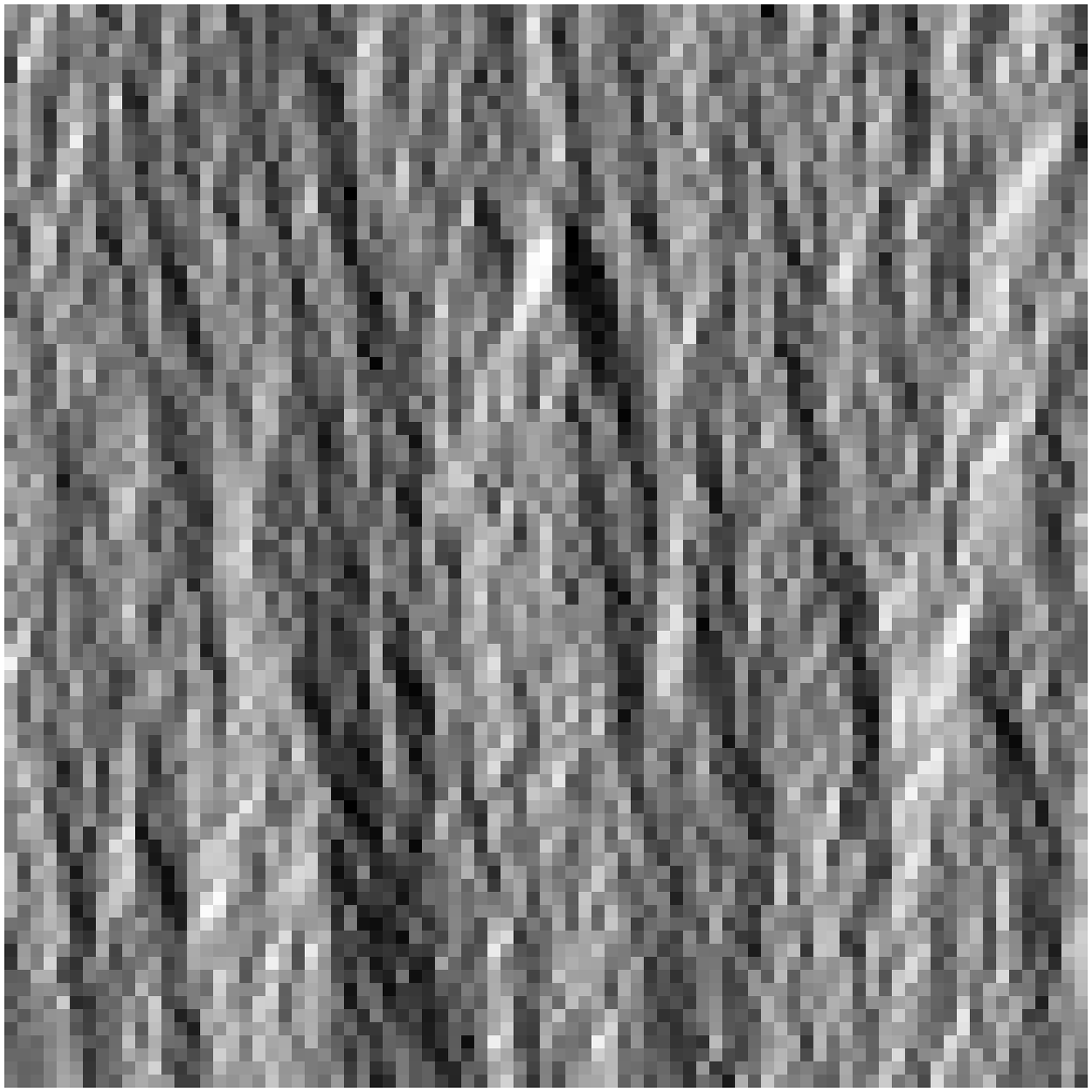}}\\
	\vspace{0.15cm}
		\framebox{\includegraphics[width=0.17\textwidth]{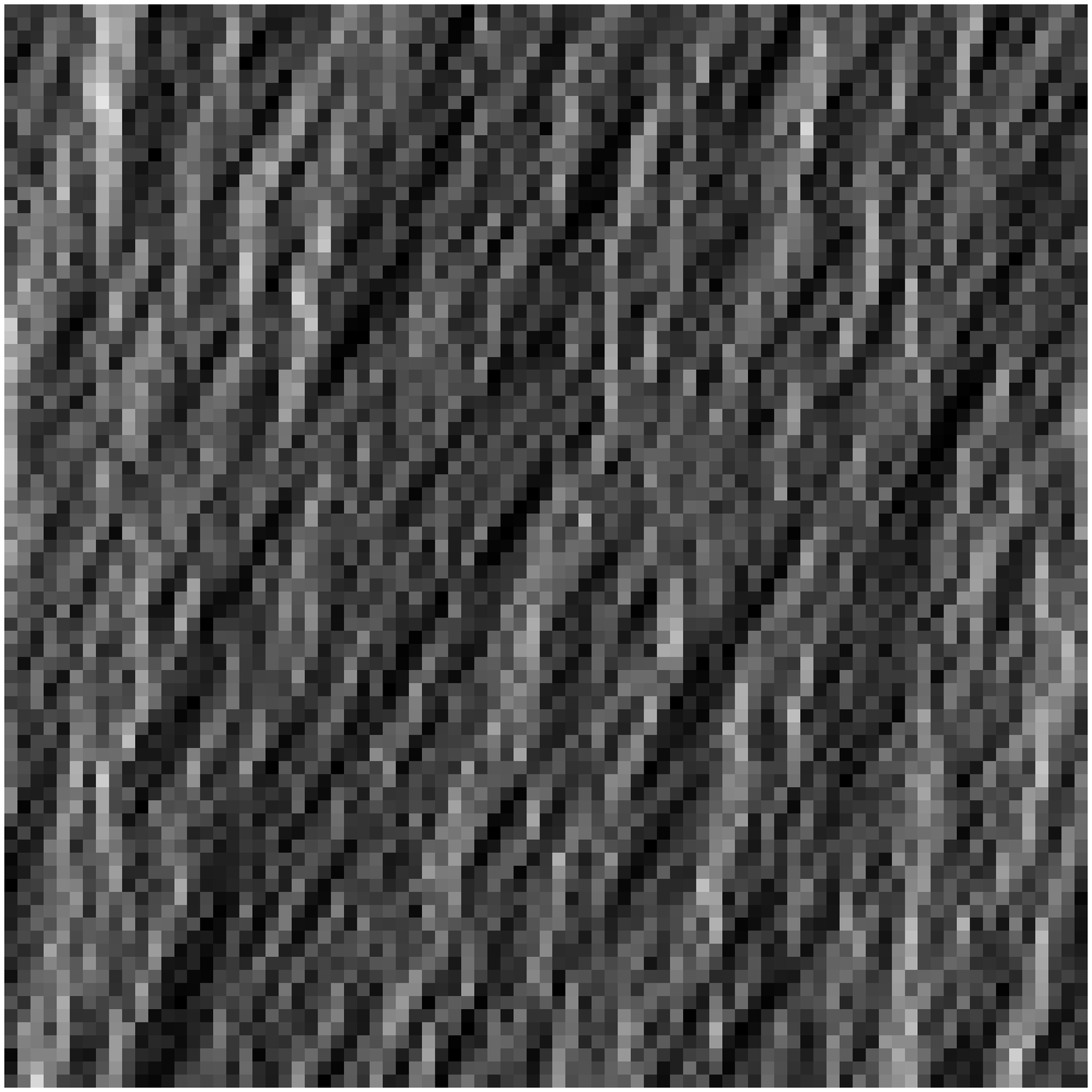}}
		\hspace{0.05cm}
		\framebox{\includegraphics[width=0.17\textwidth]{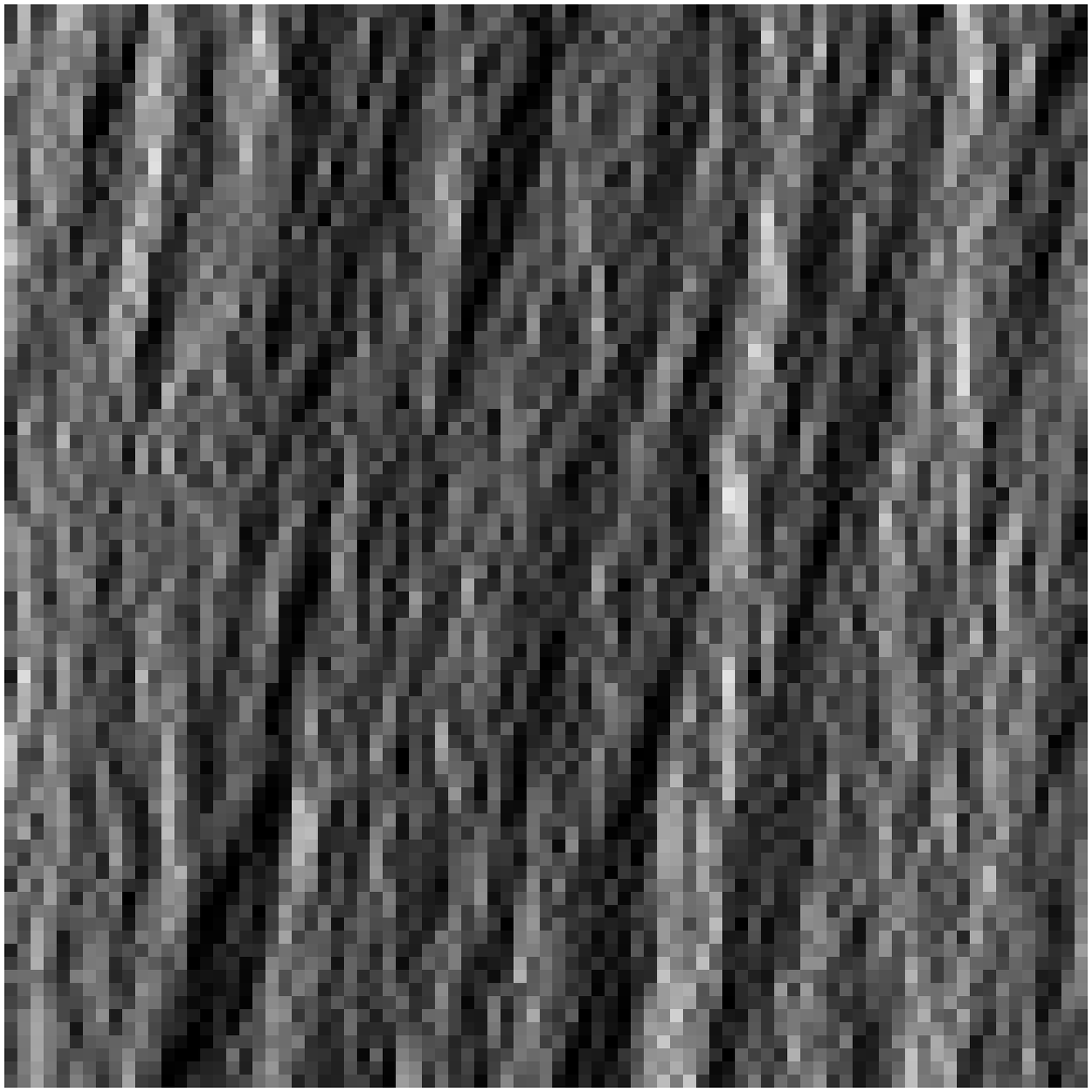}}
		\hspace{0.05cm}
		\framebox{\includegraphics[width=0.17\textwidth]{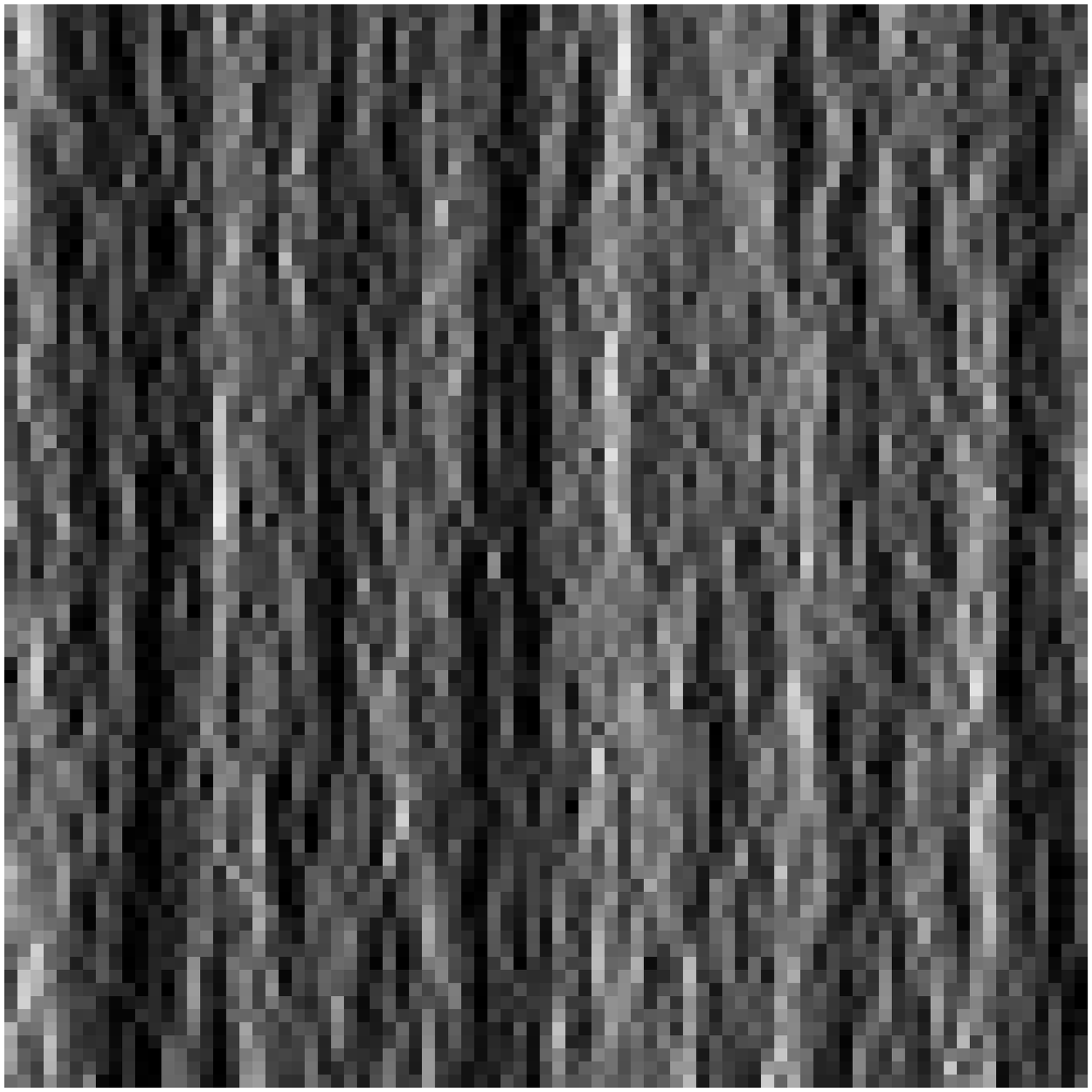}}
		\hspace{0.05cm}
		\framebox{\includegraphics[width=0.17\textwidth]{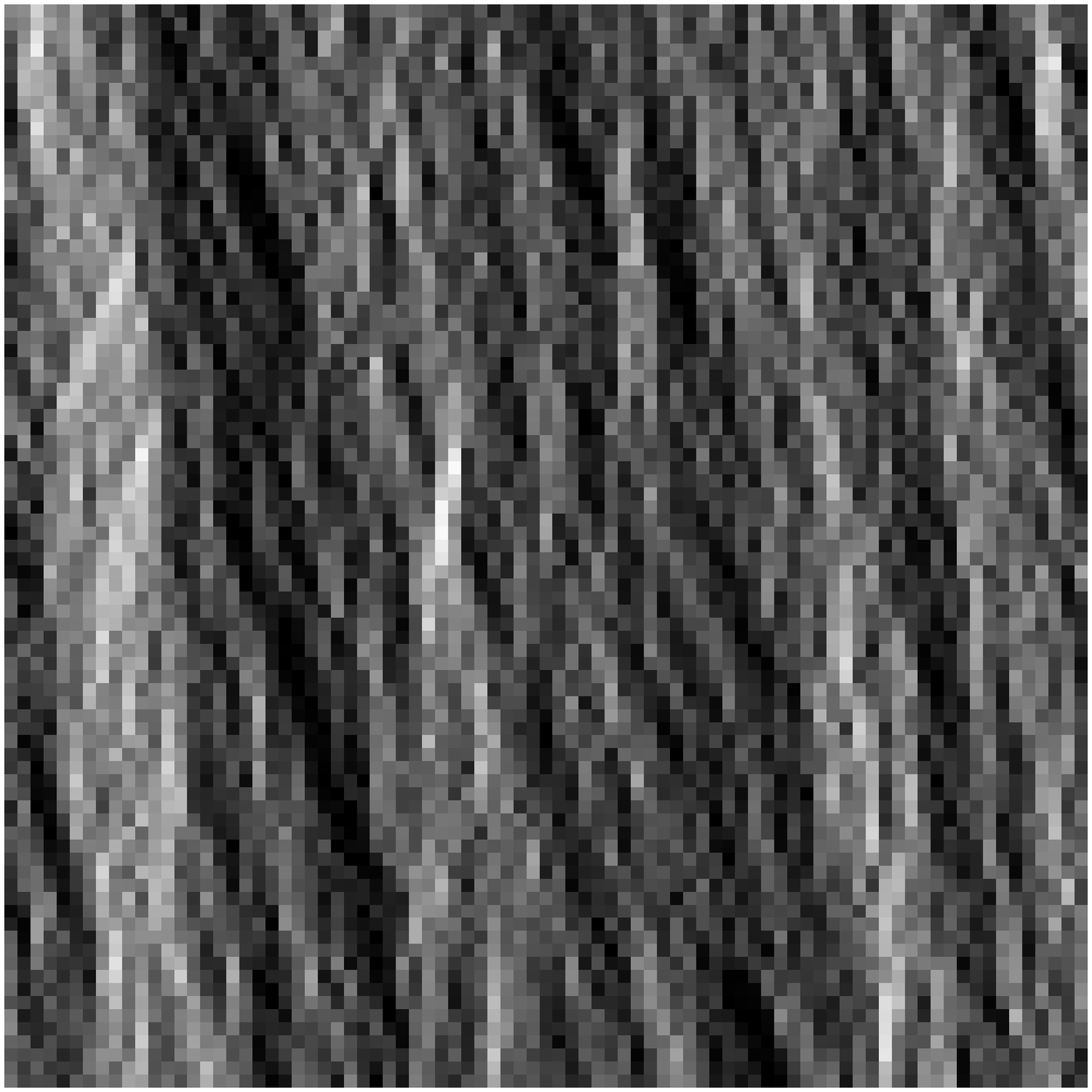}}
		\hspace{0.05cm}
		\framebox{\includegraphics[width=0.17\textwidth]{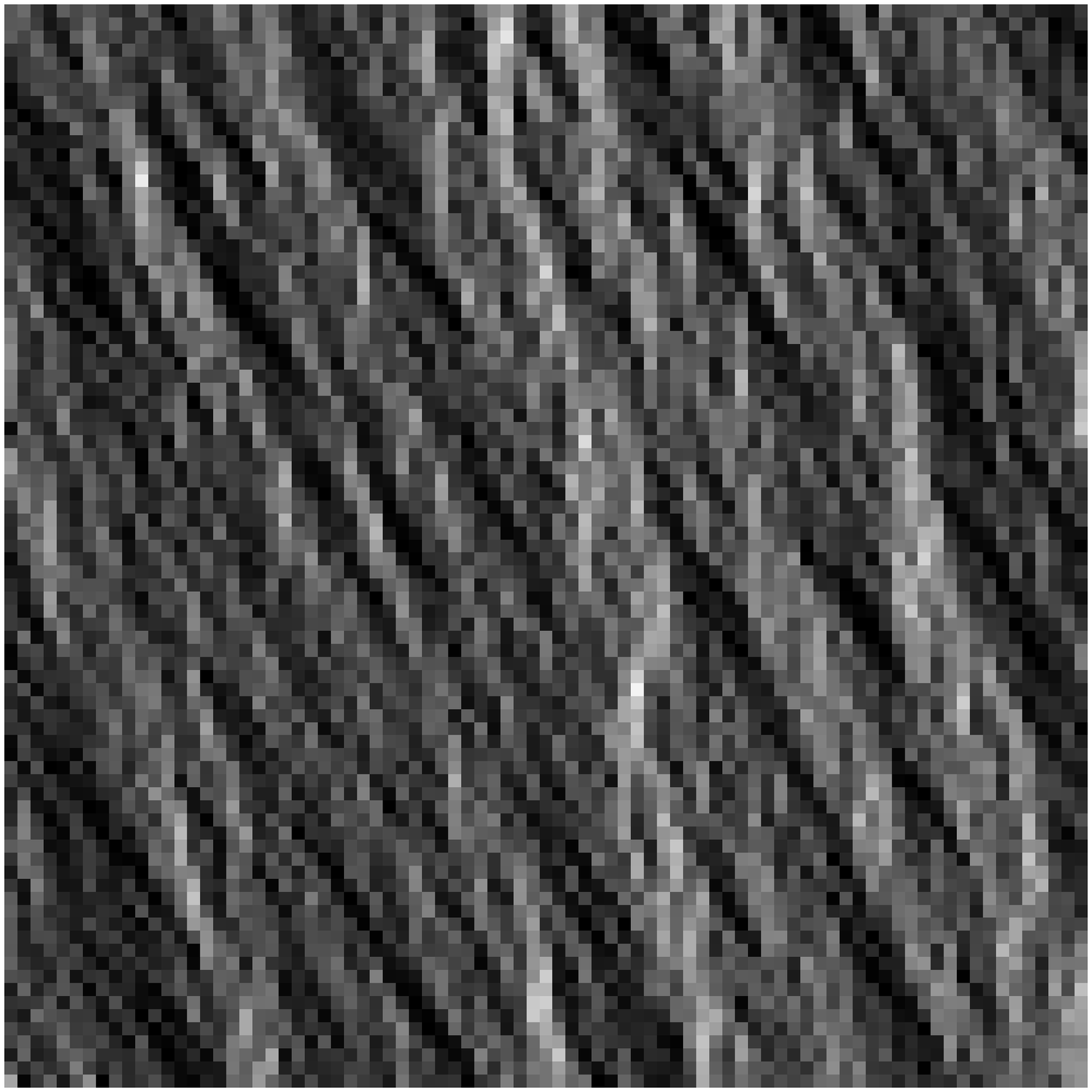}}\\
	\vspace{0.15cm}
		\framebox{\includegraphics[width=0.17\textwidth]{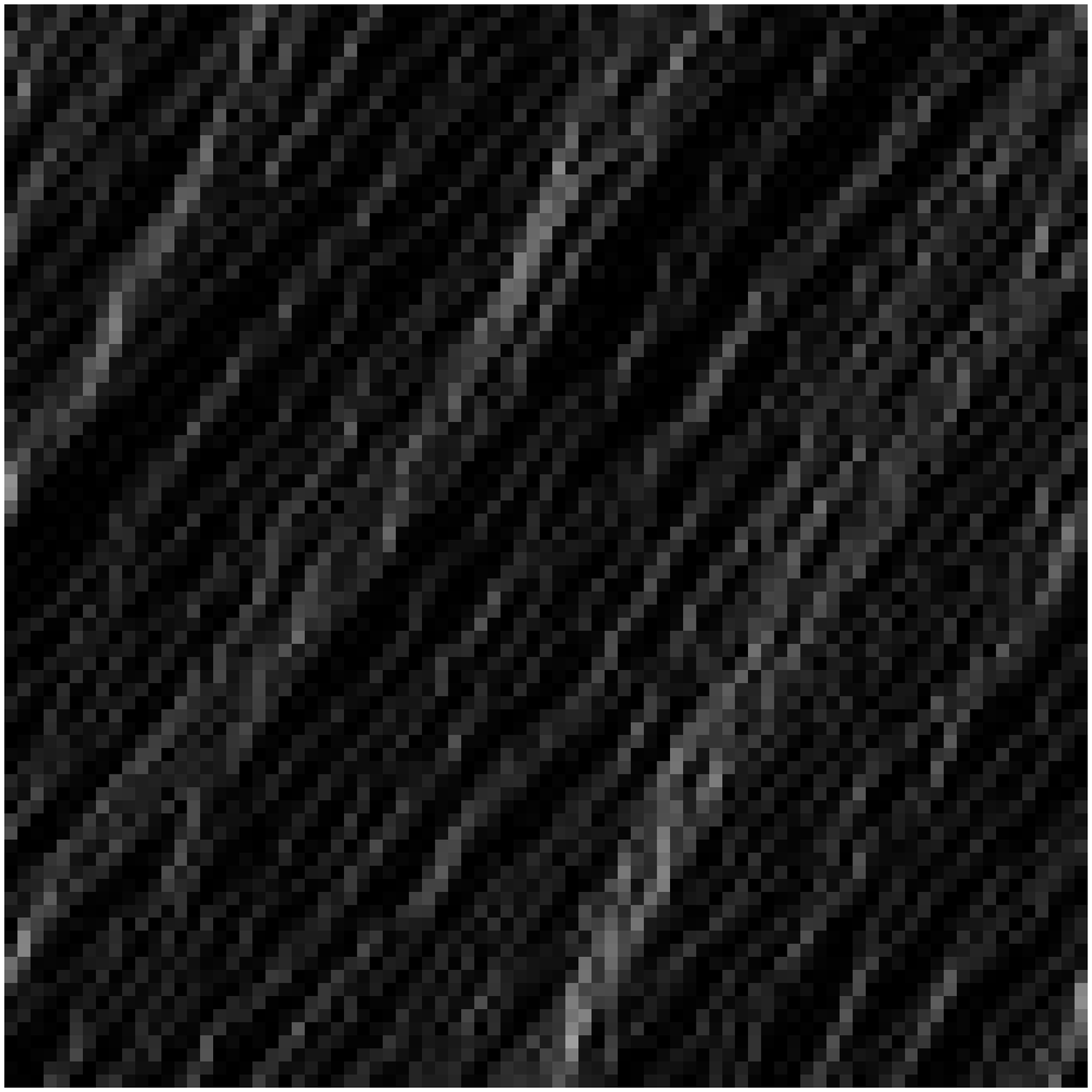}}
		\hspace{0.05cm}
		\framebox{\includegraphics[width=0.17\textwidth]{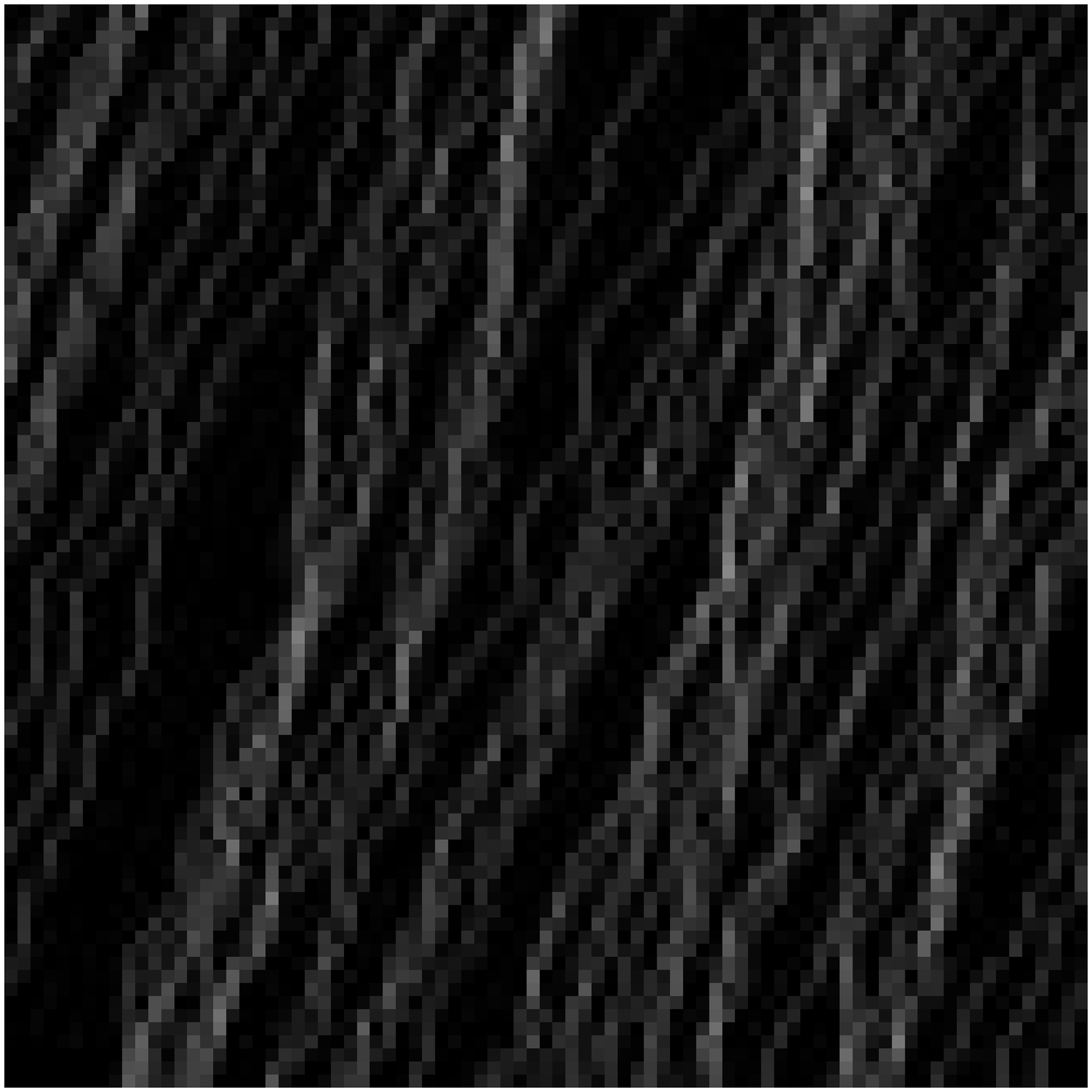}}
		\hspace{0.05cm}
		\framebox{\includegraphics[width=0.17\textwidth]{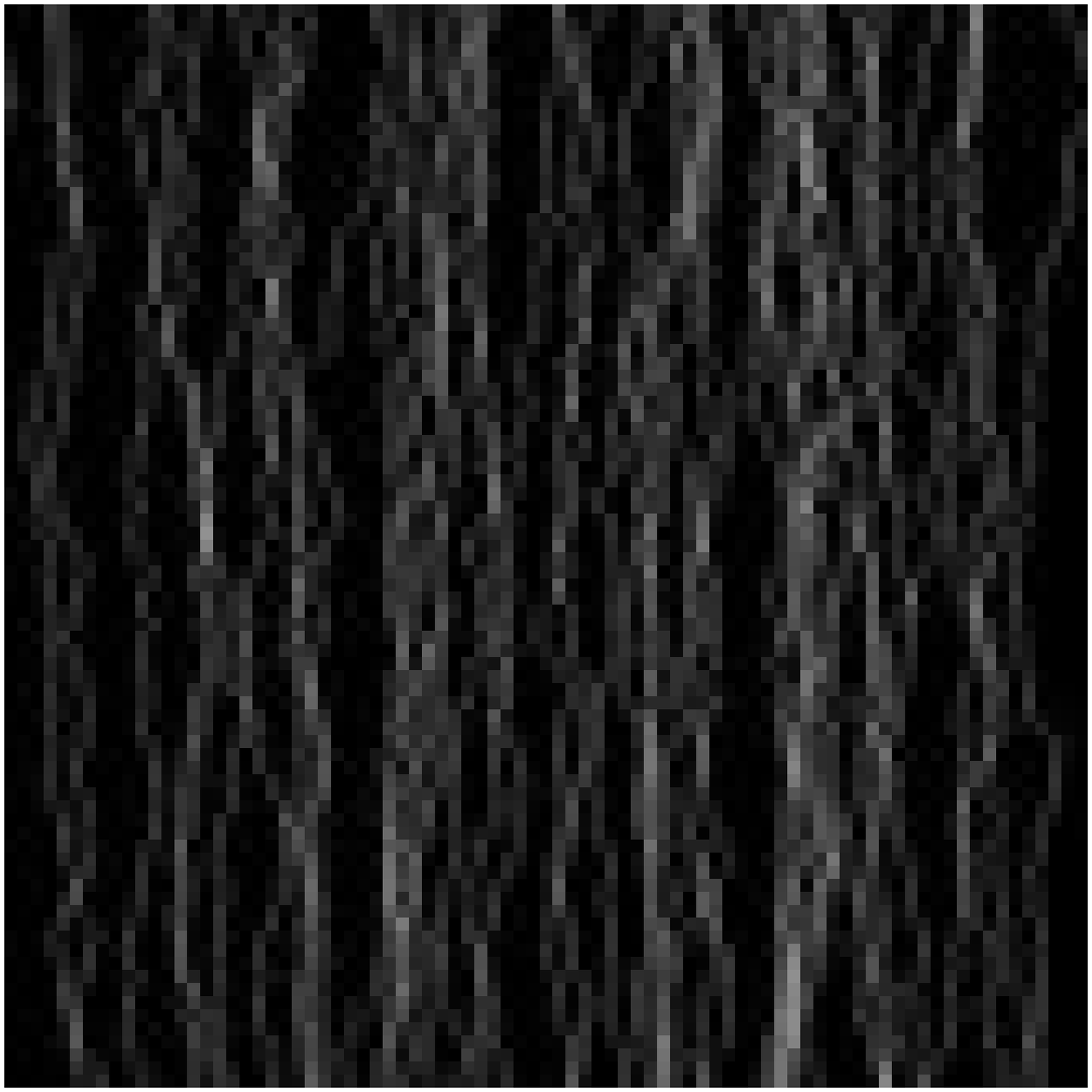}}
		\hspace{0.05cm}
		\framebox{\includegraphics[width=0.17\textwidth]{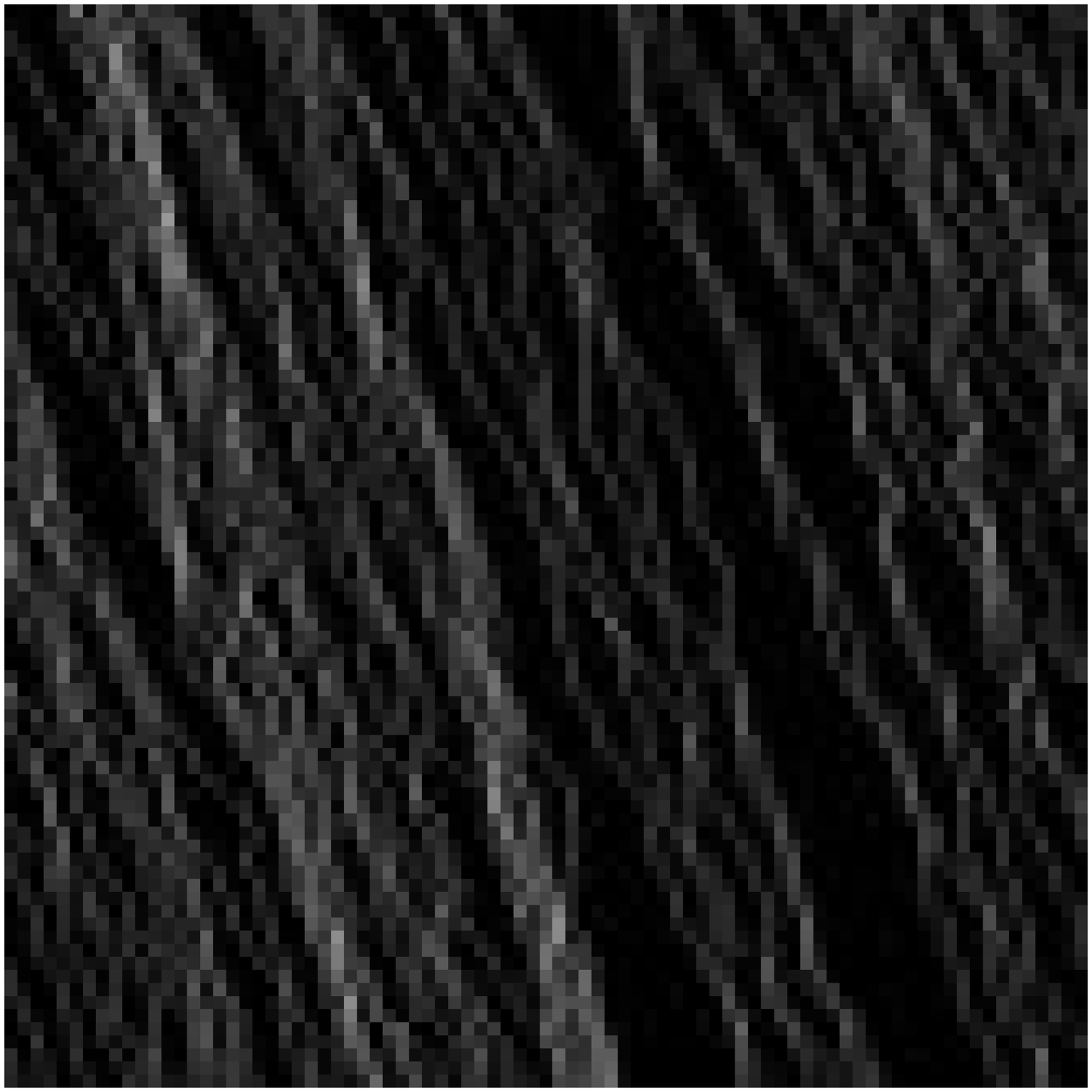}}
		\hspace{0.05cm}
		\framebox{\includegraphics[width=0.17\textwidth]{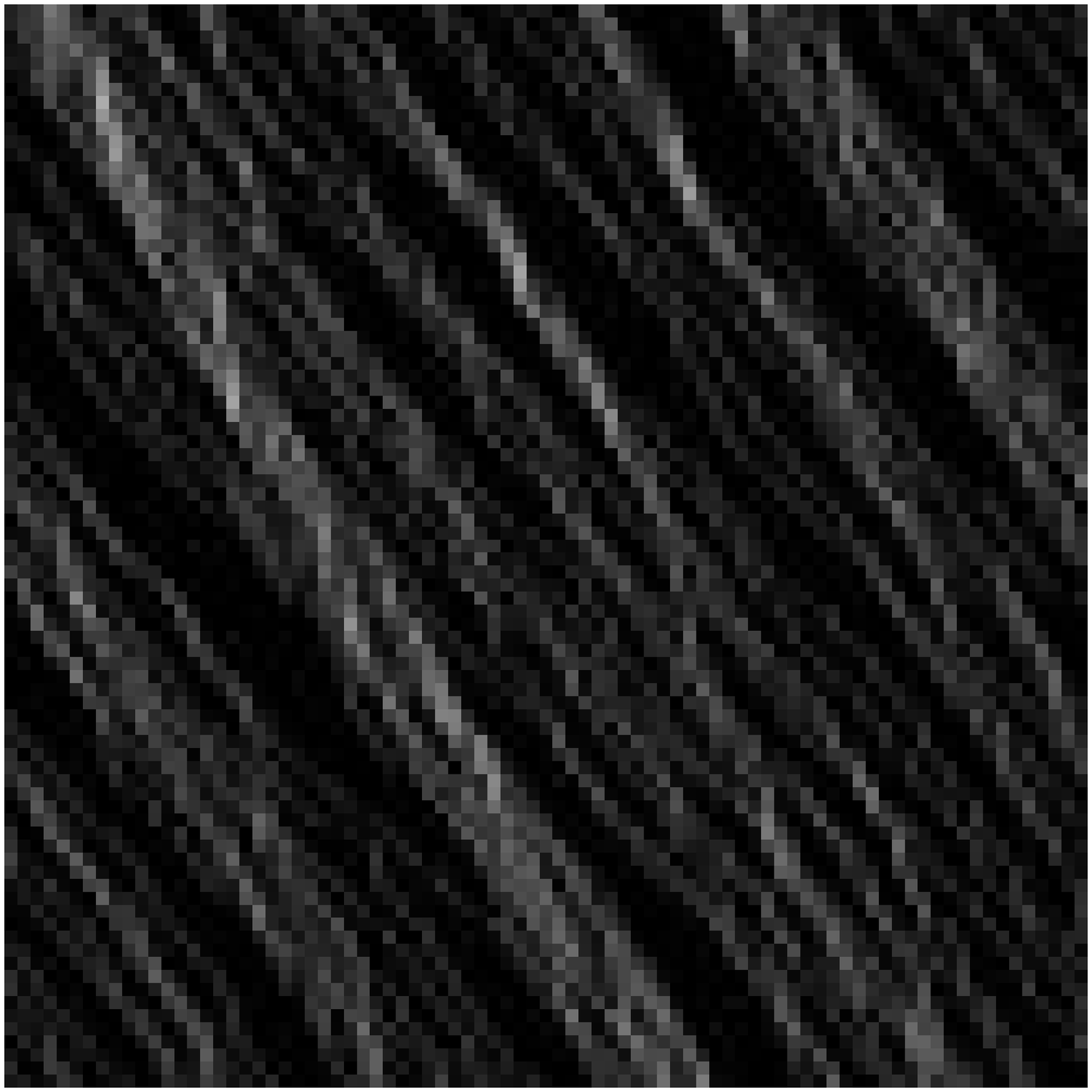}}
	\caption{
		Typical time-space diagrams of the asymmetric simple exclusion process (ASEP 
		model) with a random sequential update and $\gamma = \one - \delta$. The 
		shown lattices each contain 400 cells, with a visible period of 400 time 
		steps (note that for clarity, the space and time axes are located 
		horizontally and vertically, respectively). The global densities in the 
		systems were set for each row to $k \in \lbrace \text{0.1}, \text{0.3}, 
		\text{0.5}, \text{0.7}, \text{0.9} \rbrace$ vehicles/cell. For each column, 
		the ASEP's probability to move to the left was set to $\gamma \in \lbrace 
		\text{0.1}, \text{0.3}, \text{0.5}, \text{0.7}, \text{0.9} \rbrace$.
	}
	\label{fig:TCA:ASEPTimeSpaceDiagrams}
\end{figure*}

Updating the configuration of CA essentially amounts to updating the states of 
all its cells. In general, there are two methods for the update procedure:

\textbf{Sequential update}\\
This updating procedure considers each cell in the lattice one at a time. If all 
cells are considered consecutively, two updating directions are possible: 
\emph{left-to-right} and \emph{right-to-left}. There is also a third 
possibility, called \emph{random sequential update}. Under this scheme and with 
$N$ particles in the lattice, each time step is divided in $N$ smaller substeps. 
At each of these substeps, a random cell (or vehicle) is chosen and the CA rules 
are applied to it. As a consequence of the updating procedure, each particle is 
on average updated after $N$ smaller substeps, which introduces a certain amount 
of noise in the system. We have depicted several typical time-space diagrams for 
the ASEP with $\gamma = \one - \delta$ in 
\figref{fig:TCA:ASEPTimeSpaceDiagrams}. Furthermore note that a hidden 
assumption here is that, after completing a substep, the local information is 
immediately available to the whole system, which can violate causality (as 
information is now transmitted through the lattice at an infinite speed).\\

\textbf{Parallel update}\\
This is the classic update procedure that is used for all TCA models discussed 
in this report. For a parallel update, all cells in the system are updated in one 
and the same time step. Compared to a sequential updating procedure, this one is 
computationally more efficient (note that it is equivalent to a left-to-right 
sequential update). There is however one peculiarity associated with this 
updating scheme: because all particles are considered simultaneously, certain 
lattice configurations can not exist, i.e., the \emph{Garden of Eden} (GoE) 
states mentioned in section \ref{sec:TCA:IngredientsOfACellularAutomaton}. An 
example of such a \emph{paradisiacal state}, is two vehicles right behind each 
other, with the following having a non-zero speed. This state would imply that 
in single-lane traffic, the FIFO property was violated and consequently a 
collision occurred. Such GoE states do not exist when using a random sequential 
update.

\setlength{\fboxsep}{0pt}
\setlength{\fboxsep}{\tempfboxsep}

\begin{figure}[!htbp]
	\centering
	\includegraphics[width=\halffigurewidth]{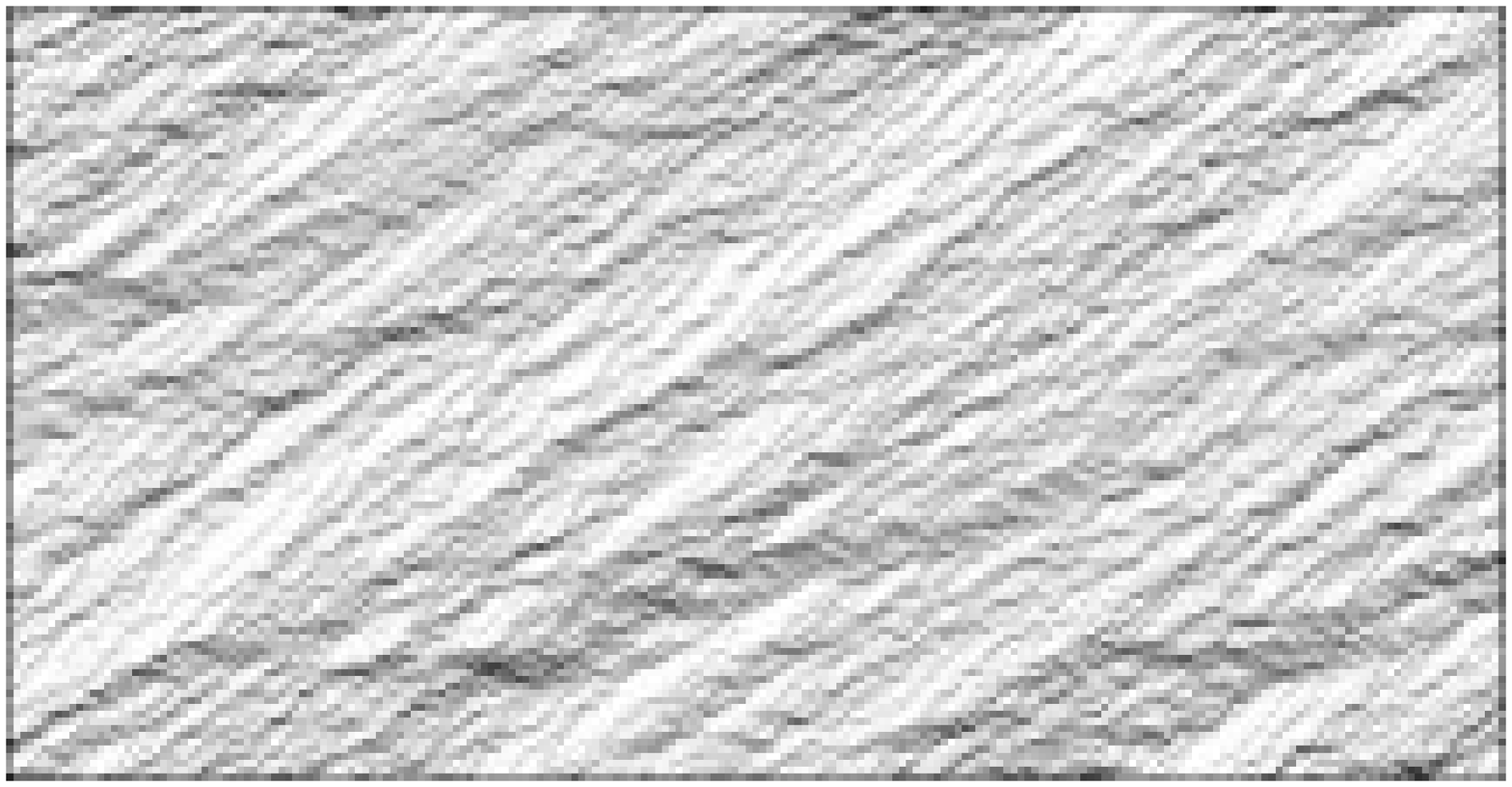}
	\hspace{\figureseparation}
	\includegraphics[width=\halffigurewidth]{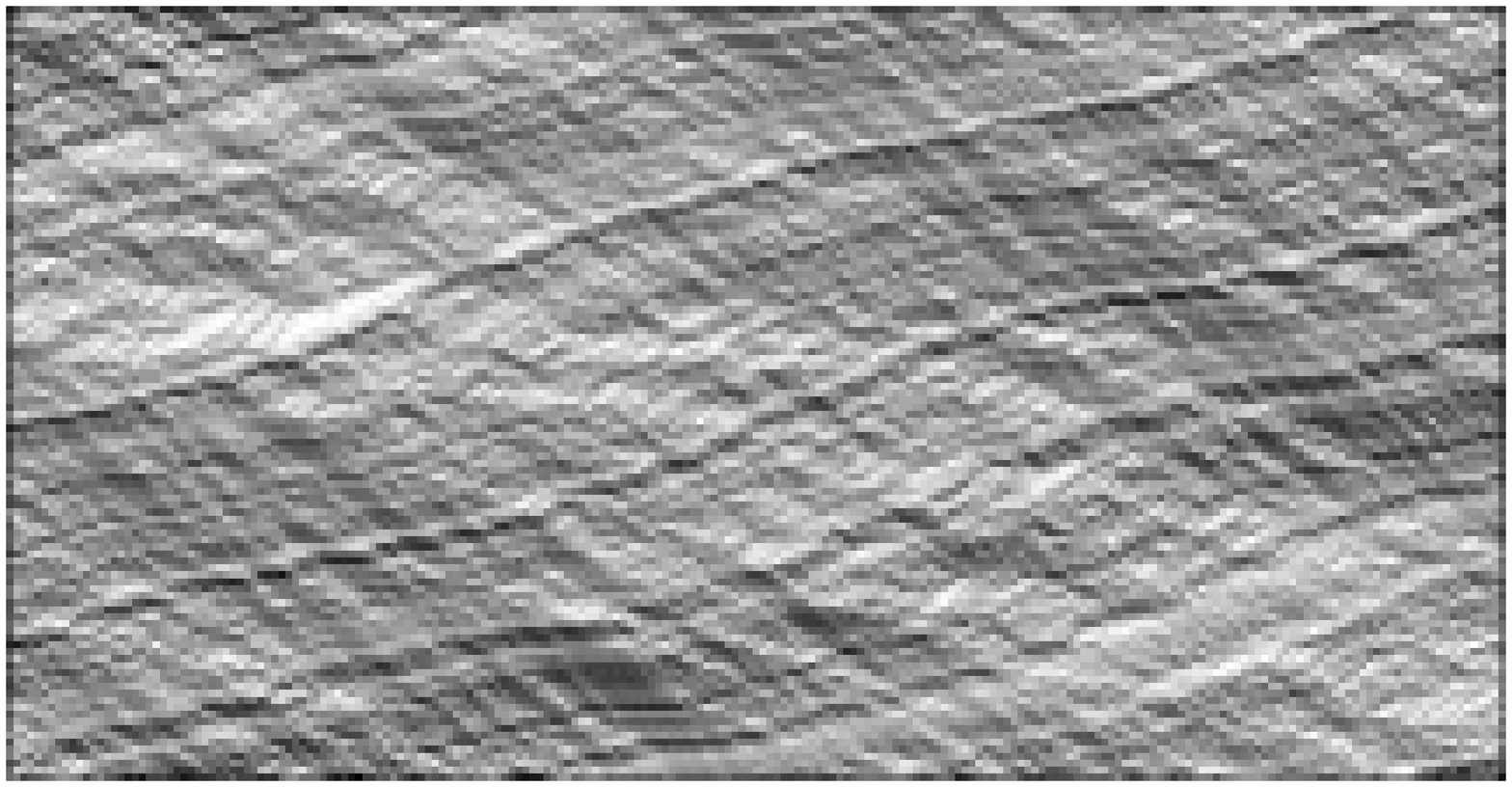}
	\caption{
		Typical time-space diagrams of the TASEP model with a random sequential 
		update. The shown lattices each contains 300 cells, with a visible period of 
		580 time steps. The global density in the system was set to $k = $~0.3 
		vehicles/cell (\emph{left}), and $k =$~0.7 vehicles/cell (\emph{right}). The 
		evolution of the system dynamics qualitatively looks the same in both 
		diagrams: the system is littered with mini-jams in both the low and high 
		density regimes.
	}
	\label{fig:TCA:TASEPTimeSpaceDiagrams}
\end{figure}

In \figref{fig:TCA:TASEPTimeSpaceDiagrams}, we have depicted two time-space 
diagrams for the TASEP with a random sequential updating procedure, operating on 
a closed loop. As can be seen, the diagrams qualitatively look the same, and 
have some of the same characteristic features of the time-space diagrams in 
\figref{fig:TCA:ASEPTimeSpaceDiagrams}. For the TASEP, there is no free-flow 
regime, there are no large jams in the system, and, because of the random 
sequential update, all vehicles continuously have the tendency to collide with 
each other. As a consequence, the system is littered with mini-jams in both the 
low and high density regimes \cite{NAGEL:95c,NAGEL:96}. Note that the TASEP with 
open boundary conditions exhibits a very rich behaviour, depending on the values 
for the entry and exit rates $\alpha$ and $\beta$, respectively 
\cite{KOLOMEISKY:98,SANTEN:99,SCHADSCHNEIDER:02}.

With respect to the relations between the TASEP with a random sequential update 
and other models, we mention the following two analogies: on the one hand, the 
LWR first-order macroscopic traffic flow model \cite{MAERIVOET:05c} corresponds 
to the TASEP in the hydrodynamic limit to a noisy and diffusive conservation 
law, which can be reduced to the LWR model \cite{NAGEL:95c,NAGEL:96}. On the 
other hand, the TASEP corresponds to the STCA (see section \ref{sec:TCA:STCA}), 
but now with $v_{\text{max}} = \one$ cell/time step 
\cite{CHOWDHURY:00,HELBING:01}.

\begin{figure}[!htbp]
	\centering
	\includegraphics[width=\halffigurewidth]{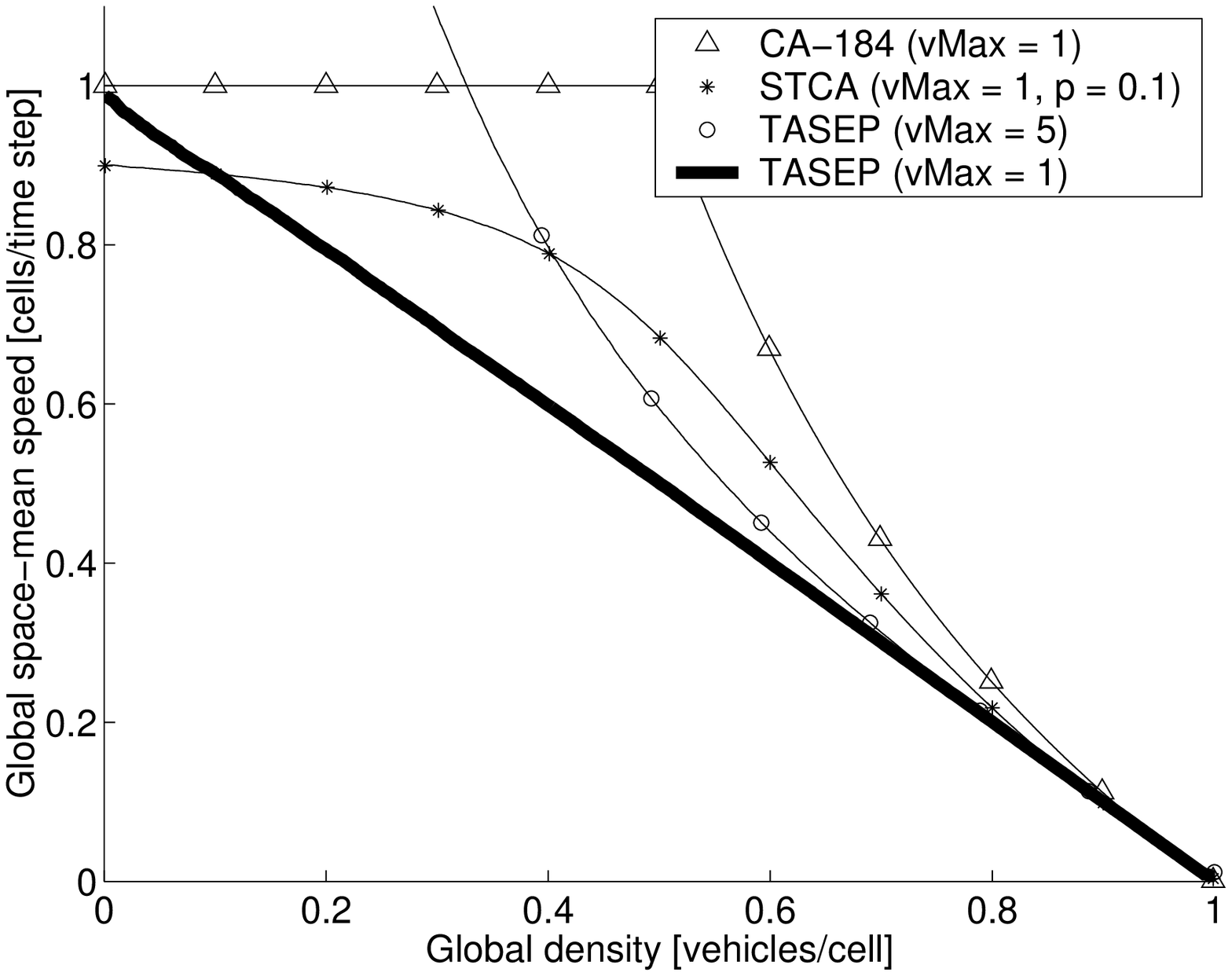}
	\hspace{\figureseparation}
	\includegraphics[width=\halffigurewidth]{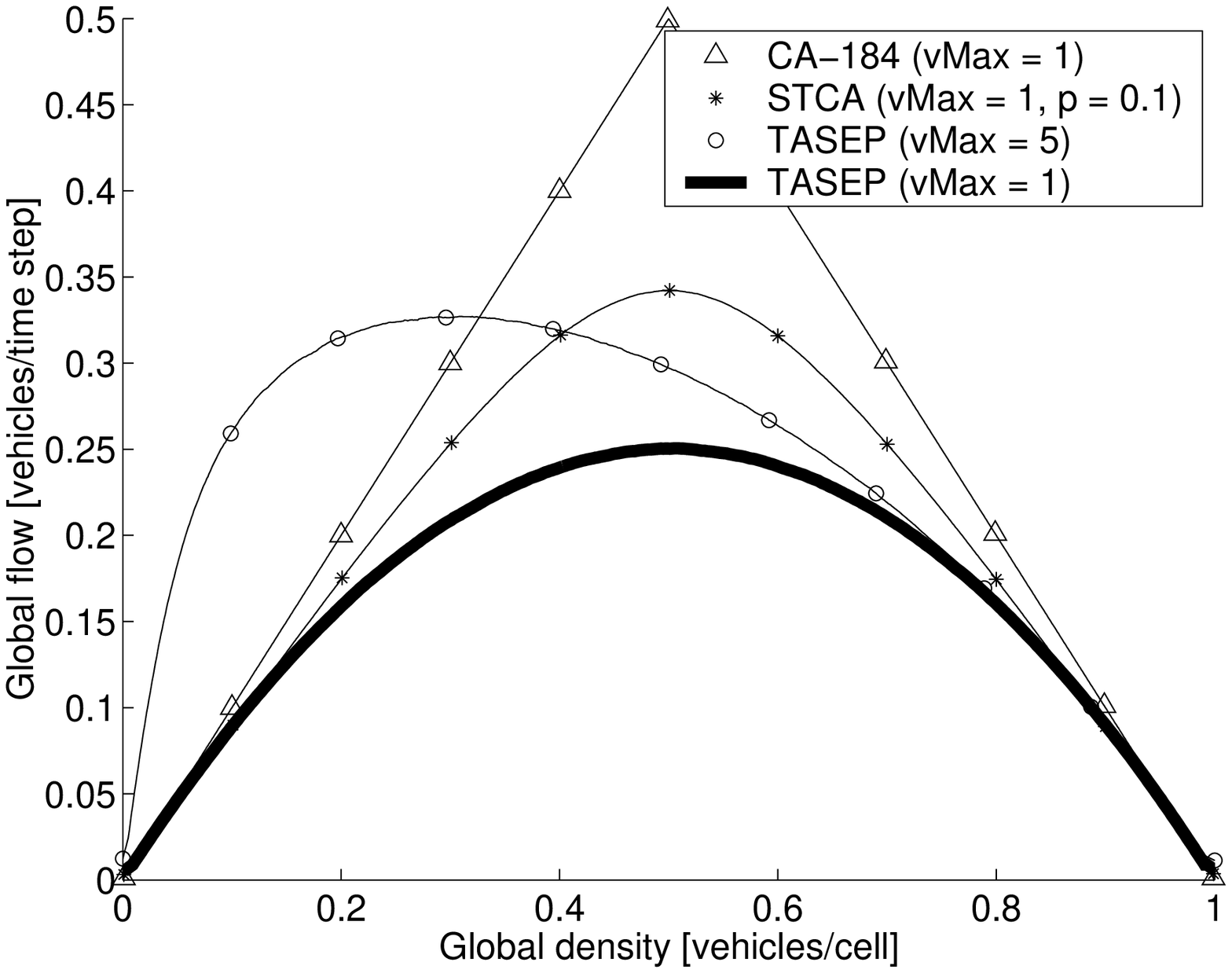}
	\caption{
		A comparison of the ($k$,$\overline v_{s}$) (\emph{left}) and ($k$,$q$) 
		(\emph{right}) diagrams for the CA-184 with $v_{\text{max}} = \one$ 
		($\Delta$), the STCA with $v_{\text{max}} = \one$ and $p =$~0.1 ($\star$), 
		the TASEP with random sequential update and $v_{\text{max}} = \five$ 
		($\circ$), and the TASEP with random sequential update and $v_{\text{max}} = 
		\one$ (thick solid line). Some distinct characteristics of the TASEP are the 
		absence of a free-flow regime, and for $v_{\text{max}} = \one$ cells/time 
		step, the exact correspondence with Greenshields linear relation between the 
		density and the mean speed.
	}
	\label{fig:TCA:TASEPDensityFlowSMSFundamentalDiagrams}
\end{figure}

To gain more insight into the macroscopic behaviour of the TASEP with random 
sequential update, we provide its ($k$,$\overline v_{s}$) and ($k$,$q$) diagrams 
in \figref{fig:TCA:TASEPDensityFlowSMSFundamentalDiagrams}. Looking at the 
($k$,$\overline v_{s}$) diagram on the left part, we notice that the TASEP with 
$v_{\text{max}} = \one$ cell/time step corresponds exactly to Greenshields' 
original linear relation between the density and the mean speed 
\cite{GREENSHIELDS:35,MAERIVOET:05d}. This in fact is a further testimony of the 
close link between the TASEP and the LWR model with a triangular $q_{e}(k)$ 
fundamental diagram. Increasing the TASEP's maximum speed, leads to a more 
curved relation, intersection the vertical axis at the point 
(0,$v_{\text{max}}$). In any case, the ($k$,$\overline v_{s}$) diagram also 
reveals the absence of a distinct free-flow branch, corresponding to the 
observations of the large amount of mini-jams for all global densities, as could 
be seen in the time-space diagrams of \figref{fig:TCA:TASEPTimeSpaceDiagrams}.

Studying the ($k$,$q$) diagram in the right part of 
\figref{fig:TCA:TASEPDensityFlowSMSFundamentalDiagrams}, we can see that the 
TASEP corresponds with the STCA for $v_{\text{max}} = \one$ and an arbitrary 
slowdown probability (e.g., $p =$~0.1). The diagram also shows how the CA-184 
leads to a sharp transition between the free-flow and the congested regime, as 
opposed to the rounded peak of capacity flow at $k =$~0.5 vehicles/cell for the 
STCA. However, whereas the TASEP also has its capacity flow at the same value, 
there does not occur such a phase transition as in the other models. Finally, we 
can see that increasing the maximum speed $v_{\text{max}}$ for the TASEP 
introduces no significant qualitative changes, except for a skewing towards 
lower densities \cite{NAGEL:95c}.\\

\sidebar{
	Note that with respect to the computational complexity of the implemented TCA 
	models, most measurements in this report took a few hours to obtain, using an 
	Intel P4 2.8 GHz with 512 MB RAM, running the Java$^{\text{\tiny{TM}}}$ JDK 
	1.3.1 under Windows 2000. In sharp contrast to this, are the computations for 
	the TASEP model, which took nearly \emph{two weeks} to complete.
}\\

			\subsubsection{Emmerich-Rank TCA (ER-TCA)}
			\label{sec:TCA:ERTCA}

Whereas the classical STCA model provided a reasonable qualitative agreement 
with real-world observations, Emmerich and Rank addressed the quantitative 
discrepancies between the model and real-world data. To this end, they proposed 
a variation on the STCA, extending the influence of the space gap on a vehicles 
updated speed \cite{EMMERICH:97}.

In their work, Emmerich and Rank fundamentally modified the STCA in two steps: 
(i) they changed the parallel update procedure to a \emph{right-to-left 
sequential update procedure} (see section \ref{sec:TCA:TASEP} for more details), 
and (ii) they changed the behaviour of vehicles that are slowing down. In a 
nutshell, (i) leads to the important result that vehicles are now able to drive 
directly behind each other (i.e., with a zero space gap) at high speeds, because 
the gaps in a traffic stream are used more efficiently. The reason is that due 
to the specific sequential update, a downstream vehicle is moved first (for a 
closed loop, the vehicle with the largest space gap is chosen first), after 
which the next vehicle upstream will see a larger space gap. 

Just as the STCA can be seen as a special case of the optimal velocity model 
(OVM), based on a linear optimal velocity function (for a description of the 
OVM, we refer the reader to our overview in \cite{MAERIVOET:05c}), the ER-TCA 
model generalises this function by making a vehicle's speed dependent on a 
variable safe distance and its current speed \cite{CHOWDHURY:00}. This affects 
(ii), i.e., vehicles that are slowing down: when determining the new speed of a 
vehicle, the ER-TCA model first checks if the vehicle is within 10 cells of its 
direct frontal leader. If this is the case, then the vehicle will slow down 
according to a table lookup in a \emph{gap-speed matrix} $M_{g_{s_{i}},v_{i}}$. 
This matrix is constructed in such a way that collisions are avoided (i.e., 
$M_{i,j} \leq \text{min} \lbrace i, j \rbrace$):

\begin{equation}
\label{eq:TCA:ERTCAGapSpeedMatrix}
	M^{\text{T}} =
		\left (
			\begin{array}{ccccccccccc}
				\zero & \zero & \zero & \zero  & \zero  & \zero  & \zero  & \zero  & \zero  & \zero  & \zero\\
				\zero & \one  & \one  & \one   & \one   & \one   & \one   & \one   & \one   & \one   & \one\\
				\zero & \one  & \two  & \two   & \two   & \two   & \two   & \two   & \two   & \two   & \two\\
				\zero & \one  & \two  & \three & \three & \three & \three & \three & \three & \three & \three\\
				\zero & \one  & \two  & \three & \four  & \four  & \four  & \four  & \four  & \four  & \four\\
				\zero & \one  & \two  & \three & \four  & \four  & \four  & \four  & \four  & \four  & \five\\
			\end{array}
		\right )
\end{equation}

\begin{figure}[!htbp]
	\centering
	\includegraphics[width=\halffigurewidth]{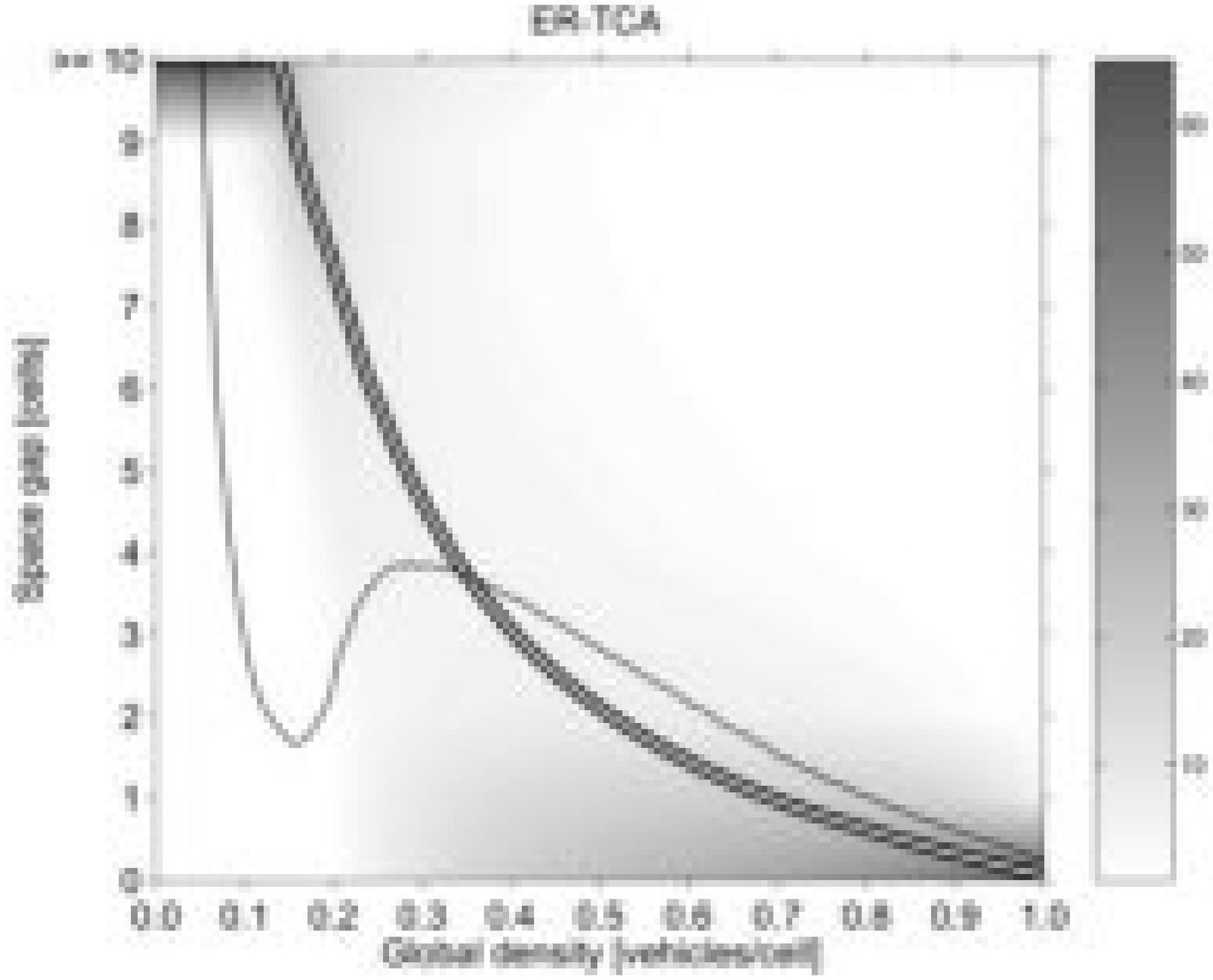}
	\hspace{\figureseparation}
	\includegraphics[width=\halffigurewidth]{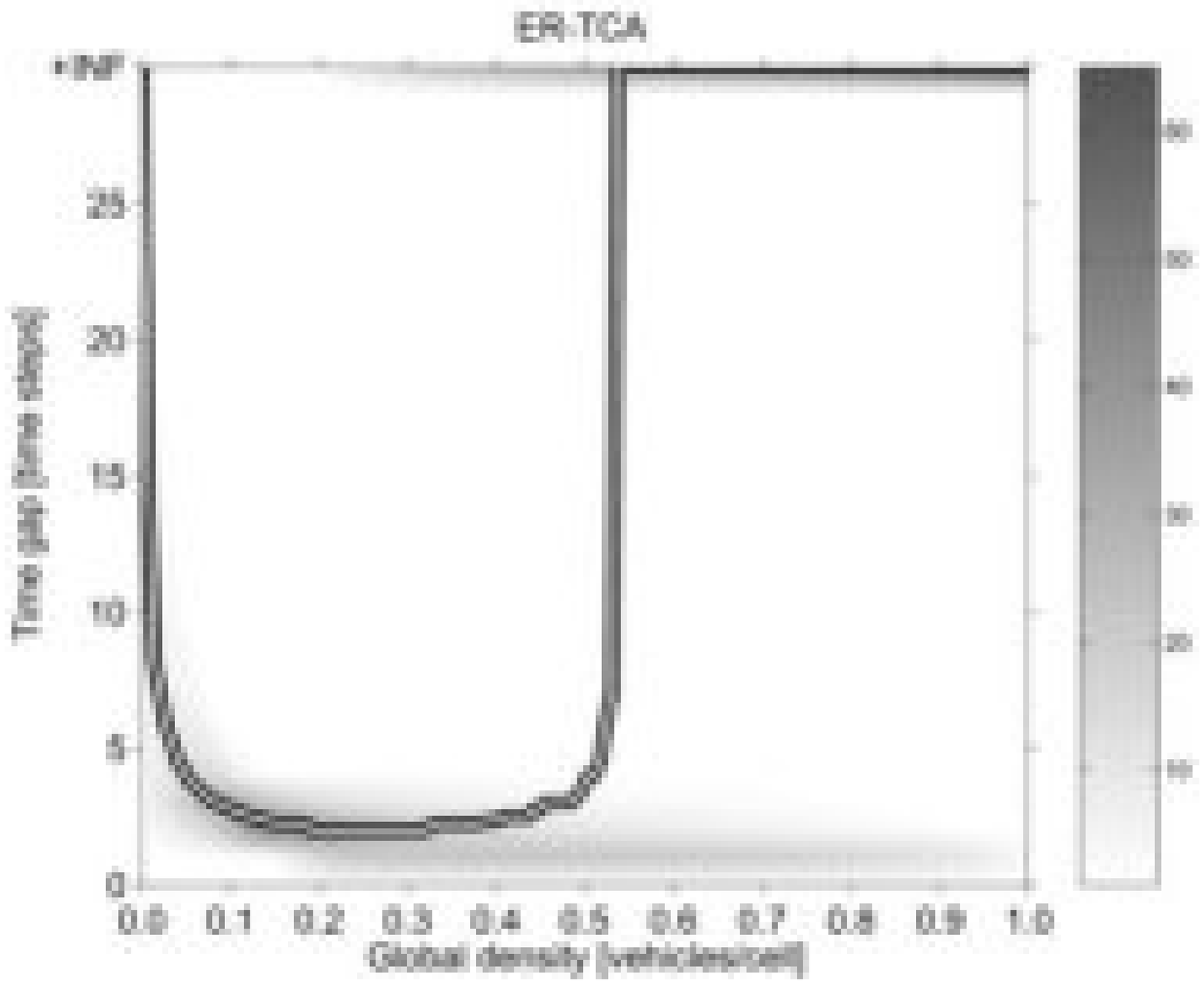}
	\caption{
		Histograms of the distributions of the vehicles' space gaps $g_{s}$ 
		(\emph{left}) and time gaps $g_{t}$ (\emph{right}), as a function of the 
		global density $k$ in the ER-STCA (with $v_{\text{max}} = \five$ cells/time 
		step and $p = $~0.35). The thick solid lines denote the mean space gap and 
		median time gap, whereas the thin solid line shows the former's standard 
		deviation. The grey regions denote the probability densities.
	}
	\label{fig:TCA:ERTCASpaceGapTimeGapHistogram}
\end{figure}

The matrix in equation \eqref{eq:TCA:ERTCAGapSpeedMatrix}, conveys the idea that 
lower speeds require lower space gaps, and that vehicles tend to keep larger 
space gaps when travelling at higher speeds. This latter effect is also visible 
in the distribution of the vehicles' space gaps, as visualised in the histograms 
in the left part of \figref{fig:TCA:ERTCASpaceGapTimeGapHistogram}, where, in 
contrast to the STCA's space gaps distribution of 
\figref{fig:TCA:STCASpaceGapHistogram}, large space gaps are observed for 
densities near the critical density. Furthermore, because of this mechanism, 
vehicles will have smoother decelerations, instead of the abrupt slowing down in 
the STCA model and some of its variations.

\begin{figure}[!htbp]
	\centering
	\includegraphics[width=\halffigurewidth]{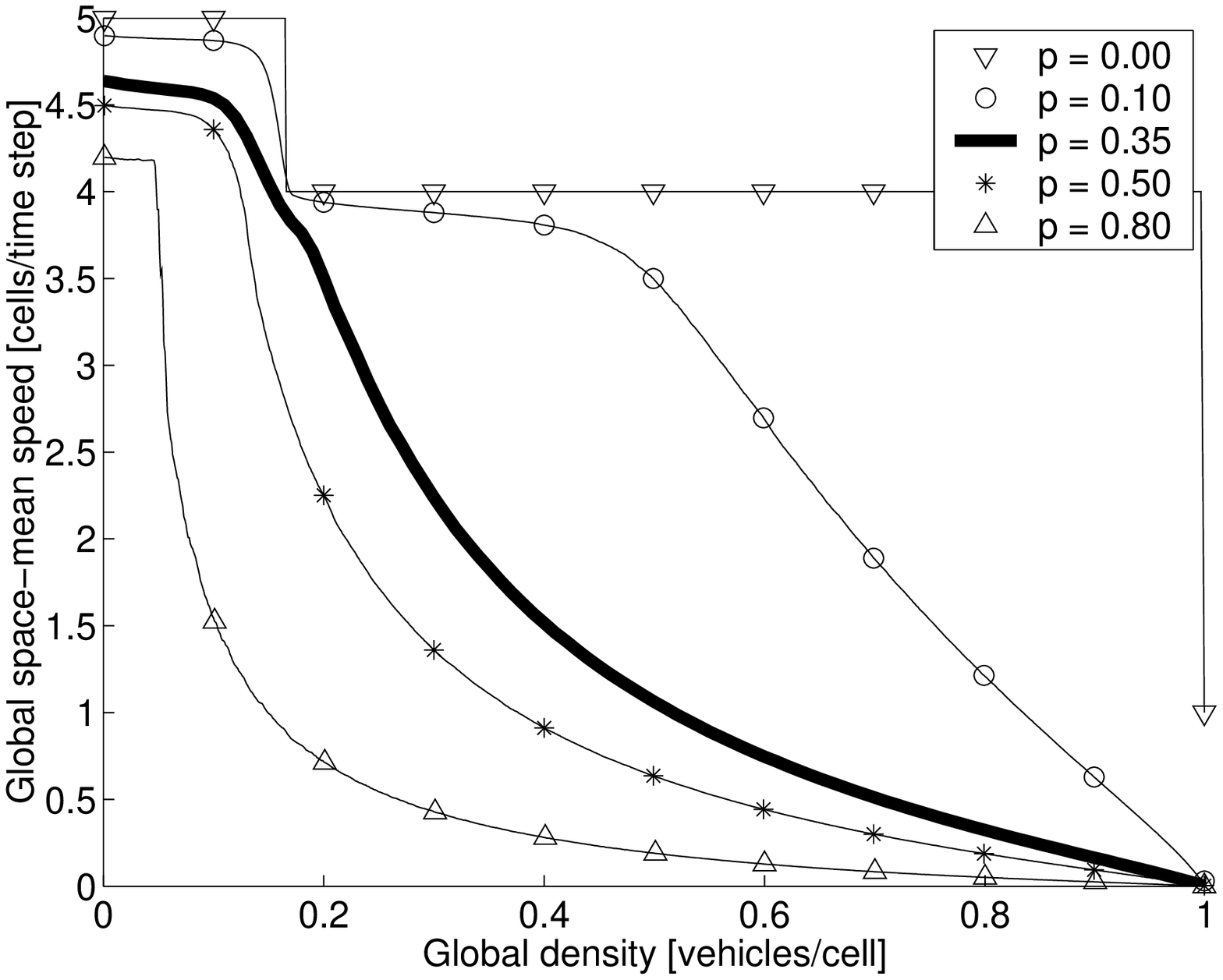}
	\hspace{\figureseparation}
	\includegraphics[width=\halffigurewidth]{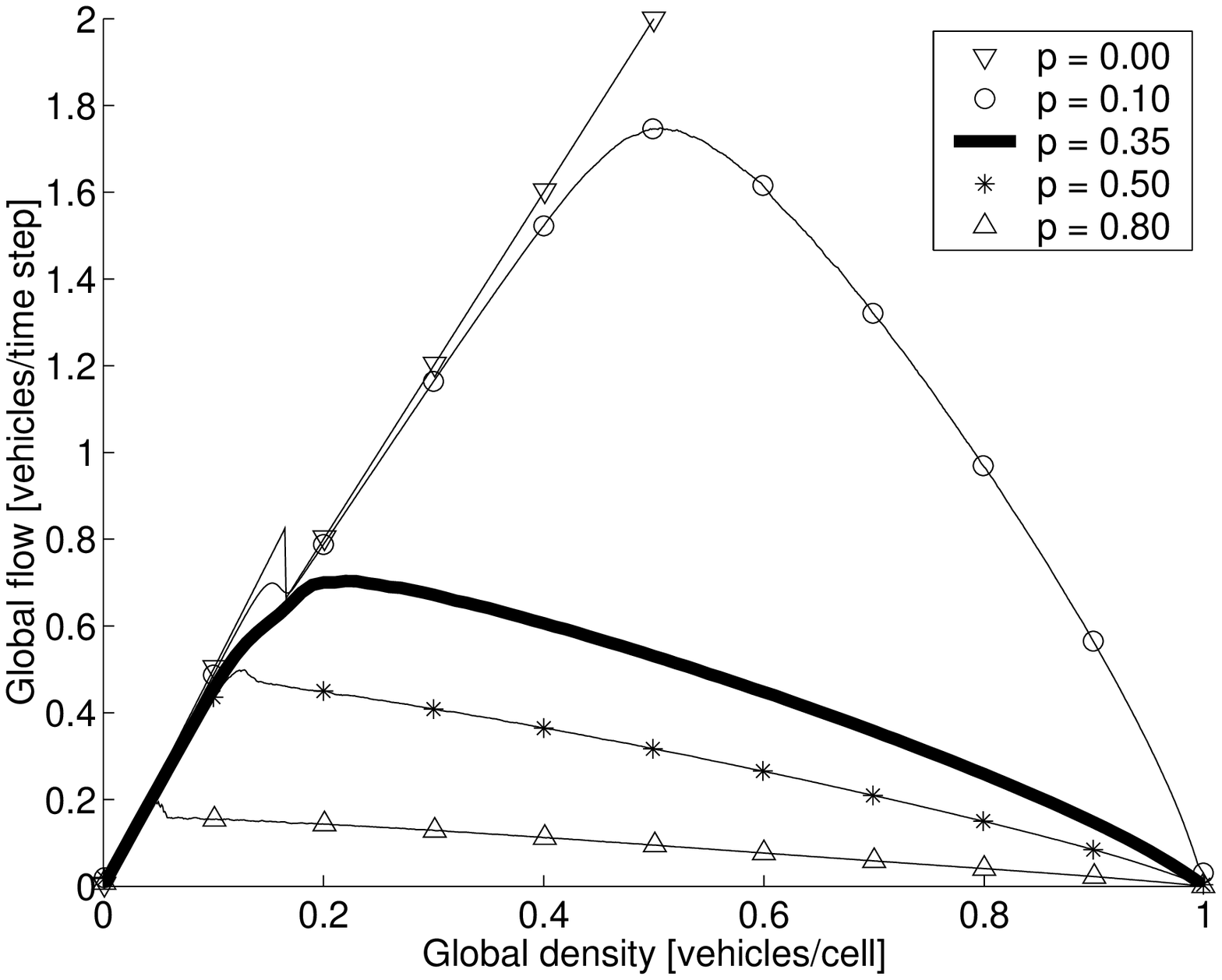}
	\caption{
		\emph{Left:} several ($k$,$\overline v_{s}$) diagrams for the ER-TCA, each 
		for a different slowdown probability $p$. It is clear from the diagram, that 
		for low values of $p$, the resulting diagrams are unrealistic, including 
		plateaus of constant space-mean speed in the congested regime. \emph{Right:} 
		several ($k$,$q$) diagrams for the same ER-TCA models as before. Due to the 
		system dynamics in the ER-TCA, very high capacity flows are possible. To 
		constrain these flows, the slowdown probability $p$ has to be quite large in 
		order to obtain realistic results. In both parts of the figure, the thick 
		solid line denotes the original model of Emmerich and Rank, who used a value 
		$p =$~0.35 as their best fit to experimental data.
	}
	\label{fig:TCA:ERTCADensityFlowSMSFundamentalDiagrams}
\end{figure}

To understand some of the system dynamics of the ER-TCA model, we have provided 
several ($k$,$\overline v_{s}$) and ($k$,$q$) diagrams in 
\figref{fig:TCA:ERTCADensityFlowSMSFundamentalDiagrams}. For $p =$~0.35, we can 
see in the ($k$,$q$) diagram in right part, that the free-flow branch gets 
\emph{curved}, implying that vehicles travel at a slightly lower speed when they 
approach the capacity-flow regime. Because vehicles can travel at high speeds in 
dense platoons, the ER-TCA model can achieve very high capacity flows, even 
leading to $q > \one$ vehicle/time step. In order to constrain these flows to 
realistic values, the ER-TCA model needs a quite high slowdown probability, 
e.g., $p =$~0.35.

These two effects, i.e., a curving of the free-flow branch and an increased 
capacity flow, are basically what the ER-TCA model is all about, as there is no 
qualitative change in the congested branch of the ($k$,$q$) diagram. There are 
however some serious drawbacks to the ER-TCA model. First and foremost, the 
($k$,$q$) diagram is no longer non-monotonic for low densities when the 
sequential update is replaced by a parallel one \cite{CHOWDHURY:00,KNOSPE:04}. 
Secondly, the model exhibits too large time headways in the free-flow regime 
when compared with real-world data. This effect is also visible in the 
distribution of the vehicles' time gaps, as depicted in the histograms in the 
right part of \figref{fig:TCA:ERTCASpaceGapTimeGapHistogram}, where, in contrast 
to the STCA's time gaps distribution of 
\figref{fig:TCA:STCASpaceMeanSpeedTimeGapHistogram}, a large amount of finite 
time gaps extends well into the region of medium densities. Third, due to the 
sequential update, the ER-TCA model's downstream jam dynamics are unstable, just 
as in the STCA model \cite{KNOSPE:04}. Fourth, as can be seen from the 
($k$,$\overline v_{s}$) diagram in the left part of 
\figref{fig:TCA:ERTCADensityFlowSMSFundamentalDiagrams}, for small slowdown 
probabilities $p$, the resulting space-mean speed in the system is very 
unrealistic, even including plateaus of constant speed in the congested regime, 
e.g., the curve associated with $p =$~0.1 (we consider $p = \zero$ as a 
degenerate case).

		\subsection{Slow-to-start models}

In order to obtain a correct behavioural picture of traffic flow breakdown and 
stable jam, it is necessary that a vehicle's minimum time headway or reaction 
time should be smaller than its escape time from a jam, or equivalently, the 
outflow from a jam (i.e., the queue discharge rate) must be lower than its 
inflow \cite{EISENBLATTER:98,KRAUSS:99,KAYATZ:01,JOST:02,JOST:03,NAGEL:03}. If 
this is not the case, as in e.g., the STCA model where both times are exactly 
the same, then all jams will be unstable, as can be seen in the time-space 
diagram of \figref{fig:TCA:STCATimeSpaceDiagrams}. Because of their unstable 
jamming behaviour, the previously discussed stochastic models, experience 
neither a capacity drop nor a hysteresis loop, for which stable jams are a 
necessary prerequisite. Although the STCA-CC seems to be an exception to this 
rule, the downstream fronts of its jams are still too unstable, in the sense 
that new jams can emerge all too easily, which is unrealistic behaviour with 
respect to real-life traffic flows \cite{WOLF:99}.

As just mentioned, one mechanism that deals with this, is by leaving free-flow 
traffic undisturbed, and by \emph{significantly reducing the outflow from a jam} 
once a breakdown occurs, thereby stabilising the downstream front of a jam. 
Instead of just eliminating the noise in free-flow traffic in the STCA-CC, this 
reduced outflow can also be accomplished more intuitively, by making the 
vehicles wait a short while longer before accelerating again from stand still. 
As such, they are said to be \emph{``slow to start''}.

Note that there exists yet another mechanism that allows for the reproduction of 
the capacity drop and hysteresis phenomena (we will only briefly mention it 
here). The approach followed by Werth, is based on the premise that drivers take 
into account the \emph{speed difference} with their direct frontal leader, 
instead of just the space gap as was previously assumed. This leads to 
\emph{Galilei invariant} vehicle-vehicle interactions (i.e., the system dynamics 
remain the same if a new linear moving coordinate system is substituted in the 
equations). Interestingly, the metastability in this model is not due to cruise 
control or slow-to-start rules, but rather a result of the anticipation adopted. 
The model can exhibit stable dense platoons of fast vehicles, resulting in a 
stabilisation of the free-flow branch, and consequently leading to hysteretic 
behaviour \cite{WERTH:98,WOLF:99,CHOWDHURY:00}.\\

\sidebar{
	With respect to real-world units, we give some typical values associated with 
	the capacity drop and hysteresis phenomena (based on \cite{WOLF:99}): an 
	outflow $q_{\text{out}} \approx$~1800 vehicles/hour/lane at an associated 
	density of $k_{\text{out}} \approx$~20 vehicles/km/lane, with 
	$q_{\text{cap}}$, $k_{\text{crit}}$, and $k_{\text{jam}}$ equal to 2700 
	vehicles/hour/lane, 20 vehicles/km/lane, and 140 vehicles/km/lane, 
	respectively.
}\\

			\subsubsection{Takayasu-Takayasu TCA (T$^{\two}$-TCA)}
			\label{sec:TCA:T2TCA}

In 1993, Takayasu and Takayasu proposed a deterministic TCA model, based on the 
CA-184 (see section \ref{sec:TCA:CA184}), that incorporated a \emph{delay in 
acceleration for stopped vehicles} \cite{TAKAYASU:93}. Their motivation stems 
from the fact that high-speed vehicles are in general able to decelerate very 
quickly, but conversely, it takes them a lot longer to attain this high speed 
when they start from a stopped condition. As such, Takayasu and Takayasu 
introduced a delay, based on the rationale that a vehicle will only start to 
move when it recognises movement of its direct frontal leader. Translating this 
into a rule set, we can write the T$^{\two}$-TCA's rules based on those of the 
CA-184, but now with the following modifications (note that $v_{\text{max}} = 
\one$ cell/time step):

\begin{quote}
	\textbf{R1}: \emph{braking}\\
		\begin{equation}
		\label{eq:TCA:T2TCAR1}
			v_{i}(t - \one) > g_{s_{i}}(t - \one) \quad \Longrightarrow \quad v_{i}(t) \leftarrow g_{s_{i}}(t - \one),
		\end{equation}

	\textbf{R2}: \emph{delayed acceleration}\\
		\begin{equation}
		\label{eq:TCA:T2TCAR2}
			\begin{array}{l}
				v_{i}(t - \one) = \zero \quad \wedge \quad g_{s_{i}}(t - \one) \geq \two\\
					\qquad \Longrightarrow \quad v_{i}(t) \leftarrow \one,
			\end{array}
		\end{equation}

	\textbf{R3}: \emph{vehicle movement}\\
		\begin{equation}
		\label{eq:TCA:T2TCAR3}
			x_{i}(t) \leftarrow x_{i}(t - \one) + v_{i}(t).
		\end{equation}
\end{quote}

From this rule set it follows that a vehicle will always drive at a speed of one 
cell/time step, unless it has to brake and stop according to rule R1, equation 
\ref{eq:TCA:T2TCAR1}. Furthermore, the vehicle is only allowed to accelerate 
again to this speed of one cell/time step, on the condition that it has a 
sufficiently large space gap in front, as dictated by rule R2, equation 
\ref{eq:TCA:T2TCAR2}. As a result, the introduced delay is \emph{spatial in 
nature}, and it only affects stopped vehicles.

\begin{figure}[!htbp]
	\centering
	\includegraphics[width=\halffigurewidth]{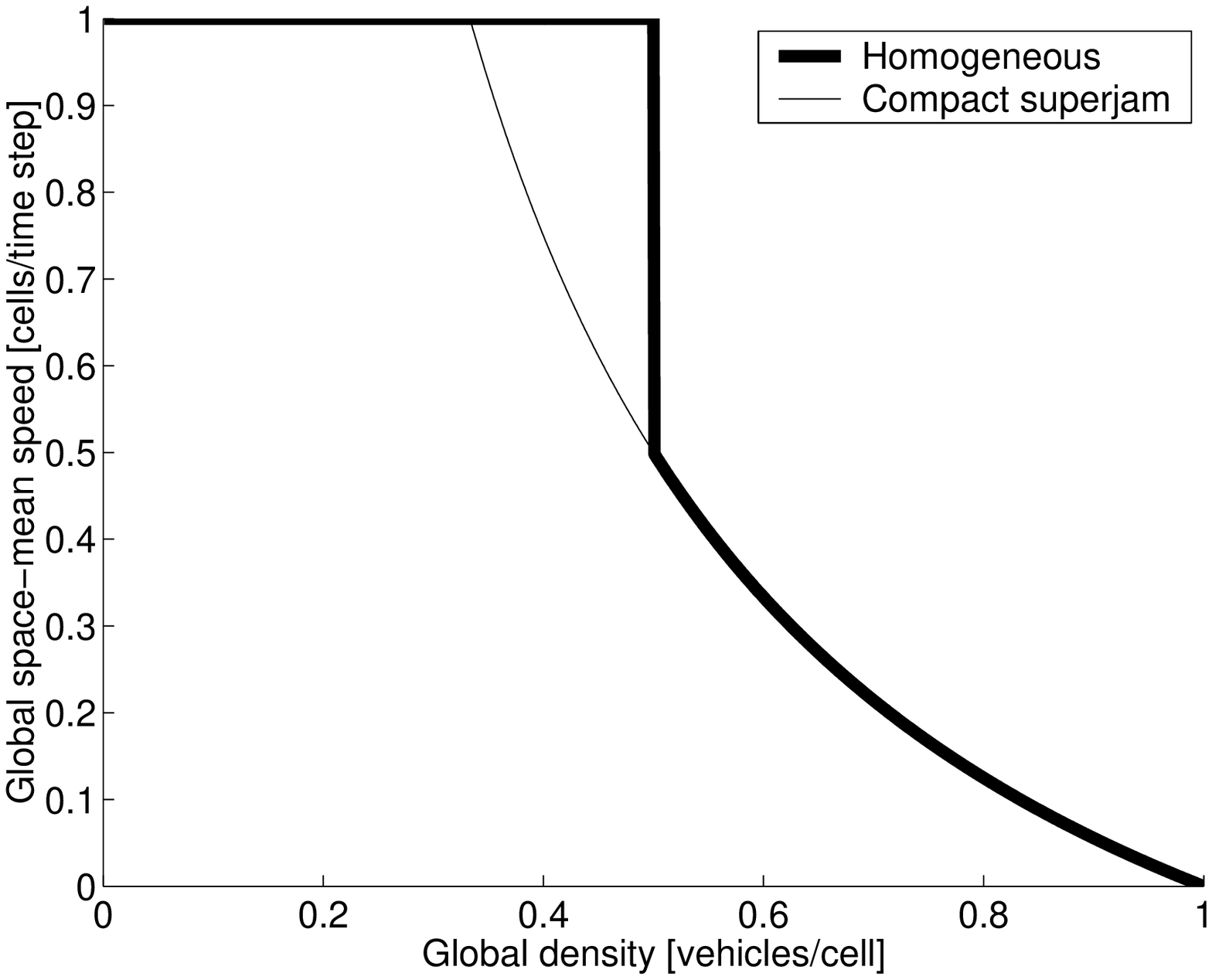}
	\hspace{\figureseparation}
	\includegraphics[width=\halffigurewidth]{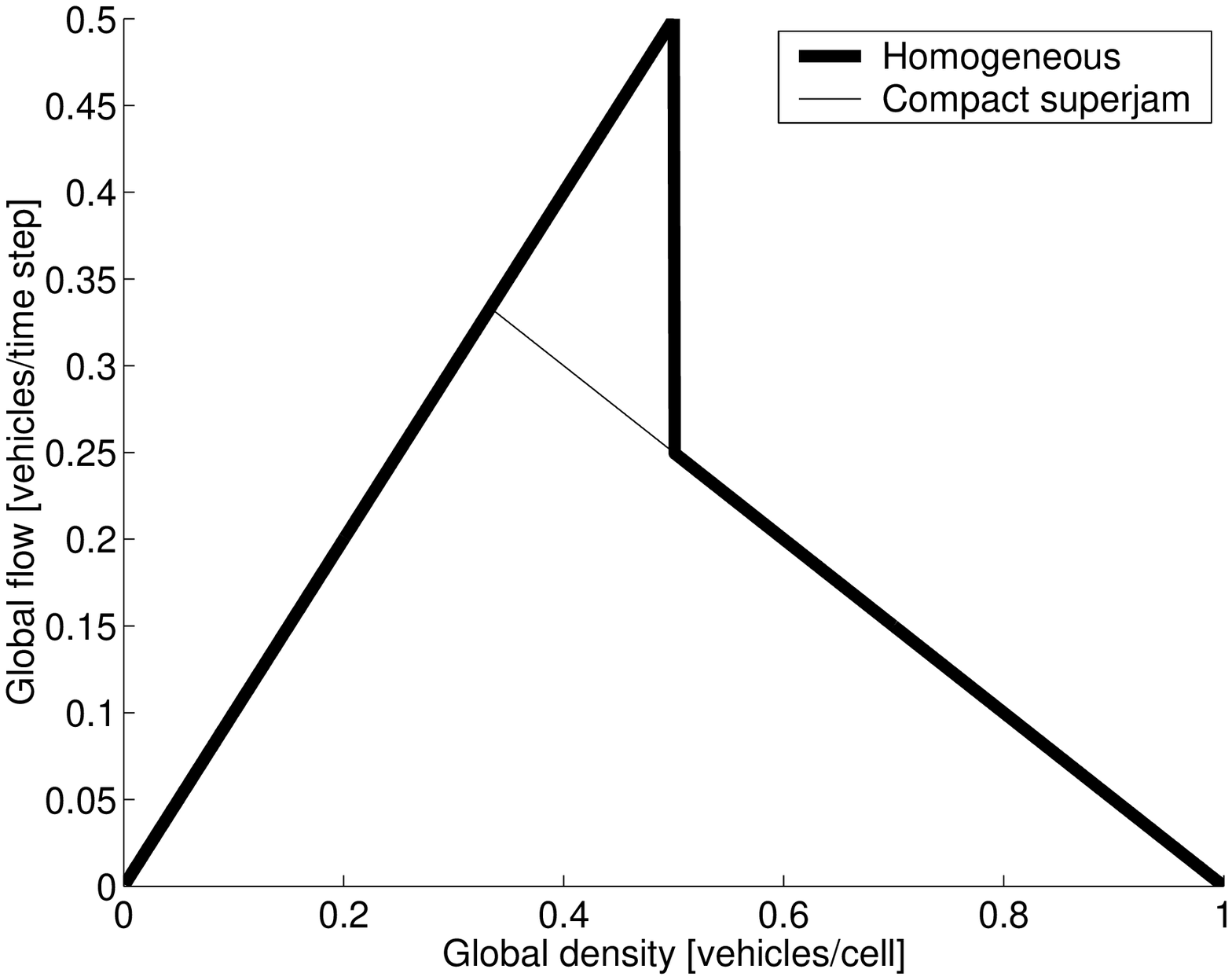}
	\caption{
		Two ($k$,$\overline v_{s}$) (\emph{left}) and ($k$,$q$) (\emph{right}) 
		diagrams for the T$^{\two}$-TCA model, with $v_{\text{max}} = \one$ 
		cells/time step. The thick solid line denotes global measurements that were 
		obtained when starting from homogeneous initial conditions; the thin solid 
		line is based on a compact superjam as the initial condition (see section 
		\ref{sec:TCA:SingleCellModels} for an explanation of these conditions). The 
		right part clearly shows a typical reversed $\lambda$ shape, which indicates 
		a capacity drop.
	}
	\label{fig:TCA:T2TCACCDensityFlowSMSFundamentalDiagrams}
\end{figure}

In \figref{fig:TCA:T2TCACCDensityFlowSMSFundamentalDiagrams}, we have depicted 
the resulting ($k$,$\overline v_{s}$) and ($k$,$q$) diagrams for the 
T$^{\two}$-TCA model. The observed behaviour is similar to that of the STCA-CC 
model in section \ref{sec:TCA:STCACC}, in that the T$^{\two}$-TCA model also 
exhibits \emph{bistability}. Starting from homogeneous initial conditions, the 
space-mean speed in the system undergoes a sharp drop once a vehicle has to 
stop. The reverse process, i.e., going from the congested to free-flow regime, 
is accompanied by a smooth continuous transition. Takayasu and Takayasu state 
that this corresponds to a second-order phase transition, because their order 
parameter (the sum of the jamming times) follows a power-law distribution, with 
jam times tending to infinity once the system goes beyond the critical density. 
With respect to the T$^{\two}$-TCA's tempo-spatial behaviour, we note that the 
critical density for the former transition is located at $k_{c} =$~0.5 
vehicles/cell, at which point all vehicles travel at a speed of one cell/time 
step with all space gaps equal to one cell. The density at which the recovery 
associated with latter transition occurs, is equal to $k = \frac{\one}{\three}$ 
vehicles/cell, at which point all vehicles travel at a speed of one cell/time 
step, but now with all space gaps equal to two cells. Fukui and Ishibashi later 
modified the delaying process, resulting in a system that always relaxes to a 
state in which the space-mean speed oscillates between two values, both smaller 
than one cell/time step \cite{FUKUI:97}.\\

\sidebar{
	The original background for Takayasu and Takayasu's work, was based on the 
	presence of so-called 1/$f$ noise (also known as \emph{pink noise} or 
	\emph{flicker noise}) in the Fourier transformed density fluctuations of 
	motorway traffic. The seemingly random stop-and-go motions of jammed vehicles, 
	could indicate a chaotic behaviour (as opposed to just statistical noise), 
	closely coupled with self-organised criticality (see also the end of section 
	\ref{sec:TCA:STCACC}) \cite{NAGEL:93}. In the free-flow regime of the 
	T$^{\two}$-TCA model, jams have a finite life time leading to a flat spectrum, 
	as opposed to the congested regime where jams have an infinite life time, 
	leading to a 1/$f$ spectrum \cite{TAKAYASU:93}.
}\\

Schadschneider and Schreckenberg later provided a generalisation of the 
T$^{\two}$-TCA model: keeping $v_{\text{max}} = \one$ cell/time step, they now 
modified the braking and acceleration behaviour of a vehicle. On the one hand, 
they kept Takayasu and Takayasu's original acceleration rule R2, equation 
\eqref{eq:TCA:T2TCAR2}, and on the other hand, they allowed a vehicle with a 
space gap of just one cell to accelerate with a \emph{slow-to-start probability} 
$\one - p_{t}$ \cite{SCHADSCHNEIDER:97c}. They furthermore also introduced a 
randomisation for moving vehicles, similar to the STCA (see section 
\ref{sec:TCA:STCA}), making vehicles stop with a slowdown probability $p$. 
Several interesting phenomena occur for certain values of both probabilities $p$ 
and $p_{t}$. The modified spatial slow-to-start rule can lead to the appearance 
of an \emph{inflection point} in the ($k$,$q$) diagram at very high densities. 
The effect gets strongly exaggerated when $p_{t} \rightarrow \one$, at which 
point a completely blocked state of zero flow appears for all global densities 
$k \geq$~0.5 vehicles/cell \cite{SCHADSCHNEIDER:97c,SANTEN:99,CHOWDHURY:00}.

			\subsubsection{The model of Benjamin, Johnson, and Hui (BJH-TCA)}
			\label{sec:TCA:BJHTCA}

Around the same time that Takayasu and Takayasu proposed their T$^{\two}$-TCA 
model, Benjamin, Johnson, and Hui (BJH) constructed another type of TCA model, 
using a slow-to-start rule that is of a \emph{temporal nature} 
\cite{BENJAMIN:96}. Their BJH-TCA model is based on the STCA (see section 
\ref{sec:TCA:STCA}), but extended it with a rule that adds a small delays to a 
stopped car that is pulling away from the downstream front of a queue. Benjamin 
et al. attribute this rule to the fact that it mimics the behaviour of a driver 
who momentarily looses attention, or when a vehicle's engine is slow to react. 
Their slow-to-start rule allows a stopped vehicle to move again with this 
\emph{slow-to-start probability} $\one - p_{s}$. If the vehicle did not move, 
then it tries to move again but this time with probability $p_{s}$. Due to this 
peculiar acceleration procedure, all vehicles require a memory that, as 
mentioned before, makes the slow-start-rule temporal in nature 
\cite{CHOWDHURY:00}. As a result of this new systematic behaviour, jams will now 
become less ravelled (as opposed to the STCA), because the slow-to-start rule 
will have the tendency to merge queues.

The BJH-TCA model was also applied to the description of a motorway with an 
on-ramp, leading to the conclusions that (i) it actually is beneficial to have 
jams on the main motorway, due to the fact that these jams homogenise the 
traffic streams as they compete for stopped vehicles, and (ii) it is desirable 
to set a maximum speed limit on this main motorway which allows to maximise the 
performance of the on-ramp. Note that in their discussion, Benjamin et al. used 
the queue length at the on-ramp as a performance measure. In our opinion, this 
is not a very good choice as it ignores e.g., the total time spent in the 
system, which we believe is a more important measure (see also the work of 
Bellemans \cite{BELLEMANS:03} and Hegyi \cite{HEGYI:01} in this respect).

To conclude, we note that the ($k$,$q$) diagrams of the BJH-TCA and 
T$^{\two}$-TCA models qualitatively look the same, with the exception the former 
does not have the possibility of an inflection point, or a density region with 
zero flow, as was the case for the latter model (see section 
\ref{sec:TCA:T2TCA}) \cite{SCHADSCHNEIDER:97c,SANTEN:99}.

			\subsubsection{Velocity-dependent randomisation TCA (VDR-TCA)}
			\label{sec:TCA:VDRTCA}

As already explained in the introduction of this section, reducing the outflow 
from a jam is responsible for the capacity drop and hysteresis phenomenon. To 
this end, Barlovi\'c et al. proposed a TCA model that generalises the STCA model 
(see section \ref{sec:TCA:STCA}) by employing an intuitive slow-to-start rule 
for stopped vehicles \cite{BARLOVIC:98,BARLOVIC:03}. Similar to the STCA-CC (see 
section \ref{sec:TCA:STCACC}), the complete rule set for the VDR-TCA is as 
follows:

\begin{quote}
	\textbf{R0}: \emph{determine stochastic noise}\\
		\begin{equation}
		\label{eq:TCA:VDRTCAR0}
			\left \lbrace
				\begin{array}{lcl}
					v_{i}(t - \one) = \zero & \Longrightarrow & p'(t) \leftarrow p_{\zero}, \\
					v_{i}(t - \one) > \zero & \Longrightarrow & p'(t) \leftarrow p, \\
				\end{array}
			\right.
		\end{equation}

	\textbf{R1}: \emph{acceleration and braking}\\
		\begin{equation}
		\label{eq:TCA:VDRTCAR1}
			v_{i}(t) \leftarrow \text{min} \lbrace v_{i}(t - \one) + \one,g_{s_{i}}(t - \one),v_{\text{max}} \rbrace,
		\end{equation}

	\textbf{R2}: \emph{randomisation}\\
		\begin{equation}
		\label{eq:TCA:VDRTCAR2}
			\xi(t) < p'(t) \Longrightarrow v_{i}(t) \leftarrow \text{max} \lbrace \zero,v_{i}(t) - \one \rbrace,
		\end{equation}

	\textbf{R3}: \emph{vehicle movement}\\
		\begin{equation}
		\label{eq:TCA:VDRTCAR3}
			x_{i}(t) \leftarrow x_{i}(t - \one) + v_{i}(t).
		\end{equation}
\end{quote}

As before, in rule R2, equation (\ref{eq:TCA:VDRTCAR2}), $\xi(t) \in 
[\zero,\one[$ denotes a uniform random number (specifically drawn for vehicle 
$i$ at time $t$) and $p'(t)$ is the stochastic noise parameter, \emph{dependent 
on the vehicle's speed} (hence the name `velocity-dependent randomisation'). The 
probabilities $p_{\zero}$ and $p$ are called the \emph{slow-to-start 
probability} and the \emph{slowdown probability}, respectively, with 
$p_{\zero},p \in [\zero,\one]$. Note that Barlovi\'c et al. only considered the 
case with two different noise parameters (i.e., $p_{\zero}$ and $p$), ignoring 
the more general case where we can have a noise parameter for each possible 
speed (i.e., $p_{\zero}$, \ldots, $p_{v_{\text{max}}}$). Their model was also 
considered for systems with open boundary conditions \cite{BARLOVIC:02b}.

\begin{figure}[!htbp]
	\centering
	\includegraphics[width=\figurewidth]{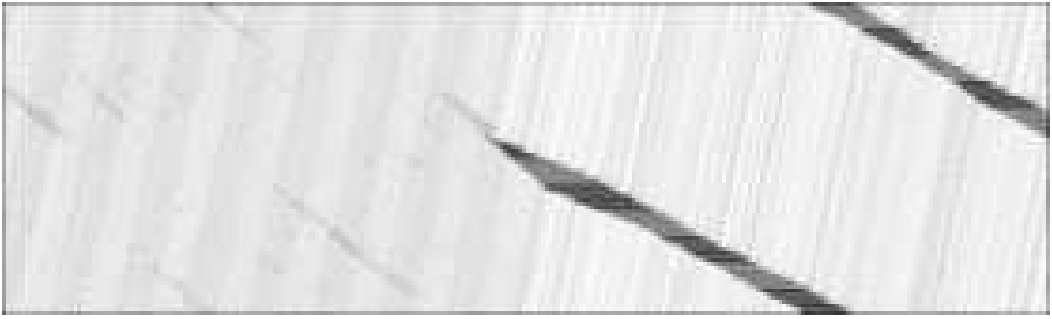}
	\caption{
		A time-space diagram of the VDR-TCA model for $v_{\text{max}} = \five$ 
		cells/time step, $p_{\zero} = $~0.5, $p =$~0.01, and a global density of $k 
		= \frac{\one}{\text{6}}$ vehicles/cell. The shown lattice contains 300 
		cells, with a visible period of 1000 time steps. We can see the breakdown of 
		an initially homogeneous traffic pattern. As the phase separation takes 
		place, a persistent compact jam is formed, surrounded by free-flow traffic. 
		The significant decrease of the density in the regions outside the jam 
		results from the jam's reduced outflow.
	}
	\label{fig:TCA:VDRTCATimeSpaceDiagram}
\end{figure}

Depending on their speed, vehicles are subject to different randomisations: 
typical metastable behaviour results when $p_{\zero} \gg p$, meaning that 
stopped vehicles have to wait longer before they can continue their journey. 
This has the effect of a reduced outflow from a jam, so that, in a closed 
system, this leads to an equilibrium and the formation of a \emph{compact jam}. 
For such a typical situation, e.g., $p_{\zero} =$~0.5 and $p =$~0.01, the 
tempo-spatial evolution is depicted in \figref{fig:TCA:VDRTCATimeSpaceDiagram}. 
We can see an initially homogeneous traffic pattern (one \emph{metastable} 
phase) breaking down and kicking the system into a \emph{phase-separated state}, 
consisting of a compact jam surrounded by free-flow traffic. In such a state, 
traffic jams in the system will absorb as many vehicles as is necessary, in 
order to have a free-flow phase in the rest of the system \cite{HELBING:01}. 
Note that the VDR-TCA can also be equipped with a cruise control, by turning of 
fluctuations for vehicles driving at the maximum speed $v_{\text{max}}$.

\begin{figure}[!htbp]
	\centering
	\includegraphics[width=\halffigurewidth]{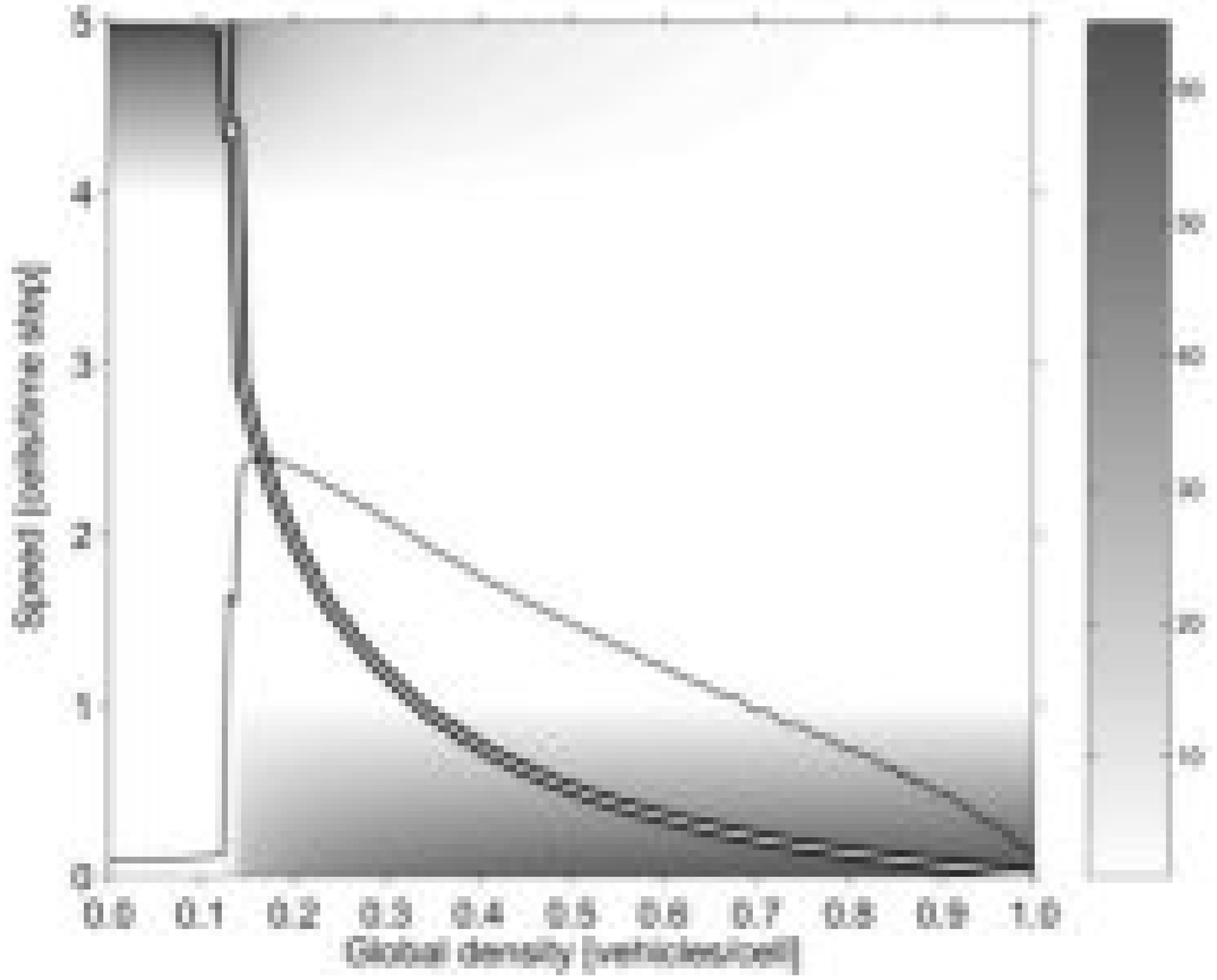}
	\hspace{\figureseparation}
	\includegraphics[width=\halffigurewidth]{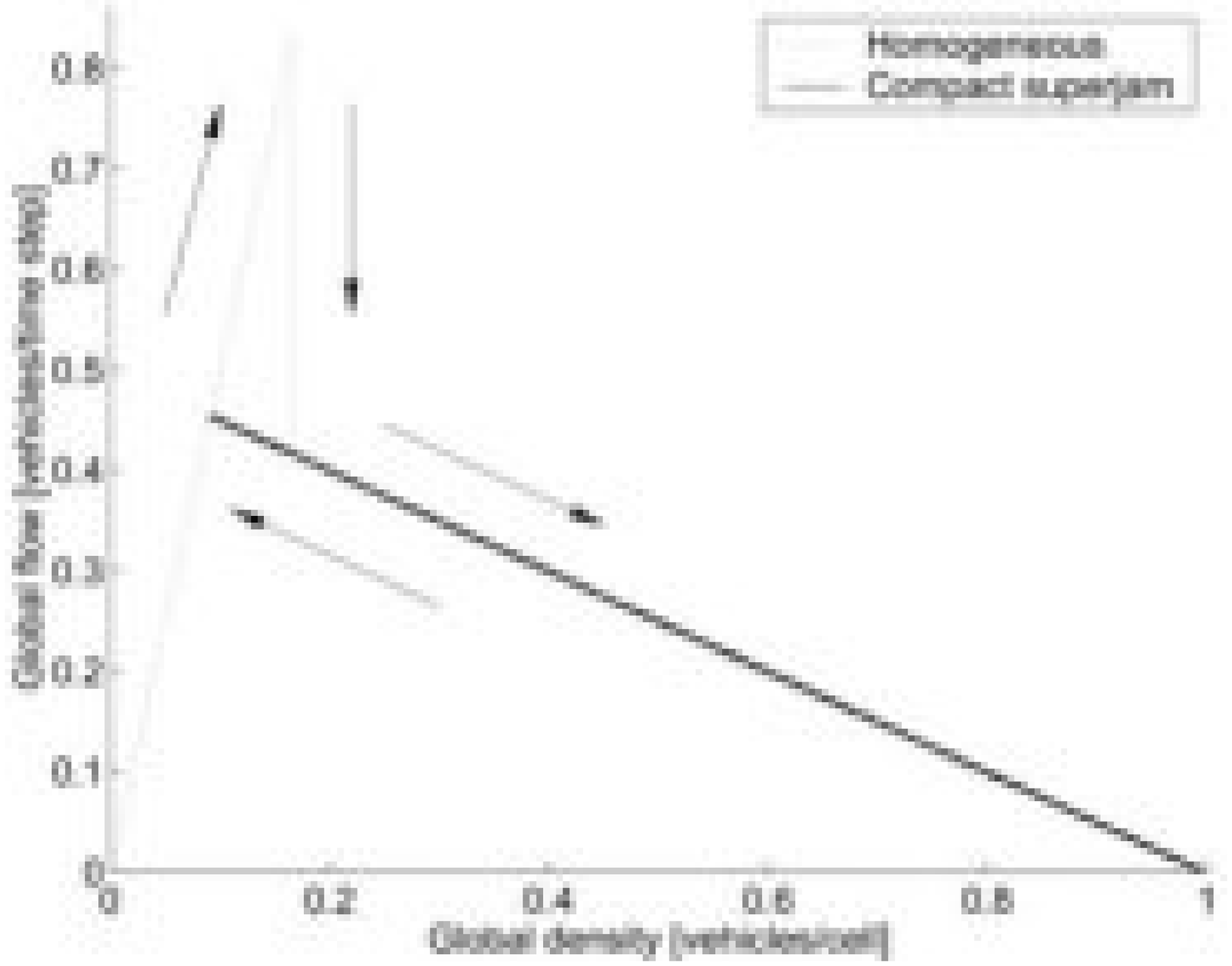}
	\caption{
		\emph{Left:} a contour plot containing the histograms of the distributions 
		of the vehicles' speeds $v$ as a function of the global density $k$ in the 
		VDR-TCA (with $v_{\text{max}} = \five$ cells/time step, $p_{\zero} =$~0.5 
		and $p =$~0.01). The thick solid line denotes the space-mean speed, whereas 
		the thin solid line shows its standard deviation. The grey regions denote 
		the probability densities. \emph{Right:} a ($k$,$q$) diagrams for the same 
		TCA model. The dotted line denotes global measurements that were obtained 
		when starting from homogeneous initial conditions; the solid line is based 
		on a compact superjam as the initial condition. The right part clearly shows 
		a typical reversed $\lambda$ shape, which indicates a capacity drop.
	}
	\label{fig:TCA:VDRTCASpeedHistogramDensityFlowFundamentalDiagram}
\end{figure}

In the left part of 
\figref{fig:TCA:VDRTCASpeedHistogramDensityFlowFundamentalDiagram}, we have 
plotted a histogram of the distributions of the vehicles' speeds, for all global 
densities $k \in [\zero,\one]$. Here we can clearly see the distinction between 
the free-flow and the congested regime: the space-mean speed remains more or 
less constant at a high value, then encounters a sharp transition (i.e., the 
capacity drop), resulting in a steady declination as the global density 
increases. Note that as the critical density is encountered, the standard 
deviation jumps steeply; this means that vehicles' speeds fluctuate wildly at 
the transition point (because they are entering and exiting the congestion 
waves). Once the compact jam is formed, the dominating speed quickly becomes 
zero (because vehicles are standing still inside the jam). Although most of the 
weight is attributed to this zero-speed, there is a non-negligible maximum speed 
present for intermediate densities. If the global density is increased further 
towards the jam density, this maximum speed disappears and the system settles 
into a state in which all vehicles either have speed zero or one (i.e., 
systemwide stop-and-go traffic).

Studying the ($k$,$q$) diagram in the right part of 
Fig.~\ref{fig:TCA:VDRTCASpeedHistogramDensityFlowFundamentalDiagram}, gives us 
another view of this phase transition. We can see a capacity drop taking place 
at the critical density, where traffic in its vicinity behaves in a metastable 
manner. This metastability is characterised by the fact that sufficiently large 
disturbances of the fragile equilibrium can cause the flow to undergo a sudden 
decrease, corresponding to a first-order phase transition. The state of very 
high flow is then destroyed and the system settles into a phase separated state 
with a large megajam and a free-flow zone. The large jam will persist as long as 
the density is not significantly lowered, thus implying that recovery of traffic 
from congestion follows a hysteresis loop. In contrast to the STCA-CC's 
bistability, the VDR-TCA model is truly \emph{metastable}, because now the 
free-flow branch in the ($k$,$q$) diagram becomes unstable for large enough 
perturbations. Furthermore, the spontaneous formation of jams in the downstream 
front that troubled the STCA, is suppressed in the VDR-TCA model.\\

\sidebar{
	Note that if $p_{\zero} \ll p$, then the behaviour of the system will be 
	drastically different. Four distinct traffic regimes emerge in the limiting 
	case where $p_{\zero} = \zero$ and $p = \one$; in this case, the model is 
	called \emph{fast-to-start} \cite{GRABOLUS:01}. In these four regimes, moving 
	vehicles can never increase their speed once the system has settled into an 
	equilibrium. Furthermore, there exists a regime which experiences forward 
	propagating density waves, corresponding to a non-concave region in the 
	system's flow-density relation. For more information, we refer to our work in 
	\cite{MAERIVOET:04i} and \cite{MAERIVOET:04d}.
}\\

			\subsubsection{Time-oriented TCA (TOCA)}
			\label{sec:TCA:TOCA}

Considering the STCA model (see section \ref{sec:TCA:STCA}), Brilon and Wu 
acknowledged the fact that it is quite capable of reproducing traffic dynamics 
in urban street networks. However, they also recognised the fact that the model 
performed rather inadequate when it comes to correctly describing the 
characteristics of traffic flows on motorways, e.g., compared to field data of a 
German motorway. Brilon and Wu blamed the unrealistic car-following behaviour of 
the STCA model for its inferior capabilities. At the core of their argument, 
they attributed this to the fact that the STCA model is exclusively based on 
spatial variables (e.g., space headways). In order to alleviate these problems, 
they proposed to use a model that was based on temporal variables (e.g., time 
headways), leading to more realistic vehicle-vehicle interactions 
\cite{BRILON:99b}. The rule set for this time-oriented TCA model (TOCA) is as 
follows:

\begin{quote}
	\textbf{R1}: \emph{acceleration}\\
		\begin{equation}
			\begin{array}{lcl}
				g_{s_{i}}(t - \one) > (v_{i}(t - \one) \cdot \overline g_{t_{s}}) \quad \wedge\\
				\xi_{\one} (t) < p_{\text{acc}}\\
					\quad \Longrightarrow \quad v_{i}(t) \leftarrow \text{min} \lbrace v_{i}(t - \one) + \one,v_{\text{max}} \rbrace,
			\end{array}
		\end{equation}

	\textbf{R2}: \emph{braking}\\
		\begin{equation}
			v_{i}(t) \leftarrow \text{min} \lbrace v_{i}(t), g_{s_{i}}(t - \one) \rbrace,
		\end{equation}

	\textbf{R3}: \emph{randomisation}\\
		\begin{equation}
			\begin{array}{lcl}
				g_{s_{i}}(t - \one) < (v_{i}(t - \one) \cdot \overline g_{t_{s}}) \quad \wedge\\
				\xi_{\two} (t) < p_{\text{dec}}\\
			  	\quad \Longrightarrow \quad v_{i}(t) \leftarrow \text{max} \lbrace v_{i}(t) - \one, \zero \rbrace,
			\end{array}
		\end{equation}

	\textbf{R4}: \emph{vehicle movement}\\
		\begin{equation}
			x_{i}(t) \leftarrow x_{i}(t - \one) + v_{i}(t).
		\end{equation}
\end{quote}

In the above rules, $\xi_{\one}(t), \xi_{\two}(t) \in [\zero,\one[$ are random 
numbers drawn from a uniform distribution, $\overline g_{t_{s} \geq \Delta T}$ 
is the \emph{safe time gap}, $p_{\text{acc}}$ is the \emph{acceleration 
probability}, and $p_{\text{dec}}$ is the \emph{deceleration probability}. 
Because all interactions between vehicles in the STCA are bounded by the update 
time step, their speeds will never oscillate, leading to a rigid and stable 
system. As a consequence of the TOCA's temporal rules however, vehicles will now 
behave more \emph{elastically}, taking a safe time gap into account that allows 
them to adapt their speeds with a relaxation. In this case, a vehicle will 
resort to emergency braking (i.e., an instantaneous deceleration) only if it 
gets too close to its direct frontal leader \cite{NAGEL:03}. Typical parameter 
values for the TOCA are $\overline g_{t_{s}} =$~1.2 time steps and 
$p_{\text{acc}} = p_{\text{dec}} =$~0.9. Brilon and Wu also extended their model 
with rudimentary rules that allowed for lane changes on unidirectional 
multi-lane roads.

\begin{figure}[!htbp]
	\centering
	\includegraphics[width=\halffigurewidth]{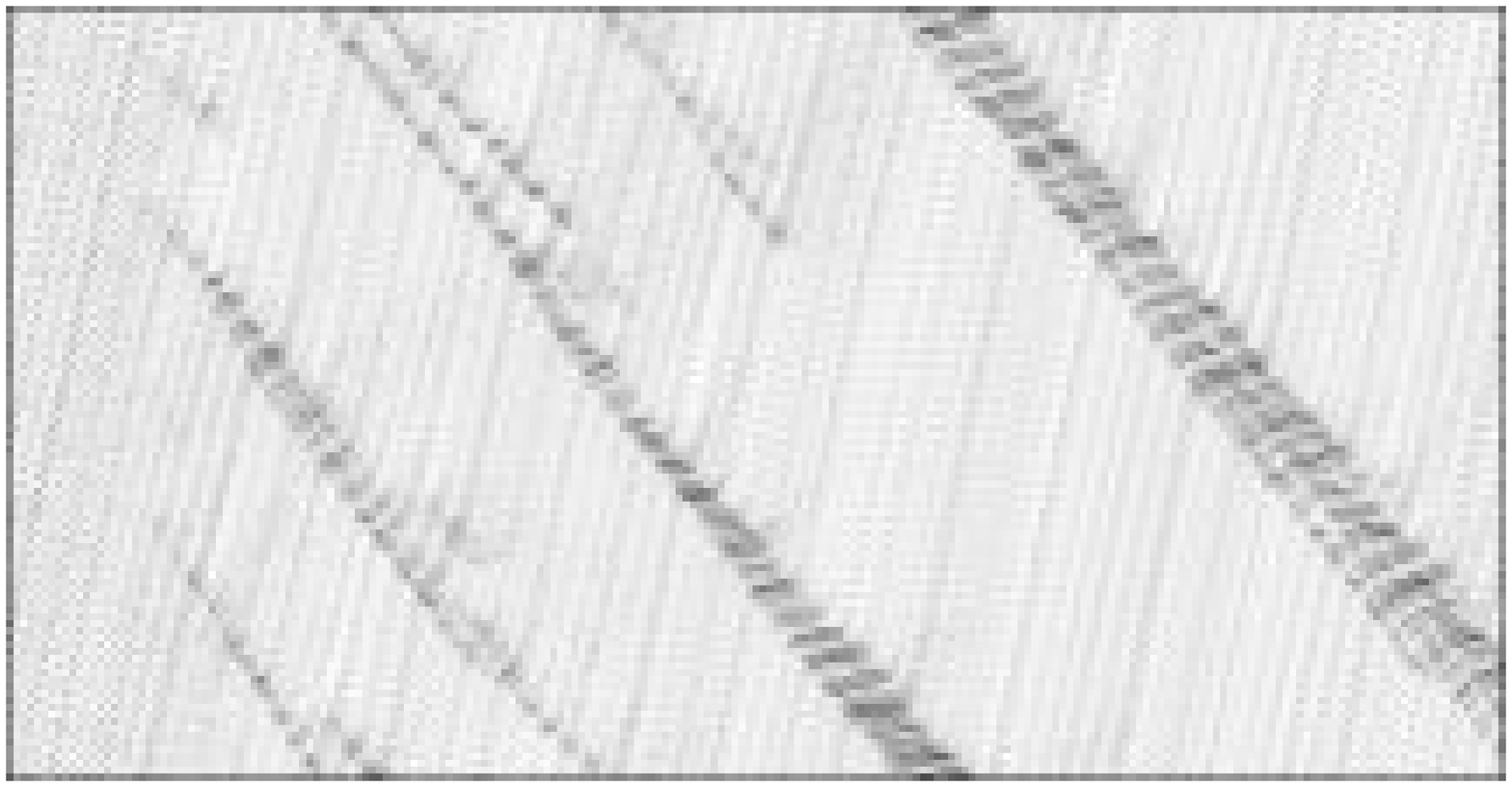}
	\hspace{\figureseparation}
	\includegraphics[width=\halffigurewidth]{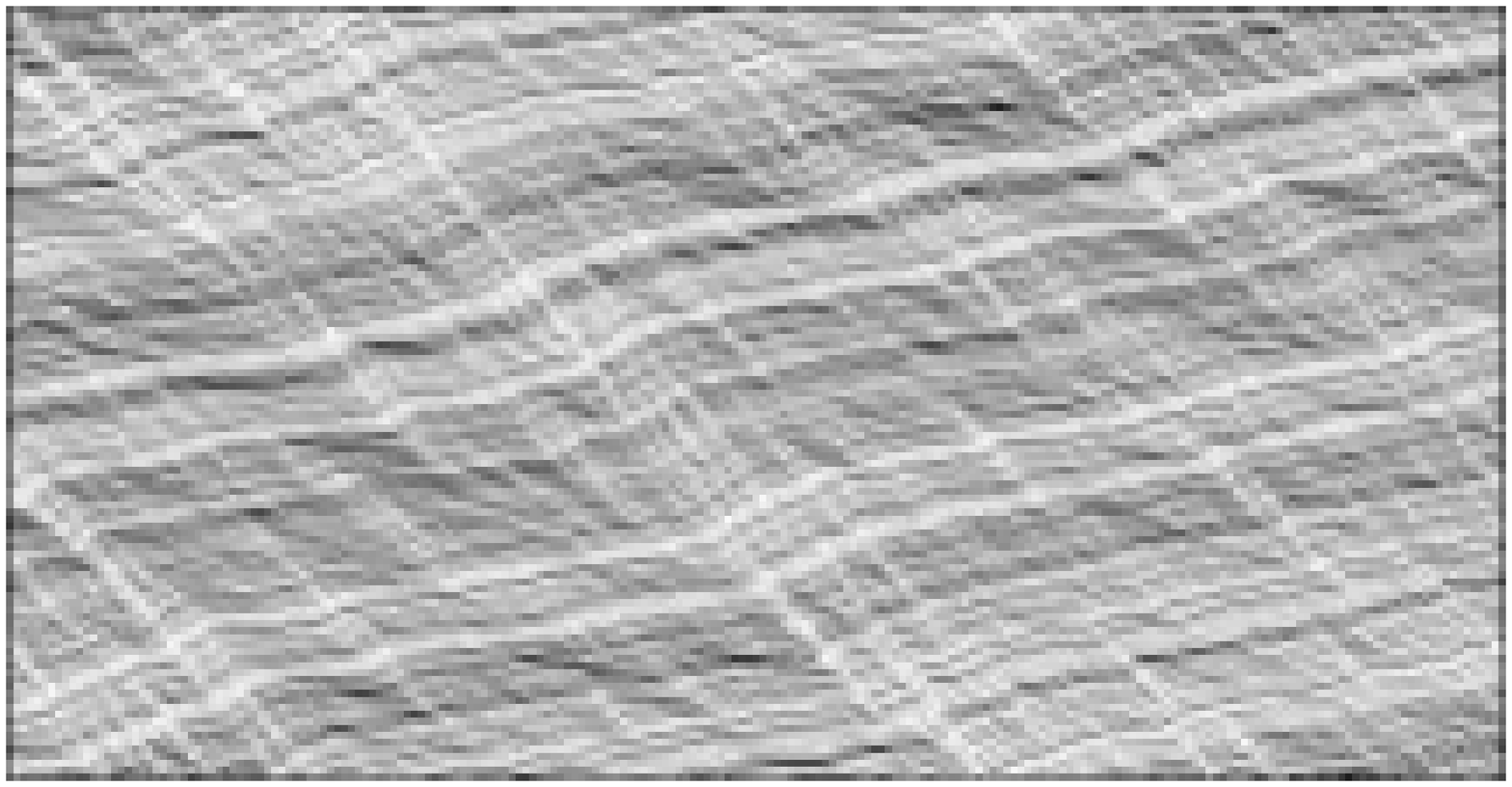}
	\caption{
		Typical time-space diagrams of the TOCA model for $v_{\text{max}} = \five$ 
		cells/time step, $\overline g_{t_{s}} =$~1.2 time steps, and $p_{\text{acc}} 
		= p_{\text{dec}} =$~0.9. The global density was set to $k = 
		\frac{\one}{\text{6}}$ vehicles/cell (\emph{left}) and $k =$~0.5 
		vehicles/cell (\emph{right}). The shown lattices each contain 300 cells, 
		with a visible period of 580 time steps. In the left part, we can see the 
		breakdown of an initially homogeneous traffic pattern, resulting in dilute 
		jam that is surrounded by free-flow traffic. In the right part, we see a 
		fully developed jam, dominating the entire system. As can be seen, for 
		moderately light densities, the jams in the TOCA model contain moving 
		vehicles.
	}
	\label{fig:TCA:TOCATimeSpaceDiagram}
\end{figure}

In the left part of \figref{fig:TCA:TOCATimeSpaceDiagram}, we can see a similar 
tempo-spatial behaviour as with the VDR-TCA (see section \ref{sec:TCA:VDRTCA}), 
in that an initially homogeneous traffic pattern breaks down, resulting in 
\emph{dilute jam} that is surrounded by free-flow traffic. The major difference 
between jamming in the VDR-TCA and TOCA models however, is that in the former 
model, vehicles come to a complete stop when entering a jam (see 
\figref{fig:TCA:VDRTCATimeSpaceDiagram}). They remain stationary until they can 
leave the downstream front of the queue. In contrast to this, the jams in the 
TOCA model contain moving vehicles. Pushing the global density even further to 
$k =$~0.5 vehicles/cell as was done in the right part of 
\figref{fig:TCA:TOCATimeSpaceDiagram}, results in a fully developed jam that 
dominates the entire system and contains temporarily stopped vehicles.

\begin{figure}[!htbp]
	\centering
	\includegraphics[width=\halffigurewidth]{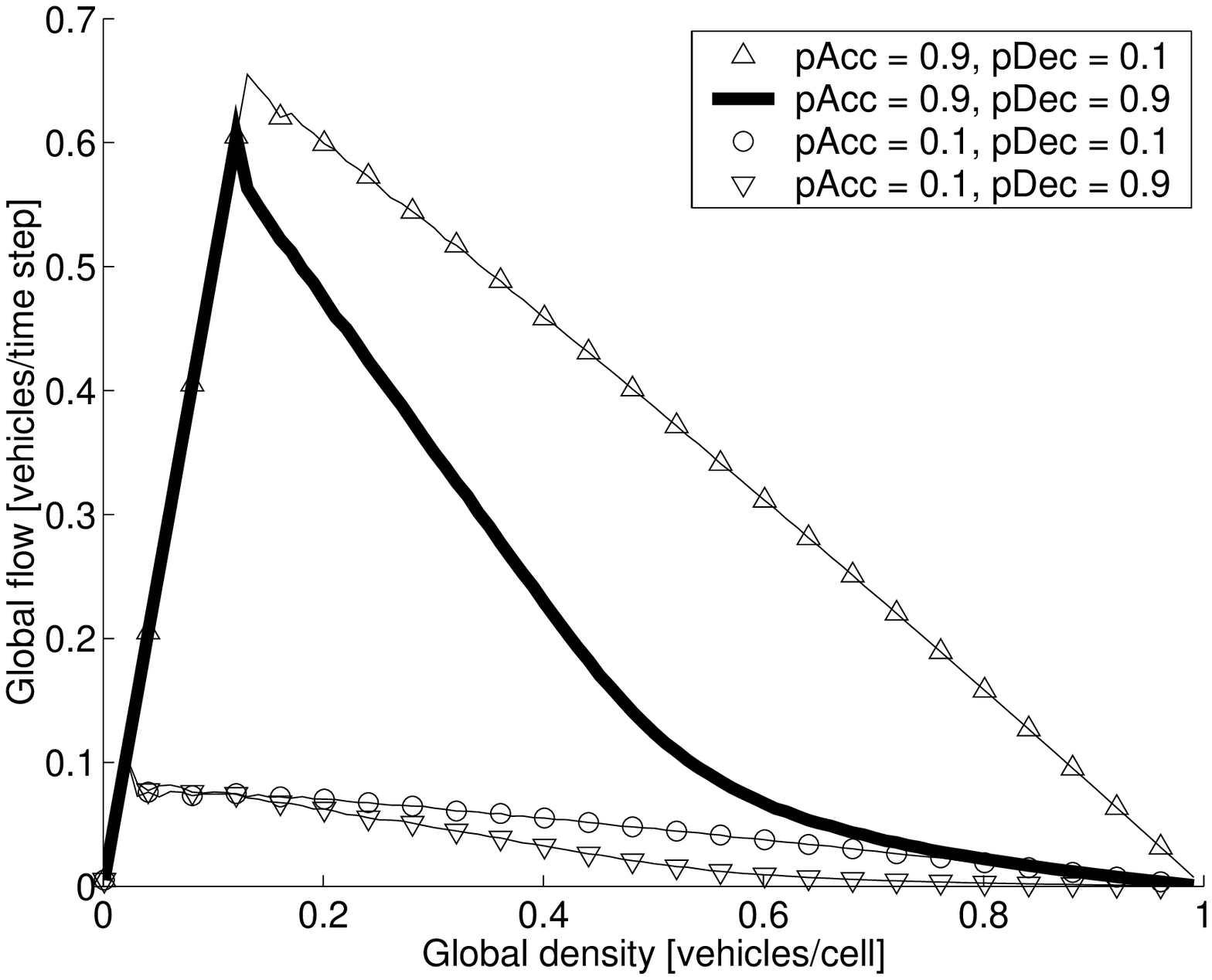}
	\hspace{\figureseparation}
	\includegraphics[width=\halffigurewidth]{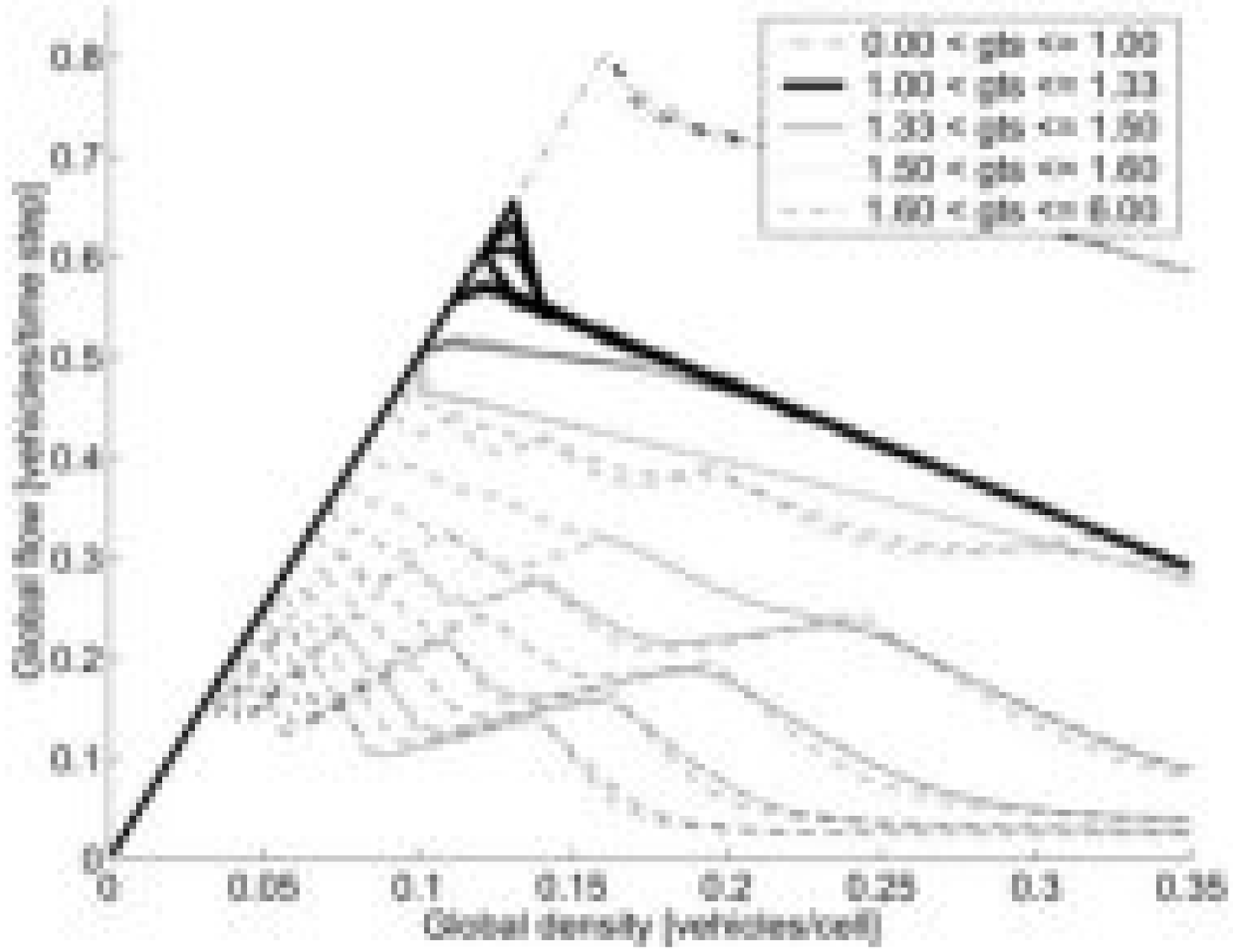}
	\caption{
		Two groups of ($k$,$q$) diagrams for the TOCA model, with $v_{\text{max}} = 
		\five$ cells/time step. \emph{Left:} four diagrams for different 
		combinations of $p_{\text{acc}}$ and $p_{\text{dec}}$, with $\overline 
		g_{t_{s}} =$~1.2 time steps. \emph{Right:} a large amount of diagrams for 
		different $g_{t_{s}}$ with $p_{\text{acc}} = p_{\text{dec}} =$~0.9. For 
		$\overline g_{t_{s}} < \Delta T$, the resulting density-flow curves are 
		non-monotonic. Note that the seemingly small capacity drops at the end of 
		each free-flow branch are in fact finite-size effects 
		\cite{NAGEL:95b,KRAUSS:97b}.
	}
	\label{fig:TCA:TOCADensityFlowFundamentalDiagrams}
\end{figure}

\figref{fig:TCA:TOCADensityFlowFundamentalDiagrams} depicts two groups of 
($k$,$q$) diagrams for the TOCA model, with $v_{\text{max}} = \five$ cells/time 
step. The left part shows four diagrams for different combinations of 
$p_{\text{acc}}$ and $p_{\text{dec}} \in \lbrace (\text{0.9},\text{0.1}), 
(\text{0.9},\text{0.9}), (\text{0.1},\text{0.1}), (\text{0.1},\text{0.9}) 
\rbrace$, each time with $\overline g_{t_{s}} =$~1.2 time steps. As can be seen, 
the default case with $p_{\text{acc}} = p_{\text{dec}} =$~0.9 leads to an 
inflection point at a moderately high density of $k =$~0.5 vehicles/cell, 
resulting in two different slopes for the congested branch of the TOCA's 
($k$,$q$) diagram. At this point, vehicles will have average space gaps less 
than one cell, and because $p_{\text{dec}}$ is rather high, vehicles will have 
the tendency to slow down (and $p_{\text{acc}}$ is smaller then one, so their 
acceleration is somewhat inhibited). As a result, a large jam, comparable to the 
system's size, will dominate tempo-spatial evolution. Furthermore, the 
acceleration probability $p_{\text{acc}}$ should take on rather high values, 
otherwise the global flow in the system is too low because vehicles are not 
accelerating anymore. In the right part of 
\figref{fig:TCA:TOCADensityFlowFundamentalDiagrams}, we have shown a large 
amount of diagrams for different $g_{t_{s}}$ with $p_{\text{acc}} = 
p_{\text{dec}} =$~0.9. Here we can see that, for $\overline g_{t_{s}} < \Delta 
T$, the resulting density-flow curves are non-monotonic. Higher values for 
$\overline g_{t_{s}}$ in more vehicles that drive more cautiously, apparently 
leading to higher values for the critical density and the capacity flow. Note 
that the seemingly small capacity drops at the end of each free-flow branch are 
in fact finite-size effects \cite{NAGEL:95b,KRAUSS:97b}.

In their original paper, Brilon and Wu claim that their TOCA model results in a 
better agreement with empirical data, a fact which is based on a qualitative 
comparison of the ($q$,$\overline v_{s}$) diagrams \cite{BRILON:99b}. Note that, 
after personal communication with the authors, it seems they performed a 
minimisation of the square errors in the ($k$,$\overline v_{s}$) diagram. 
However, in order to get the correct values for calibrating the TOCA's 
parameters, they just manually guessed, without performing a thorough numerical 
optimisation. Despite this optimistic view, Knospe et al. later investigated the 
TOCA model's capabilities more thoroughly. Their conclusions state that a quite 
large value for the deceleration probability $p_{\text{dec}}$ is necessary in 
order to obtain realistic capacity flows. Although the time headway distribution 
of a jam's downstream front in the TOCA model is correct with respect to 
real-life observations, its downstream front moves too fast due to the large 
deceleration probability. As a result, the jams in the TOCA model are more 
dilute, as could be seen in \figref{fig:TCA:TOCATimeSpaceDiagram} 
\cite{KNOSPE:04}.

			\subsubsection{TCA models incorporating anticipation}
			\label{sec:TCA:TCAModelsIncorporatingAnticipation}

One of the models related to anticipative driving (i.e., only taking a leaders' 
reactions into account, without predicting them), can be found in the work of 
Krau\ss~et al., who derived a collision-free model based on the STCA (see 
section \ref{sec:TCA:STCA}), but which uses \emph{continuous} vehicle speeds. 
Their model can be considered as a simplified version of the Gipps model 
\cite{MAERIVOET:05c}. Although the model restricts vehicles' deceleration 
capabilities, it is still able to correctly reproduce the capacity drop and 
hysteresis phenomena \cite{KRAUSS:97b}.

Another model with anticipation was proposed by Eissfeldt and Wagner 
\cite{EISSFELDT:03}. Their model is based on Krau\ss's work 
\cite{MAERIVOET:05c}, and employs a next-nearest-neighbour interaction, which 
stabilises dense flows and results in a non-unique flow-density relation.

Recently, L\'arraga et al. introduced a TCA model that includes a driver's 
\emph{anticipation} of the leading vehicle's speed \cite{LARRAGA:04}. In 
contrast to the STCA model (see section \ref{sec:TCA:STCA}), the acceleration 
and braking rules are decoupled. As a first rule, the standard acceleration 
towards the maximum speed is applied, after which the randomisation is performed 
by means of a second rule. Only then, the model considers braking in its third 
rule; however, the deceleration is not only based on the space gap between both 
vehicles, but also on an anticipation of the leading vehicle's speed:

\begin{quote}
	\textbf{R3}: \emph{anticipation and braking}\\
		\begin{equation}
			\begin{array}{l}
				v_{i}(t) \leftarrow\\
				\text{min} \left \lbrace v_{i}(t),
					\underbrace{
						g_{s_{i}}(t - \one) +
						\left [ (\one - \alpha_{i}) \cdot v_{i + \one}(t - \one) + \frac{\one}{\two} \right ]
					}_{\text{safe distance}} \right \rbrace,
			\end{array}
		\end{equation}
\end{quote}

with $v_{i}(t)$ on the right-hand side corresponding to the computed speed after 
applying rule R2, $[x]$ denoting $x$ rounded to the nearest integer, $v_{i + 
\one}(t - \one)$ the speed of the leading vehicle at the current time step, and 
$\alpha_{i} \in [\zero,\one]$ an anticipatory driving parameter for the \ith 
vehicle. In their work, L\'arraga et al. considered all $\alpha_{i}$ to be 
equal.

The interesting aspect of this anticipatory TCA model, is that for certain 
values of $\alpha$, it can result in \emph{dense platoons of vehicles}, 
travelling coherently and thereby leading to forward propagating density 
structures. In the free-flow regime, the ($k$,$q$) diagram also exhibits a 
slight curvature near the capacity flow, similar to the ER-TCA model (see 
section \ref{sec:TCA:ERTCA}). Del Ri\'o and L\'arraga later also extended the 
model to accommodate for multi-lane traffic flows \cite{DELRIO:05}.

			\subsubsection{Ultra discretisation, slow-to-accelerate, and driver's perspective}
			\label{sec:TCA:BCA}

It is also possible to derive a cellular automaton model, based on the 
discretisation of a partial differential equation. Starting from a PDE (e.g., 
the Burgers equation \cite{MAERIVOET:05c}, we can obtain an finite difference 
equation by discretising the spatial and temporal dimensions, resulting in a 
model that still has continuous state variables. As a further step, we can now 
also discretise these state variables, using a process called the 
\emph{ultra-discretisation method} (UDM) \cite{TOKIHIRO:96}. The result of the 
UDM can be interpreted as a cellular automaton in the \emph{Euler 
representation}. The latter means that for a TCA model, a road is considered to 
be a field, whereby the individual cars are not distinguished 
\cite{NISHINARI:01b}. The interesting part of this type of CA is that its cells 
are allowed to hold multiple vehicles, which makes it possible to implicitly 
model multi-lane traffic in a simplified sense (because the effects of lane 
changes are neglected) \cite{CHOWDHURY:00}. As a next step, this obtained CA can 
be cast in its \emph{Lagrangian representation}, by means of an 
\emph{Euler-Lagrange transformation} \cite{NISHINARI:01b,MATSUKIDAIRA:03}. The 
resulting Lagrange representation treats the positions of all vehicles 
individually, thus leading to the well-know position-based rule sets of the TCA 
models discussed in this report.

Nishinari proposed an interesting TCA model, based on the above UDM scheme. 
Their discretisation leads to the so-called \emph{Burgers cellular automaton} 
(BCA), which is for single-lane traffic equivalent to the CA-184 TCA model (see 
section \ref{sec:TCA:CA184}) \cite{NISHINARI:99,NISHINARI:01}. Emmerich et al. 
also provided a TCA model, by applying the UDM scheme to a Korteweg-de Vries 
equation. In contrast to the BCA model, their work resulted in a second-order 
TCA model because the CA's global map not only needs the configuration at the 
previous time step $t - \one$, but also the configuration at time step $t - 
\two$ \cite{EMMERICH:98,CHOWDHURY:00}.

\begin{figure}[!htbp]
	\centering
	\includegraphics[width=0.75\figurewidth]{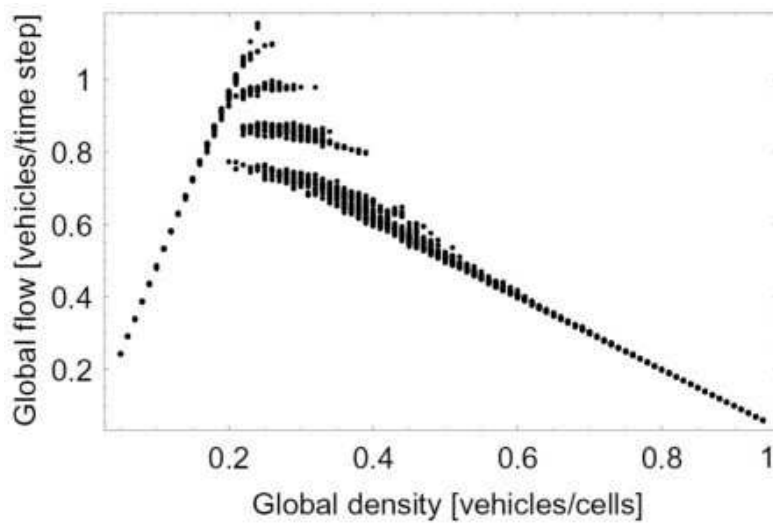}
	\caption{
		A ($k$,$q$) diagram of Nishinari et al.'s extended BCA model, with 
		$v_{\text{max}} = \five$ cells/time step, $\Delta T =$~1.3~s, $\Delta X 
		=$~7.5~m, and a driver's perspective of two vehicles ahead. The resulting 
		diagram exhibits multiple metastable branches. Vehicles inside jams come to 
		a complete stop only for the lowest metastable branch; for the higher 
		branches, vehicles inside jams are still able to move forward. Depending on 
		the strength of a local perturbation, traffic will shift from the highest 
		branch to one of the lower branches (image reproduced after 
		\cite{NISHINARI:04}).
	}
	\label{fig:TCA:S2ATCADensityFlowFundamentalDiagram}
\end{figure}

Nishinari et al. recently extended the BCA model, thereby allowing for 
slow-to-start effects with $v_{\text{max}} > \one$ cell/time step 
\cite{NISHINARI:04}. Their model contains a rule similar to the classic notion 
of slow-to-start rules, but now generalised for moving vehicles, leading to the 
terminology of a \emph{slow-to-accelerate} rule. Taking the idea of anticipation 
one step further, they also incorporated a \emph{driver's perspective}, meaning 
that a vehicle will base its acceleration and braking decisions not only on the 
basis of its space gap and the anticipated speed of the vehicle ahead, but also 
on the space gap with the \emph{next} leading vehicle (or even a vehicle located 
more downstream). As a result, the model exhibits \emph{multiple metastable 
branches} in the ($k$,$q$) diagram, as can be seen in 
\figref{fig:TCA:S2ATCADensityFlowFundamentalDiagram}. For the lowest metastable 
branch, vehicles inside jams will come to a complete stop. In contrast to this, 
vehicles will still be able to move forward inside jams for the higher branches. 
Note that depending on the strength of a local perturbation, traffic will shift 
from the highest branch to one of the lower branches. Finally, Nishinari et al. 
also combined the model with the classic STCA (see section \ref{sec:TCA:STCA}), 
thereby allowing for stochasticity in both the acceleration and braking rules. 

	\section{Multi-cell models}

Whereas all the previously discussed TCA models were based on a single-cell 
setup, this section introduces some of the existing multi-cell TCA models (still 
for single-lane traffic). In a multi-cell model, a vehicle is allowed to span a 
number of consecutive cells in the longitudinal direction, i.e., $l_{i} \geq 
\one$ cell.

In the subsequent sections, we discuss several multi-cell TCA models encountered 
in literature. We first start with an overview of the artifacts that can be 
introduced when switching to a multi-cell setup. Subsequently, we describe three 
multi-cell TCA models, which have more intricate rule sets than the simple 
models of section \ref{sec:TCA:SingleCellModels}:

\begin{itemize}
	\item Helbing-Schreckenberg TCA (HS-TCA)
	\item Brake-light TCA (BL-TCA)
	\item The model of Kerner, Klenov, and Wolf (KKW-TCA)
\end{itemize}

Note that with respect to the measurements performed on the TCA models' 
lattices, we assume homogeneous traffic flows, i.e., all vehicles have the same 
length. This allows us, after suitable adjustment with the average vehicle 
length $\overline l = l_{i}$, to express the global density as $k_{g} \in 
[\zero,\one]$.

		\subsection{Artifacts of a multi-cell setup}
		\label{sec:TCA:ArtifactsOfAMultiCellSetup}

It might seem that a translation of the classic STCA model (see section 
\ref{sec:TCA:STCA}) into a multi-cell version would be straightforward. However, 
using a finer discretisation introduces a very specific artifact, i.e, 
\emph{hysteresis}. In order to investigate this phenomenon, we have performed 
several experiments based on a multi-cell translation of the STCA model (now 
called the MC-STCA). In what follows, we assume a closed-loop lattice consisting 
of $\ten^{\five}$ cells. The simulations ran each for $\five \times 
\ten^{\five}$ time steps, with $\Delta T = \one$~s.

\begin{figure}[!htbp]
	\centering
	\includegraphics[width=\halffigurewidth]{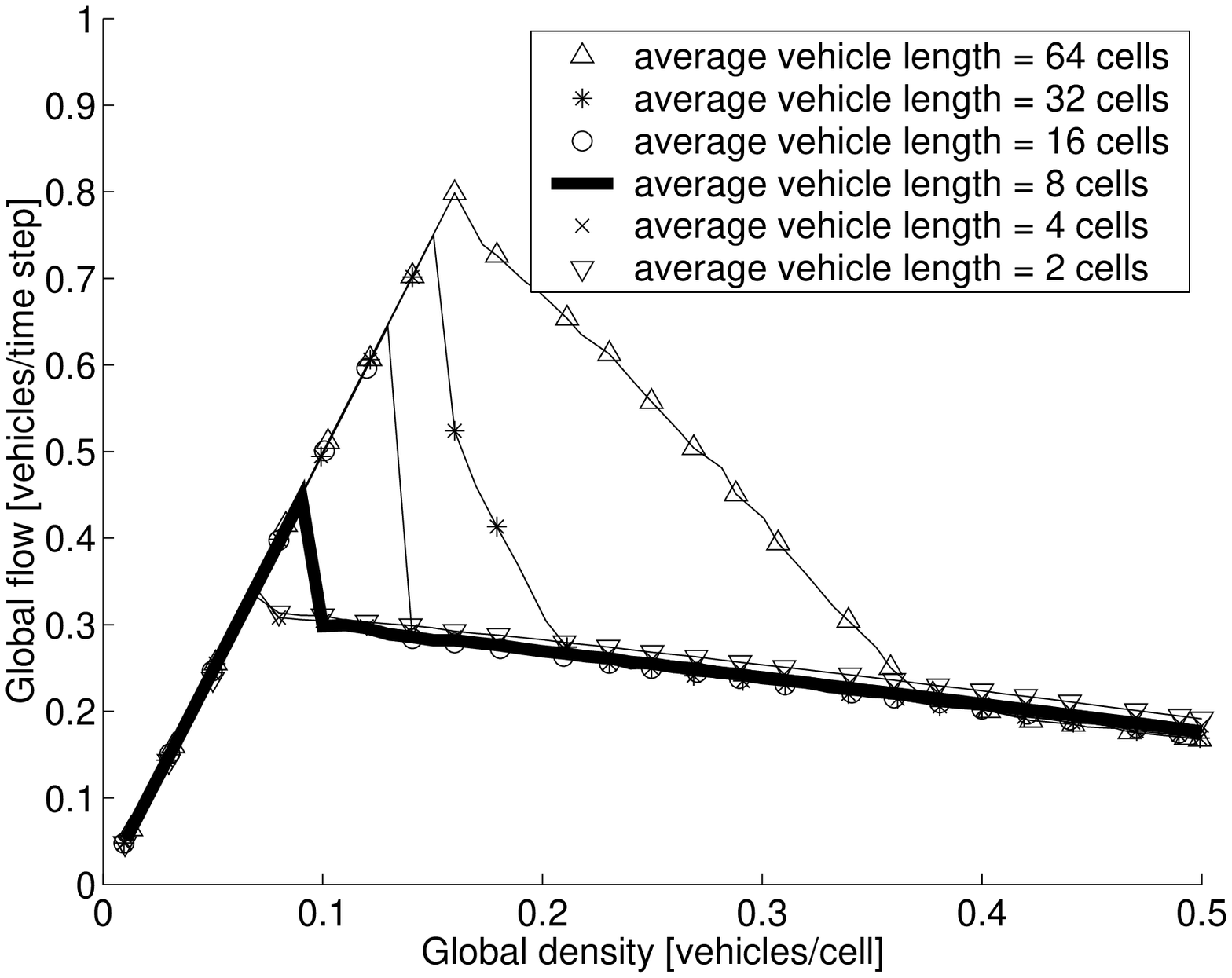}
	\hspace{\figureseparation}
	\includegraphics[width=\halffigurewidth]{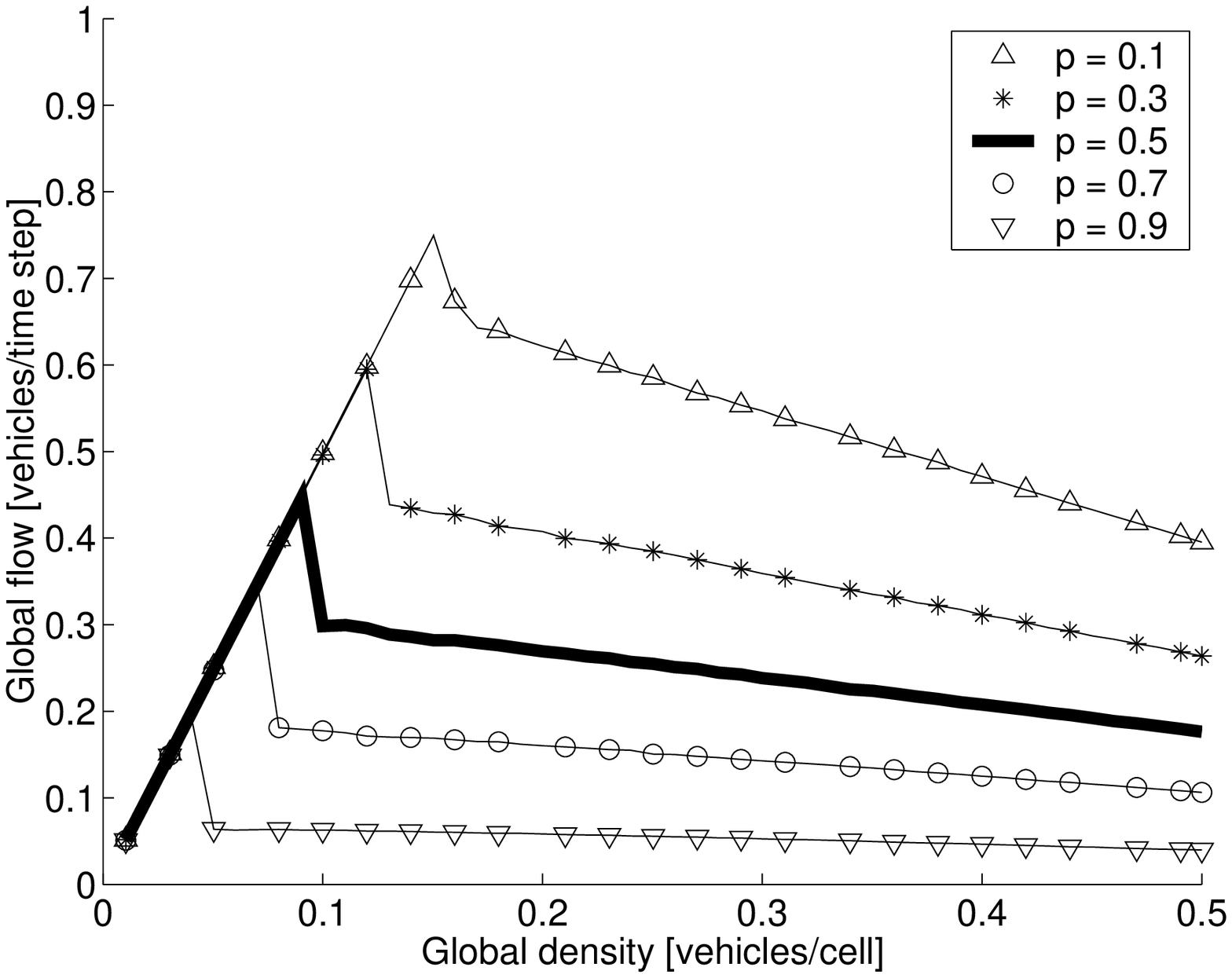}
	\caption{
		\emph{Left:} several ($k$,$q$) diagrams of the MC-STCA, for $\overline l \in 
		\lbrace \two, \four, \text{8}, \text{16}, \text{32}, \text{64} \rbrace$ 
		cells and $p =$~0.5. As can be seen, an increase of the average vehicle 
		length apparently results in a higher critical density, with an associated 
		higher capacity flow (followed by a capacity drop). \emph{Right:} the same 
		setup for the MC-STCA, but now with a fixed $\overline l =$~8 cells and 
		$v_{\text{max}} = \five \times \text{8} =$~40 cells/time step. The ($k$,$q$) 
		diagrams depict the results of changing the slowdown probability $p \in 
		\lbrace \text{0.1}, \text{0.3}, \text{0.5}, \text{0.7}, \text{0.9} \rbrace$: 
		an increase of $p$, leads to decrease of both the critical density and the 
		capacity flow.
	}
	\label{fig:TCA:MCSTCAFundamentalDiagrams}
\end{figure}

Setting the slowdown probability to $p =$~0.5, the left part of 
\figref{fig:TCA:MCSTCAFundamentalDiagrams} shows the resulting ($k$,$q$) 
diagrams for different spatial discretisations, each time for homogeneous 
initial conditions. The average vehicle length was set to $\overline l \in 
\lbrace \two, \four, \text{8}, \text{16}, \text{32}, \text{64} \rbrace$ cells. 
In these experiments, we also scaled the maximum speed $v_{\text{max}}$ 
correspondingly (e.g., if $\overline l = \four$ cells, then $v_{\text{max}}$ 
would become $\five \times \four =$~20 cells/time step), as can be seen from the 
coinciding free-flow branches in the left part in 
\figref{fig:TCA:MCSTCAFundamentalDiagrams}. We also notice that an increase of 
the average vehicle length apparently results in a higher critical density, with 
an associated higher capacity flow. Furthermore, the flow seems to encounter a 
\emph{capacity drop} at this critical dense.

What causes this capacity drop~? To answer this question, we must first consider 
what happens in the deterministic case where $p = \zero$. Here, our experiments 
have shown that there is no difference between a single-cell and a multi-cell 
setup. Setting $p > \zero$, the randomisation rule R2, equation 
\eqref{eq:TCA:STCAR2}, introduces fluctuations in the high speeds of vehicles in 
free-flow traffic. However, these speed fluctuations are actually small compared 
to the vehicles' speeds themselves. Because of this limited influence, the 
free-flow branch of the ($k$,$q$) diagrams remains \emph{very stable}. The 
smaller the discretisation, i.e., the larger the average vehicle length, the 
more stable the free-flow branch becomes for larger densities (note however that 
the capacity drop gets less pronounced for increasing average vehicle lengths). 
This capacity drop behaviour due to a stabilisation effect, is akin to the 
observations in the STCA's cruise-control limit (see section 
\ref{sec:TCA:STCACC}), and thus different from the VDR-TCA (see section 
\ref{sec:TCA:VDRTCA}), where a reduced outflow from a jam causes the drop in 
flow \cite{KNOSPE:04}. In contrast to this, random initial conditions or a 
superjam to start the simulations with, will always lead to the congested 
branch, thereby indicating a hysteretic phase transition. As the left part of 
\figref{fig:TCA:MCSTCAFundamentalDiagrams} indicates, changing the 
discretisation level of the STCA, by adjusting the average vehicle length and 
relatively keeping the same maximum speed, has only an effect on the length of 
the free-flow branch; the traffic dynamics in the congested regime remain the 
same.

Holding $\overline l$ fixed at 8 cells and $v_{\text{max}} = \five \times 
\text{8} =$~40 cells/time step, the right part of 
\figref{fig:TCA:MCSTCAFundamentalDiagrams} shows the resulting ($k$,$q$) 
diagrams for different values of the slowdown probability $p \in \lbrace 
\text{0.1}, \text{0.3}, \text{0.5}, \text{0.7}, \text{0.9} \rbrace$. It is clear 
that an increase of $p$, leads to a decrease of both the critical density and 
the capacity flow. Note that the size of the capacity drop remains approximately 
the same for the different $p$.

To conclude, we mention the work of Grabolus who performed extensive numerical 
studies on the STCA. He also noted that it is possible to translate any 
multi-cell STCA variant into an \emph{equivalent} single-cell STCA model, by 
suitably adjusting the values of the density and the maximum speed 
\cite{GRABOLUS:01}.\\

\sidebar{
	Interestingly, the use of a smaller discretisation was already considered by 
	Barrett et al. in the early course of the TRANSIMS project 
	\cite{BARRETT:95,MAERIVOET:05c}. In their work, they introduce the terminology 
	of \emph{multi-resolution} TCA models, corresponding to our multi-cell setup. 
	Although they discuss several methods for integral refinements of the TCA's 
	lattice, they do not make any mention of the observed hysteresis phenomenon 
	introduced by a finer discretisation.
}\\

		\subsection{Advanced multi-cell models}

Having discussed the repercussions of switching to a multi-cell setup, we now 
illustrate three TCA models that have more complex rule sets. We discuss their 
properties by means of time-space diagrams, fundamental diagrams of global and 
local measurements, and histograms of the distributions of the space and time 
gaps.

			\subsubsection{The model of Helbing and Schreckenberg (HS-TCA)}
			\label{sec:TCA:HSTCA}

Similar in spirit as the STCA (see section \ref{sec:TCA:STCA}) and the ER-TCA 
(see section \ref{sec:TCA:ERTCA}), Helbing and Schreckenberg proposed their 
HS-TCA model in analogy with the optimal velocity model \cite{HELBING:99c}. In 
fact, their model can be seen as a direct discretisation of the OVM, with the 
following rule set:

\begin{quote}
	\textbf{R1}: \emph{acceleration and braking}\\
		\begin{equation}
		\label{eq:TCA:HSTCAR1}
			v_{i}(t) \leftarrow v_{i}(t - \one) + \lfloor \alpha~(V(g_{s_{i}}(t - \one)) - v_{i}(t - \one)) \rfloor,
		\end{equation}

	\textbf{R2}: \emph{randomisation}\\
		\begin{equation}
		\label{eq:TCA:HSTCAR2}
			\xi(t) < p \Longrightarrow v_{i}(t) \leftarrow \max \lbrace \zero,v_{i}(t) - \one \rbrace,
		\end{equation}

	\textbf{R3}: \emph{vehicle movement}\\
		\begin{equation}
		\label{eq:TCA:HSTCAR3}
			x_{i}(t) \leftarrow x_{i}(t - \one) + v_{i}(t).
		\end{equation}
\end{quote}

The function $V(g_{s_{i}})$ in rule R1, equation \eqref{eq:TCA:HSTCAR1}, is the 
discrete version of the optimal velocity function; it is specified in the form 
of a lookup table, containing speed entries for each space gap (see 
\tableref{table:TCA:HSTCAOVM}) and has the following meaning: higher values for 
the parameter $\alpha$ indicate an almost instantaneous adaptation of the 
vehicle's speed to the OVF, whereas lower values denote an increasing inertia 
and longer adaptation times \cite{HELBING:99c}. However, as stated by Chowdhury 
et al. and Knospe et al., the role of $\alpha$ is a bit unclear as it does not 
exactly correspond to the timescale of the adaptation to the OVF (which is the 
case for the original optimal velocity model) \cite{CHOWDHURY:00,KNOSPE:04}. 
Furthermore, certain values for $\alpha$ can, in combination with the OVF, lead 
to collisions between vehicles (because $\alpha$ reduces a vehicle's braking 
capability). Knospe et al. later provided the necessary conditions that 
guarantee collision-free driving, and avoid the possible backward moving of 
vehicles \cite{KNOSPE:04}. Note that, similar to the Fukui-Ishibashi models (see 
sections \ref{sec:TCA:DeterministicFITCA} and \ref{sec:TCA:SFITCA}), vehicles 
are allowed to accelerate instantaneously in the HS-TCA model. The model is 
stochastic, in that it introduces randomisation by means of rule R2, equation 
\eqref{eq:TCA:HSTCAR2}, with $\xi(t) \in [\zero,\one[$ a random number drawn 
from a uniform distribution. 

\begin{table}[!htbp]
	\centering
	\begin{tabular}{|c|c||c|c|}
		\hline
		$g_{s_{i}}$ & $V(g_{s_{i}})$ & $g_{s_{i}}$ & $V(g_{s_{i}})$\\
		\hline
		\hline 
		0, 1 & 0 & 11        & 8\\
		\hline
		2, 3 & 1 & 12        & 9\\
		\hline
		4, 5 & 2 & 13        & 10\\
		\hline
		6    & 3 & 14, 15    & 11\\
		\hline
		7    & 4 & 16 -- 18  & 12\\
		\hline
		8    & 5 & 19 -- 23  & 13\\
		\hline
		9    & 6 & 24 -- 36  & 14\\
		\hline
		10   & 7 & $\geq$~37 & 15\\
		\hline
	\end{tabular}
	\caption{
		A possible optimal velocity function (OVF) for the TCA model of Helbing and 
		Schreckenberg (HS-TCA). The OVF is represented as a table, giving the 
		optimal speed $V(g_{s_{i}})$ (expressed as cells/time step) associated with 
		each possible space gap $g_{s_{i}}$ (expressed as a number of cells).
	}
	\label{table:TCA:HSTCAOVM}
\end{table}

\begin{figure}[!htbp]
	\centering
	\framebox{\includegraphics[width=\halffigurewidth]{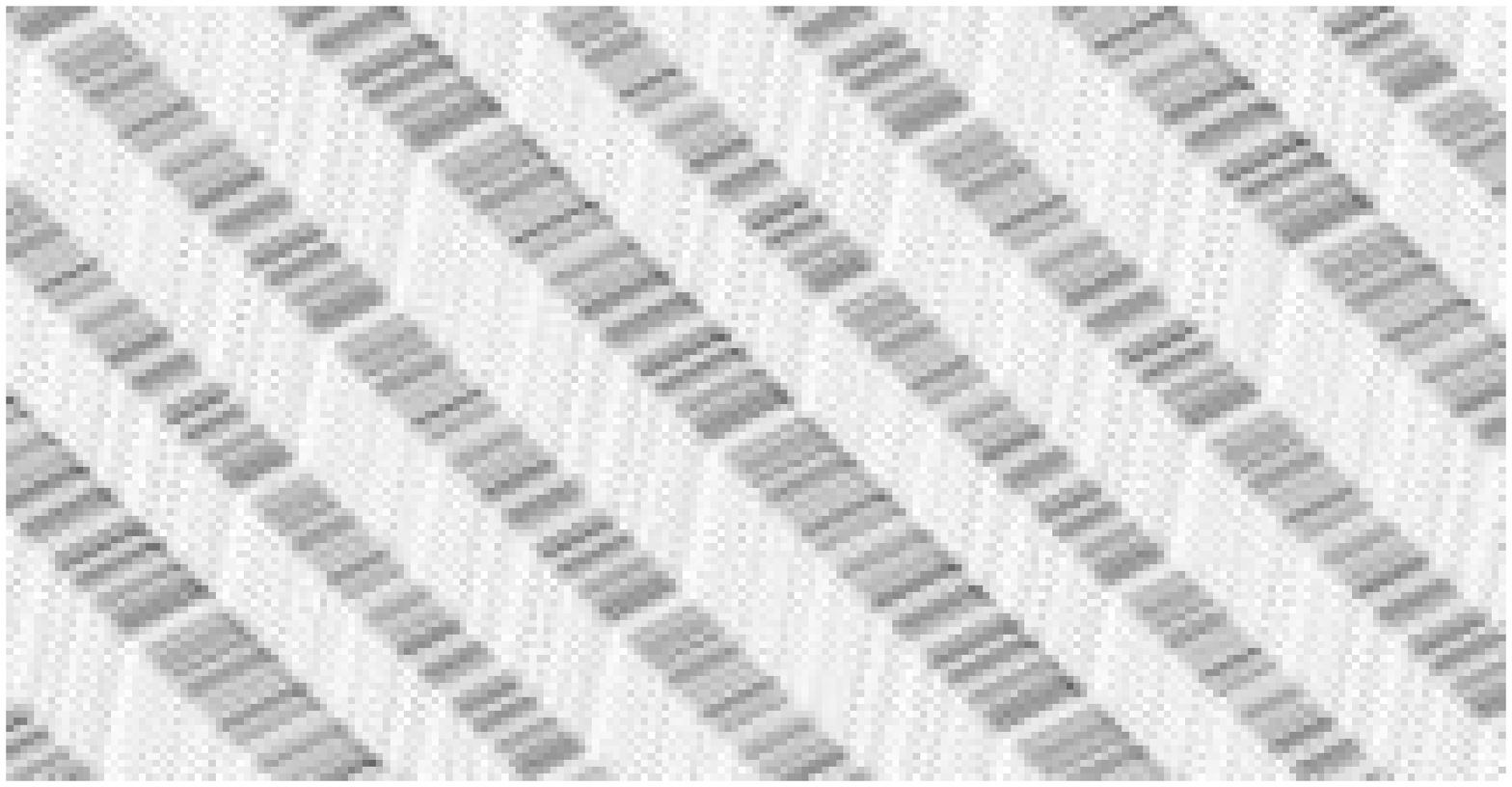}}
	\hspace{\figureseparation}
	\framebox{\includegraphics[width=\halffigurewidth]{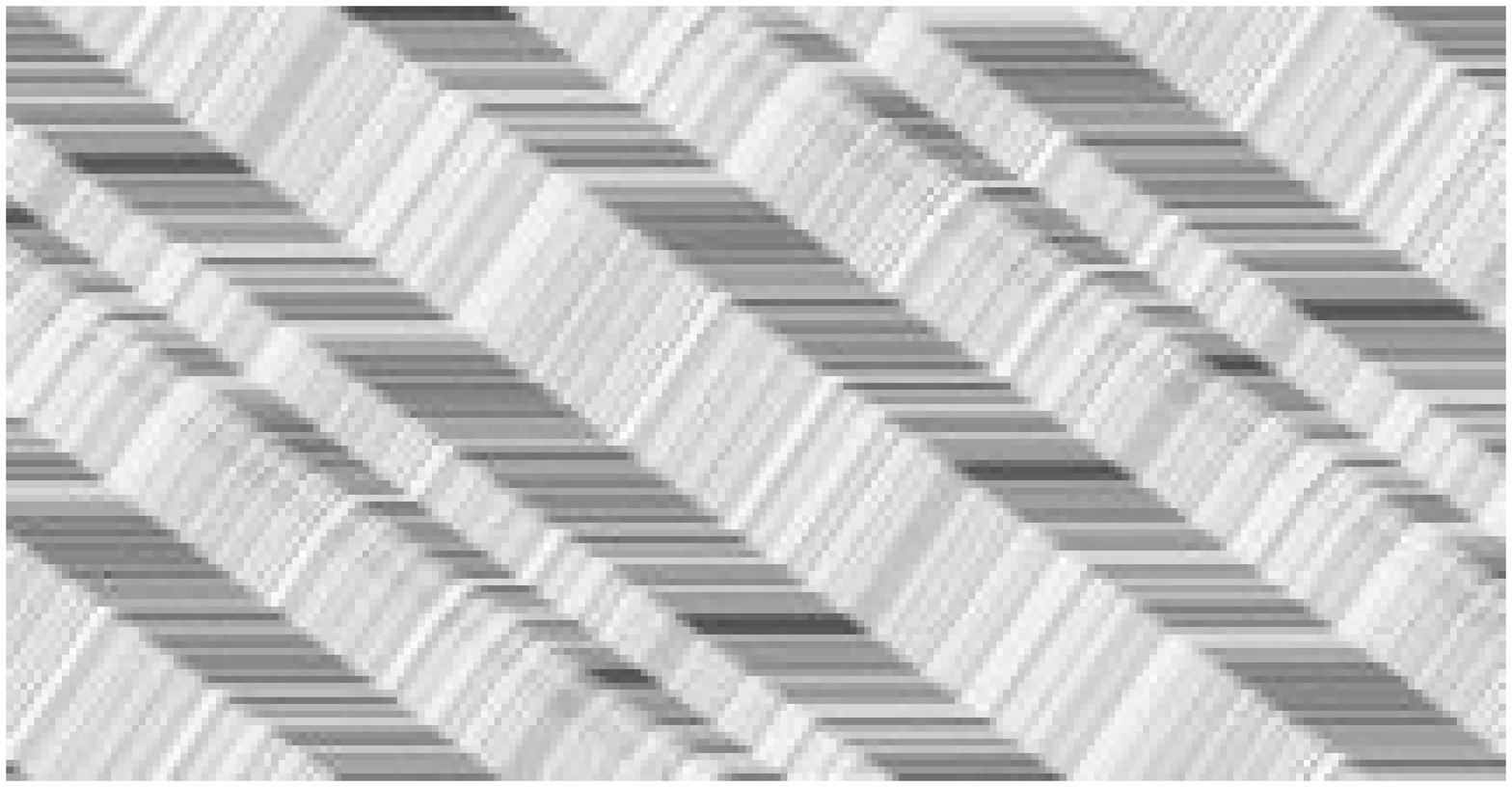}}
	\caption{
		Typical time-space diagrams of the HS-TCA model, with $l = \two$ cells, $p 
		=$~0.001 $\alpha = \one \div$~1.3, and $v_{\text{max}} =$~15 cells/time 
		step. The shown closed-loop lattices each contain 300~$\times \two =$~600 
		cells, with a visible period of 580 time steps. The global density $k$ was 
		set to 0.25 vehicles/cell (\emph{left}) and 0.40 vehicles/cell 
		(\emph{right}). The formation of congestion waves leads to dense, compact 
		jams containing stopped vehicles. Vehicles strive to decelerate smoothly, 
		but are allowed to accelerate instantaneously when exiting jams fronts.
	}
	\label{fig:TCA:HSTCATimeSpaceDiagrams}
\end{figure}

In \figref{fig:TCA:HSTCATimeSpaceDiagrams}, we have given two time-space 
diagrams of the HS-TCA for global densities $k =$~0.25 and $k =$~0.40 
vehicles/cell. The length of a vehicle was $l = \two$ cells, $p =$~0.001, 
$\alpha = \one \div$~1.3, $v_{\text{max}} =$~15 cells/time step, $\Delta T = 
\one$~s, and $\Delta X =$~2.5~m. Due the small slowdown probability, the system 
dynamics are strongly deterministic, totally dependent on the initial 
(homogeneous) conditions. In the left diagram we can observe how vehicles can 
accelerate instantaneously when exiting a jam. Note that for higher densities, 
all jams become dense and compact, always containing stopped vehicles, as is 
depicted in the right diagram. Because of the non-linearity introduced by the 
discretised optimal velocity function, all tempo-spatial patterns in the system 
are of a chaotic nature (i.e., nonlinear with stochastic noise) 
\cite{KNOSPE:04}.

\setlength{\fboxsep}{0pt}
\setlength{\fboxsep}{\tempfboxsep}

\begin{figure}[!htbp]
	\centering
	\includegraphics[width=\halffigurewidth]{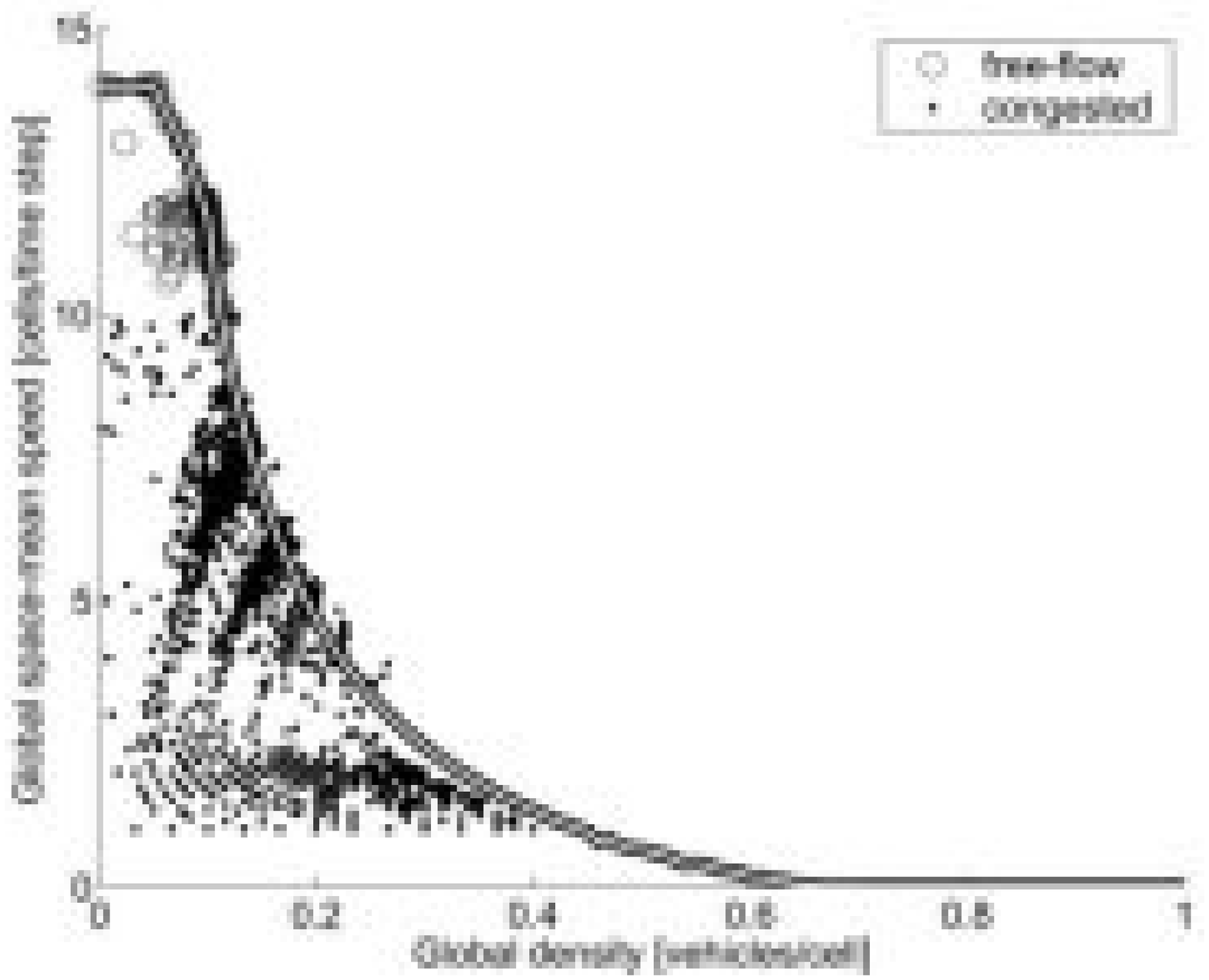}
	\hspace{\figureseparation}
	\includegraphics[width=\halffigurewidth]{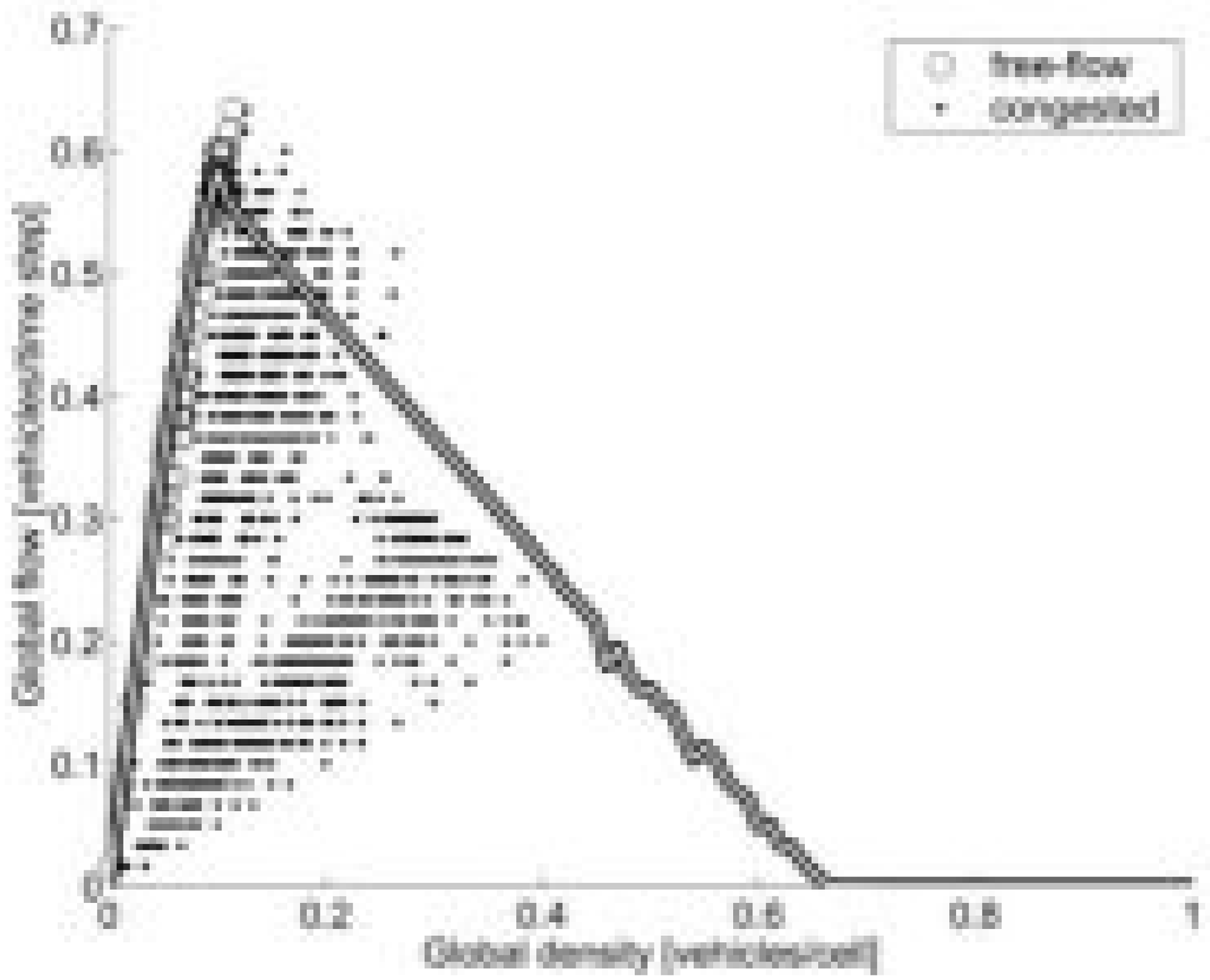}
	\caption{
		The ($k$,$\overline v_{s}$) (\emph{left}) and ($k$,$q$) (\emph{right}) 
		diagrams  for the HS-TCA, obtained by local and global measurements. The 
		local measurements tend to form clusters around certain space-mean speeds, 
		corresponding to the speeds dictated by the discretised optimal velocity 
		function of \tableref{table:TCA:HSTCAOVM}. These clusters are visible in the 
		right diagram as branches with different slopes. Remarkably, from a certain 
		finite density $k \ll \one$ vehicle/cell on, all vehicles always come to a 
		full stop and the flow in the system becomes zero.
	}
	\label{fig:TCA:HSTCAFundamentalDiagrams}
\end{figure}

The ($k$,$\overline v_{s}$) and ($k$,$q$) diagrams in 
\figref{fig:TCA:HSTCAFundamentalDiagrams} are based on local and global 
measurements. A feature of these diagrams is that the local measurements tend to 
form clusters around certain space-mean speeds (see the left part of 
\figref{fig:TCA:HSTCAFundamentalDiagrams}): these clusters correspond to the 
speeds dictated by the discretised optimal velocity function of 
\tableref{table:TCA:HSTCAOVM}, each time associated with an average space gap 
corresponding to the inverse of the locally measured density. As a result, the 
($k$,$q$) diagram in the right part of \figref{fig:TCA:HSTCAFundamentalDiagrams} 
shows several branches, each one with a different OVF speed. The lowest branch 
corresponds to the speed of the backward propagating waves, i.e., the jam speed. 
Even more striking, is that from a certain finite density $k \ll \one$ 
vehicle/cell on, all vehicles always come to a full stop and the flow in the 
system becomes zero \cite{HELBING:99c}.

\begin{figure}[!htbp]
	\centering
	\includegraphics[width=\halffigurewidth]{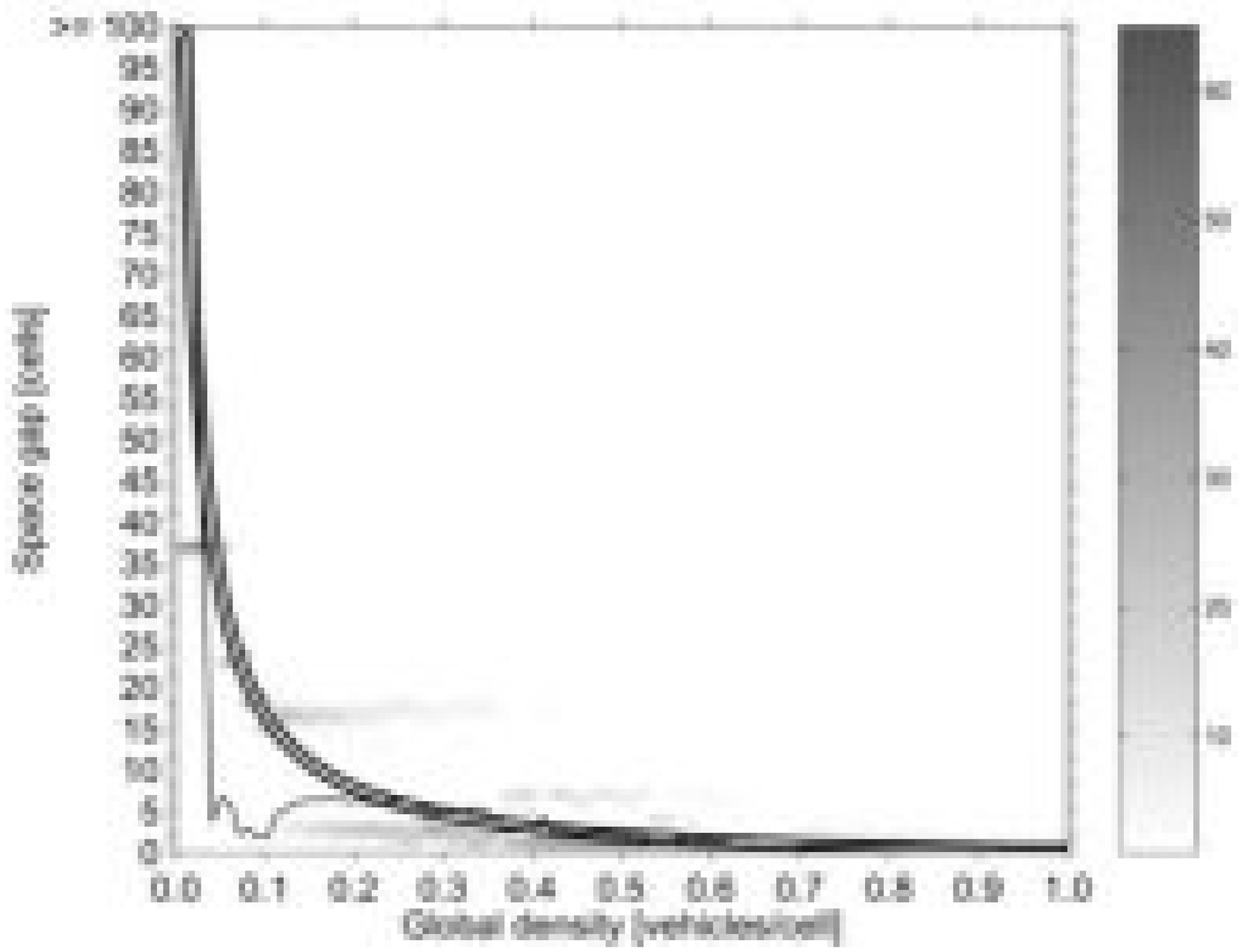}
	\hspace{\figureseparation}
	\includegraphics[width=\halffigurewidth]{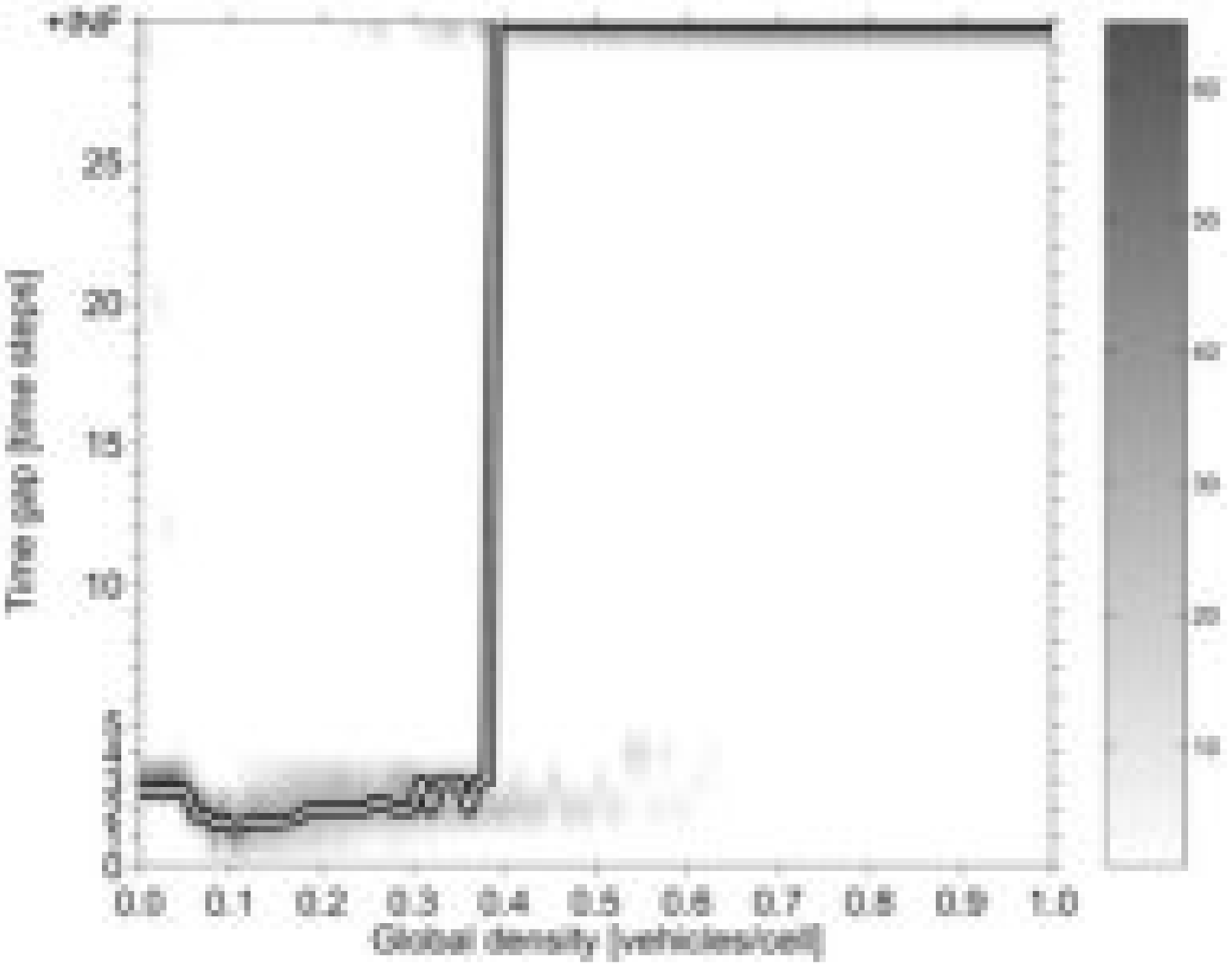}
	\caption{
		Histograms of the distributions of the vehicles' space gaps $g_{s}$ 
		(\emph{left}) and time gaps $g_{t}$ (\emph{right}), as a function of the 
		global density $k$ in the HS-TCA. The thick solid lines denote the mean 
		space gap and median time gap, whereas the thin solid line shows the 
		former's standard deviation. The grey regions denote the probability 
		densities.
	}
	\label{fig:TCA:HSTCASpaceTimeGapHistograms}
\end{figure}

To conclude our discussion of the HS-TCA, we give the histograms of the 
distributions of the space and time gaps in the left and right parts, 
respectively, of \figref{fig:TCA:HSTCASpaceTimeGapHistograms}. The most 
prominent features of these histograms, are that (i) there exist small clusters 
of probability mass between certain space gaps (i.e., 15 -- 20, 25 -- 25, and 35 
-- 40 cells), corresponding to groups of vehicles, (ii) for higher densities, we 
can observe a spread-out cluster of probability mass, corresponding to the 
lowest local measurements in the left part of 
\figref{fig:TCA:HSTCAFundamentalDiagrams}, and (iii) in contrast to the previous 
TCA models, the median of the time gap for the HS-TCA is already very small for 
densities $k <$~0.1.\\

\sidebar{
	The HS-TCA might seem an interesting improvement, as its being based on a 
	discretisation of the optimal velocity model. But although its authors state 
	that it \emph{``reproduces many of empirically observed features''} 
	\cite{HELBING:99c}, Knospe et al. showed several shortcomings in the model 
	\cite{KNOSPE:04}: care must be taken to avoid collisions, and the model fails 
	to reproduce the synchronised-flow regime entirely. This latter can be 
	understood by looking at the dense, compact structure of jams in the 
	time-space diagrams of \figref{fig:TCA:HSTCATimeSpaceDiagrams}, and the 
	occurrence of branches with distinct speeds as in the right part of 
	\figref{fig:TCA:HSTCAFundamentalDiagrams}.
}\\

			\subsubsection{Brake-light TCA (BL-TCA)}
			\label{sec:TCA:BLTCA}

Recently, an interesting idea was pursued by Knospe et al.; their TCA model 
includes \emph{anticipation} effects, introduced by equipping the vehicles with 
\emph{brake lights} \cite{KNOSPE:00b}. The focus of this (and the following) TCA 
model lies in a correct reproduction of the three phases of traffic as 
introduced by Kerner et al. \cite{KERNER:04,MAERIVOET:05c}. In a sense, the 
BL-TCA incorporates many of the features encountered in previously discussed 
single-cell TCA models. First of all, the BL-TCA has randomisation for 
spontaneous braking. Secondly, it has slow-to-start behaviour for the capacity 
drop and hysteresis phenomena. Moreover, it incorporates anticipation which can 
lead to a stabilisation of the free-flow branch. Finally, it includes elements 
for reproducing synchronised traffic. These latter two aspects clearly go beyond 
the standard incentive if drivers to avoid collisions. As such, it is the desire 
for smooth and comfortable driving (which resembles \emph{human behaviour}), is 
responsible for the occurrence of traffic states like e.g., synchronised traffic 
\cite{KNOSPE:02b}. To achieve all this, the rule set of the BL-TCA becomes quite 
complex, in comparison with some of the more standard single-cell TCA models of 
section \ref{sec:TCA:SingleCellModels}:

\begin{quote}
	\textbf{R0}: \emph{determine stochastic noise}\\
		\begin{equation}
		\label{eq:TCA:BLTCAR0}
			\left \lbrace
				\begin{array}{l}
					b_{i + \one}(t - \one) = \one \quad \wedge \quad t_{h_{i}}(t - \one) < t_{s_{i}}(t - \one)\\
						\qquad \Longrightarrow p(t) \leftarrow p_{b}, \\
					v_{i}(t - \one) = \zero\\
						\qquad \Longrightarrow p(t) \leftarrow p_{\zero}, \\
					\text{else}\\
						\qquad \Longrightarrow p(t) \leftarrow p_{d},\\
					b_{i}(t) \leftarrow \zero,
				\end{array}
			\right.
		\end{equation}

	\textbf{R1}: \emph{acceleration}\\
		\begin{equation}
			\begin{array}{lcl}
				(b_{i}(t - \one) = \zero \quad \wedge \quad b_{i + \one}(t - \one) = \zero) \quad\\
					\quad \vee \quad t_{h_{i}}(t) \geq t_{s_{i}}(t)\\
						\qquad \Longrightarrow v_{i}(t) \leftarrow \text{min} \lbrace v_{i}(t + \one), v_{\text{max}} \rbrace,
			\end{array}
		\end{equation}

	\textbf{R2a}: \emph{determine effective space gap}\\
		\begin{equation}
		\label{eq:TCA:BLTCAR2a}
			\begin{array}{lcl}
				g_{s_{i}}^{*}(t) \leftarrow\\
					\quad g_{s_{i}}(t - \one) +\\
					\quad \text{max} \lbrace \underbrace{\text{min} \lbrace v_{i + \one}(t - \one),
						g_{s_{i + \one}}(t - \one) \rbrace}_{\text{anticipated speed of leading vehicle}} -
							g_{s_{\text{security}}}, \zero \rbrace,
			\end{array}
		\end{equation}

	\textbf{R2b}: \emph{braking}\\
		\begin{equation}
		\label{eq:TCA:BLTCAR2b}
			\begin{array}{lcl}
				v_{i}(t)      \leftarrow \text{min} \lbrace v_{i}(t), g_{s_{i}}^{*}(t) \rbrace,\\
				v_{i}(t) < v_{i}(t - \one)\\
					\quad \Longrightarrow b_{i}(t) \leftarrow \one,
			\end{array}
		\end{equation}

	\textbf{R3}: \emph{randomisation}\\
		\begin{equation}
		\label{eq:TCA:BLTCAR3}
			\begin{array}{lcl}
				\xi(t) < p(t) \Longrightarrow\\
					\quad p(t) = p_{b} \quad \wedge \quad v_{i}(t) = v_{i}(t - \one) + \one\\
					\qquad \Longrightarrow b_{i}(t) \leftarrow \one,\\
					\quad v_{i}(t) \leftarrow \max \lbrace \zero,v_{i}(t) - \one \rbrace,
			\end{array}
		\end{equation}

	\textbf{R4}: \emph{vehicle movement}\\
		\begin{equation}
			x_{i}(t) \leftarrow x_{i}(t - \one) + v_{i}(t),
		\end{equation}
\end{quote}

where $b_{i}(t)$ denotes the state (0 or 1) of the brake light of the \ith 
vehicle at time step $t$, $t_{h_{i}} = g_{s_{i}} / v_{i}$ and $t_{s_{i}} = 
\text{min} \lbrace v_{i}, h \rbrace$ with $h$ the interaction range of the brake 
light. As such, $t_{h_{i}}$ is the time to reach the leading vehicle, which gets 
compared with an \emph{interaction horizon} $t_{s_{i}}$ that depends on the 
speed $v_{i}$ and is constrained by $h$. If the leading vehicle is far away, its 
brake light should not influence the following vehicle. Furthermore, rule R0 
also takes into account that drivers are more alert when they are travelling at 
high speeds. The slowdown probability $p$ in rule R0, equation 
\eqref{eq:TCA:BLTCAR0}, corresponds to either the \emph{braking probability} 
$p_{b}$, the slow-to-start probability $p_{\zero}$, or the classic slowdown 
probability $p_{d}$ for decelerations. Finally, $g_{s_{i}}^{*}(t)$ in rules R2a 
and R2b, equations \eqref{eq:TCA:BLTCAR2a} and \eqref{eq:TCA:BLTCAR2b}, 
respectively, denotes the \emph{effective space gap}, based on the 
\emph{anticipated speed} of the leading vehicle and taking into account a 
\emph{security constraint} $g_{s_{\text{security}}}$. Just as the previous TCA 
models, the BL-TCA is stochastic, in that it introduces randomisation by means 
of rule R3, equation \eqref{eq:TCA:BLTCAR3}, with $\xi(t) \in [\zero,\one[$ a 
random number drawn from a uniform distribution. If a vehicle was in the process 
of braking due to the previous rules, then its brake light $b_{i}$ is turned on. 
Note that Knospe also extended the BL-TCA with rules that allow asymmetric lane 
changing on a two-lane road (unidirectional), incorporating a right-lane 
preference as well as an overtaking prohibition on the right lane. As such, the 
model correctly reflects the density inversion phenomenon (see also section 
\ref{sec:TCA:MultiLaneTraffic}) \cite{KNOSPE:02b,KNOSPE:02c}.

In the remainder of this discussion, we set $p_{b} =$~0.94, $p_{\zero} =$~0.5, 
$p_{d} =$~0.1, $h =$~6 time steps, $g_{s_{\text{security}}} =$~7 cells, 
$v_{\text{max}} =$~20 cells/time step, with a vehicle length of $l =$~5 cells, 
$\Delta T = \one$~s, and $\Delta X =$~1.5~m \cite{KNOSPE:00b,KNOSPE:04}. With 
respect to the calibration of the BL-TCA model's parameters, Knospe et al. 
provide a nice overview, giving intuitive analogies for each of these parameters 
(e.g., $p_{0}$ is associated with the speed of the backward propagating waves) 
\cite{KNOSPE:04}.

\begin{figure}[!htbp]
	\centering
	\framebox{\includegraphics[width=\halffigurewidth]{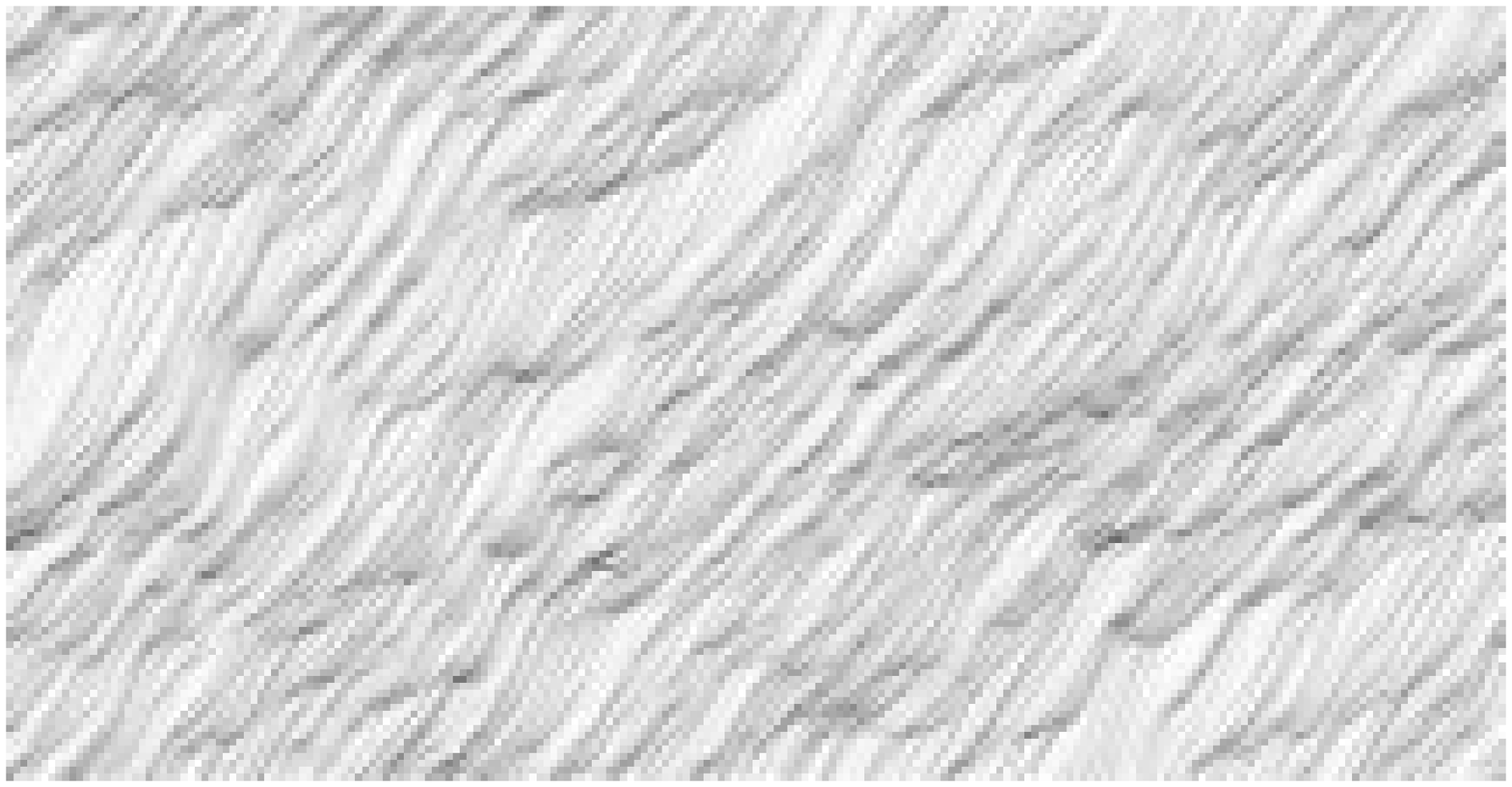}}
	\hspace{\figureseparation}
	\framebox{\includegraphics[width=\halffigurewidth]{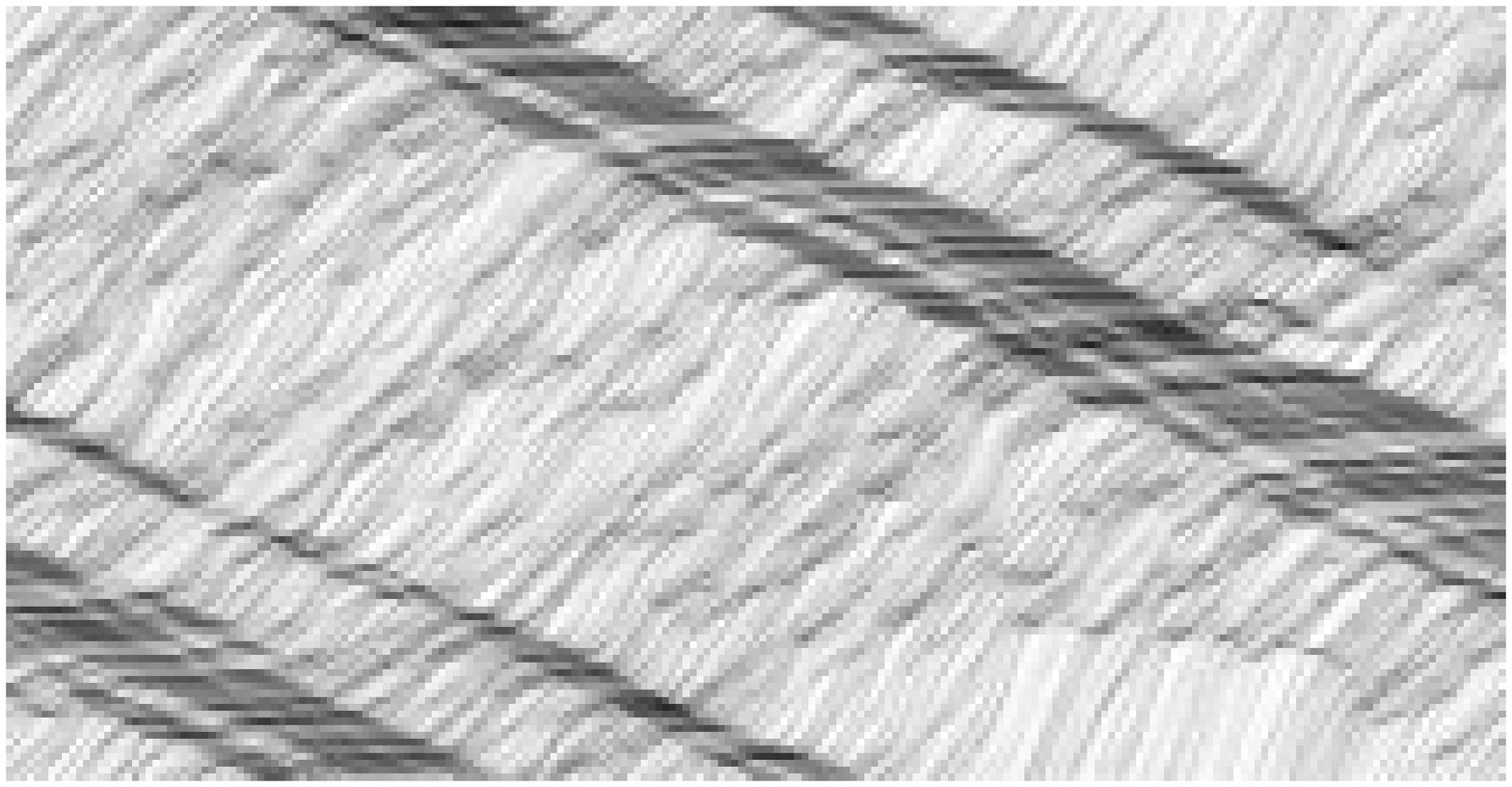}}
	\caption{
		Typical time-space diagrams of the BL-TCA model (refer to the text for the 
		used parameter values). The shown closed-loop lattices each contain 
		300~$\times \five =$~1500 cells, with a visible period of 580 time steps. 
		The global density $k$ was set to 0.25 vehicles/cell (\emph{left}) and 0.40 
		vehicles/cell (\emph{right}). The visible forward propagating density waves 
		are a result of the anticipation and synchronisation phenomena. At higher 
		densities, stable jams occur, indicative of the wide-moving jam phase.
	}
	\label{fig:TCA:BLTCATimeSpaceDiagrams}
\end{figure}

In \figref{fig:TCA:BLTCATimeSpaceDiagrams}, we have given two time-space 
diagrams of the BL-TCA for global densities $k =$~0.25 and $k =$~0.40 
vehicles/cell. As can be seen in the time-space diagram in the left part, the 
anticipation and synchronisation phenomena lead to forward propagating density 
waves, where vehicles carry the density downstream. Going to higher densities, 
we can see stable jams, indicative of the wide-moving jam phase (see also 
Kerner's three-phase traffic theory \cite{KERNER:04,MAERIVOET:05c}).

\setlength{\fboxsep}{0pt}
\setlength{\fboxsep}{\tempfboxsep}

\begin{figure}[!htbp]
	\centering
	\includegraphics[width=\halffigurewidth]{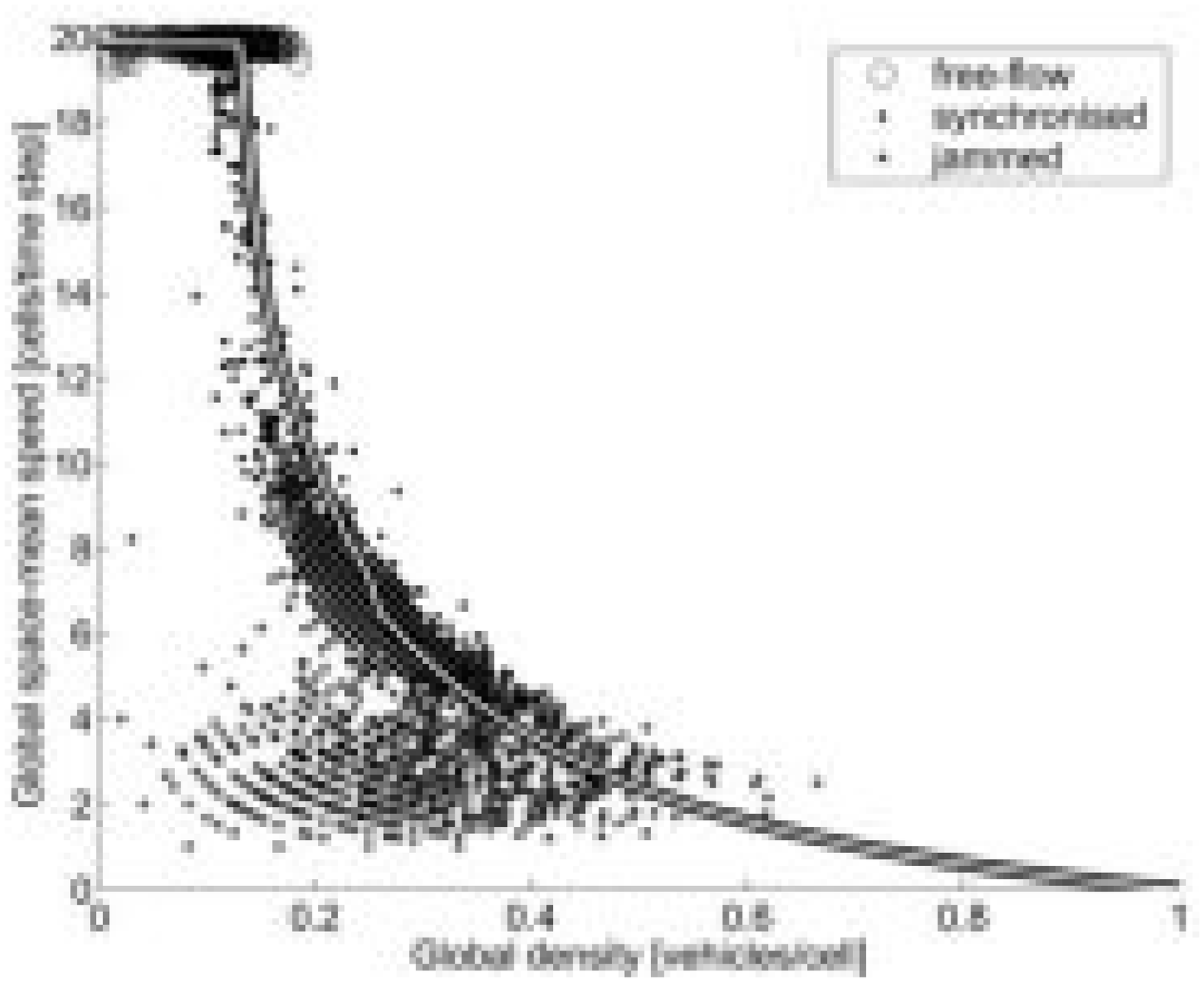}
	\hspace{\figureseparation}
	\includegraphics[width=\halffigurewidth]{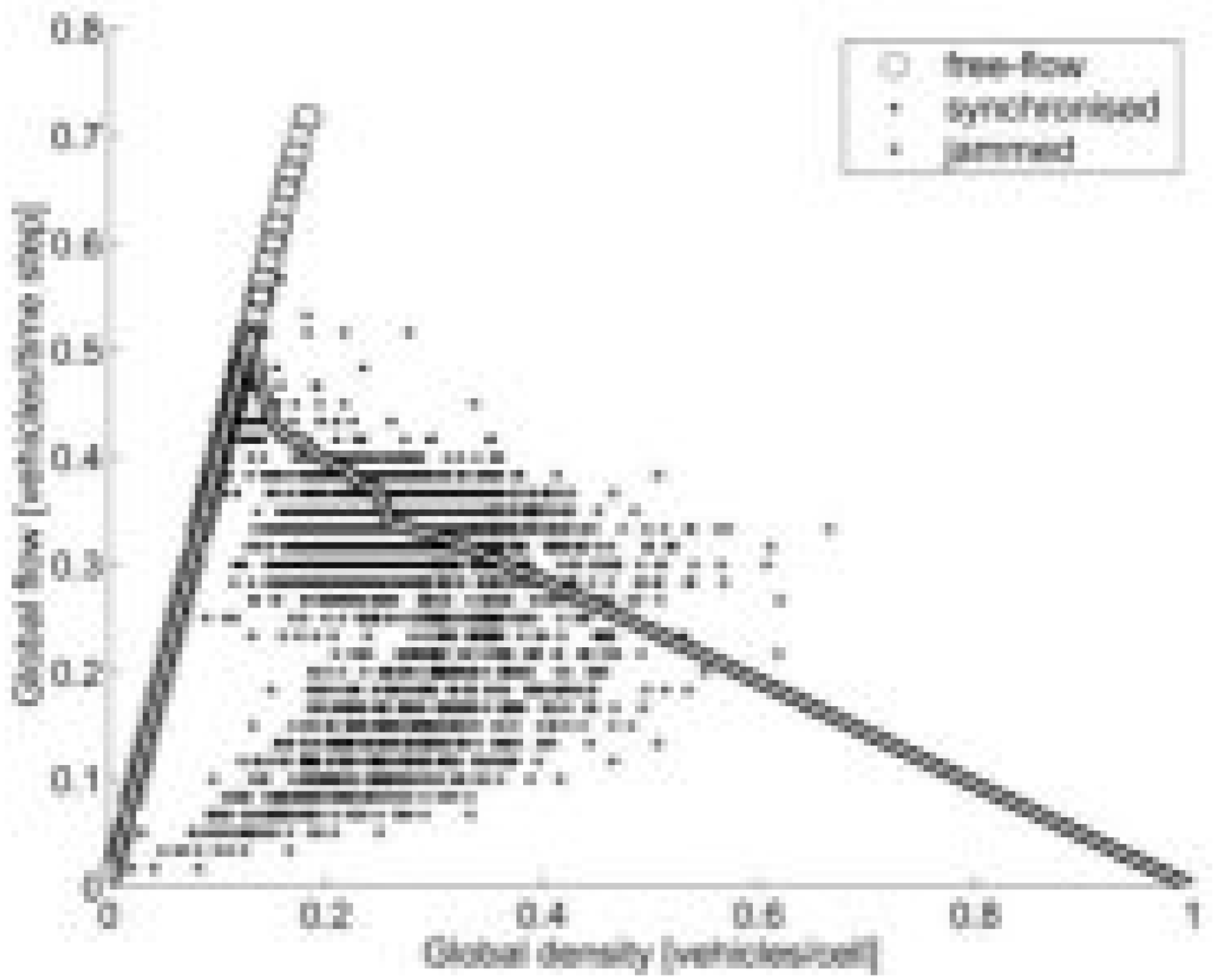}
	\caption{
		The ($k$,$\overline v_{s}$) (\emph{left}) and ($k$,$q$) (\emph{right}) 
		diagrams for the BL-TCA model, obtained by local and global measurements. 
		The local measurements discriminate between the free-flow ($\circ$), 
		synchronised-flow ($\cdot$), and jammed regimes ($\star$). The synchronised 
		regime is visible as a wide scatter in the data points, having various 
		speeds but relatively high flows. The data points in the wide-moving jam 
		correspond to Kerner's line $J$.
	}
	\label{fig:TCA:BLTCAFundamentalDiagrams}
\end{figure}

Looking at the ($k$,$\overline v_{s}$) and ($k$,$q$) diagrams in 
\figref{fig:TCA:BLTCAFundamentalDiagrams}, we can use the local measurements to 
discriminate between the free-flow ($\circ$), synchronised-flow ($\cdot$), and 
jammed regimes ($\star$). The synchronised regime is visible as a wide scatter 
in the data points, having various speeds but relatively high flows. The data 
points in the wide-moving jam correspond to Kerner's so-called line $J$ 
\cite{KERNER:04,MAERIVOET:05c}. The use of a finer discretisation can lead to 
metastable states (see section \ref{sec:TCA:ArtifactsOfAMultiCellSetup}), but as 
Knospe et al. note, the slow-to-start behaviour in rule R0, equation 
\eqref{eq:TCA:BLTCAR0}, is necessary in order to produce the correct speed of 
the backward propagating wave, as a result of a reduced outflow from a jam 
\cite{KNOSPE:04}.

\begin{figure}[!htbp]
	\centering
	\includegraphics[width=\halffigurewidth]{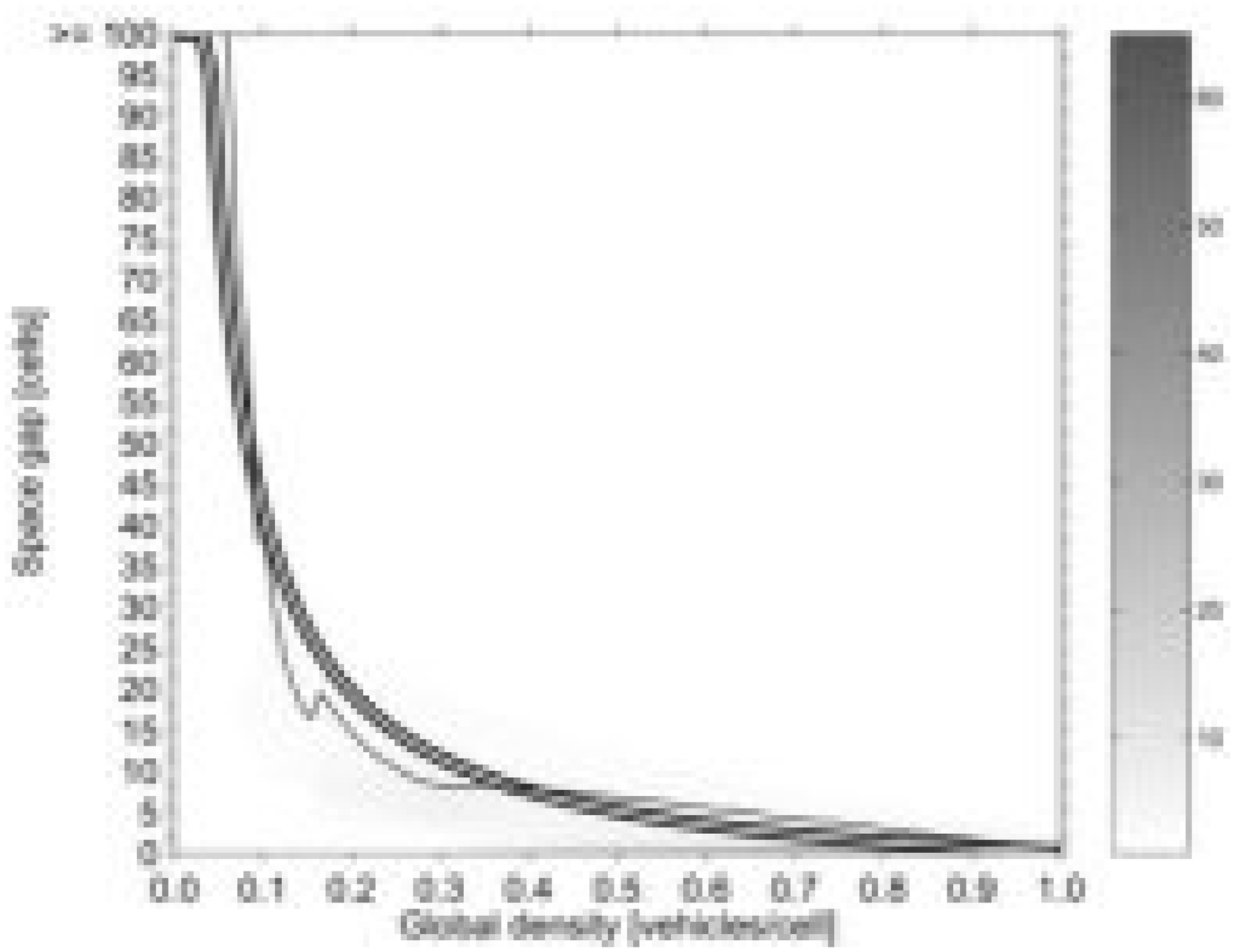}
	\hspace{\figureseparation}
	\includegraphics[width=\halffigurewidth]{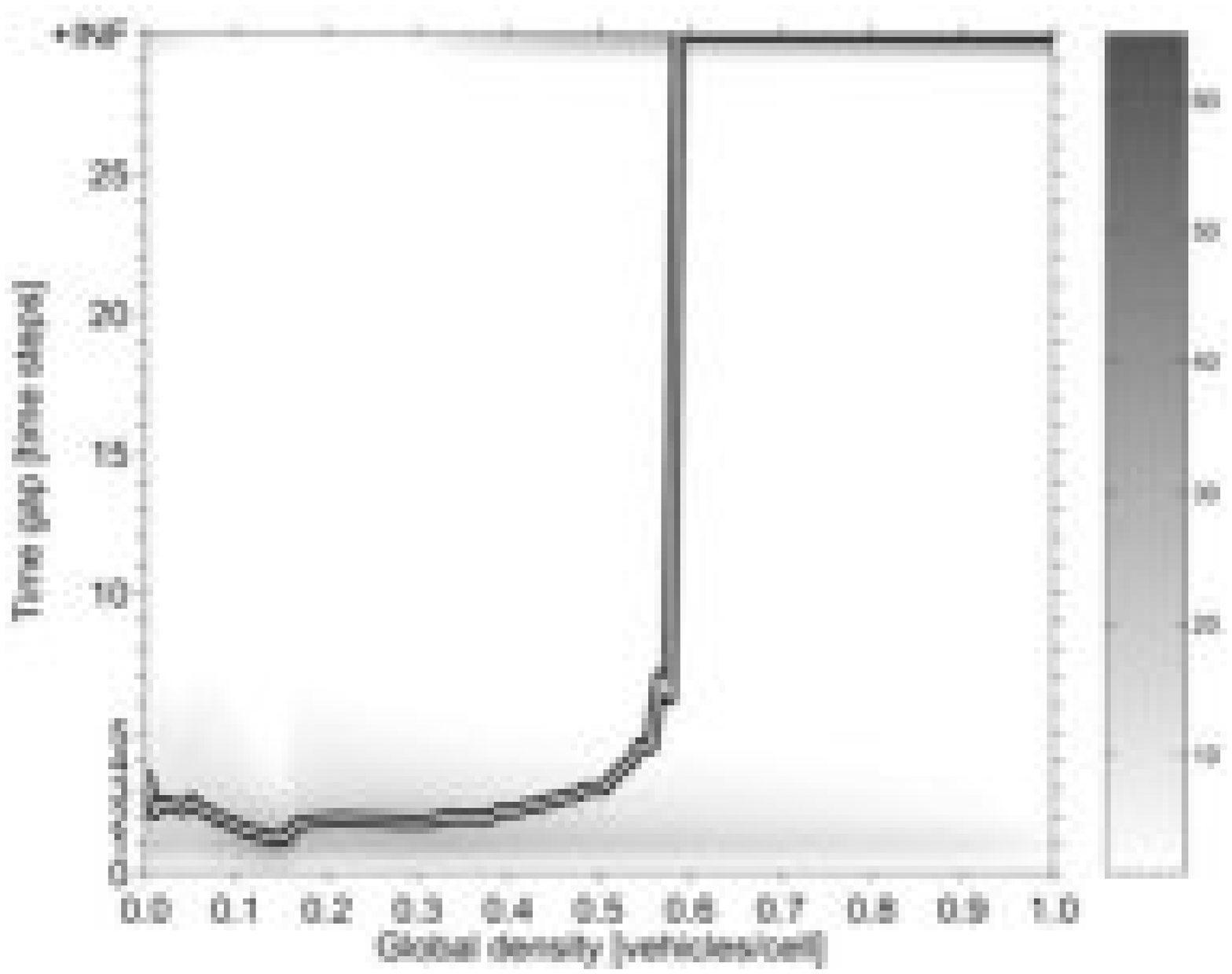}
	\caption{
		Histograms of the distributions of the vehicles' space gaps $g_{s}$ 
		(\emph{left}) and time gaps $g_{t}$ (\emph{right}), as a function of the 
		global density $k$ in the BL-TCA model. The thick solid lines denote the 
		mean space gap and median time gap, whereas the thin solid line shows the 
		former's standard deviation. The grey regions denote the probability 
		densities.
	}
	\label{fig:TCA:BLTCASpaceTimeGapHistograms}
\end{figure}

Finally, \figref{fig:TCA:BLTCASpaceTimeGapHistograms} depicts the histograms of 
the distributions of the space and time gaps in the left and right parts, 
respectively. In contrast to the HS-TCA, there are no more clusters for the 
space gap (see left part of \figref{fig:TCA:HSTCASpaceTimeGapHistograms}), but 
rather a smooth region of probability mass: as the global density of the system 
increases, the average space gap diminishes continuously and monotonically. The 
observations for the distributions of the time gaps correspond to those 
encountered in literature \cite{KNOSPE:00b,KNOSPE:04}: from the right part of 
\figref{fig:TCA:BLTCASpaceTimeGapHistograms}, we can see a wide range of 
probability mass at low densities (free-flow traffic), corresponding to a wide 
distribution of time gaps. At intermediate densities (synchronised flow), the 
distribution tends to peak, leading to a small dense cluster at approximately $k 
=$~0.15 vehicles/cell, with a median time gap of 1 time step. Finally, at higher 
densities (jammed traffic), the distribution of the time gaps gets more peaked, 
as is illustrated by the narrowing of the grey region of probability mass.

			\subsubsection{The model of Kerner, Klenov, and Wolf (KKW-TCA)}
			\label{sec:TCA:KKWTCA}

Based upon the BL-TCA of Knospe et al., Kerner, Klenov, and Wolf (KKW) refined 
this approach by extending it. Their work resulted in a family of models that 
incorporate the notion of a \emph{synchronisation distance} for individual 
vehicles \cite{KERNER:03}. Derived from this model class, Kerner et al. proposed 
discretised versions in the form of traffic cellular automata models. In this 
report, we consider the KKW-1 TCA model, of which the complex rule set is as 
follows \cite{KERNER:02}:

\begin{quote}
	\textbf{R1a}: \emph{determine synchronisation distance}\\
		\begin{equation}
			D_{i}(t) \leftarrow D_{\zero} + D_{\one} v_{i}(t - \one),\\
		\end{equation}

	\textbf{R1b}: \emph{determine acceleration and deceleration}\\
		\begin{equation}
			\left \lbrace
				\begin{array}{lcl}
					v_{i}(t - \one) < v_{i + \one}(t - \one) \Longrightarrow 
						\Delta_{\text{acc}_{i}}(t) \leftarrow a,\\
					v_{i}(t - \one) = v_{i + \one}(t - \one) \Longrightarrow 
						\Delta_{\text{acc}_{i}}(t) \leftarrow \zero,\\
					v_{i}(t - \one) > v_{i + \one}(t - \one) \Longrightarrow 
						\Delta_{\text{acc}_{i}}(t) \leftarrow -b,\\
				\end{array}
			\right.
		\end{equation}

	\textbf{R1c}: \emph{determine desired speed}\\
		\begin{equation}
			\left \lbrace
				\begin{array}{lcl}
					g_{s_{i}}(t - \one) > (D_{i}(t) - l_{i})\\
					\qquad \Longrightarrow v_{\text{des}_{i}}(t) \leftarrow v_{i}(t - \one) + a,\\
					g_{s_{i}}(t - \one) \leq (D_{i}(t) - l_{i})\\
					\qquad \Longrightarrow v_{\text{des}_{i}}(t) \leftarrow v_{i}(t - \one) + \Delta_{\text{acc}_{i}}(t),\\
				\end{array}
			\right.
		\end{equation}

	\textbf{R1d}: \emph{determine deterministic speed}\\
		\begin{equation}
			v_{i}(t) \leftarrow \text{max} \lbrace \zero,
				\text{min} \lbrace v_{\text{max}}, g_{s_{i}}(t), v_{\text{des}_{i}}(t) \rbrace \rbrace,
		\end{equation}

	\textbf{R2a}: \emph{determine acceleration probability}\\
		\begin{equation}
			\left \lbrace
				\begin{array}{lcl}
					v_{i}(t) < v_{p} \Longrightarrow p_{a}(t) \leftarrow p_{a_{\one}},\\
					v_{i}(t) \geq v_{p} \Longrightarrow p_{a}(t) \leftarrow p_{a_{\two}},\\
				\end{array}
			\right.
		\end{equation}

	\textbf{R2b}: \emph{determine braking probability}\\
		\begin{equation}
			\left \lbrace
				\begin{array}{lcl}
					v_{i}(t) = \zero \Longrightarrow p_{b}(t) \leftarrow p_{\zero},\\
					v_{i}(t) > \zero \Longrightarrow p_{b}(t) \leftarrow p_{d},\\
				\end{array}
			\right.
		\end{equation}

	\textbf{R2c}: \emph{determine stochastic noise}\\
		\begin{equation}
			\left \lbrace
				\begin{array}{lcl}
					\xi(t) < p_{a}(t)                          & \Longrightarrow & \eta_{i}(t) \leftarrow a,\\
					p_{a}(t) \leq \xi(t) < p_{a}(t) + p_{b}(t) & \Longrightarrow & \eta_{i}(t) \leftarrow -b,\\
					\xi(t) \geq p_{a}(t) + p_{b}(t)            & \Longrightarrow & \eta_{i}(t) \leftarrow \zero,\\
				\end{array}
			\right.
		\end{equation}

	\textbf{R2d}: \emph{determine stochastic speed}\\
		\begin{equation}
			v_{i}(t) \leftarrow \text{max} \lbrace \zero,
				\text{min} \lbrace v_{\text{max}}, v_{i}(t) + \eta_{i}(t), v_{i}(t) + a \rbrace \rbrace,
		\end{equation}

	\textbf{R3}: \emph{vehicle movement}\\
		\begin{equation}
			x_{i}(t) \leftarrow x_{i}(t - \one) + v_{i}(t).
		\end{equation}
\end{quote}

As can be seen from this overview, the KKW-TCA model's rule set is mainly 
composed of a \emph{deterministic} part (rules R1a -- R1d) and a 
\emph{stochastic} part (rules R2a -- R2d). In the deterministic part, the 
synchronisation distance $D_{i}$ is computed first with rule R1a, which uses a 
linear function (other forms, e.g., quadratic functions, are also possible). The 
parameters $D_{\zero}$ and $D_{\one}$ need to be estimated. Rule R1c determines 
the desired speed $v_{\text{des}_{i}}$: the first part of the rule allows the 
vehicle to accelerate, whereas the second part of the rule uses an acceleration 
$\Delta_{\text{acc}_{i}}$ defined by rule R1b ($a$ and $b$ are parameters 
denoting the acceleration, and respectively braking, capabilities). As such, a 
vehicle will tend to adapt its speed to that of its direct frontal leader, 
whenever the vehicle is within a zone of interaction (i.e., the synchronisation 
distance). The deterministic speed is then computed by means of rule R1d, which 
takes into account the maximum speed $v_{\text{max}}$, the space gap $g_{s_{i}}$ 
to avoid a collision, and the previously computed desired speed of rule R1c.

In the stochastic part for computing the speed, a randomisation is introduced in 
rule R2d by means of a stochastic acceleration $\eta_{i}$. The values of 
$\eta_{i}$ are obtained in rule R2c with probability $p_{a}$ for accelerating, 
and probability $p_{b}$ for braking. The former is dependent on the vehicles 
computed deterministic speed and the parameters $v_{p}$, $p_{a_{\one}}$, and 
$p_{a_{\two}}$ with $p_{a_{\one}} > p_{a_{\two}}$ and $p_{a_{\one}} + 
p_{a_{\two}} \leq \one$. The latter, $p_{b}$ is dependent on the vehicles 
computed deterministic speed and the slowdown probability $p_{d}$ and the 
slow-to-start probability $p_{\zero}$ with $p_{\zero} > p_{d}$.

In the remainder of this discussion, we set $D_{\zero} =$~60, $D_{\one} =$~2.55, 
$a = b = \one$, $v_{p} =$~28, $p_{a_{\one}} =$~0.2, $p_{a_{\two}} =$~0.052, 
$p_{\zero} =$~0.425, $p_{d} =$~0.04, $v_{\text{max}} =$~60 cells/time step, with 
a vehicle length of $l =$~15 cells, $\Delta T = \one$~s, and $\Delta X =$~0.5~m 
\cite{KNOSPE:04}.

\begin{figure}[!htbp]
	\centering
	\framebox{\includegraphics[width=\halffigurewidth]{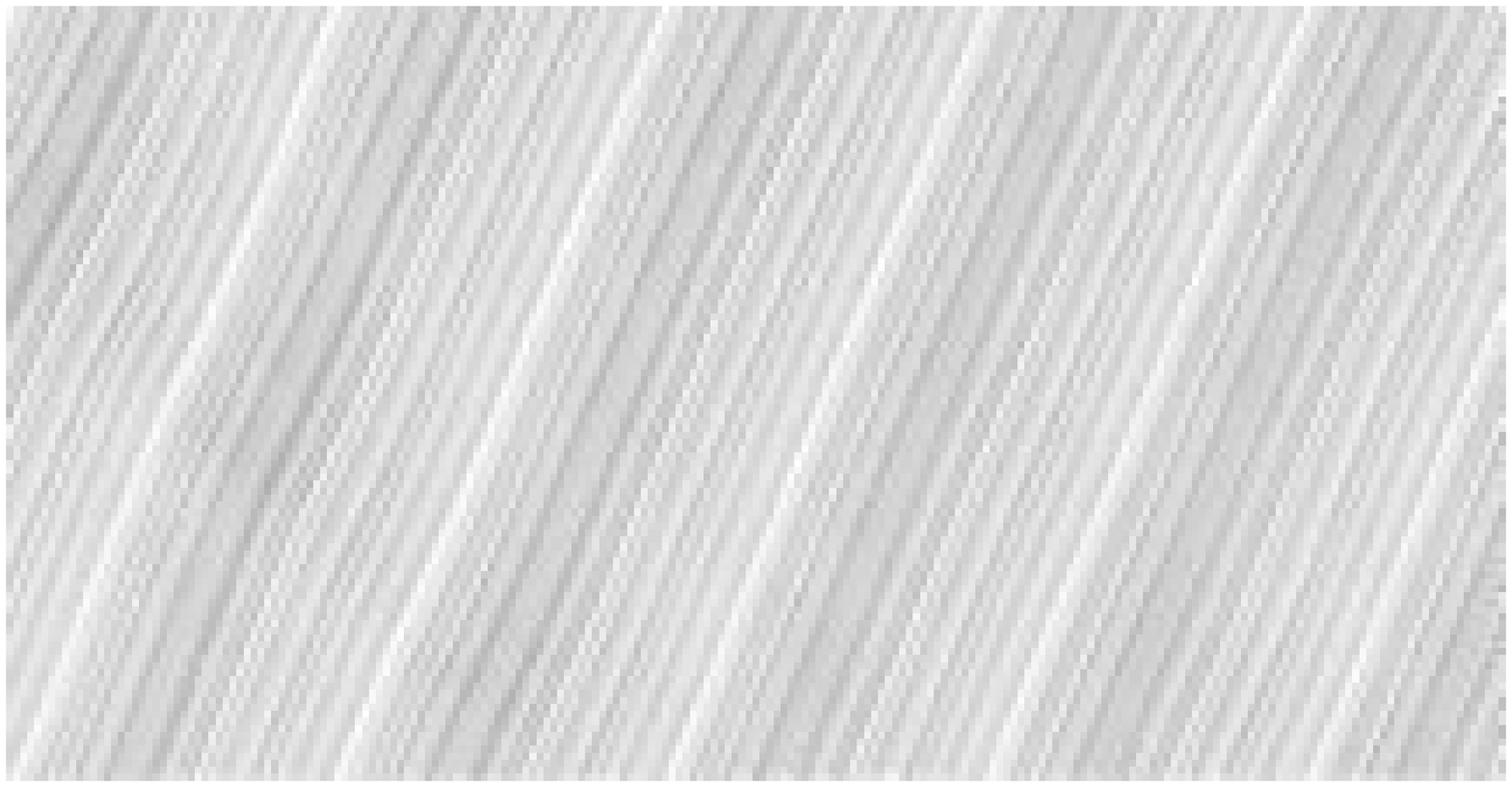}}
	\hspace{\figureseparation}
	\framebox{\includegraphics[width=\halffigurewidth]{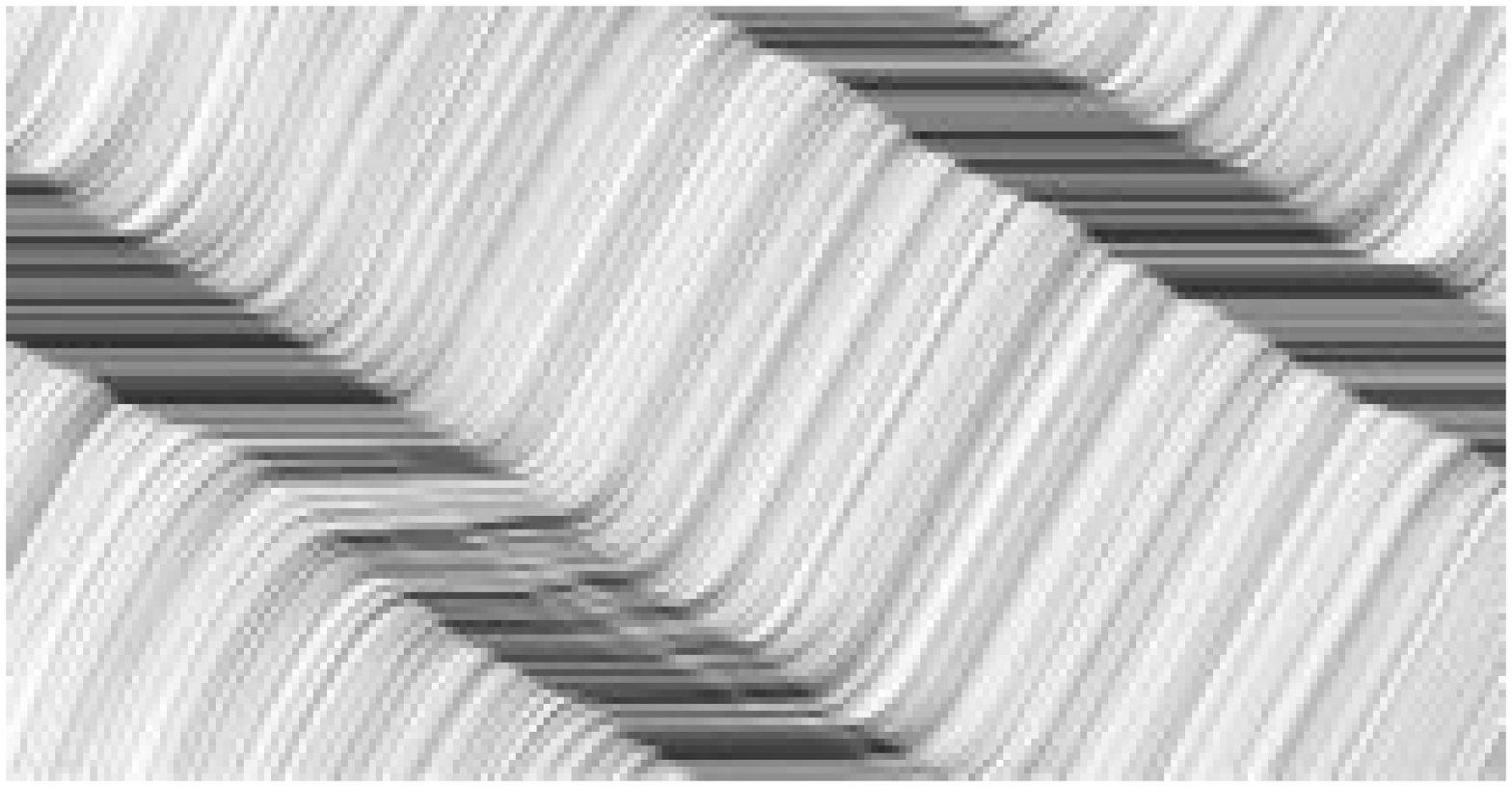}}
	\caption{
		Typical time-space diagrams of the KKW-TCA model (refer to the text for the 
		used parameter values). The shown closed-loop lattices each contain 
		300~$\times \text{15} =$~4500 cells, with a visible period of 580 time 
		steps. The global density $k$ was set to 0.25 vehicles/cell (\emph{left}) 
		and 0.40 vehicles/cell (\emph{right}). Note the stable flow of vehicles 
		surrounding the dense and compact superjams.
	}
	\label{fig:TCA:KKWTCATimeSpaceDiagrams}
\end{figure}

Considering the KKW-TCA models' time-space diagrams in 
\figref{fig:TCA:KKWTCATimeSpaceDiagrams}, we can see that, in contrast to the 
BL-TCA (see section \ref{sec:TCA:BLTCA}), there are less spontaneous formations 
of small traffic jams. The forward propagating density waves in 
\figref{fig:TCA:BLTCATimeSpaceDiagrams} are absent in the KKW-TCA model. 
However, the two models show good correspondence with respect to the speed of 
the backward propagating waves.

\setlength{\fboxsep}{0pt}
\setlength{\fboxsep}{\tempfboxsep}

\begin{figure}[!htbp]
	\centering
	\includegraphics[width=\halffigurewidth]{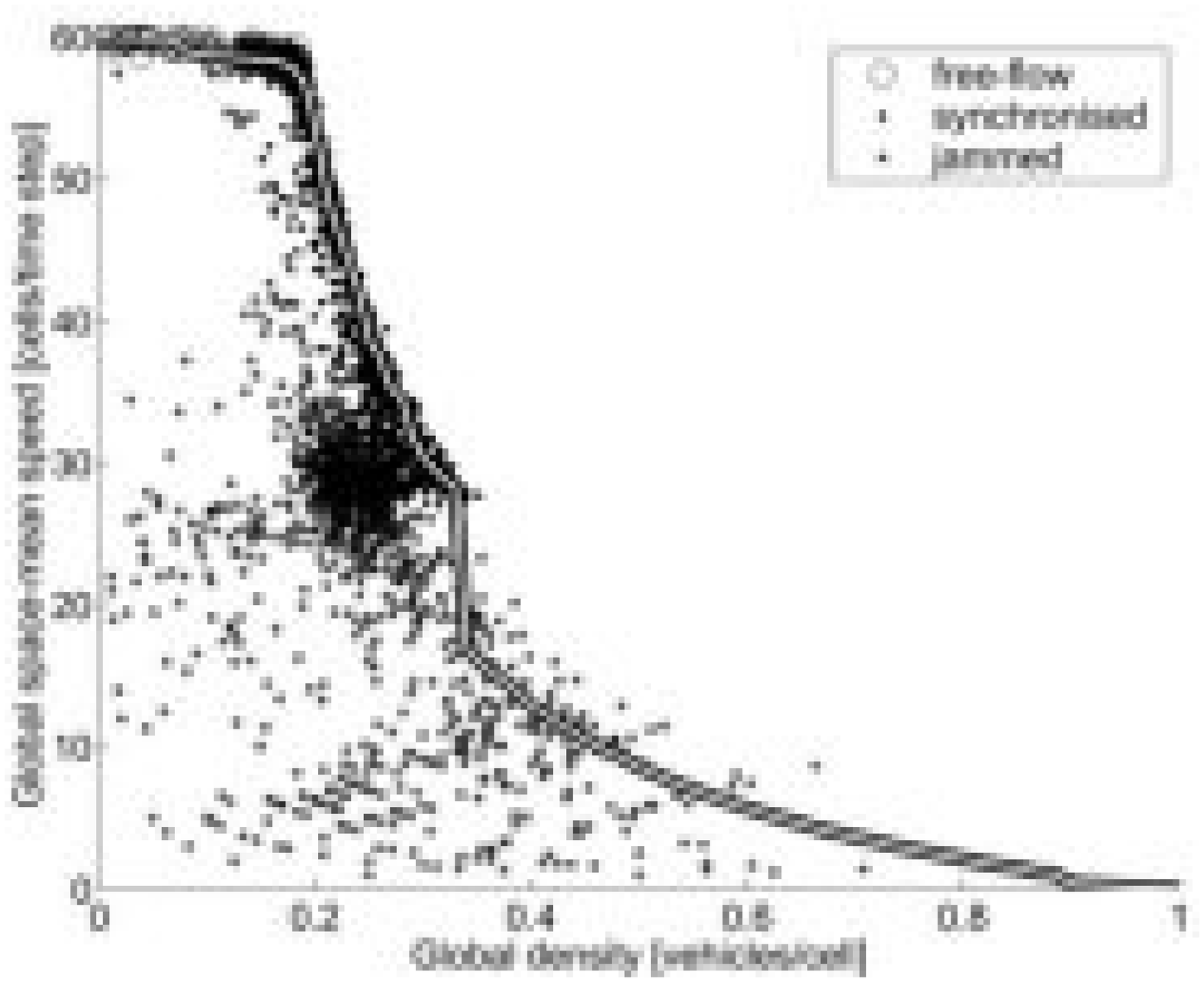}
	\hspace{\figureseparation}
	\includegraphics[width=\halffigurewidth]{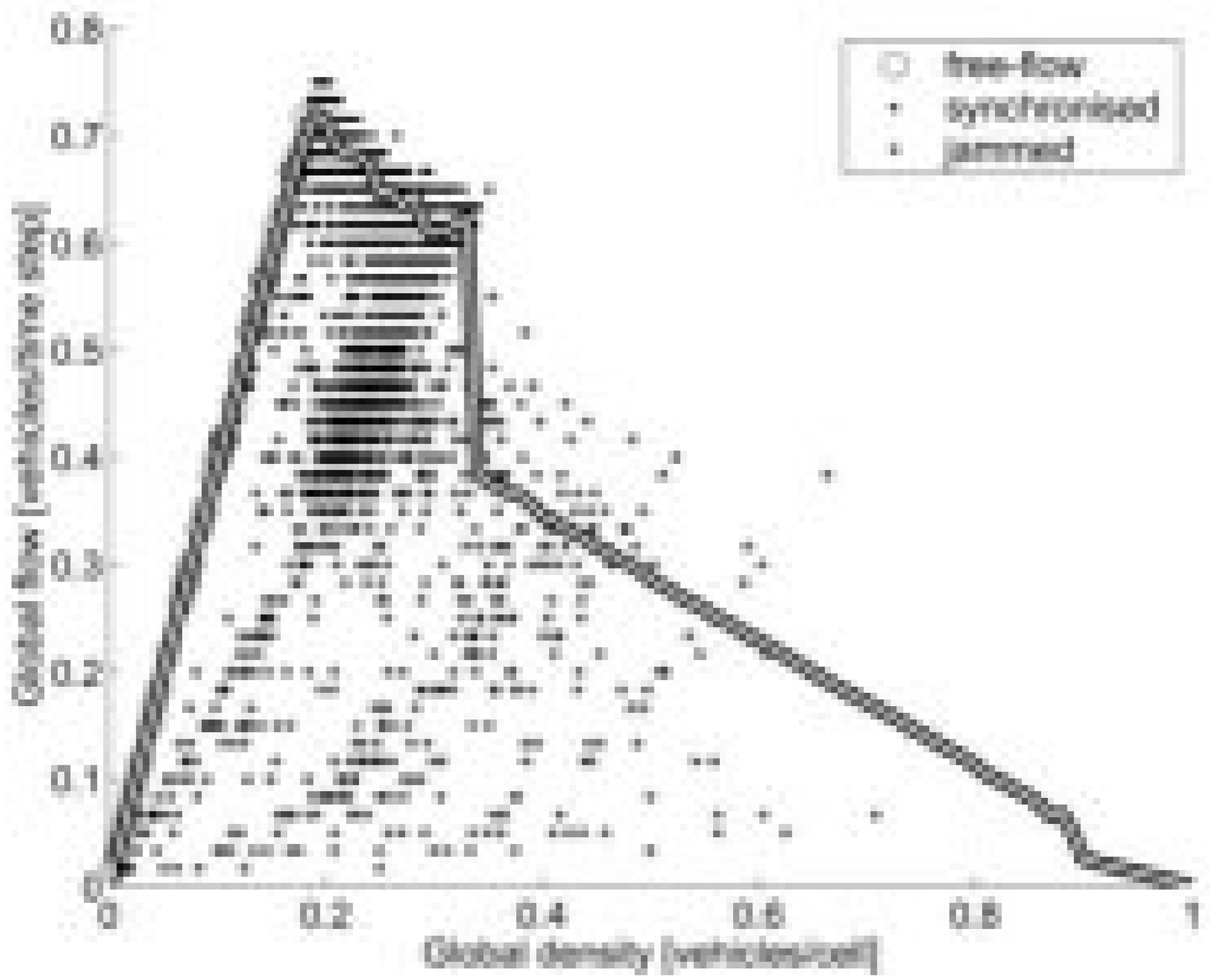}
	\caption{
		The ($k$,$\overline v_{s}$) (\emph{left}) and ($k$,$q$) (\emph{right}) 
		diagrams for the KKW-TCA model, obtained by local and global measurements. 
		The local measurements discriminate between the free-flow ($\circ$), 
		synchronised-flow ($\cdot$), and jammed regimes ($\star$). The synchronised 
		regime is visible as a wide scatter in the data points, having various 
		speeds but flows comparable to the capacity flow.
	}
	\label{fig:TCA:KKWTCAFundamentalDiagrams}
\end{figure}

Similar as in the BL-TCA model's effective space gap $g_{s_{i}}^{*}(t)$, the 
synchronisation distance $D$ is responsible for producing the typical 
two-dimensional scatter in the ($k$,$\overline v_{s}$) and ($k$,$q$) diagrams in 
\figref{fig:TCA:KKWTCAFundamentalDiagrams}. When a driver who is within the 
synchronisation distance adapts the vehicle's speed, the only factors taken into 
account are the current speed of the direct frontal leader and a safety 
criterion (in the form of the current space gap); it is this effect that 
produces the scatter in the data, because the exact specification of this speed 
is absent. In both diagrams of \figref{fig:TCA:KKWTCAFundamentalDiagrams}, the 
local measurements discriminate between the free-flow ($\circ$), 
synchronised-flow ($\cdot$), and jammed regimes ($\star$). One of the major 
differences between these two models, is that the flow in the synchronised 
regime is almost a factor two larger for the KKW-TCA than the BL-TCA. The 
KKW-TCA also experiences a capacity drop similar as in the BL-TCA, but also 
undergoes an abrupt transition when going from the synchronised-flow to the 
wide-moving jam regime around a global density of some 0.4 vehicles/cell (see 
the left part of \figref{fig:TCA:KKWTCAFundamentalDiagrams}). Because the model 
is built around the assumption that vehicles tend to approximate the behaviour 
of their direct leader within a certain synchronisation distance, the resulting 
traffic regimes correspond well to Kerner's empirical observations 
\cite{KERNER:04,MAERIVOET:05c}.

In \figref{fig:TCA:BLTCASpaceTimeGapHistograms}, we have depicted the histograms 
of the distributions of the space and time gaps in the left and right parts, 
respectively. The distributions are similar to those of the BL-TCA, but there 
are some important differences. With respect to the space gaps in the left part 
of \figref{fig:TCA:BLTCASpaceTimeGapHistograms}, there is a high variance in the 
jammed regime, due to the fact that there are vehicles in free-flow traffic, as 
well as inside the wide-moving jams (although most of the probability mass is 
assigned to the zero space gap inside the dense jams). Considering the time gaps 
in the right part of \figref{fig:TCA:BLTCASpaceTimeGapHistograms}, we can see 
that they always form a tight cluster around the median of the distribution, 
indicating very narrow distributions with an pronounced peak. This is completely 
different behaviour than in the BL-TCA model (see the right part of 
\figref{fig:TCA:BLTCASpaceTimeGapHistograms}). The main reason is probably due 
to the lack of an anticipation effect in the KKW-TCA model. Even more severe, is 
the fact that the KKW-TCA model, despite its elaborate construction based on a 
synchronisation distance, completely fails to describe the microscopic structure 
of motorway traffic. The BL-TCA model however succeeds in having a good fit on 
both macroscopic and macroscopic scales, as stated according to Knospe et al. 
\cite{KNOSPE:02c,KNOSPE:04}.

\begin{figure}[!htbp]
	\centering
	\includegraphics[width=\halffigurewidth]{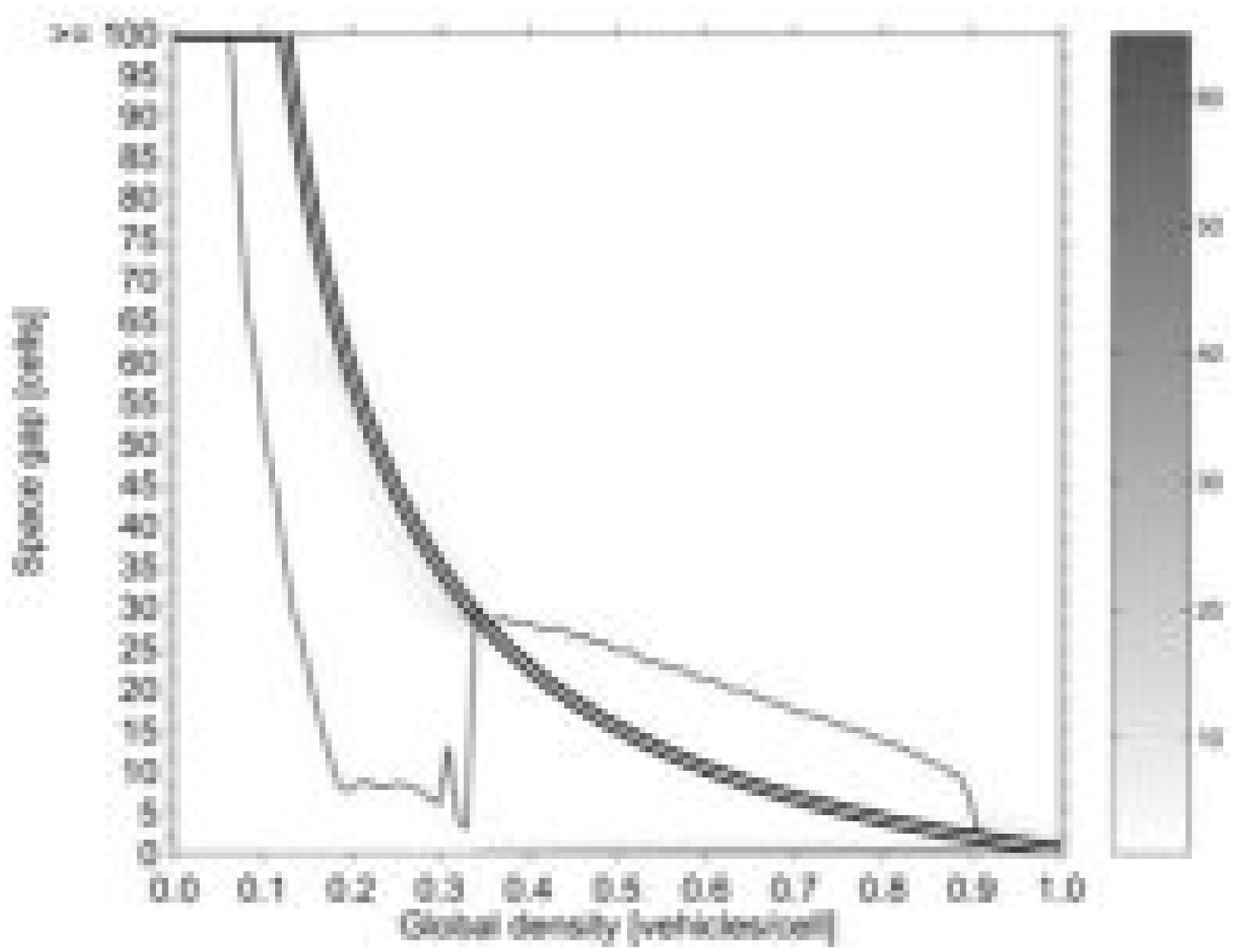}
	\hspace{\figureseparation}
	\includegraphics[width=\halffigurewidth]{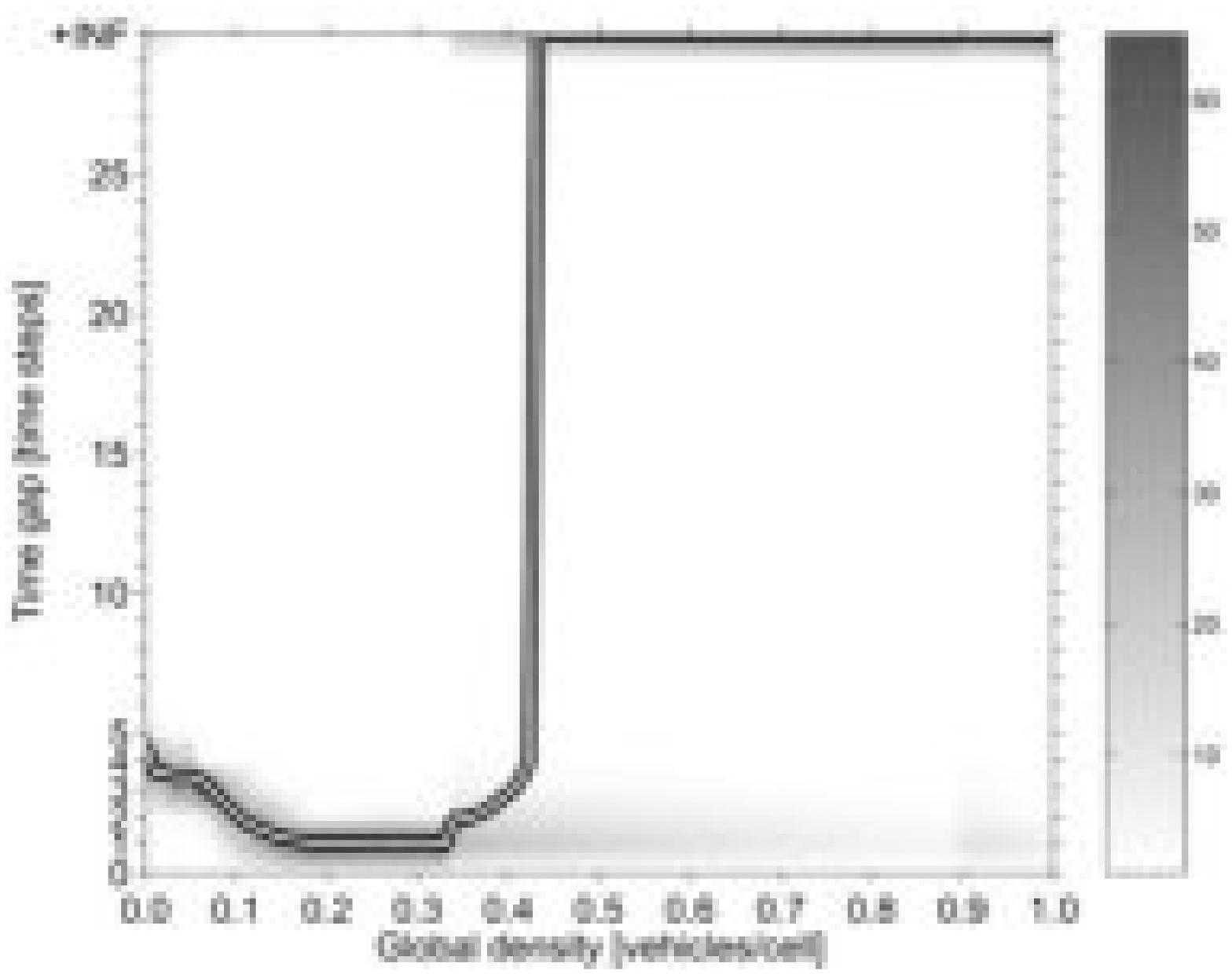}
	\caption{
		Histograms of the distributions of the vehicles' space gaps $g_{s}$ 
		(\emph{left}) and time gaps $g_{t}$ (\emph{right}), as a function of the 
		global density $k$ in the KKW-TCA model. The thick solid lines denote the 
		mean space gap and median time gap, whereas the thin solid line shows the 
		former's standard deviation. The grey regions denote the probability 
		densities.
	}
	\label{fig:TCA:KKWTCASpaceTimeGapHistograms}
\end{figure}

	\section{Multi-lane traffic, city traffic, and analytical results}

In this final section on traffic cellular automata models, we take a look at 
some other aspects related to TCA models. We first discuss some properties and 
methodologies for modelling multi-lane traffic in the context of a cellular 
automaton, after which we briefly consider several approaches for dealing with 
city traffic. The final part of the section concludes with an overview of 
different analytical treatments of TCA models.

		\subsection{Multi-lane traffic}
		\label{sec:TCA:MultiLaneTraffic}

In this section, we briefly discuss some properties and methodologies for 
modelling multi-lane traffic in the context of a cellular automaton. To this 
end, we illustrate the types of lane changes that are possible, then discuss the 
general setup for a lane-changing model. We conclude with a short overview on 
the implementation of lane-change rules and explain the phenomenon of ping-pong 
traffic, an artifact introduced by an inferior implementation.

			\subsubsection{Types of lane changes}

In general, there are two types of lane changes identified: \emph{mandatory lane 
changes} (MLC) and \emph{discretionary lane changes} (DLC) \cite{MAERIVOET:05c}. 
In the former case, a vehicle is obliged to execute a lane change, e.g., because 
it needs to exit the motorway at an off-ramp, or because the vehicle is by law 
obliged to drive in the right shoulder lane. In the latter case, a vehicle 
changes a lane at its own discretion, e.g., when approaching and overtaking a 
slow-moving leading vehicle.

With respect to the rules for lane changing, there are also two approaches: 
\emph{symmetric} and \emph{asymmetric}. In the US, the symmetric approach is 
more applicable: this is embodied by the fact that motorways have a large number 
of lanes (i.e., more than three), with vehicles driving at lower speeds (e.g., 
60 miles/hour, corresponding to some 100 kilometres/hour), effectively using all 
lanes more homogeneously. Such a system is typically called 
\emph{``keep-your-lane''}, as frequent lane changes are discouraged. In contrast 
to this, people in most European countries are obliged by law to drive on the 
outer right shoulder lane whenever possible. Motorways have fewer lanes 
(typically either two or three, unidirectional), operating at higher speeds of 
e.g., 120 kilometres/hour. In addition, most of these countries have instituted 
an overtaking prohibition on the right lane, with large trucks restricted to the 
two most right lanes.

With respect to this latter system of asymmetric lane changes, the phenomenon of 
\emph{density} or \emph{lane inversion} plays an important role, especially on 
the numerous 2x2 motorways in Europe (see also the beginning of section 
\ref{sec:TCA:SingleCellModels} for a discussion of this phenomenon). Another 
aspect that has a significant influence, is the change of driver behaviour, 
e.g., near on-ramps. Here, drivers might avoid the shoulder lane to allow 
traffic to enter, or because of their increased attention, they might induce a 
more subtle effect such as the capacity funnel (see also our discussion in 
\cite{MAERIVOET:05d} for more details on this phenomenon).

			\subsubsection{General setup for lane changing}
			\label{sec:TCA:GeneralSetupForLaneChanging}

Deciding on whether or not to perform a lane change, is typically split in two 
separate steps: first, a vehicle checks if it is \emph{desirable} to change 
lanes, i.e., making the distinction between a mandatory or discretionary lane 
change. If a lane change is indeed desirable, then the second step proceeds to 
check whether or not such a lane change can be performed at all with respect to 
safety and collision avoidance. Thus, there is a check for \emph{gap 
acceptance}.

One of the first approaches to model such lane-changing behaviour an a two-lane 
road in a TCA model, is due to Nagatani. His work was based on the deterministic 
CA-184 model (see section \ref{sec:TCA:CA184}) \cite{NAGATANI:93}. One of the 
artifacts of his lane-changing rules, was the existence of states in which 
blocks of vehicles alternated from one lane to another, without moving at all. 
To circumvent this problem, Nagatani randomised the lane-changing behaviour 
\cite{NAGATANI:94}. Rickert et al. later applied this lane-changing methodology, 
by extending the STCA model (see section \ref{sec:TCA:STCA}) to handle two-lane 
unidirectional traffic \cite{RICKERT:96}. Wagner et al. later assessed the 
previous work of Rickert et al., concluding that it did not capture certain 
aspects (e.g., density inversion) of traffic flows very well \cite{WAGNER:97}. 
To this end, they built upon the previous work, adding a more specialised 
security constraint that takes into account the fact that vehicles should also 
consider the following vehicles in the target lane, thereby avoiding severe 
disruptions. As a final comment, they state that the lane-changing rules in a 
TCA model typically do not provide a realistic microscopic model, but they 
rather lead to a good correspondence with respect to observed macroscopic 
features (e.g., the frequency of lane changes).

In order to address the correct reproduction of the density inversion 
phenomenon, Nagel et al. artificially introduced a \emph{slack parameter}, 
capturing the inclination of a driver to change back to the right lane. They 
furthermore also provided an extensive classification of some 10 lane changing 
rules and criteria encountered in literature \cite{NAGEL:98}. Another excellent 
overview of multi-lane traffic is given by Chowdhury et al. \cite{CHOWDHURY:00}. 

As all the previous work dealt with unidirectional roads, it seems logical to 
consider \emph{bidirectional traffic}, i.e., traffic with adjacent but opposing 
lanes. Simon and Gutowitz were among the first to consider a TCA model of such 
traffic, with vehicles driving on two lanes \cite{SIMON:98b}. Central to their 
approach, is the notion of a \emph{local density} that each driver must assess 
before attempting to complete an overtaking manoeuvre. When a driver encounters 
a slower moving vehicle, a check is made whether or not there is enough space 
\emph{in front} of this leading vehicle (this is the local density). If the 
check is positive, then a lane change can be performed (under the condition of 
course that there is a safe gap in the opposing lane). With this scheme in mind, 
high density traffic thus excludes such overtaking manoeuvres, due to the fact 
that the local density is too low to complete them.\\

\sidebar{
	Note that some authors, e.g., Gundaliya et al. \cite{GUNDALIYA:04}, 
	Mallikarjuna and Ramachandra Rao \cite{MALLIKARJUNA:05}, use a peculiar 
	variant of a multi-lane setup. Their models have essentially a multi-cell 
	structure, but now the multi-cell concept is extended in the lateral 
	direction. So cells not only get smaller, but also `thinner', allowing 
	\emph{variable-width vehicles}, e.g., motor cycles that can more easily pass 
	other vehicles in the same lane. In our opinion, this leads to unnecessary 
	complexity, giving little benefits. In fact, we believe that such a scheme 
	directly opposes the idea behind a CA model, as explained at the introduction 
	of this report. We strongly feel that heterogeneity in a TCA model should 
	\emph{only} be incorporated by means of different lengths, maximum speeds, 
	acceleration characteristics, anticipation levels, and stochastic noise for 
	distinct classes of vehicles and/or drivers. Any other approach would be 
	better off with a continuous microscopic model.
}\\

			\subsubsection{Implementation of lane-changing rules and the phenomenon of ping-pong traffic}

The basic implementation of a lane-changing model in a TCA setting, leads to two 
substeps that are consecutively executed at each time step of the CA:

\begin{itemize}
	\item first, the lane-changing model is executed, exchanging vehicles between 
	\emph{laterally} adjacent lanes,
	\item then, all vehicles are moved forward (i.e., \emph{longitudinal}) by 
	applying the car-following part of the TCA model's rules.
\end{itemize}

One immediate result from this approach, is that a lane change in a TCA model is 
completed within one time step (i.e., $\Delta T$). This is in contrast to 
real-life traffic, where lane changes have a duration of several seconds 
\cite{NAGEL:98}.

For more than two lanes, care must be taken to avoid so-called \emph{scheduling 
conflicts} during the first substep. Consider for example three lanes, with two 
vehicles driving in the outer left, respectively outer right, lane at the same 
longitudinal position. If the cell in the middle lane is empty, then the 
vehicles may decide to move to this location, resulting in a lateral collision. 
In order to compensate this, one possibility is to choose a vehicle at random 
(or by preference), thereby allowing it to perform its requested lane change. 
Another possibility is to perform left-to-right lane changes in even time steps, 
and right-to-left lane changes in odd time steps.

As hinted earlier, the `correctness' of a lane-change model should be judged on 
the basis of certain macroscopic observations. Examples of these are the 
frequency of lane changes with respect to different densities, the capacity 
flows for all lanes separately and combined, the critical density at which a 
breakdown occurs in each of the lanes, \ldots Good indicators can be found in 
the many small fluctuations typically exhibited by multi-lane TCA models, 
instead of the large jams in single-lane traffic. Traffic flows get more fluid 
if vehicles are allowed to pass moving bottlenecks \cite{WAGNER:97,NAGEL:98}. 
However, under certain conditions, Helbing and Huberman have shown the existence 
of coherent states, where vehicles' speeds are synchronised across adjacent 
lanes. For heterogeneous traffic flows, this can lead to a moving `solid block' 
of vehicles \cite{HELBING:98c}.

When implementing lane-change rules in a TCA model, care must however be taken 
that the implementation does not introduce any unrealistic artifacts. A 
prominent example of this, plaguing many TCA models, is a phenomenon called 
\emph{ping-pong traffic}. Nagatani was among the first to observe this peculiar 
behaviour of vehicles in traffic flows (see section 
\ref{sec:TCA:GeneralSetupForLaneChanging}). In ping-pong traffic, vehicles 
typically alternate between lanes during successive time steps. As explained 
earlier, one way to resolve this behaviour is by randomising the lane-change 
decision, thereby quickly destroying any such artificial patterns 
\cite{NAGATANI:94,RICKERT:96}.

		\subsection{City traffic and intersection modelling}

When modelling city traffic, essentially two approaches can be followed: either 
the entire road network is considered as a two-dimensional lattice (i.e., a 
\emph{grid}), or each road in the network is a single longitudinal lattice 
(single- or multi-lane) with explicitly modelled intersections. The former was 
historically used in the context of phase transitions in a CA, whereas the 
latter is more applicable to describe real-life traffic flows in populated 
cities.

In this section, we illustrate both approaches, starting with a classic grid 
layout as embodied by the Biham-Middleton-Levin (BML) and 
Chowdhury-Schadschneider (ChSch) TCA models, after which we briefly comment on 
explicit descriptions of intersections in TCA models.

		\subsubsection{Grid traffic}

The first model of `city traffic' was proposed by Biham, Middleton, and Levine 
(BML). It was developed around the same time Nagel and Schreckenberg presented 
their STCA (see section \ref{sec:TCA:STCA}). The BML-TCA, is a two-dimensional 
model that describes traffic on a square grid in a toroidal setup (i.e., 
opposing sides are identified), with vehicles distributed randomly over the 
lattice \cite{BIHAM:92}. The model is in fact a very simplistic model, in that 
it assumes that all vehicles either move from the south to the north direction, 
or from the west to the east. Each cell of the lattice is assumed to contain a 
traffic light, in the sense that all west-east vehicles try to move during even 
time steps, and all south-north vehicles during odd time steps (thus 
$v_{\text{max}} = \one$ cell/time step for all vehicles). The BML-TCA 
constitutes a fully deterministic model, where the only randomness is introduced 
through the initial conditions. Note that its one-dimensional version 
corresponds to the CA-184 and the TASEP (see sections \ref{sec:TCA:CA184} and 
\ref{sec:TCA:TASEP}).

\begin{figure}[!htbp]
	\centering
	\includegraphics[height=0.82\halffigurewidth]{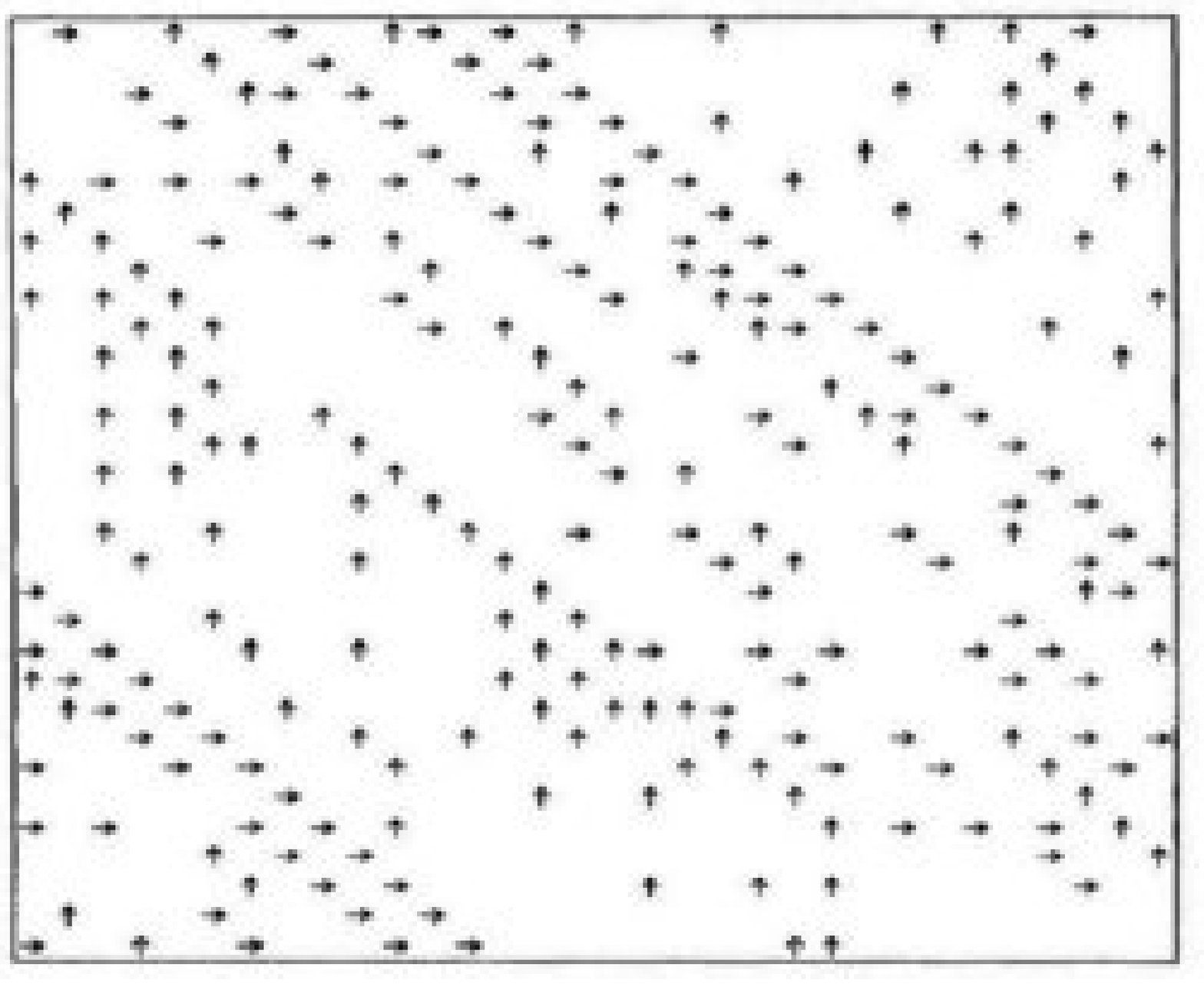}
	\hspace{\figureseparation}
	\includegraphics[height=0.82\halffigurewidth]{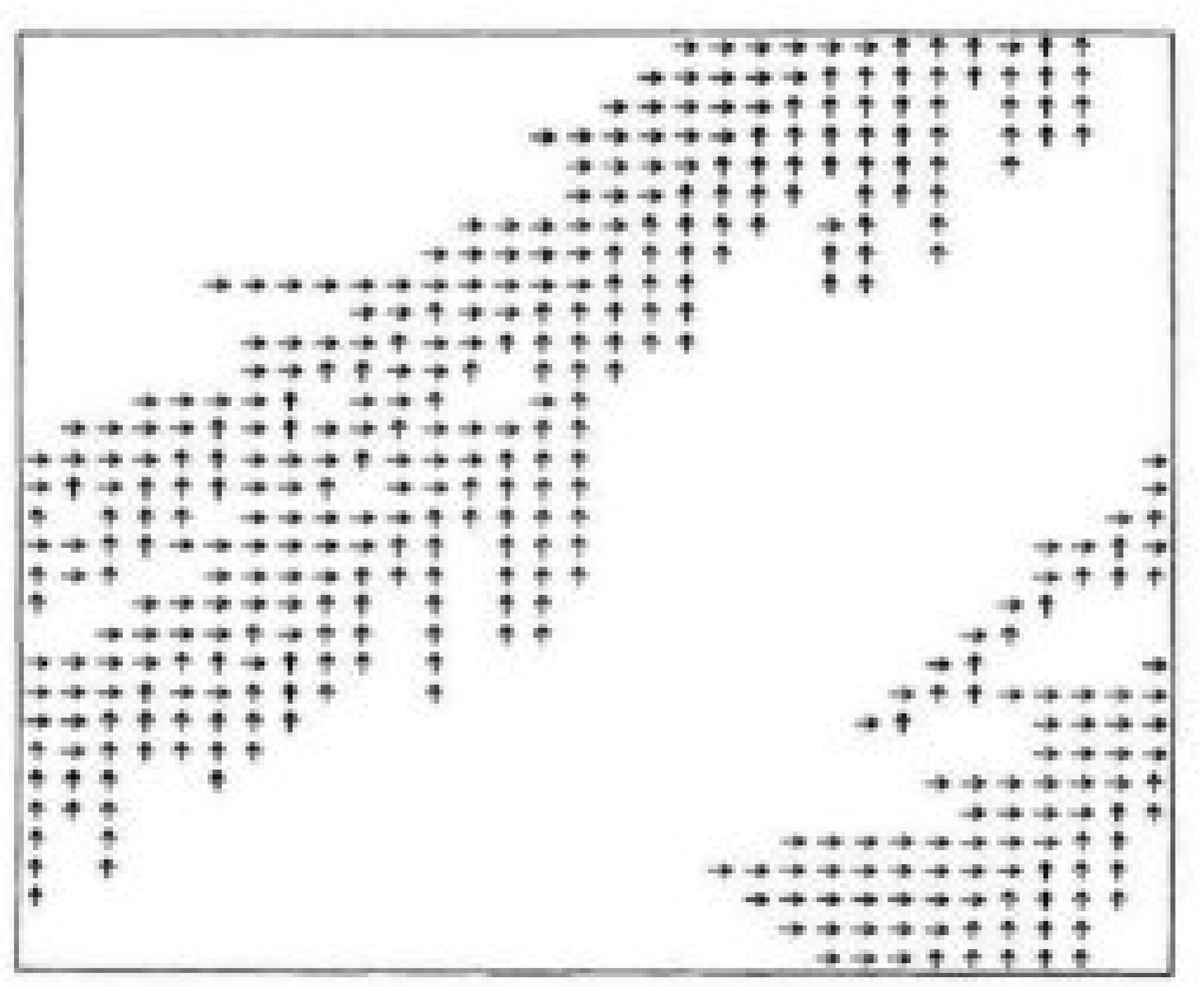}\\
	\vspace{\figureseparation}
	\includegraphics[height=0.75\figurewidth]{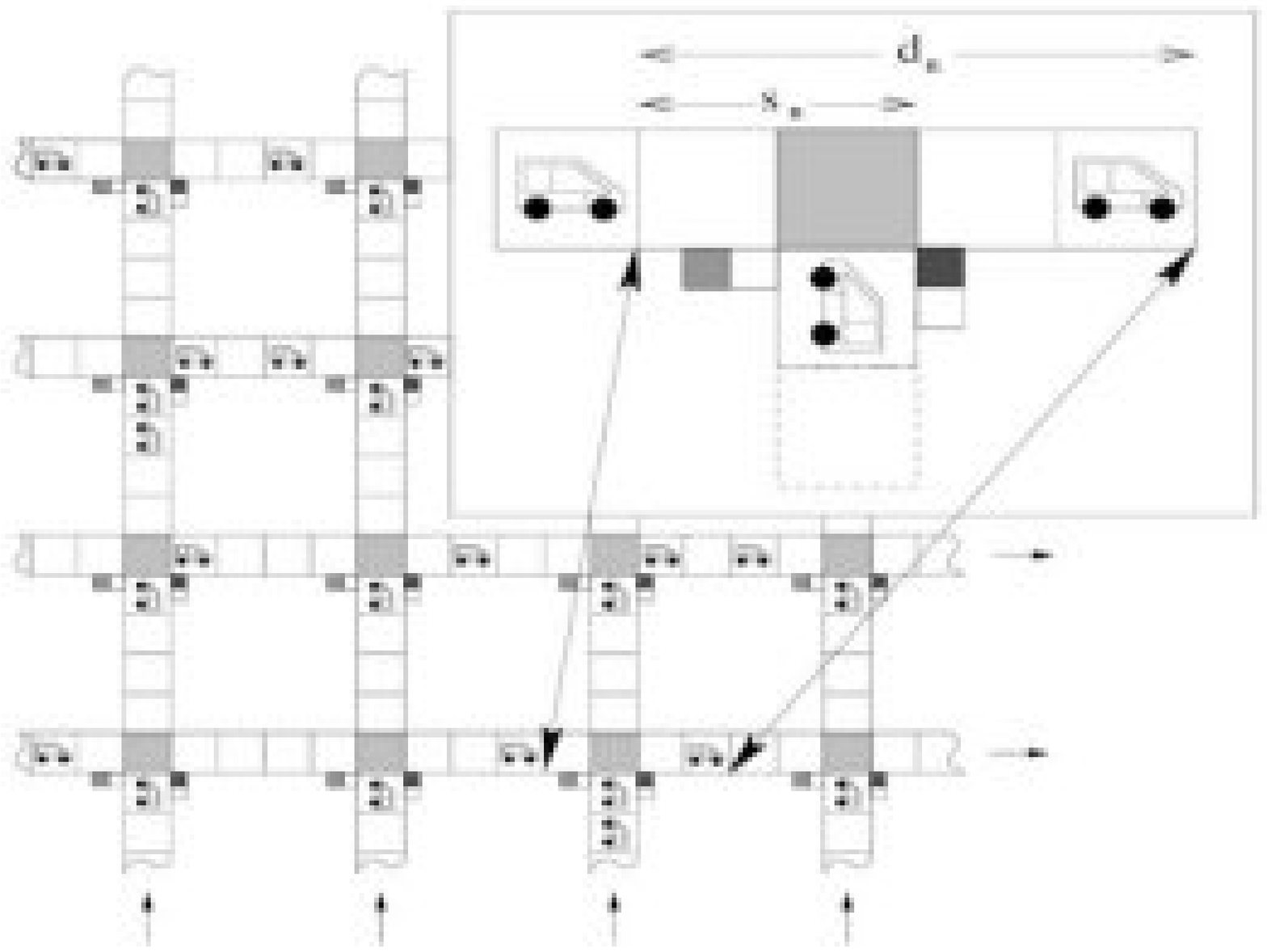}
	\caption{
		\emph{Left:} snapshot of the spatial structure in the BML-TCA for $k 
		=$~0.25. In this free-flow regime, all vehicles move alternatingly, with the 
		right-oriented arrows denoting west-east travelling vehicles, and the 
		upward-oriented arrows denoting south-north travelling vehicles. 
		\emph{Right:} same setup as before, but now for $k \approx$~0.4082. In this 
		congested regime, a global cluster emerges, completely composed of blocked 
		vehicles. \emph{Bottom:} an overview of the ChSch-TCA, showing the street 
		segments of finite length between the BML-TCA's original intersections. The 
		first two images are reproduced after \cite{BIHAM:92}, the third after 
		\cite{BARLOVIC:03}.
	}
	\label{fig:TCA:CityTraffic}
\end{figure}

Depending on the global density of vehicles in the lattice, the model results in 
two distinct traffic regimes, with a \emph{sharp first-order phase transition} 
between them. The first regime, i.e., free-flow traffic, corresponds to a state 
with alternate moving vehicles (i.e., west-east and south-north moving); an 
example is depicted in the left part of \figref{fig:TCA:CityTraffic}. In the 
congested regime, a self-organised global cluster emerges, completely composed 
of blocked vehicles (see e.g., the right part of \figref{fig:TCA:CityTraffic}). 
When the phase transition between both regimes occurs, the space-mean speed 
changes abruptly from one to zero cells/time step \cite{BIHAM:92,ANGEL:05}. 
Fukui and Ishibashi studied the repercussions of a local disruption in the 
lattice (e.g., a crashed vehicle that remains stopped for an eternal period), 
and found that it provides the seed of a growing global cluster \cite{FUKUI:93}. 
Biham et al. also considered a less restrictive version of the above model, in 
which now all vehicles try to move at each time step. In case of conflicts 
between a west-east and a south-north vehicle, one of them is chosen at random. 
Another variation considers also opposing traffic, which can lead to 
\emph{gridlocked} situations where no vehicles are able to move at all. A 
generalisation of the BML-TCA, was provided by Freund and Poschel who consider a 
similar setup, but now with traffic moving in all four directions 
\cite{FREUND:95}. Finally, Shi is able to obtain analytical expressions for the 
critical densities at which the previously mentioned phase transitions occur 
\cite{SHI:99}.

In the work of Chowdhury et al., a comprehensive overview is given, describing 
extensions to the BML-framework  \cite{CHOWDHURY:00}. This overview includes 
asymmetric distributions of the west-east and south-north vehicles, unequal 
maximum speeds, two-level crossings (where two vehicles can share the same 
cell), faulty traffic lights (here, either a west-east or south-north vehicle is 
chosen at random to occupy a cell, irrespective of the current time step), road 
blocks, line- and point-defects (i.e., a crowded `street' of the model, 
corresponding to a dense horizontal or vertical row of cells), random turning of 
vehicles, cut-off streets (similar to a row of two-level crossings), and so 
forth and so on.

Chowdhury and Schadschneider later extended the BML-TCA model to incorporate 
randomisation effects like in the STCA model, having the result that jamming can 
now occur spontaneously \cite{CHOWDHURY:99c}. Their model furthermore contains 
street segments of finite length between the cells, with vehicles driving 
according to the STCA's rules on these streets. The original cells in the 
BML-TCA model form the signallised intersections of the Chowdhury-Schadschneider 
model (ChSch-TCA), as can be seen in the bottom part of 
\figref{fig:TCA:CityTraffic}. At sufficiently large densities, a transition can 
occur that leads to a self-organising state of completely gridlocked traffic. 
Barlovi\'c later provided a solution to this problem, making the model 
well-suited for assessing the results of different traffic light control 
policies in a city \cite{BARLOVIC:03}.

		\subsubsection{Explicit intersection modelling}

In contrast to the previous section were all traffic operations were essentially 
defined on a two-dimensional lattice, it is also possible to consider a complete 
road network, consisting of \emph{separate links} that are connected to each 
other by means of \emph{intersections}. These intersections can either be 
signallised, or unsignallised, turning priorities can be defined, as well as 
different geometrical layouts (e.g., roundabouts).

Road networks based on the above assumptions, typically combine a set of basic 
building blocks. As such, the network is logically decomposed in a set of 
\emph{nodes} and \emph{links}. The former denote the intersections, whereas the 
latter can, depending on the implementation, refer to individual lanes, a group 
of adjacent lanes, or even a road with two-way traffic. In general, traffic 
operations on motorways are primarily influenced by the behaviour of vehicles on 
links, i.e., their car-following and lane-changing behaviour. Conversely, 
traffic operations in cities and denser street networks, are primarily defined 
by the behaviour of vehicles at intersections, i.e., queueing delays at traffic 
lights, priority turns, \ldots In many cases, the intersection logic is 
simplified, such that all decisions (conflict resolving et cetera) are taken 
\emph{before} a vehicle enters the intersection \cite{HELBING:04}.

Several non-exhaustive examples include the work of Esser and Schreckenberg with 
applications to the city of Duisburg \cite{ESSER:97}, the work of Simon and 
Nagel who primarily focussed on single-lane traffic in combination with several 
setups for controlling traffic lights, applying their work to the city of Dallas 
(different links have different slowdown probabilities associated with them, 
thus enabling to model different street capacities) \cite{SIMON:98}, the work of 
Diedrich et al. who consider the effects of various implementations of on- and 
off-ramps in the classic STCA model \cite{DIEDRICH:00}, and all the references 
on TRANSIMS, the travel behaviour in Switzerland, the region of Dallas, the city 
of Portland, and the city of Geneva (where all intersections are replaced by 
generalised roundabouts), mentioned in our discussion in \cite{MAERIVOET:05c}.\\

\sidebar{
	All these examples have in common that they are based on simple building 
	blocks. Despite this elegance, most of them however, do not provide 
	satisfactory information regarding the calibration and validation of their 
	underlying models (this for example with respect to the correct observed 
	queueing delays at intersections). A popular technique is to use 
	\emph{sources} and \emph{sinks}, where vehicles are added and removed, 
	allowing tuning of the simulator in order to agree with incoming on-line 
	measurements.	Clearly, we feel that besides a need for elaborate descriptions 
	of the employed models, there is perhaps even a bigger need for correct 
	information with respect to these models' fidelity and accuracy.
}\\

		\subsection{Analytical results}
		\label{sec:TCA:AnalyticalTCA}

Because most studies based on TCA models heavily rely on numerical simulations, 
this creates the danger of introducing artifacts (e.g., finite-size effects) 
that obscure the true dynamics of the systems under consideration. Although most 
of these problems should resolve in the so-called \emph{thermodynamic limit} 
where $K_{\mathcal{L}}, T_{\text{mp}} \rightarrow +\infty$ (i.e., a lattice with 
infinite length considered over an infinite time period), resorting to this 
approach is computationally not feasible. As a result, researchers have focussed 
on analytical methods. Except for the most trivial cases with a deterministic 
(i.e., noiseless) TCA model, these analytical methods most of the time provide 
approximations at best.

In this section, we illustrate several of these analytical methods encountered 
in literature. Our discussion focusses on the concept of a mean-field theory, 
after which we elaborate on some of its improvements that lead to better 
agreement with numerical results.\\

\sidebar{
	Note that other avenues for analytical treatments of CA models, and TCA models 
	in particular, are also explored. In this section, we will however not go into 
	detail about them. For more information, we refer the reader to the 
	interesting work of Fuk\'s and Boccara 
	\cite{FUKS:98,FUKS:99,FUKS:01,FUKS:04,BOCCARA:00}.
}\\

			\subsubsection{Mean-field theory}

As mentioned in the introduction of this section, for the case of arbitrary 
$v_{\text{max}}$ and $p = \zero$ or $p = \one$, or for $v_{\text{max}} = \one$ 
cell/time step, the analytical solution of the resulting TCA model is exactly 
known. This solution, expressed as its ($k$,$q$) diagram, corresponds to the set 
of diagrams as depicted in 
\figref{fig:TCA:DFITCADensityFlowSMSFundamentalDiagrams} (see section 
\ref{sec:TCA:DeterministicFITCA}) for the DFI-TCA.

The problem is to find an analytical description of how the system evolves in 
time through the state space, i.e., what are the occurring configurations~? The 
evolution of a system, can be described by what is called a \emph{master 
equation}. For cellular automata, this equation is a first-order differential 
equation, describing the change in probability of a system's lattice to be in a 
certain configuration. The downside is that, in general, this master equation 
can not be solved exactly.

For the TASEP model (see section \ref{sec:TCA:TASEP}) with open boundary 
conditions and random sequential update, the master equation can be solved 
exactly \cite{RAJEWSKY:96,SANTEN:99}. In a first step, the master equation is 
elegantly written in vector form, comprising a \emph{transfer matrix} that 
contains the time-evolution of the probabilities. By assuming the 
\emph{matrix-product ansatz} (MPA) formalism, the transfer matrix can be 
rewritten as a product of local transfer matrices, operating on sets of cells. 
This provides a algebra that can be solved exactly, thereby solving the TASEP 
analytically. Note that for the TASEP with a parallel update however, obtaining 
the exact solution is difficult, because no simple MPA decomposition into local 
matrices is possible.

In contrast to this promising result, obtaining an analytical solution becomes 
harder to even intractable for the STCA model (see section \ref{sec:TCA:STCA}) 
with $v_{\text{max}} > \one$ cell/time step and $\zero < p < \one$. In the 
master equation, probabilities of cluster of cells will occur, making its 
solution very hard \cite{SCHADSCHNEIDER:02}. One well-known method that is 
suitable for dealing with many-particle systems in statistical mechanics, is the 
construction of a \emph{mean-field theory} (MFT) of the model. Such a MFT can 
provide an approximation of the master equation; in some cases, the MFT turns 
out to be an exact solution.

The idea behind a MFT, is that all correlations between neighbouring cells are 
neglected. For TCA models, such a \emph{site-oriented mean-field theory} (SOMF) 
assumes that all cluster probabilities are replaced by single cell 
probabilities. The MFT now replaces the effects of these individual cells with 
an average effect (the `mean field'), which simplifies computations 
considerably. When translating the STCA's rules R1 -- R3, i.e., equations 
\eqref{eq:TCA:STCAR1} -- \eqref{eq:TCA:STCAR3}, R1 is decoupled into separate 
acceleration and braking rules R1a and R1b, after which their order is changed 
to R1b, R3, R4, R1a. The upshot of this is that there are no stopped vehicles in 
the system, thereby reducing the number of possible states for a cell by one. If 
$v_{\text{max}} = \one$ cell/time step, then the system can be fully described 
by cell occupancies. Applying this SOMF theory to the STCA model, results in 
considerably underestimation of the flow in the system (even for the restricted 
case of $v_{\text{max}} = \one$ cell/time step) 
\cite{SCHRECKENBERG:95,SCHADSCHNEIDER:99b,SCHADSCHNEIDER:02}.

			\subsubsection{Improving the SOMF theory}

As mentioned in the previous section, setting $v_{\text{max}} = \one$ cell/time 
step leads to an underestimation of the flow. However, when switching from a 
parallel update procedure to a random sequential one, the resulting SOMF theory 
becomes exact~! It turns out that the reason for the underestimation, can be 
traced back to its neglecting of all correlations between cells (which are a 
consequence of the parallel update procedure). As explained in the beginning of 
section \ref{sec:TCA:TASEP}, using a parallel update excludes certain Garden of 
Eden states. However, the SOMF theory naively includes these paradisiacal 
states. As a solution, these GoE states can be eliminated, resulting in a 
\emph{paradisiacal mean-field theory} (pMFT). In systems with higher maximum 
speeds, more GoE states occur, making it difficult to derive a pMFT. Even then, 
the theory still remains an approximation (albeit a better one) when using a 
parallel update procedure 
\cite{SCHADSCHNEIDER:98,SCHADSCHNEIDER:99b,SCHADSCHNEIDER:02}.

Taking into account short-range correlations, can be done by considering a 
\emph{car-oriented mean-field theory} (COMF). Instead of dealing with cells and 
their occupancies, the COMF theory computes the probabilities $P_{n}(v)$ of 
finding a space gap of $n$ cells for a vehicle driving with speed $v$ 
\cite{SCHADSCHNEIDER:97}. In a sense, the COMF theory approximates the problem 
by neglecting the correlations between space gaps of successive vehicles 
\cite{CHOWDHURY:00}. As such, it gives qualitatively good approximations for $p 
\rightarrow \zero$; in all other cases, the COMF theory starts to fail, because 
there are also correlations between the space gaps 
\cite{SANTEN:99,CHOWDHURY:00}. Note that the COMF theory has also been applied 
to the BJH-TCA and VDR-TCA models (see sections \ref{sec:TCA:BJHTCA} and 
\ref{sec:TCA:VDRTCA}, respectively) \cite{SCHADSCHNEIDER:97c}.

Another approach to analytically solve the master equation, is to explicitly 
take into account the correlations between neighbouring cells, by considering 
\emph{clusters} composed of $n$ consecutive cells 
\cite{SCHADSCHNEIDER:99b,SCHADSCHNEIDER:02}. Such a \emph{site-oriented 
cluster-theoretic approach} proves to perform better than the COMF theory from 
the previous section \cite{SCHRECKENBERG:95}. The improvement of the 
approximation is even better when considering larger clusters; it is exact for 
$n \rightarrow +\infty$ \cite{SCHADSCHNEIDER:97b,SANTEN:99,CHOWDHURY:00}.

	\section{Summary and outlook}

This report gave an elaborate and understandable review of traffic cellular 
automata (TCA) models, which are a class of computationally efficient 
microscopic traffic flow models. TCA models arise from the physics discipline of 
statistical mechanics, having the goal of reproducing the correct macroscopic 
behaviour based on a minimal description of microscopic interactions.

We began with an overview of cellular automata (CA) models, their background and 
physical setup. Applying this technique to the modelling of traffic flows, we 
discretise a road into a number of small cells (a procedure called coarse 
graining), having a width of e.g., $\Delta X =$~7.5~m. Time is also discretised 
into units of approximately $\Delta T =$~1~s. After introducing the mathematical 
notations, we showed how to perform measurements on a TCA model's lattice of 
cells, and how to convert these quantities into real-world units and vice versa.

Subsequently, we gave an extensive account of the behavioural aspects of several 
TCA models encountered in literature. Already, several reviews of TCA models 
exist, but none of them consider all the models exclusively from the behavioural 
point of view. In this respect, our overview fills this void, as it focusses on 
the behaviour of the TCA models, by means of time-space diagrams, ($k$,$q$) 
diagrams and the like, and histograms showing the distributions of vehicles' 
speeds, space, and time gaps. In the report, we have distinguished between 
single-cell and multi-cell models, whereby in the latter vehicles are allowed to 
span a number of consecutive cells. We concluded with a concise overview of TCA 
models in a multi-lane setting, and some of the TCA models used to describe city 
traffic as a two-dimensional grid of cells, or as a road network with explicitly 
modelled intersections. The final part of the report illustrated some of the 
more common analytical approximations to single-cell TCA models.

Considering the state-of-the-art in using TCA models, our analysis indicates 
that the field has evolved rapidly over the last decade. Starting from initial 
attempts based on rather crude models, the past few years have seen an increase 
in the computational complexity as well as the available computational power. 
More complex models are developed, of which we believe the brake-light TCA model 
of section \ref{sec:TCA:BLTCA} is the most promising: it is able to faithfully 
reproduce the correct real-life empirical observations, and quite some work has 
been done at calibrating the model, see e.g., the recent work of Knospe et al. 
\cite{KNOSPE:04}. To conclude, we note an evolving trend of using these TCA 
models as the physical models underlying multi-agent systems, in part describing 
the behaviour of individual people in large-scale road networks 
\cite{MAERIVOET:05c}.



\appendix
%

\section{TCA+ Java$^{\text{\tiny{TM}}}$ software}
\label{sec:TCA+Software:Appendix}

As already briefly mentioned in the paper, all simulations were performed by 
means of our \emph{Traffic Cellular Automata +} software \cite{MAERIVOET:04d}. 
It was developed for the Java\trademark Virtual Machine (JVM), and can be 
downloaded\footnote{From May 2002 until June 2005, the software has been 
downloaded some 800 times, of which we suspect one third to be traffic coming 
from search engines' indexing robots.} from:

\begin{quote}
	\texttt{http://smtca.dyns.cx}
\end{quote}

The software is also referenced on the \emph{Traffic 
Forum}\footnote{\texttt{http://www.trafficforum.org}} (see section \emph{Links}, 
subsection \emph{Online Traffic Simulation or Visualization (Java Applets)}, 
item \emph{Java (Swing) application for several cellular automata models}). 

In this appendix, we summarise our rudimentary TCA+ software. We start with an 
overview of its features, explain how to run the software, and conclude with 
some technical details with respect to the implementation of its code base.

	\subsection{Overview and features}
	\label{sec:TCA+Software:OverviewAndFeatures}

The TCA+ software package's goal is two-fold: on the one hand, it provides an 
\emph{intuitive didactical tool} for getting acquainted with the concept of 
single-lane traffic cellular automata models. On the other hand, it provides a 
rich enough code base to perform hand-tailored \emph{simulation experiments}, as 
well as giving insight into the details of programming TCA models.

In a nutshell, our software considers one-dimensional traffic cellular automata 
with periodic boundary conditions, i.e., vehicles driving on a unidirectional 
circular road. Different sets of rules can be chosen, and for each set its 
parameters (e.g., stochastic noise) can be changed at run time. Both local and 
global measurements can be performed on the lattice by means of artificial loop 
detectors. A traffic light with cyclical red and green phases was also added, 
allowing to study elementary queueing behaviour. In the software, we have 
implemented the TCA models listed in 
\tableref{table:TCA+Software:ImplementedTCAModels}.

\begin{table}[!htb]
	\centering
	\begin{tabular}{|l|c|}
		\hline
		\emph{TCA model}             & \emph{Refer to section}\\
		\hline
		\hline 
		CA-184                       & \ref{sec:TCA:CA184}\\
		\hline
		DFI-TCA                      & \ref{sec:TCA:DeterministicFITCA}\\
		\hline
		STCA                         & \ref{sec:TCA:STCA}\\
		\hline
		STCA-CC                      & \ref{sec:TCA:STCACC}\\
		\hline
		SFI-TCA                      & \ref{sec:TCA:SFITCA}\\
		\hline
		TASEP                        & \ref{sec:TCA:TASEP}\\
		\hline
		ER-TCA                       & \ref{sec:TCA:ERTCA}\\
		\hline
		deterministic T$^{\two}$-TCA & \ref{sec:TCA:T2TCA}\\
		\hline
		stochastic T$^{\two}$-TCA    & \ref{sec:TCA:T2TCA}\\
		\hline
		VDR-TCA                      & \ref{sec:TCA:VDRTCA}\\
		\hline
		VDR-CC-TCA                   & \ref{sec:TCA:VDRTCA}\\
		\hline
		TOCA                         & \ref{sec:TCA:TOCA}\\
		\hline
		MC-STCA                      & \ref{sec:TCA:ArtifactsOfAMultiCellSetup}\\
		\hline
		HS-TCA                       & \ref{sec:TCA:HSTCA}\\
		\hline
		BL-TCA                       & \ref{sec:TCA:BLTCA}\\
		\hline
		KKW-TCA                      & \ref{sec:TCA:KKWTCA}\\
		\hline
	\end{tabular}
	\caption{
		All TCA models implemented in our TCA+ software, accompanied by references 
		to the respective sections in the paper where they are extensively 
		discussed.
	}
	\label{table:TCA+Software:ImplementedTCAModels}
\end{table}

\begin{figure*}[!htbp]
	\centering
	\includegraphics[width=\textwidth]{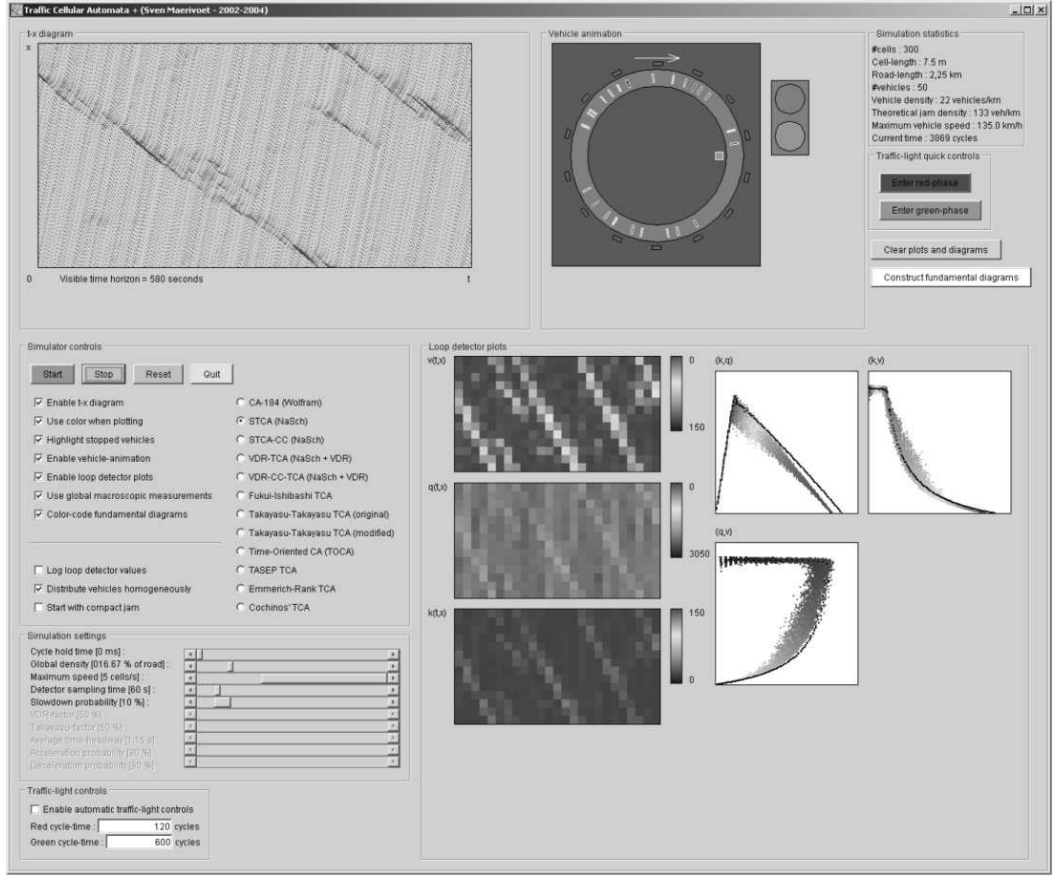}
	\caption{
		A screenshot of the TCA+'s main graphical user interface (GUI) for 
		single-cell TCA models. The GUI is rather huge, spanning approximately 
		1400x1200 pixels, consisting of several panels: a scrolling time-space 
		diagram containing vehicle trajectories, an animation of the road situation, 
		a panel containing some simulation statistics, several simulator controls, 
		scrolling loop detector plots and plots of the ($k$,$q$), ($k$,$\overline 
		v_{s}$), and ($q$,$\overline v_{s}$) diagrams.
	}
	\label{fig:TCA+Software:GUIScreenshot}
\end{figure*}

In \figref{fig:TCA+Software:GUIScreenshot}, we show a screenshot of the main 
graphical user interface (GUI). As can be judged from the image, the TCA+'s GUI 
is rather huge, spanning approximately 1400x1200 pixels (scrollbars are 
automatically placed if it does not fit on the screen). It consists of several 
panels:

\begin{itemize}
	\item a scrolling time-space diagram containing vehicle trajectories and an 
		animation of the road situation,
	\item a panel containing some simulation statistics,
	\item several simulator controls,
	\item and scrolling loop detector plots and plots of the ($k$,$q$), 
		($k$,$\overline v_{s}$), and ($q$,$\overline v_{s}$) diagrams.
\end{itemize}

In the following paragraphs, we describe each of these features in more detail. 
Note that there currently are two versions of the GUI: a standard version for 
all the single-cell TCA models, and a modified multi-cell TCA version with 
limited functionality (mainly for creating coloured tempo-spatial diagrams).\\

\textbf{Vehicle animation}

Looking at the time-space diagram in the upper-left panel, we can discern the 
individual vehicle trajectories, as well as the typical backwards-travelling 
shock waves of congestion. In this scrolling diagram, the time axis goes from 
the left to the right, while the space axis goes from the bottom to the top (and 
is a one-to-one mapping of the consecutive cells on the ring road). Each pixel 
here corresponds to a unique cell of the simulator and each vehicle is coloured 
with a certain shade of yellow (in order to easily distinguish between different 
neighbouring vehicles). There is also a setting available that allows stopped 
vehicles to be coloured red. In the upper-middle panel, the actual geometrical 
configuration of the ring road is depicted. This allows us to view the current 
physical situation on the road, i.e., the positions of all the vehicles. Each 
vehicle can be coloured with a certain shade of yellow (the same as in the 
time-space diagram). The current phase of the traffic light is also shown, as 
well as the positions of all the loop detectors: their positions are indicated 
by the small purple boxes alongside the road. The small green box indicates the 
position of the traffic light, with vehicles travelling in clockwise fashion.\\

\textbf{Simulation statistics}

In the upper-right panel, we can find the length of the ring road (expressed in 
the number of cells), the number of vehicles currently in the simulator, the 
global vehicle density, and the current time step. There is also a small panel 
that allows to quickly set the status of the traffic light to either red or 
green.\\

\textbf{Simulator controls and settings}

The middle-left panel contains buttons for starting, stopping (i.e., pausing), 
resetting, and quitting the simulator. Several preferences can also be 
specified, i.e., whether or not to activate several panels containing the 
simulator's output. There is also the possibility to log the measurements from 
the loop detectors to a default file (called \emph{detector-values.data}). And 
finally, the type of traffic cellular automaton (i.e., its rule set) can also be 
selected from a list, specified by radio control buttons.

Note that there are several initial conditions possible for each density level: 
it is possible to start with a homogeneous state (all vehicles are spaced 
evenly), with a compact superjam of vehicles that are all stopped, or with a 
random initialisation (see also the introduction of section 
\ref{sec:TCA:SingleCellModels}).

If the simulation goes (visually) too fast, the cycle hold time can be 
increased, thereby freezing the simulation for a while between two consecutive 
time steps. Besides this, the ring road's global density and the vehicles' 
maximum speed can be specified. The sampling time for the artificial loop 
detectors can be adjusted (to increase or smooth out fluctuations). And finally, 
all probabilities can be adjusted between 0\% and 100\% in incremental steps of 
1\%.

The red and green cycle times for the traffic light can be specified, such that 
the light can operate automatically, thereby inducing artificial queues at 
regular intervals. One can also control the traffic light manually (enabling the 
red or green phase) using the small upper-right panel; but if applied, the 
traffic-light controls override these manual settings.\\

\textbf{Plots of macroscopic measurements}

The software has the ability to extract both local and global macroscopic flow 
measurements from several uniformly road-side placed loop detectors which record 
flows, densities, and space-mean speeds.

The three large coloured regions in the middle panel represent the measured (and 
averaged) values of the local flows, local densities, and local space-mean 
speeds of the loop detectors. Pair-wise correlating these values, results in the 
plots of the ($k$,$q$), ($k$,$\overline v_{s}$), and ($q$,$\overline v_{s}$) 
diagrams in the lower-right panel. The coloured dots indicate locally obtained 
measurements, whereas the black dots represent globally obtained ones.

Note the small button that allows to construct these diagrams: when it is 
pressed, the global density is incrementally increased from 0\% to 100\%, each 
time adding a single vehicle to the ring road. The simulation is then ran for a 
certain amount of time and the measurements from all the loop detectors are 
recorded. When all densities are processed (an indicator of the total time left 
is shown), the diagrams should be clearly visible in the loop detector plots in 
the lower-right panel.

	\subsection{Running the software}

When visiting the website mentioned in the introduction of this appendix, there 
are two options for downloading the software. One is by downloading the 
\emph{compiled classes}, whereas the other is to download the programme's 
\emph{source code}. Once the compiled software has been downloaded, it is 
relatively easy to start the graphical user interface. Considering the 
single-cell setup GUI, the software is ran by executing the following command:

\begin{quote}
	\texttt{java -jar tca.jar}
\end{quote}

Note that a Java\trademark Development Kit (JDK) (preferably 
Sun's\footnote{\texttt{http://java.sun.com}}) should be installed. Furthermore, 
due to a change in the threading of the Java\trademark Swing\trademark API, it 
appears that only JDK/JRE 1.3.1 is suitable~!

	\subsection{Technical implementation details}

It should be noted that the software is not implemented as an applet, but 
instead as a full Java\trademark application because it uses Swing\trademark 
components that are not standard supported by most browsers (at least not 
without installing a necessary plugin). The source itself logically consists of 
three different parts:

\begin{itemize}
	\item the TCA engine with different rule sets, 
	\item the graphical user interface,
	\item and a whole range of predefined experiments. 
\end{itemize}

The geometrical configuration used in the single-cell TCA engine is a 
unidirectional ring road with a single lane. Vehicles are located in cells of 
$\Delta X =$~7.5~m and can have speeds of 0 to 5 cells/time step (corresponding 
to a maximum speed of 135 km/h). One iteration in the simulation corresponds to 
a time step of $\Delta T = \one$~s.

A number of artificial loop detectors are uniformly placed alongside the road, 
aggregating various macroscopic traffic measurements (i.e., flows, densities and 
space-mean speeds). In the GUI, global measurements on the entire lattice are 
performed according to the methodology explained in section 
\ref{sec:TCA:GlobalMeasurements}, whereas local measurements are performed 
according to section \ref{sec:TCA:LocalMeasurementsFiniteLength}. Note that for 
the TCA software itself, it is also possible to perform local measurements using 
a detector of unit length, according to the methodology explained in section 
\ref{sec:TCA:LocalMeasurementsUnitLength}.

Besides the standard single-cell GUI and the limited multi-cell GUI, there also 
exist some predefined experiments. These allow to create the ($k$,$q$), 
($k$,$\overline v_{s}$), and ($q$,$\overline v_{s}$) diagrams, histograms of the 
vehicles' speeds, space gaps, and time gaps, as well as several order parameters 
(density correlations, nearest neighbours, and an inhomogeneity measure that 
compares the locally recorded densities to the current global density).

Inside the TCA+ software, several packages are available:

\begin{itemize}
	\item \texttt{tca.base} containing the definitions of cells, global states, 
		loop detectors, and the traffic cellular automaton's lattice,
	\item \texttt{tca.automata} containing implementations of all the TCA models 
		mentioned in section \ref{sec:TCA+Software:OverviewAndFeatures},
	\item \texttt{tca.simulator} containing the classes related to the single-cell 
		and multi-cell GUIs,
	\item \texttt{tca.experiments.fundamentaldiagrams},\\
		\texttt{tca.experiments.histograms},\\
		and \texttt{tca.experiments.orderparameters} containing setups for the 
		previously mentioned experiments.
\end{itemize}

%

\section{Glossary of terms}
\label{appendix:Glossary}

	\subsection{Acronyms and abbreviations}

\begin{tabular}{ll}
	ASEP           & asymmetric simple exclusion process\\
	BCA            & Burgers cellular automaton\\
	BJH            & Benjamin, Johnso, and Hui\\
	BJH-TCA        & Benjamin-Johnson-Hui traffic cellular\\
	               & automaton\\
	BL-TCA         & brake-light traffic cellular automaton\\
	BML            & Biham, Middleton, and Levine\\
	BML-TCA        & Biham-Middleton-Levine traffic cellular\\
	               & automaton\\
	CA             & cellular automaton\\
	CA-184         & Wolfram's cellular automaton rule 184\\
	ChSch-TCA      & Chowdhury-Schadschneider traffic\\
	               & cellular automaton\\
	CML            & coupled map lattice\\
	COMF           & car-oriented mean-field theory\\
	DFI-TCA        & deterministic Fukui-Ishibashi traffic\\
	               & cellular automaton\\
	DLC            & discretionary lane change\\	
	ECA            & elementary cellular automaton\\
	ER-TCA         & Emmerich-Rank traffic cellular\\
	               & automaton\\
	GoE            & Garden of Eden state\\
	HS-TCA         & Helbing-Schreckenberg traffic cellular\\
	               & automaton\\
	JDK            & Java\trademark Development Kit\\
	KKW-TCA        & Kerner-Klenov-Wolf traffic cellular\\
	               & automaton\\
\end{tabular}

\begin{tabular}{ll}
	LGA            & lattice gas automaton\\
	LWR            & Lighthill, Whitham, and Richards\\
	MC-STCA        & multi-cell stochastic traffic cellular\\
	               & automaton\\
	MFT            & mean-field theory\\
	MLC            & mandatory lane change\\	
	MPA            & matrix-product ansatz\\
	NaSch          & Nagel and Schreckenberg\\
	NCCA           & number conserving cellular automaton\\
	OVF            & optimal velocity function\\
	OVM            & optimal velocity model\\
	PCE            & passenger car equivalent\\
	PCU            & passenger car unit\\
	pMFT           & paradisiacal mean-field theory\\
	SFI-TCA        & stochastic Fukui-Ishibashi traffic\\
	               & cellular automaton\\
	SMS            & space-mean speed\\
	SOC            & self-organised criticality\\
	SOMF           & site-oriented mean-field theory\\
	SSEP           & symmetric simple exclusion process\\
	STCA           & stochastic traffic cellular automaton\\
	STCA-CC        & stochastic traffic cellular automaton\\
	               & with cruise control\\
	T$^{\two}$-TCA & Takayasu-Takayasu traffic cellular\\
	               & automaton\\
	TASEP          & totally asymmetric simple exclusion\\
	               & process\\
	TCA            & traffic cellular automaton\\
	TMS            & time-mean speed\\
\end{tabular}

\begin{tabular}{ll}
	TOCA           & time-oriented traffic cellular\\
	               & automaton\\
	TRANSIMS       & TRansportation ANalysis and SIMulation\\
	               & System\\
	UDM            & ultra-discretisation method\\
	VDR-TCA        & velocity-dependent randomisation traffic\\
	               & cellular automaton\\
\end{tabular}

	\subsection{List of symbols}

\renewcommand{\arraystretch}{1.2}

\begin{tabular}{ll}
	$\mathcal{C}(\zero)$                           & a CA's initial configuration\\
	$\mathcal{C}(t)$                               & a CA's global configuration at time step $t$\\
	$\delta$                                       & a CA's local transition rule\\
	$G$                                            & a CA's global map\\
	$G^{-\one}$                                    & a reversible CA's inverse global map\\
	$K_{\mathcal{L}}$                              & the number of cells in one lane of a TCA's lattice\\
	$\mathcal{L}$                                  & a CA's lattice (e.g., $\mathbb{Z}^{\two}$)\\
	$\mathcal{N}_{i}$                              & the (partially) ordered set of cells in the\\
	                                               & neighbourhood of the \ith cell\\
	$| \mathcal{N} |$                              & the number of cells in the neighbourhood\\
	                                               & of each cell\\
	$\mathcal{O}_{\mathcal{C}(t) | G^{-\one}}^{-}$ & the backward orbit of the configuration $\mathcal{C}(t)$\\
	                                               & under $G^{-\one}$\\
	$\mathcal{O}_{\mathcal{C}(\zero) | G}^{+}$     & the forward orbit of the initial configuration\\
	                                               & $\mathcal{C}(\zero)$ under $G$\\
	$\sigma_{i}(t)$                                & the state of the \ith cell at time step $t$\\
	$\Sigma$                                       & the set of all possible states a CA's cells can\\
	                                               & be in (e.g., $\mathbb{Z}_{\two}$)\\
	$\Sigma^{\mathcal{L}}$                         & the set of all possible global configurations\\
	                                               & of a CA\\
	$\Sigma^{\mathcal{N}}$                         & the set of all possible configurations of a\\
	                                               & cell's neighbourhood\\
	$| \Sigma^{\Sigma^{\mathcal{N}}} |$            & the number of all possible rules of a CA\\
	$\mathcal{T}_{\mathcal{C}(\zero) | G}$         & the trajectory/orbit of the initial configuration\\
	                                               & $\mathcal{C}(\zero)$ under $G$\\
	$\alpha$                                       & the entry rate of particles in the TASEP model\\
	$\alpha_{i}$                                   & the anticipatory driving parameter of vehicle $i$\\
	$a$                                            & the acceleration capability of a vehicle in the\\
	                                               & KKW-TCA model\\
	$\beta$                                        & the exit rate of particles in the TASEP model\\
	$b$                                            & the deceleration capability of a vehicle in the\\
	                                               & KKW-TCA model\\
	$b_{i}(t)$                                     & the state of the brake light of vehicle $i$ at time\\
	                                               & $t$ in the BL-TCA model\\
	$\delta$                                       & the probability for a particle to move to the\\
	                                               & right in the TASEP model\\
\end{tabular}

\begin{tabular}{ll}
	$\Delta_{\text{acc}_{i}}$                      & the deterministic acceleration of vehicle $i$ in\\
	                                               & the KKW-TCA model\\
	$\Delta T$                                     & a TCA's temporal discretisation\\
	$\Delta V$                                     & a TCA's speed discretisation\\
	$\Delta X$                                     & a TCA's spatial discretisation\\
	$D_{\zero}$                                    & a parameter for the synchronisation distance\\
	                                               & in the KKW-TCA model\\
	$D_{\one}$                                     & a parameter for the synchronisation distance\\
	                                               & in the KKW-TCA model\\
	$D_{i}$                                        & the synchronisation distance of vehicle $i$ in\\
	                                               & the KKW-TCA model\\
	$\eta_{i}$                                     & the stochastic acceleration of vehicle $i$ in\\
	                                               & the KKW-TCA model\\
	$\gamma$                                       & the probability for a particle to move to the\\
	                                               & left in the TASEP model\\
	$\overline g_{s}$                              & the average space gap\\
	$g_{s_{i}}^{*}(t)$                             & the effective space gap of vehicle $i$ at time $t$\\
	                                               & in the BL-TCA model\\
	$g_{s_{\text{security}}}$                      & a security constraint for the space gap in the\\
	                                               & BL-TCA model\\
	$\overline g_{t}$                              & the median time gap\\
	$\overline g_{t_{s}}$                          & the safe time gap in the TOCA model\\
	$h$                                            & the upper limit to the interaction horizon in\\
	                                               & the BL-TCA model\\
	$\xi(t)$                                       & a random number in $[\zero,\one[$ drawn at time $t$\\
	                                               & from a uniform distribution\\
	$k_{g}$                                        & the global density of a TCA's\\
	                                               & lattice\\
	$k_{l}$                                        & the local density of a TCA's lattice\\
	$K_{\mathcal{L}}$                              & the number of cells in one lane of a TCA's\\
	                                               & lattice\\
	$\mathcal{L}$                                  & a TCA's lattice\\
	$l_{i}$                                        & the length of vehicle $i$\\
	$\overline l$                                  & the average length of all vehicles on a TCA's\\
	                                               & lattice\\
	$L_{\mathcal{L}}$                              & the number of lanes in a TCA's lattice\\
	$M_{g_{s_{i}},v_{i}}$                          & the gap-speed matrix of the ER-TCA model\\
\end{tabular}

\begin{tabular}{ll}
	$p$                                            & the slowdown probability in $[\zero,\one]$\\
	$p_{\zero}$                                    & the slow-to-start probability in $[\zero,\one]$\\
	$p_{a}$                                        & the acceleration probability in $[\zero,\one]$ in the\\
	                                               & KKW-TCA model\\
	$p_{a_{\one}}$                                 & a parameter for the acceleration probability in\\
	                                               & the KKW-TCA model\\
	$p_{a_{\two}}$                                 & a parameter for the acceleration probability in\\
	                                               & the KKW-TCA model\\
	$p_{\text{acc}}$                               & the acceleration probability in $[\zero,\one]$ in the\\
	                                               & TOCA model\\
	$p_{b}$                                        & the braking probability in $[\zero,\one]$ in the\\
	                                               & BL-TCA model\\
	                                               & the deceleration probability in $[\zero,\one]$ in the\\
	                                               & KKW-TCA model\\
	$p_{d}$                                        & the slowdown probability in $[\zero,\one]$ in the\\
	                                               & BL-TCA model\\
	$p_{\text{dec}}$                               & the deceleration probability in $[\zero,\one]$ in\\
	                                               & the TOCA model\\
	$p_{s}$                                        & the slow-to-start probability in $[\zero,\one]$ in\\
	                                               & the BJH-TCA model\\
	$p_{t}$                                        & the slow-to-start probability in $[\zero,\one]$ in\\
	                                               & the T$^{\two}$-TCA model\\
	$P_{n}(v)$                                     & the probabilities of finding a space gap of\\
	                                               & $n$ cells\\
	                                               & for a vehicle driving with speed $v$\\
	$q_{g}$                                        & the global flow of a TCA's lattice\\
	$q_{l}$                                        & the local flow of a TCA's lattice\\
	$t_{s_{i}}$                                    & the interaction horizon in the BL-TCA model\\
	$v_{\text{des}_{i}}$                           & the desired speed of vehicle $i$ in the\\
	                                               & KKW-TCA model\\
	$v_{p}$                                        & a parameter for the acceleration probability\\
	                                               & in the KKW-TCA model\\
	$\overline v_{s_{\text{ff}}}$                  & the space-mean speed in the free-flow regime\\
	$\overline v_{s_{g}}$                          & the global space-mean speed of a TCA's lattice\\
	$\overline v_{s_{l}}$                          & the local space-mean speed of a TCA's lattice\\
	$x_{i}^{l,b}$                                  & the longitudinal position of vehicle $i$'s\\
	                                               & left-back neighbour\\
\end{tabular}

\begin{tabular}{ll}
	$x_{i}^{l,f}$                                  & the longitudinal position of vehicle $i$'s\\
	                                               & left-front neighbour\\
	$x_{i}^{r,b}$                                  & the longitudinal position of vehicle $i$'s\\
	                                               & right-back neighbour\\
	$x_{i}^{r,f}$                                  & the longitudinal position of vehicle $i$'s\\
	                                               & right-front neighbour\\
\end{tabular}

\section*{Acknowledgements}

Dr. Bart De Moor is a full professor at the Katholieke Universiteit Leuven, Belgium.
\noindent
Our research is supported by:
\textbf{Research Council KUL}: GOA AMBioRICS, several PhD/post\-doc
\& fellow grants,
\textbf{Flemish Government}:
\textbf{FWO}: PhD/post\-doc grants, projects, G.0407.02 (support vector machines),
G.\-0197.02 (power islands), G.0141.03 (identification and cryptography), G.0491.03
(control for intensive care glycemia), G.0120.\-03 (QIT), G.0452.04 (new quantum algorithms),
G.0499.04 (statistics), G.0211.05 (Nonlinear), research communities (ICCoS, ANMMM, MLDM),
\textbf{IWT}: PhD Grants, GBOU (McKnow),
\textbf{Belgian Federal Science Policy Office}: IUAP P5/\-22 (`Dynamical Systems and
Control: Computation, Identification and Modelling', 2002-2006), PODO-II (CP/40:
TMS and Sustainability),
\textbf{EU}: FP5-Quprodis, ERNSI,
\textbf{Contract Research/agreements}: ISMC/IPCOS, Data4s,TML, Elia, LMS,
Mastercard.

\bibliography{paper}

\end{document}